\documentclass[a4paper,12pt]{article}

\usepackage{jheppub} 
   \usepackage[T1]{fontenc}                  
\usepackage{hyperref}
\hypersetup{
    bookmarks=true,         
    unicode=false,          
    colorlinks=true,       
    linkcolor=red,          
    citecolor=violet,        
    filecolor=magenta,      
    urlcolor=blue,           
    linktocpage=true
}
\usepackage{graphics}
\usepackage{rotating}
\usepackage{subfigure}
\usepackage{pstricks}
\usepackage{color}
\usepackage{amsfonts}
\usepackage{mathrsfs}
\usepackage{epsfig}
\usepackage{amsmath,amssymb,amsthm,graphicx,latexsym}
\usepackage{ulem}

\newcommand{\be}{\begin{equation}}
\newcommand{\ee}{\end{equation}}
\newcommand{\bea}{\begin{eqnarray}}
\newcommand{\eea}{\end{eqnarray}}
\newcommand{\bml}{\begin{subequations}}
\newcommand{\eml}{\end{subequations}}
\newcommand{\bfig}{\begin{figure}}
\newcommand{\efig}{\end{figure}}

\newcommand{\slashed}{\hspace{-1.1ex}/}

\begin{document}
$~~~~~~~~~~~~~~~~~~~~~~~~~~~~~~~~~~~~~~~~~~~~~~~~~~~~~~~~~~~~~~~~~~~~~~~~~~~~~~~~~~~~$\textcolor{red}{\bf TIFR/TH/15-18}
\title{\textsc{\fontsize{85}{100}\selectfont \sffamily \bfseries Hysteresis in the Sky}}

\author[a]{Sayantan Choudhury
\footnote{\textcolor{purple}{\bf Presently working as a Visiting (Post-Doctoral) fellow at DTP, TIFR, Mumbai, \\$~~~~~$Alternative
 E-mail: sayanphysicsisi@gmail.com}. ${}^{}$}}
\author[b]{Shreya Banerjee}

\affiliation[a]{Department of Theoretical Physics, Tata Institute of Fundamental Research, Colaba, Mumbai - 400005, India
}
\affiliation[b]{Department of Astronomy and Astrophysics, Tata Institute of Fundamental Research, Colaba, Mumbai - 400005, India
}

\emailAdd{sayantan@theory.tifr.res.in, shreya.banerjee@tifr.res.in  }

\abstract{Hysteresis is a phenomenon occurring naturally in several
magnetic and electric materials in condensed matter physics. When applied to cosmology,
aka {\it cosmological hysteresis}, has interesting and vivid
implications in the scenario of a cyclic bouncy universe. Most importantly, this physical prescription 
can be treated as an alternative proposal to inflationary paradigm.
{\it Cosmological hysteresis} is caused by the asymmetry in the
equation of state parameter during expansion and contraction phase
of the universe, due to the presence of a single scalar field.
This process is purely thermodynamical in nature, results in a non-vanishing
hysteresis loop integral $(\oint pdV)$ in cosmology. When applied to variants of modified 
gravity models -1) Dvali-Gabadadze-Porrati (DGP) brane world gravity, 2) Cosmological constant dominated Einstein gravity, 3) Loop Quantum Gravity (LQG), 4) Einstien-Gauss-Bonnet brane world 
gravity and 5) Randall Sundrum single brane world gravity (RSII), under certain circumstances, this phenomenon
leads to the increase in amplitude of the consecutive cycles
and to a universe with older and larger successive cycles,
provided we have physical mechanisms to make the universe bounce and
turnaround. This inculcates an arrow of time in a dissipationless
cosmology. Remarkably, this phenomenon appears to be widespread
in several cosmological potentials in variants of modified gravity background, which we explicitly study for- i) Hilltop, ii) Natural and iii) Colemann-Weinberg potentials, in this paper.
Semi-analytical analysis of these models, for different potentials
with minimum/minima, show that the conditions which creates
a universe with an ever increasing expansion, depend on the
signature of the hysteresis loop integral $(\oint pdV)$ as well as on the
variants of model parameters.  
}

\maketitle
\flushbottom
\section{Introduction}

Hysteresis is a phenomenon that arises in systems with a lag between its input and output.
When this lag is dynamic i.e it changes with time, we get hysteresis loops. It can be evaluated
by purely thermodynamical expressions where the output depends on the current and past inputs.
Such phenomenon naturally occurs in many laboratory systems like ferromagnetic and ferroelectric
materials and is often incorporated artificially in several electrical systems.

In analogy with hysteresis, in cosmology, we have the phenomenon of cyclic universe in which
the universe re-borns repeatedly after each cycle. Just as in hysteresis, where the material
undergoes through the same process over and over again, in a cyclic universe, the universe
starts from big bang and ends in big crunch repeatedly. Several models of cyclic universe
have been proposed in literature \cite{eliade,jaki,starobinsky}. Such models also arise
naturally as exact solutions of Einstein equations for a closed universe filled with perfect fluid.
However in most of these models, all the cycles are identical to one another.  Also all these
models does not provide any prescription for avoiding singularity.  Hence these models remain
unsuccessful in solving some of the major problems of big bang model i.e the flatness and
horizon problem, and avoidance of singularity. However Tolman \cite{tolman} in his paper used
a radically different approach by which one could get an oscillating universe with increasing
expansion maximum after each cycle~\footnote{The same phenomenon occurs naturally in other cyclic universe models as proposed in refs.~\cite{Steinhardt:2001st,Lehners:2008vx}.}. He postulated the presence of a viscous fluid which gave
rise to asymmetric, irreversible equation of motion. This created an inequality between the
pressures at the time of expansion and contraction phases which resulted in the growth of
both energy and entropy. Thus he showed a novel way of linking thermodynamical principles
to the model of cyclic universe. This unusual approach helped in solving the horizon and
flatness problem. In later years, theory of inflation \cite{Baumann:2014nda,Baumann:2009ds,Lyth:1998xn} was developed which addressed both
the horizon and flatness problem. But none of these models were able to avoid big bang
singularity. Also the model proposed by Tolman led to an inevitable increase in entropy with each cycle. 

However, in \cite{Kanekar:2001qd,Sahni:2012er}, the authors have proposed a method of
avoiding both the presence of singularity and increasing entropy by using a
cosmological analog model of hysteresis. The basic idea of generating the cyclic
universe with an increasing maximum remains the same and in the words of Tolman is
``if the pressure tends to be greater during a compression than during a previous
expansion, as would be expected with a lag behind equilibrium conditions, an element
of fluid can return to its original volume with increased energy..". The authors created
the asymmetry in pressure using the scalar field dynamics generated during inflationary
paradigm, thereby maintaining the symmetric nature of the equation of motion hence
avoiding entropy production.  Originally proposed in \cite{Kanekar:2001qd} and later
extended in \cite{Sahni:2012er}, the authors demonstrated that ``a universe filled
with a scalar field possess the intriguing property of {\it`hysteresis'}." The central
idea was to show that the presence of a massive scalar field "under certain reasonable
conditions at the bounce \cite{Cai:2013kja,Cai:2013vm,Cai:2012ag,Cai:2012va,Li:2014era,Brandenberger:2012zb,Cai:2011zx,Lilley:2015ksa,Falciano:2008gt,
Lilley:2010av,Lilley:2011ag,Battefeld:2014uga,Graham:2011nb,Koehn:2013upa}, gives rise to growing expansion cycles, the increase in
expansion amplitude being related to the work done by/on the scalar field during
the expansion/contraction of the universe." This leads to the production of hysteresis
loop defined as $\oint pdV$, during each oscillatory cycle. The loop area is largest
in case of inflationary potentials since they give rise to largest asymmetry between
expansion and contraction pressures. But the phenomenon of hysteresis is generic i.e.
independent of the nature of potential. Any potential with a proper minima which
randomizes the phase of the scalar field as it oscillates around the minima during
expansion, thereby making all possible values of $\dot\phi$ probable at turnaround,
is cable of causing the phenomenon of hysteresis. However potentials without any
proper minimum will result in a unique value of $\dot\phi$ which will make
\be p_{exp} = p_{cont},\ee thereby making: \be \oint pdV = 0.\ee Such potentials are not
suitable candidates for causing hysteresis. In order to avoid big bang and big crunch,
the authors have made used of the presence of existing models like brane world
scenario in the early universe, and the presence of negative density or phantom
like density in the late universe. These models replaces the big bang singularity
by bounce and the big crunch by re-collapse or turnaround. 

Our present paper is based on the analysis done by  \cite{Sahni:2012er}. We have
further investigated the phenomenon of hysteresis in different models like the
variants of cosmological constant model including $\Lambda$CDM, higher dimensional models like Dvali-Gabadadze-Porrati (DGP)
brane world gravity model, Loop Quantum Gravity model and Einstein Gauss-Bonnet brane world gravity model in brane world, and
in models where the dynamics of the scalar field gets modified which can be achieved
by making the cosmological constant field dependent. Our aim is to study not only
the phenomenon of hysteresis in different models but also to constrain the
parameters of the model using hysteresis. We have mostly studied the models
which can give rise to both the phenomena of bounce in the early universe and
turnaround in the late universe. We have also investigated the equivalent
conditions required to achieve such bouncing and re-collapsing scenarios. We
have shown that our analysis holds true for any general form of the potential
of the scalar field with a proper minimum. We have also shown that the phenomenon
of hysteresis or the asymmetry in pressure can be achieved irrespective of whether
the slow roll conditions of inflation are satisfied or not. A notable feature
of this analysis is that an increase in expansion maximum after each cycle now
depends not only on the sign of $\oint pdV$ but also on the parameters of
the models that we have considered. Thus we see that using the remarkable
cosmological effect of hysteresis as proposed by \cite{Kanekar:2001qd,Sahni:2012er},
there are numerous methods and models in which a cyclic universe with
an ever increasing amplitude maximum can be achieved. 

The plan of the paper is as follows:
\begin{itemize}
 \item  In section \ref{sq1}, we have discussed the mathematical formulation leading
to the phenomenon of hysteresis in cosmological scenario. 

\item In section \ref{sq2} we have tried
to draw an analogy between the hysteresis in ferromagnetic materials and
in a cyclic universe. This simple analysis leads to the conclusion that
the equation of state parameter w plays the role of magnetic field $H$
in cosmology and the scale  factor mimics the behavior of magnetization $M$.
 
 \item In section \ref{sq3}, \ref{sq4}, \ref{sq5}, \ref{sq6} ,\ref{sq8}, we have studied the behavior of each of the above mentioned
models for flat, closed and open universe and have discussed various unexplored important cosmological features.

\item In section \ref{sq3}, \ref{sq5}, \ref{sq6} ,\ref{sq8}, we have explicitly studied cosmological hysteresis in the context of higher dimensional theories, where we have
shown the results for both space-like and time-like extra dimensions. For the sake of simplicity we restrict ourselves up to five dimensions ($D=5$). But one can extend 
the computation for dimensions, $D>5$.
\end{itemize}
In this paper, we have drawn various physical conclusions by explicitly solving the
equations governing the dynamics of the system using the semi-analytical techniques. Though the
analysis is perfectly true for any kind of cosmological potential with a proper minimum/minima, we have
the studied the detailed features for three different potentials - hilltop potential, natural potential
and Colemann-Weinberg potential. All these potentials have well defined minimum/minima
and have free parameters which can be adjusted to get the required results.
Thus this analysis helps us to put stringent constraints on the characteristic parameters of these
models in the bouncing scenario along with cosmological hysteresis. Though the analysis that
we have performed holds good under certain physically acceptable approximations and limiting cases,
but we can at least show mathematically if there are any limiting cases
in which these potentials combined with the models can give rise to the
phenomenon of cosmological hysteresis i.e. make $\oint pdV$ non-zero. In this paper we
have also explicitly derived the expression for work done in one complete
cycle of expansion and contraction, and have shown it to be non zero.
But the sign of the integral depends on how we have chosen the sign
and magnitudes of the parameters of our models. The most interesting result of our analysis is that there are several
models which can give rise to a cyclic universe with an
increasing amplitude of expansion.

\section{Basics of cosmological hysteresis}
\label{sq1}
Before going to the technical details of the ``Cosmological hysteresis'' let us mention clearly the underlying assumptions. It is important to note that  
throughout the analysis in this paper we assume that during a specified period in the time line of the universe, it is described by a single massive scalar field which is minimally interacting with the gravity sector.
The presence of this scalar field is responsible to generate the required asymmetry in the pressure of the universe and this leads to an overall increase in the energy of the universe and hence
an increase in its amplitude of the expansion rate.

The action governing the dynamics of this scalar field $\phi$ with potential $V(\phi)$ within effective field theory description, is given by
\begin{equation}
S = \int d^{4}x\sqrt{-g}\left[\frac{M^2_{p}}{2}R + \frac{1}{2}g^{\mu\nu}\partial_{\mu}\phi \partial_{\nu}\phi -V(\phi)\right] = S_{EH} + S_{\phi}
\end{equation}  
where the signature of the metric throughout our analysis is (-, +, +, +). The total action $S$ can be written as the sum of the standard Einstein Hilbert action ($S_{EH}$) and the action for the scalar field ($S_{\phi}$).
 The energy-momentum tensor for the scalar field can be computed from the matter part of the action \be S_{\phi}=\int d^{4}x \sqrt{-g}~{\cal L}_{\phi}\ee as:
\begin{eqnarray}
T_{\mu\nu}^{\phi}  = -\frac{2}{\sqrt{-g}}\frac{\partial \left(\sqrt{-g}~{\cal L}_{\phi}\right)}{\partial g_{\mu\nu}}=\partial_{\mu}\phi \partial_{\nu}\phi -g_{\mu\nu}\left(\frac{1}{2}\partial^{\beta}\phi\partial_{\beta}\phi+V(\phi)\right)
\label{eq:lagr}
\end{eqnarray}
and for a homogeneous and isotropic spatially flat ($k=0$) FLRW cosmological background, the energy density and pressure for a scalar field can be computed from the energy momentum tensor as:
\bea
\rho = \frac{1}{2} \dot{\phi}^2 + V (\phi) \, ,\\   p =
\frac{1}{2} \dot{\phi}^2 - V (\phi) \, , 
\label{eq:scalar_rho}
\eea
and the resulting equation of state parameter $w$ can be written as:
\bea
w=\frac{p}{\rho}=\frac{\dot{\phi}^2 - 2V (\phi)}{\dot{\phi}^2 + 2V (\phi)},
\eea
where we assume that the energy momentum tensor for scalar field can be approximated via perfect fluid.
Also in the spatially flat FLRW cosmological background the scalar field equation of motion is given by:
\begin{equation}
\Box\phi+\frac{dV}{d\phi} = \ddot \phi + 3 H \dot \phi + \frac{dV}{d\phi} = 0 \, , 
\label{eq:scalar field}
\end{equation}
where $\Box$ is the d'Alembertian operator four dimension defined as:
\bea 
\Box=\frac{1}{\sqrt{-g}}\partial_{\mu}\left(\sqrt{-g}g^{\mu\nu}\partial_{\nu}\right)=\frac{d^{2}}{dt^{2}}+3H\frac{d}{dt}.
\eea
Here $H$ is the Hubble parameter defined as: \be H=\frac{1}{a(t)}\frac{da(t)}{dt},\ee where $a(t)$ is the scale factor.\\
\textbf{Origin of cosmic hysteresis:} If we closely look at Eq~(\ref{eq:scalar field}), then we can conclude that when the universe expands i.e. $H>0$ the second term $3 H \dot \phi$ mimics the role of friction and opposes the motion of the scalar field, thus serves the purpose of 
damping during its motion. This lowers the kinetic energy of the scalar field compared to its potential energy, giving rise to a soft equation of state ($P=-\rho$ in case $\dot{\phi}^{2}/2<<V(\phi)$ i.e. slow roll regime).
By contrast, in a contracting ($H<0$) phase of the universe, the term $3 H \dot \phi$ behaves like 
anti-friction and favors the motion of the scalar field and hence accelerates it. This makes the kinetic energy of the scalar field much larger than the potential energy, giving rise to stiff equation of state ($P=\rho$ in case $\dot{\phi}^{2}/2>>V(\phi)$). 
As a result, from second law of thermodynamics, we can convey that a net asymmetry in the pressure (during expansion and contraction cycle) leads to a net non-zero
work done by/on the scalar field. In addition, if we now postulate the presence of bouncing and recollapsing mechanisms during contraction and expansion
respectively, one can expect that a non-zero work done during a given oscillatory cycle to be converted into expansion energy, resulting in the growth
in the maximum amplitude and hence maximum volume of the universe of each successive expansion cycle. Thus producing older and larger cycles. 
In \cite{Sahni:2012er}, using simple thermodynamic arguments, the authors have developed the equations which relate the change in maximum amplitude
of the scale factor after successive cycles to the work done. The authors have shown that these equations have a universal form which is independent of
the scalar field potential responsible for hysteresis. As has been discussed in \cite{Sahni:2012er}, though the process is independent of the potential,
``the presence of hysteresis is closely linked to the ability of the field $\phi$ to oscillate''. This urges the presence of potential minimum/minima. This
is because only potentials with well defined minimum can make the field oscillate. As a result Oscillations of the scalar during expansion makes its
phase arbitrary, thereby making all values of $\dot{\phi}$ equally likely at turnaround. Thus assuring that the values of $\phi$ and $\dot{\phi}$, when
the universe turns around and contracts, are nearly uncorrelated with its phase space value when the field $\phi$ began oscillating. As a result,
the field almost always rolls up the potential along the different phase space trajectory compared to the one along which it had descended during expansion. 
This gives rise to unequal pressure during expansion and contraction hence to non-zero work done. 

In the present work, we have reconsidered the above phenomenon of hysteresis and applied it to models or physical mechanisms which can make the universe bounce and turnaround in the presence of potentials having well defined minimum/minima.
Further solving Eq.~(\ref{eq:scalar_rho}) and Eq.~(\ref{eq:scalar field}) simultaneously, along with the Friedmann equations derived from various cosmological background model, 
and applying the proper bounce and turnaround conditions, which will be discussed in the following sections, we get the explicit expressions for the scalar field and scale factor as a function of time.

It has been first pointed out in ref.\cite{Kanekar:2001qd} that when we plot the equation of state given by $w = p/\rho$ vs the scale factor from a specified cosmological model, we get a hysteresis loop whose area contributes
to the work done by/on the scalar field during expansion and contraction of the Universe. The general expression for the work done by/on the scalar field during one cycle is given by 
\begin{eqnarray}
\oint pdV &=& \int_{cont} pdV + \int_{exp} pdV
\end{eqnarray}
where the contributions $\int_{cont} pdV$ and $\int_{exp} pdV$ represent the work done by/on the scalar field during the phase of contraction and expansion of the Universe respectively.
 The signature of the integral depends on the pressure during the phase of contraction and expansion i.e if the pressure corresponding to the contraction phase is greater than the 
 pressure due to expansion phase i.e. \be p_{cont} > p_{exp}\ee then the overall signature of the $p-dV$ work done is negative i.e. \be \oint pdV < 0.\ee On the other hand, if the pressure corresponding to the contraction
 phase is smaller than the 
 pressure due to expansion phase i.e. \be p_{cont} < p_{exp}\ee then the overall signature of the $p-dV$ work done is positive i.e. \be \oint pdV > 0.\ee
 For a universe characterized by a scale factor $a(t)$, the total volume at any given time is given by $a^{3}(t)$ (neglecting the overall constant factor). Hence using this input, 
 the area of the cosmological hysteresis loop or equivalently 
 the $p-dV$ work done is given by:
 \begin{eqnarray}
\oint pdV &=& 3\left[\int_{cont} pa^{2}da+ \int_{exp} pa^{2}da\right]
\end{eqnarray}
Hence we follow the following algorithm:
\begin{itemize}
 \item In the context of various cosmological frameworks i.e. DGP branewold gravity, Loop Quantum gravity (LQG), Einsitein-Gauss-Bonnet (EGB) gravity and in presence of pure and field dependent 
 cosmological constant within Einstein gravity 
 we explicitly derive an expression for this characteristic
integral in terms of the scale factor and the scalar field degrees of freedom.
\item Then we solve the equations of
motions for the scalar field under various limiting approximations and hence elaborately study the physical conditions under which the
above integral gives non zero value.
\item Further we repeat this mentioned two step process for three different cosmological potentials i.e. for Hilltop potentials, Natural potential and Coleman-Weinberg potential
within the framework of effective field theory prescription, which will be discussed in the next section in detail.
\item We also analyze the whole process graphically to see whether we get a net increase in the amplitude of the scale factor after one cycle of expansion and contraction.
\end{itemize}
Though the phenomenon of hysteresis is commonly attached to magnetic and electric systems, its appearance for different cosmological models, almost naturally, makes us appreciate and acknowledge its importance in the field of cosmology. The various advantages which a cyclic universe along with the phenomenon of hysteresis, which is our main focus in this paper, have are:
\begin{itemize}
\item The phenomenon of hysteresis is important due to the simplicity with which it can be generated. Only a thermodynamic interplay between the pressure and density, creating an asymmetry during expansion
and contraction phase of the universe, succeeds in causing cosmological hysteresis.
\item As the phenomena of cosmological hysteresis deals with the bouncing as well as the re-collapsing phase of the universe, one can avoid the appearance of Big Bang Singularity as well as the Big Crunch at early and late times.
\item Hysteresis can be generated by the presence of a single massive scalar field, which has already been studied for a wide variety of physical situations. The most exciting issue is, it do not require any other
fields for its occurrence. Hence it is very easy to handle and its properties can be studied extensively in cosmological literature.
\item A cyclic universe having the conditions to cause hysteresis, can solve all the Big Bang puzzles, hence acts as an alternative proposal to inflation \cite{Choudhury:2014hja,Choudhury:2014sxa,Choudhury:2013woa,Choudhury:2013zna,Choudhury:2011jt}.
\item In this scenario, we can always start with a closed or open universe and after allowing the universe go through a number of cycles, we get the present observable
flat universe. In this paper, we will extensively deal with several models that can give rise to such cyclic universe. Through analytical calculations,
we will show that irrespective of the nature of the universe, we can get a cyclic model with increasing amplitude of the scale factor for a wide variety of models.
\item It results in dissipative cosmology 
\cite{Kanekar:2001qd,Sahni:2012er,Vilenkin:2013rza}, which makes
the whole process irreversible, which finally causes an arrow of
time according to widespread belief. But recently in the ref.~\cite{Sahni:2015kga} 
the authors have explicitly shown that for cosmological hysteresis phenomena such an arrow of time can appear
even if equations describing cosmological evolution are dissipationless, which makes the cumulative process reversible provided they possess 
cosmological attractors in the expanding and contracting phase of the universe. Additionaly, it is important to note that 
for flat cosmological potential the increament in the expansion cycles are routinely observed. But for the steep potentials two fold cyclic pattern 
with lesser expansion cycles is observed, which being nested in the larger expansion cylces. This phenomena is exactly analogous to the 
`beat' formation in acoustics systems. In ref.~\cite{Kanekar:2001qd,Sahni:2012er} the authors have studied the cosmological beat formation
in the context of chaotic potential $V(\phi)=\frac{1}{2}m^2 \phi^2$.
\item Cosmology in this scenario has not been explored in a very wide sense. Earlier it has been studied by the authors of refs.~\cite{Kanekar:2001qd,Sahni:2012er,Sahni:2015kga}. In this paper, we have
further explored this phenomenon for several other models which have not been discussed earlier in refs.~\cite{Kanekar:2001qd,Sahni:2012er,Sahni:2015kga}. 
\item In future we plan to connect these analyses with CMB observations, by rigorous study of the cosmological perturbation theory \cite{Biswas:2015kha,Biswas:2012bp,Battarra:2014tga} in various orders of metric fluctuations
and computation of two point correlations to get the expressions for scalar and tensor power spectrum in this context. Hence we extend the study of this paper to compute the primordial non-Gaussianity in CMB from
three and four point correlations \cite{Choudhury:2015yna,Choudhury:2014uxa,Gao:2014hea,Gao:2014eaa,Maldacena:2002vr,Maldacena:2011nz,Arkani-Hamed:2015bza,Mata:2012bx,Ghosh:2014kba,Kundu:2014gxa}. 
We also plan to derive the explicit expression for various modified consistency relations between the non-Gaussian as well as other cosmological parameters in the present context.
\item Also we plan to setup the formalism of magnetogenesis from the present setup which can be treated as the alternative version of the inflationary magnetogenesis \cite{Choudhury:2014hua,Choudhury:2015jaa,Subramanian:2015lua}.
\item We also carry forward our analysis in the development of density inhomogeneities, which is the prime component to form large scale structures at late times. Also 
the specific role of cosmological hysteresis in the study of cosmological perturbations i.e. for interacting/decoupled dark matter and dark energy have not been explored at all earlier. We have some future plan to do
some computations from this setup.
\item Further using the reconstruction techniques \cite{Choudhury:2014kma,Choudhury:2013iaa,Choudhury:2014sua,Lidsey:1995np} we want to study the generic features of scalar field potentials in the framework of cosmological hysteresis.
\item Last but not the least, we also plan to further study this outstanding cosmological phenomenon for different modified gravity pictures
i.e. for variants of $f(R)$ gravity \cite{Sotiriou:2008rp,DeFelice:2010aj,Faraoni:2008mf,Capozziello:2011et}, two brane-world model in presence of the Einsitein-Hilbert term and the 
Einstein-Hilbert-Gauss-Bonnet gravity setup \cite{Choudhury:2012yh,Choudhury:2013yg,Choudhury:2013eoa,Choudhury:2013qza,Choudhury:2013aqa,Choudhury:2014hna,Choudhury:2015wfa}. 
\end{itemize}

\section{Analogy with magnetic hysteresis}
\label{sq2}
\begin{figure*}[htb]
\centering
\subfigure[ An illustration of hysteresis in ferromagnetic materials. The loop has been plotted in the $B-H$ plane.]{
    \includegraphics[width=14.2cm,height=9cm] {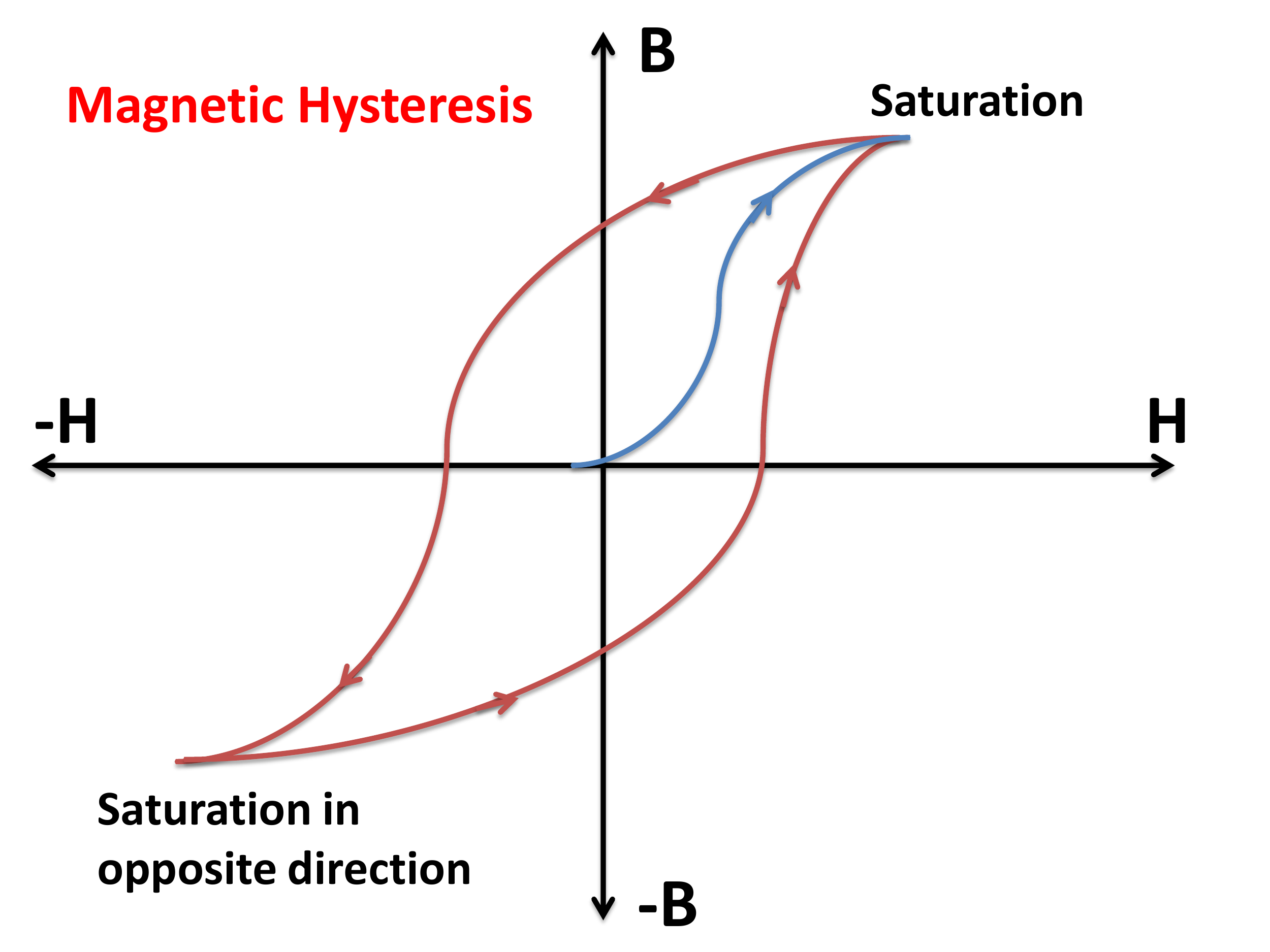}
    \label{fig1a}
}
\subfigure[An idealised illustration of cosmological hysteresis. The loop has been plotted in the $w-a$ plane. ]{
    \includegraphics[width=14.2cm,height=9cm] {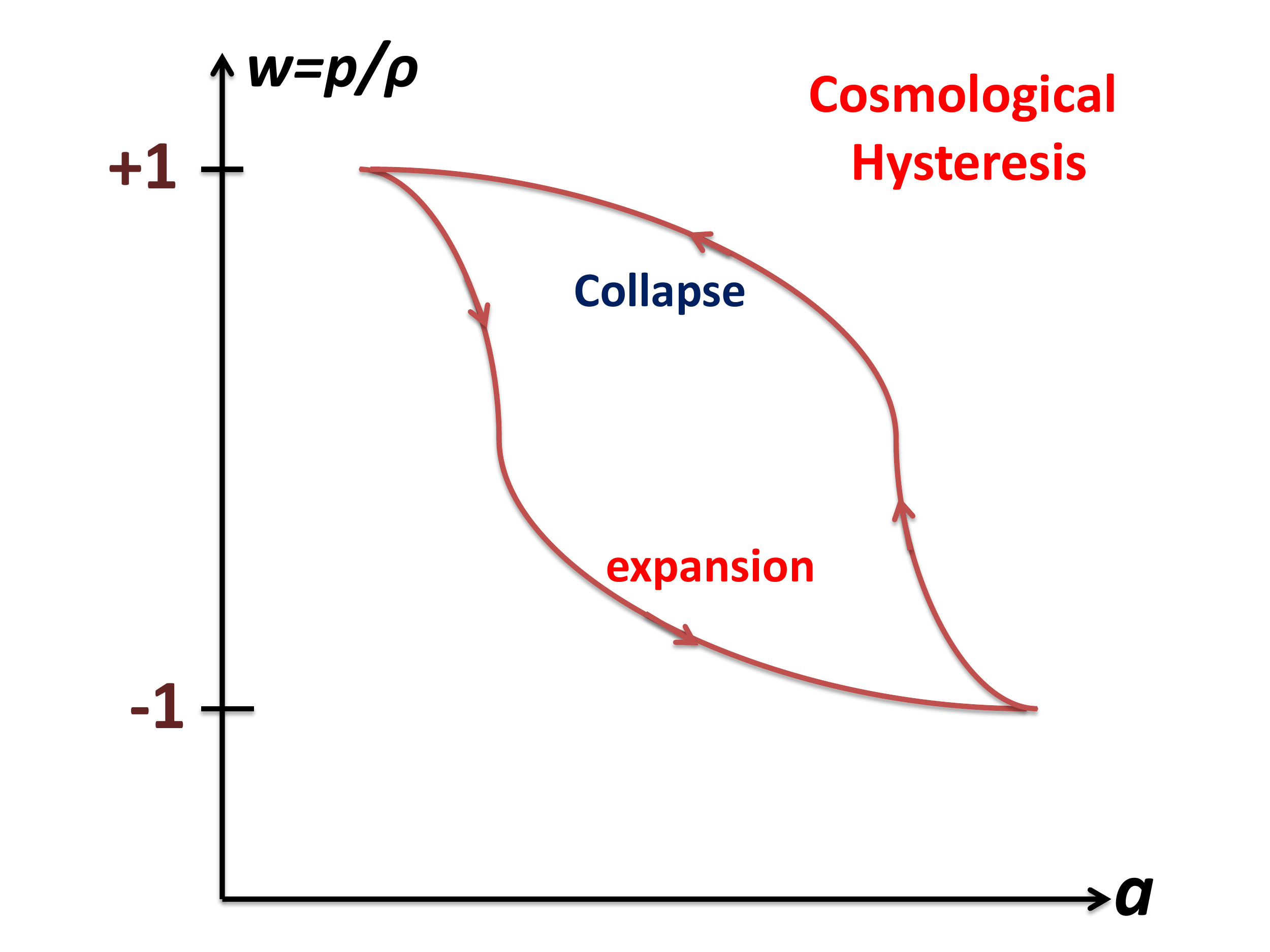}
    \label{fig1b}
}
\caption[Optional caption for list of figures]{ Analogy between magnetic hysteresis in ferromagnetic materials and cosmological hysteresis.} 
\label{fig1}
\end{figure*}

 The phenomenon of lagging of magnetic induction B or magnetization M behind the magnetic field H when a specimen of a magnetic material (such
 as iron) is subjected to a cycle of magnetization is called ``magnetic hysteresis''. The closed loop that is traced by the material in the $B-H$
 or $M-H$ plane is known as the hysteresis loop. It is related to the change in the alignment of the magnetic dipoles, as one varies the magnetic field
 which leads to magnetization and  demagnetization of the material. It is a beautiful way of depicting the effect of varying H on the system.
 Fig. \ref{fig1a} is a representative schematic diagram in which we have explicitly shown one such hysteresis loop in ferromagnetic material.
 Since we want to draw an analogy between the phenomenon of
 hysteresis in magnetism and cosmology, we have also shown Fig. \ref{fig1b} to illustrate how an ideal
 hysteresis loop looks in the case of a cyclic universe. In a cyclic universe, the hysteresis loop is traced by the universe in the $a-w$ plane where a
 is the scale factor of the universe and $w=p/ \rho$ is the equation of state parameter. Closely following  the two figures we can say that  just as
 in a magnetic material, we also vary the magnetic field and try to find the behavior of the magnetization of the material, in this cyclic model,
 we vary the pressure or the equation of state in order to find the variation of the scale factor or expansion of the universe. This motivates
 us to consider that the parameter playing the role of H in cosmological hysteresis is $w$ and that of $M/B$ is $a$.  Just as in magnetic material,
 where the magnetization oscillates within a specified maximum and minimum values, analogously in cyclic model, the scale factor  goes through maximum and
 minimum values which has been clearly shown in \cite{Kanekar:2001qd,Sahni:2012er} for few cosmological models and will again be shown in
 this paper for some other cosmological models. While the primary cause of magnetic hysteresis is the asymmetry in the behavior of the alignment of the
 magnetic dipoles with increase and decrease of $H$, the cause for hysteresis in cosmology is the asymmetry in pressure during expansion and
 contraction phases of the universe. In magnetic hysteresis loop, the minimum and maximum corresponds to the state when all the magnetic dipoles
 are aligned in reverse and along the direction of $H$ respectively. In cosmological hysteresis, the maximum is reached when the scale
 factor reaches its maximum value and the density of the scalar field reached at its minimum i.e when the condition for re-collapse is generated, and the corresponding
 minimum is reached at bounce when scale factor becomes minimum and density of the scalar field reached at its maximum value. But unlike in magnetic hysteresis where the parameters
 can take all kind of values i.e positive, negative  and zero,  in cyclic universe, one of our primary goal is to avoid singularity,
 i.e. $a\neq$0, so that singularity is replaced by bounce. While repeated cycles of magnetic hysteresis loop results in loss of finite amount of energy,
 hence decrease in the area of the loop, in cosmological hysteresis, repeated cycles may be either larger or smaller in area compared to the previous one.
 Thus, we find that a closer look at these two phenomena draws lots of similarities in their behavior. In table. \ref{fig:hysteresis} we have depicted the analogy and the parallelism between the 
 different features of these two kinds of hysteresis.
 
\begin{table*}
\centering
\small\begin{tabular}{|c|c|c|
}
\hline
\hline
Characteristics & Magnetic Hysteresis & Cosmological Hysteresis
\\
\hline\hline\hline
{\bf  Loop parameters } &  $B, M$ &  $a$
\\
 &  $H$ &  $w=p/\rho$
\\
\hline
{\bf  Hysteresis loop} & $\oint H~dB$ &  $\oint w~da~(\sim \oint p~dV)$
\\
\hline
{\bf  Largest value of } & Given by hard &  Cyclic Universe\\
{\bf  hysteresis loop } &  ferromagnetic materials & with inflationary conditions
\\ \hline
{\bf  Upper and lower  } & No limit on maximum &   $w_{max}=+1$\\
{\bf  limits of the loop } &  and minimum values of $H$ & $w_{min}=-1$
\\
\hline
{\bf  Characteristic equation } & $B=\mu(H+M)$ & $w=p/\rho$ \\
{\bf  for the loop } &  & 
\\
\hline
{\bf  Nature of } & Soft ferromagnetic materials &  Universe with softer equation\\
{\bf  the loop } & have smaller loop area & of state have smaller loop area \\
\hline
\hline
\end{tabular}
\caption{ Table showing the analogy between magnetic hysteresis and cosmological hysteresis.}\label{fig:hysteresis}
\vspace{.4cm}
\end{table*}
In this section we have shown the complete analysis of different models and studied the
 conditions under which we get an increase in the amplitude of the scale factor after each cycle.
 We have studied open closed and flat universe. We have also explicitly calculated the work done in one cycle for three different potentials.
 
\section{Hysteresis from Dvali-Gabadadze-Porrati (DGP) brane world gravity model}
\label{sq3}
The DGP model is a model of modified gravity theory proposed by Gia Dvali, Gregory Gabadadze, and Massimo Porrati. 
The model consists of a 4D Minkowski brane embedded in a 5D Minkowski bulk like Randall Sundrum (RS) II model. But unlike in RSII model \cite{Randall:1999vf,Maartens:2010ar}~\footnote{See also the details of Randall Sundrum (RS) I model 
studied in ref.~\cite{Randall:1999ee} for completeness.}, here the infinitely large 5th extra dimension is flat.
The Newton's law can be recovered by adding a 4D
Einstein--Hilbert action sourced by the brane curvature to the 5D action.
While the DGP model recovers the standard 4D gravity for small distances,
the effect from the 5D gravity manifests itself for large distances. 

The DGP model is described by the following action \cite{Dvali:2000hr}:
\begin{equation}
S=\frac{1}{2\kappa_{(5)}^{2}}\int d^{5}X\sqrt{-\tilde{g}}\,
\tilde{R}+\frac{1}{2\kappa^{2}}\int d^{4}x\sqrt{-g}\,R
+\int d^{4}x\sqrt{-g}\,{\cal L}_{M}^{\mathrm{brane}}\,,
\label{DGPaction}
\end{equation}
where $\tilde{g}_{AB}$ is the metric in the 5D bulk and 
\be g_{\mu\nu}=\partial_{\mu}X^{A}\partial_{\nu}X^{B}\tilde{g}_{AB}\ee
is the induced metric on the brane with $X^{A}(x^{c})$ being the
coordinates of an event on the brane labeled by $x^{c}$.
The first and second terms in Eq.~(\ref{DGPaction}) correspond to
Einstein--Hilbert actions in the 5D bulk and on the brane, respectively.
Note that $\kappa_{(5)}^{2}$ and 
$\kappa^{2}$ are 5D and 4D gravitational constants, 
respectively, which are related with 5D and 4D Planck masses,
$M_{5}$ and $M_{4}$, via 
\bea \kappa_{(5)}^{2}&=&\frac{1}{M_{5}^3},\\ 
\kappa^{2}&=&\frac{1}{M_{4}^2}=\frac{1}{M^{2}_{p}}= 8\pi G.\eea
The Lagrangian ${\cal L}_{M}^{\mathrm{brane}}$ describes 
matter localized on the 3-brane.

The equations of motion read
\begin{equation}
G^{(5)}_{AB}=0\,,
\end{equation}
where $G^{(5)}_{AB}$ is the 5D Einstein tensor.
The Israel junction conditions on the brane, under which 
a $Z_2$ symmetry is imposed:
\begin{equation}
\label{branejun}
G_{\mu\nu}-\frac{1}{r_c}(K_{\mu\nu}-g_{\mu\nu} K)
=\kappa_{(4)}^2T_{\mu\nu}\, ,
\end{equation}
where $K_{\mu\nu}$ is the extrinsic curvature calculated on the brane,
 $T_{\mu \nu}$ is the energy-momentum tensor of localized matter and $r_{c}$ is the crossover length scale, \be r_{c} = \frac{M_{4}^{2}}{2M_{5}^{3}},\ee because it sets the scale above which the effect of extra dimension becomes important.

The modified Friedmann equations in this model are given by \cite{Copeland:2006wr}:
\bea
H^2 + \frac{k}{a^2}\,&=&\,
 \Bigg(\sqrt{ \frac{\kappa^2 \rho}{3} + \frac{1}{4r_{c}^{2}} }  + \frac{1}{2r_{c}}
 \Bigg)^2,\\  
 \label{DGPfr1}
\dot{H} + H^2 \:&=& \:-\frac{\kappa^2}{6}(\rho + p) \left[ 1+
\left(\kappa^2 \frac{\rho}{3}   + \frac{1}{4r_{c}^2}\right)^{-1/2}
\frac{1}{2r_{c}} \right] + \left[ \sqrt{ \kappa^2
\frac{\rho}{3} + \frac{1}{4r_{c}^2}  } + \frac{1}{2r_{c}}
\right]^2 .  
 \label{DGPfr2}
\eea
In the present analysis, we have set the extra dimension as time-like which is necessary for getting late time acceleration in DGP model. but in generalized prescription one can comsider both space and time like extra dimensions. 
Also it is important to note that, in Eq.~(\ref{DGPfr2}) the spatial curvature $k$ can take values $0$ or $\pm 1$.
 Now following the proposal of  \cite{Kanekar:2001qd,Sahni:2012er}, we know that in order to get a cyclic universe, the condition for bounce and turn around are given by:
 \begin{itemize}
  \item \underline{\bf Bounce:}  \be H=\frac{\dot{a}}{a}=\frac{1}{a(t)}\frac{da(t)}{dt}=0\ee and \be \ddot a=\frac{d^{2}a(t)}{dt^2} > 0,\ee
  
  \item \underline{\bf Turn around:} \be H=\frac{\dot{a}}{a}=\frac{1}{a(t)}\frac{da(t)}{dt}=0\ee and \be\ddot a=\frac{d^{2}a(t)}{dt^2} < 0.\ee
 \end{itemize}
In the following subsections we will discuss all of these possibilities in detail for DGP brane world gravity framework. 

\subsection{Condition for bounce}

In the early universe, at high energy or equivalently in the high density regime of the brane world, $\rho r_c^2/M_4^2 \gg 1$, one can expand Eq.~(\ref{DGPfr1}) as \cite{Gumjudpai:2003vv}
\begin{equation}
H^2 + \frac{k}{a^{2}} \;=\; \frac{\rho}{3M_4^2} \left[\: 1 + \frac{1}{2}
\left(\frac{3 M_4^2 }{4 \rho r_{c}^2}\right) + \;\;\: \cdots
\;\;\: + \frac{1}{2r_c}\sqrt{\frac{3 M_4^2}{\rho}}
\:\right]^2
\label{dgpexp}
\end{equation}
At first order approximation, Eq.~\ref{dgpexp} reduces to the following expression:
\begin{equation}
H^2 + \frac{k}{a^{2}} \;=\; \frac{\rho}{3M_4^2}\left(1+\frac{1}{r_{c}}\sqrt{\frac{3M_{4}^{2}}{\rho}}\right)
\label{dgp3}
\end{equation}
This is obviously a valid assumption because at the time of bounce, the density of the matter content of the Universe (in our case scalar field) is at its maximum, hence we can neglect the contribution from the 
other higher order terms in Eq.~ (\ref{dgpexp}). Now applying the $\rho r_c^2/M_4^2 \gg 1$ at bounce, and setting $H = 0$ in Eq.~ (\ref{dgp3}), we get:
\begin{equation}
\rho_{b} = 3M_{4}^{2}\left(\frac{\sqrt{k}}{a_{b}}-\frac{1}{2r_{c}}\right)^{2}
\label{k1bounce}
\end{equation}
where $\rho_{b}$ and $a_{b}$ are the density and scale factor at bounce, respectively.
Similarly the mass content at bounce (neglecting the constant factor) is given by, \be M_{b} = \rho_{b} a_{b}^{3} =  3M_{4}^{2}\left(\frac{\sqrt{k}}{a_{b}}-\frac{1}{2r_{c}}\right)^{2}a_{b}^{3}.\ee
Hence applying the energy conservation in the present context, we get: \be \delta M + \delta W = 0,\ee where $\delta W$ is the work done during each 
expansion-contraction cycle which is given by  $\oint pdV$ which includes contribution from the area of the hysteresis loop.
By setting, \be \delta M = -\delta W = -\oint pdV,\ee we get the expression for change in amplitude of the scale factor at each successive cycle as 
\begin{equation}
\delta a_{min} = -\frac{\oint pdV}{3M_{4}^{2}\left(\frac{\sqrt{k}}{a_{b}}-\frac{1}{2r_{c}}\right)a_{b}\left[3\left(\frac{\sqrt{k}}{a_{b}}-\frac{1}{2r_{c}}\right)a_{b}-2\right]}
\end{equation}
Therefore we clearly observe that the change in amplitude of the scale factor after each cycle depends on the sign of the integral, the
curvature parameter and the cross over length scale for DGP brane world. It is independent of the density of the matter content of the universe. During our computation we also observe that the
if we fix the spatial curvature parameter $k=-1$, this makes the energy density of the scalar field imaginary, which is not at all physically possible. Hence, for $k=-1$ the cosmological bounce from DGP brane world 
model is not at all possible.
But if we neglect all the higher order terms, Eq.~(\ref{dgpexp}) reduces to the standard Friedmann equation
for which the condition for bounce becomes \be \rho_{b}=\frac{3kM_{4}^{2}}{a_{b}},\ee which is possible for both $k=\pm 1$ but not
for $k=0$. Hence we see that the bouncing condition depends largely on the order of terms which we are including because
our analysis is possible only under approximations. But since our present observed universe is flat ($k=0$), we will be
interested to study the first case in more detail. Hence in our further analysis for DGP model, we will show the
results for Eq.~(\ref{dgp3}) only i.e exclude the $k=-1$ possibility. Here we have the following expression for $\delta a_{min}$: 

\be\begin{array}{lll}\label{rk9}
 \displaystyle\delta a_{min} =\left\{\begin{array}{ll}
                    \displaystyle   -\frac{\oint pdV}{3M_{4}^{2}\left(\frac{1}{a_{b}}-\frac{1}{2r_{c}}\right)a_{b}\left[3\left(\frac{1}{a_{b}}-\frac{1}{2r_{c}}\right)a_{b}-2\right]} &
 \mbox{\small {\bf for {$k=+1$}}}  \\ \\
         \displaystyle  -\frac{\oint pdV}{\frac{3M_{4}^{2}a_{b}}{2r_{c}}\left(\frac{3}{2r_{c}}+2\right)} & \mbox{\small {\bf for {$k=0$}}}.
          \end{array}
\right.
\end{array}\ee
Let us now briefly mention the characteristic feature of the results for cosmological bounce for DGP brane world model in the following:
\begin{enumerate}
 \item For a closed universe, depending on whether the quantity appearing in the denominator for $k=+1$ is positive or negative, an increase in the scale factor after each
 cycle is possible if:
 \begin{itemize}
  \item The hysteresis loop integral \be\oint pdV < 0\ee or equivalently \be p_{exp} < p_{cont},\ee
  \item The hysteresis loop integral \be \oint pdV > 0\ee or equivalently \be p_{exp} > p_{cont}.\ee 
 \end{itemize}
\item On the other hand, for the case $k=0$, the quantity in the denominator is always positive
and finally we get an increase in the scale factor only if the hysteresis loop integral \be \oint pdV<0.\ee
\end{enumerate}

\subsection{Condition for acceleration}

From Eq.~ (\ref{DGPfr2}), at bounce or at high energy, contribution from $1/r_{c}$ term is small compared to the density of the scalar field and consequently one finally gets the standard cosmology from the condition
for acceleration as:
\begin{equation}
p < -\frac{\rho}{3}
\label{dgpaccel}
\end{equation}
which clearly implies that the cosmological bounce can be obtained by violating the strong energy condition.

Substituting the expression for $\rho_{b}$ for $k =+1,0$, we get the conditions for acceleration at bounce as:
\be\begin{array}{lll}\label{dgpbounce}
 \displaystyle\delta p_{b} <\left\{\begin{array}{ll}
                    \displaystyle   -M_{4}^{2}\left(\frac{1}{a_{b}}-\frac{1}{2r_{c}}\right)^{2} &
 \mbox{\small {\bf for {$k=+1$}}}  \\ \\
         \displaystyle  -\frac{M_{4}^{2}}{4r_{c}^{2}} & \mbox{\small {\bf for {$k=0$}}}.
          \end{array}
\right.
\end{array}\ee

This implies that the pressure of the matter content of the Universe at the time of cosmological bounce is related to the scale factor and the cross over length scale $r_{c}$.
Also we observe that the condition for acceleration depends on whether we have considered a possibility of a closed
or flat universe. For flat universe, we see that the condition for acceleration is independent of the scale factor at cosmological bounce.

Finally substituting the expressions for the energy density $\rho$ and pressure $p$ from Eq.~(\ref{eq:scalar_rho}) into Eq.~(\ref{dgpaccel}), we get the condition for the expansion of the Universe in terms of the scalar field 
degrees of freedom as:
\begin{equation}
\dot\phi^{2} < V(\phi)
\end{equation}
which is same as that we get for the standard cosmological inflationary scenario. Therefore, we
can conclude that the conditions on the dynamics of scalar which is needed for causing acceleration
in standard case remains unchanged for the DGP model at the time of cosmological bounce. This implies that the cosmological potentials
in the standard cosmological scenario, capable of satisfying the above condition, will also be able to cause
acceleration in a universe described by the DGP brane world model.

\begin{figure*}[htb]
\centering
\subfigure[ An illustration of the bouncing condition for a universe with an equation of state $w=0,\ k=0,\ \mid r_{c}\mid= 1$.]{
    \includegraphics[width=7.2cm,height=8.2cm] {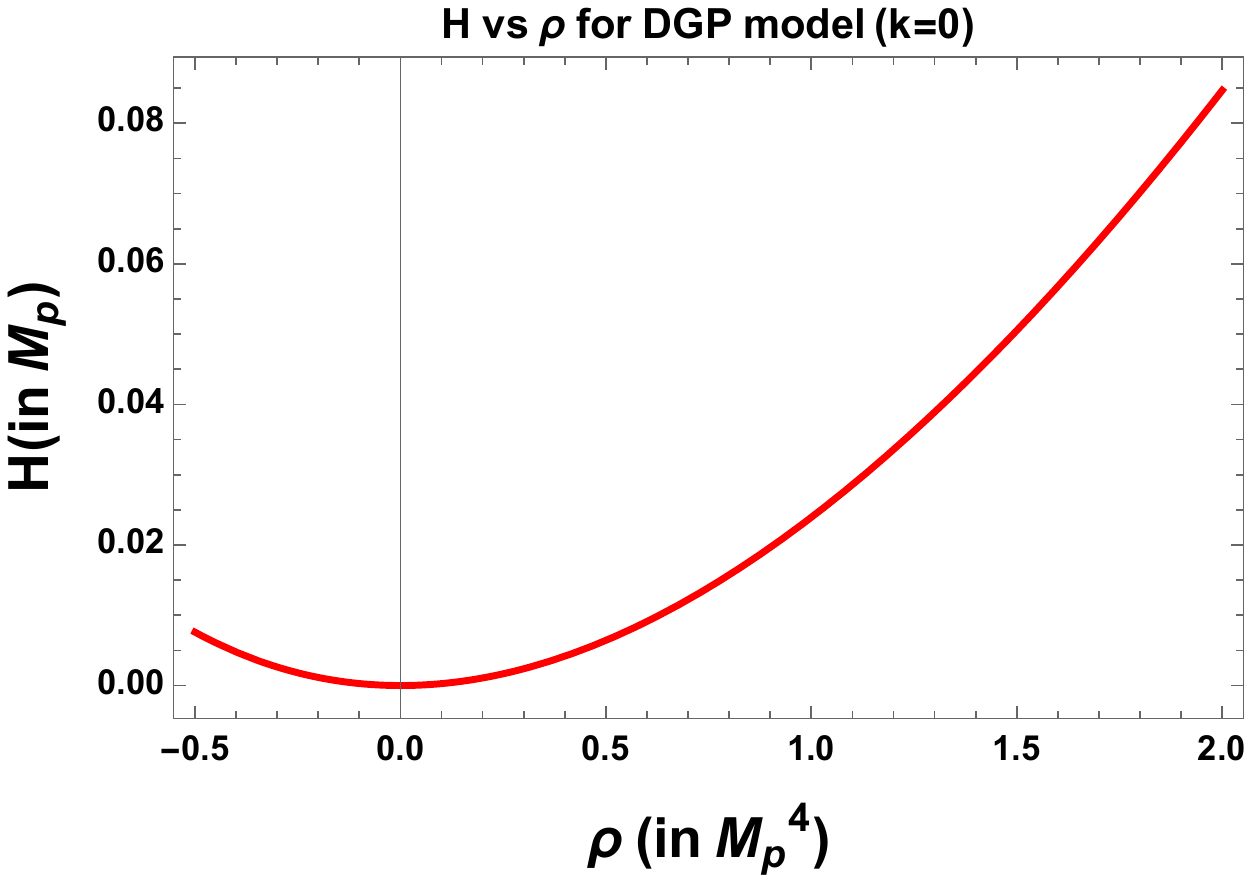}
    \label{dgp1}
}
\subfigure[ An illustration of the bouncing condition for a universe with an equation of state $w=0,\ k=1,\ \mid r_{c}\mid= 1.44$.]{
    \includegraphics[width=7.2cm,height=8.2cm] {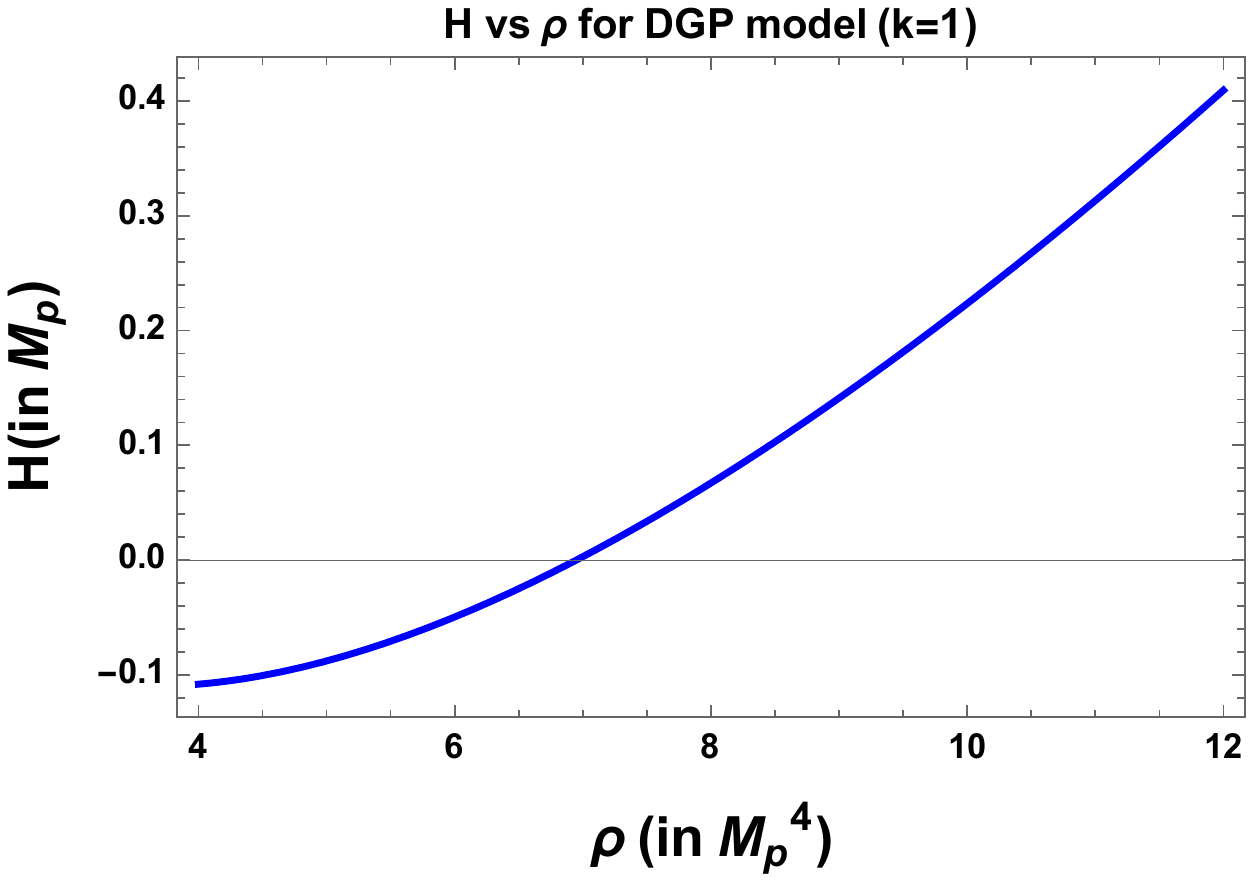}
    \label{dgp2}
}
\subfigure[An illustration of the acceleration condition at the time of bounce for a universe with an equation of state $w=0,\ k=0,\ \mid r_{c} \mid=1$.]{
    \includegraphics[width=7.2cm,height=8.2cm] {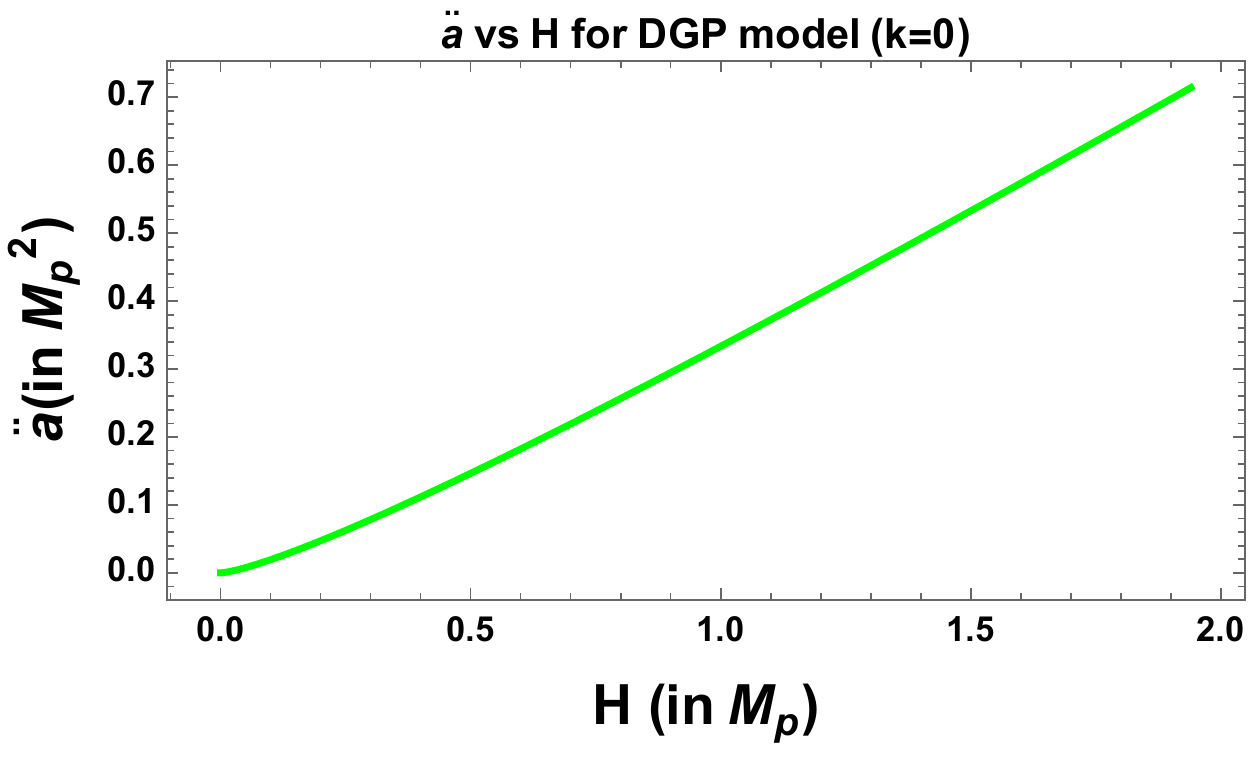}
    \label{dgp3}
}
\subfigure[An illustration of the acceleration condition at the time of bounce for a universe with an equation of state $w=0,\ k=1,\ \mid r_{c} \mid=1.44$.]{
    \includegraphics[width=7.2cm,height=8.2cm] {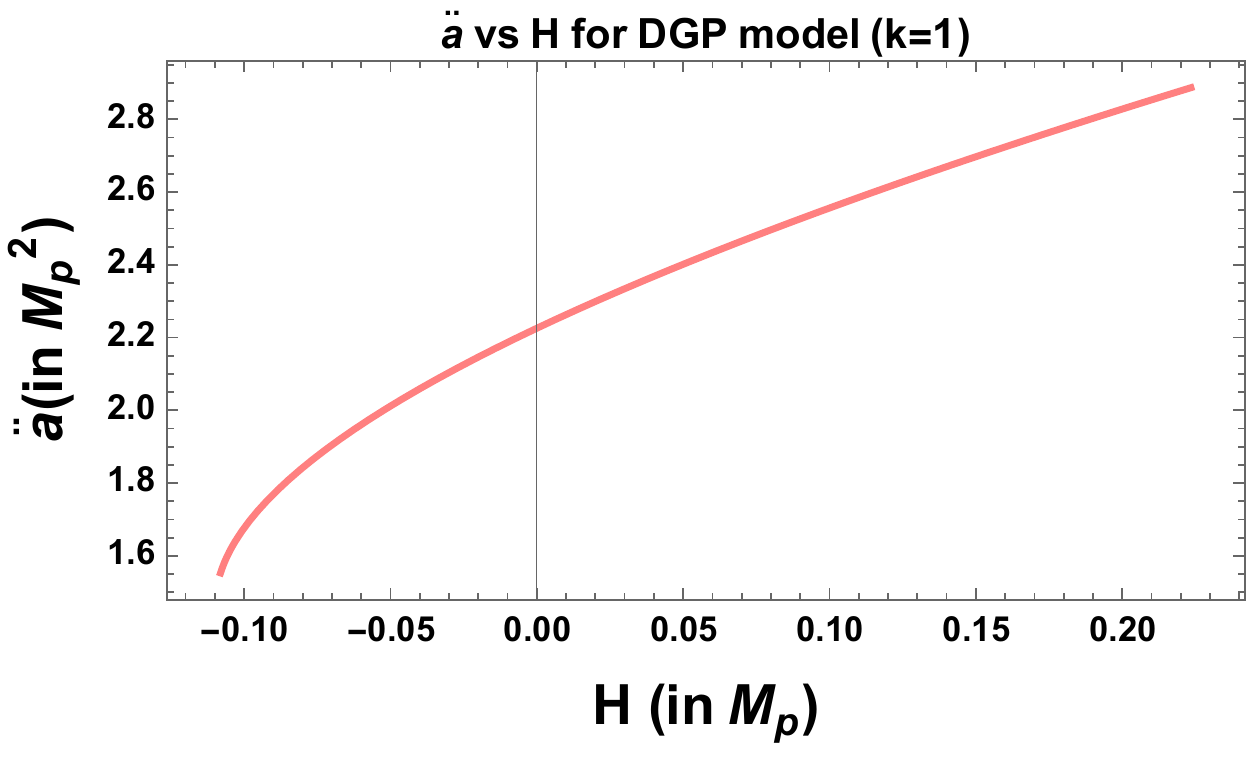}
    \label{dgp4}
}
\caption[Optional caption for list of figures]{ Graphical representation of the phenomena of bounce and acceleration for DGP model.} 
\label{fig15}
\end{figure*}

In Fig.~\ref{fig15}, we have shown the phenomena of bounce and acceleration in the DGP model. We can draw the following conclusions from the above figures:
\begin{itemize}
\item In Fig.~\ref{dgp1} and Fig.~\ref{dgp2}, we have plotted the r.h.s of Eq.~(\ref{DGPfr1}) using the relation $\rho=a^{-3(1+w)}$. The graphs have
been plotted only for the cases when $k=0,\ 1$. This is because, Eq.~(\ref{k1bounce}) shows that density becomes imaginary for the case when $k=-1$, which is unphysical. 
\item We have used $w=0$, since we require a soft equation of state for causing the acceleration and expansion. The case $w=1/3$ made $H$ only
approximately zero. The value of $r_{c}$ has been chosen such that we get proper bounce and acceleration. Thus we find that we require a larger value of $r_{c}$ for causing bounce in a closed universe as compared to a flat universe.
\item From Fig.~\ref{dgp1}, we get the bounce at $\rho=\rho_{b}=0$. Similarly from Fig.~\ref{dgp2}, we get $\rho=\rho_{b}=7.0M_{p}^{4}$.
\item In Fig.~\ref{dgp1} and Fig.~\ref{dgp2}, $H$ going to negative values may be interpreted as the universe changing its direction of motion at bounce.
As had been discussed in \cite{Kanekar:2001qd}, the condition of bounce/turnaround can be imposed by either the condition of making the scale factor changing sign with other quantities remaining same, or, $\dot{a}$ going to negative values with other quantities remaining same. 
\item Figs. \ref{dgp3} and \ref{dgp4} show the necessary condition of acceleration ($\ddot{a}> 0$) at the time of bounce. Here we have plotted the r.h.s of Eq.~(\ref{DGPfr2})
and Eq.~(\ref{DGPfr1}). These plots have also been obtained for the same parameter values as the earlier graphs.
\item Thus Fig.~\ref{fig15} shows graphically that the phenomenon of bounce is possible for dgp model having $w=0$.
\end{itemize}
\subsection{Condition for turnaround}   

To establish the condition for turnaround we can again rewrite Eq.~(\ref{DGPfr1}) as \cite{Gumjudpai:2003vv}:
\begin{eqnarray}
H^2 + \frac{k}{a^{2}}\: &=& \:\frac{1}{4r_{c}^2} \Bigg[ \sqrt{1 + \frac{4\rho
r_{c}^2}{3M_{4}^2}} + 1  \Bigg]^2 \nonumber\\
   \: &=&\: \frac{1}{4r_{c}^2} \Bigg[ \bigg( 1+ \frac{4\rho r_{c}^2}{3M_{4}^2} \bigg)
    + 2 \sqrt{1+ \frac{4\rho r_{c}^2}{3M_{4}^2}} +
   1
   \Bigg]\label{dgp4}
\end{eqnarray}
In late time universe, $\rho r_{c}^{2}/M_{4}^{2} << 1$, hence keeping terms upto the first order we get:
\begin{equation}
H^{2} + \frac{k}{a^{2}} \approx \frac{1}{r_{c}^{2}} + \frac{2\rho}{3M_{4}^{2}}
\label{dgp5}
\end{equation}
This is also a valid assumption because at the time of re-collapse, the density of the
Universe is low or reaches at its minimum but not negligible. Hence we need to keep terms upto the first order in the binomial series expansion.
Now by setting $H = 0$, we get:
\begin{equation}
\rho_{t} = \frac{3M_{4}^{2}}{2}\left(\frac{k}{a_{t}^{2}}-\frac{1}{r_{c}^{2}}\right)
\end{equation}
where $\rho_{t}$ and $a_{t}$ are the density and scale factor at turnaround respectively.

Similar to the Cosmological bounce case, here we also can equate the change in energy or mass content of the Universe to the work done after each expansion-contraction cycle and finally we get:
\begin{equation}
\delta M_{t} = \frac{3M_{4}^{2}}{2}\left(k -\frac{3a_{t}^{2}}{r_{c}^{2}}\right)\delta a_{t} = -\oint pdV
\end{equation} 
Therefore the change in the scale factor after each successive cycle is given by the following expression:
\begin{equation}
\delta a_{max} = -\frac{2}{3M_{4}^{2}\left(k - \frac{3a_{max}^{2}}{r_{c}^{2}}\right)}\oint pdV
\label{dgp6}
\end{equation}
Therefore, just like in the case of bounce, where the change in the amplitude is dependent on the parameters of the cosmological model, in such a physical prescrption 
the change in the amplitude of the scale factor at turnaround also depends not only on the work done, but also on the cross over length scale $r_{c}$ of the DGP brane world model.

\be\begin{array}{lll}\label{dgp7}
 \displaystyle\delta a_{min} =\left\{\begin{array}{ll}
                    \displaystyle   -\frac{2}{3M_{4}^{2}\left(1 - \frac{3a_{max}^{2}}{r_{c}^{2}}\right)}\oint pdV ~~~~&
 \mbox{\small {\bf for {$k=+1$}}}  \\ \\
 \displaystyle \frac{2r_{c}^{2}}{9M_{4}^{2}a_{max}^{2}}\oint pdV &
 \mbox{\small {\bf for {$k=0$}}}  \\ \\
         \displaystyle  \frac{2}{3M_{4}^{2}\left(1 + \frac{3a_{max}^{2}}{r_{c}^{2}}\right)}\oint pdV & \mbox{\small {\bf for {$k=-1$}}}.
          \end{array}
\right.
\end{array}\ee
Let us now briefly mention the characteristic feature of the results for turnaround for DGP brane world model in the following:
\begin{enumerate}
 \item For $k=-1$ we clearly observe that for an open universe, 
 an increase in amplitude of the scale factor after each successive cycle is possible if the hysteresis loop integral
 \be \oint pdV > 0.\ee

\item For $k=0$ we observe that the increase in magnitude of the scale factor depends not only on work done, but also on the cross over length scale.
We get a positive change in the scale factor with each successive cycle if the hysteresis loop integral 
\be \oint pdV > 0.\ee Hence, this is a case where increase in expansion maximum is possible only if the work done is positive.

\item For $k=+1$ if the denominator \be \left(1 - \frac{3a_{max}^{2}}{r_{c}^{2}}\right)>0\ee or equivalently \be\frac{3a_{max}^{2}}{r_{c}^{2}}<1,\ee
we get positive $\delta a_{max}$ if the hysteresis loop integral \be\oint pdV<0\ee and vice versa. Thus depending on the relative magnitude of the scale factor and cross over
length scale at turnaround, we can get an increase in expansion for both positive and negative signature of the work done.

\end{enumerate}

\subsection{Condition for deceleration}

To establish the condition for deceleration we first take the time derivative of Eq.~ (\ref{dgp5}), and using the energy conservation or equivalently the continuity 
equation, \be \dot \rho + 3H(\rho + p) = 0,\ee we can write: 
\begin{equation}
\frac{\ddot{a}}{a} = -\frac{(\rho + p)}{M_{4}^{2}} + \frac{1}{r_{c}^{2}} + \frac{2\rho}{3M_{4}^{2}}.
\end{equation}
From the above equation we get the condition for deceleration as:
\begin{equation}
(\rho_{t} + 3p_{t}) > \frac{3M_{4}^{2}}{r_{c}^{2}}
\label{dgpdecel}
\end{equation}
Therefore, turnaround can be obtained without violating the energy condition.
And just like for acceleration, here we see that the condition for deceleration
depends on $r_{c}$, which was expected because in the late universe, the effect of $r_{c}$ will become more important. 

In place of Eq.~ (\ref{dgpbounce}), we get the conditions for deceleration at turnaround as:
\be\begin{array}{lll}\label{bounce}
 \displaystyle p_{t} =\left\{\begin{array}{ll}
                    \displaystyle  M_{4}^{2}\left(\frac{1}{r_{c}^{2}} - \frac{1}{2a_{t}^{2}}\right) ~~~~&
 \mbox{\small {\bf for {$k=+1$}}}  \\ \\
 \displaystyle\frac{M_{4}^{2}}{r_{c}^{2}} &
 \mbox{\small {\bf for {$k=0$}}}  \\ \\
         \displaystyle   M_{4}^{2}\left(\frac{1}{r_{c}^{2}} + \frac{1}{2a_{t}^{2}}\right) & \mbox{\small {\bf for {$k=-1$}}}.
          \end{array}
\right.
\end{array}\ee
Finally substituting the expressions for the density $\rho$ and pressure $p$ from Eq.~(\ref{eq:scalar_rho}) into Eq.~(\ref{dgpdecel}), we get the following 
condition for expansion of the Universe in terms of the scalar field as:
\begin{equation}
\dot\phi^{2} > V(\phi) + \frac{3M_{4}^{2}}{2r_{c}^{2}}
\end{equation}
This clearly implies that the lesser the contribution from the potential energy 
as compared to the standard cosmological inflationary case or equivalently for
the acceleration case as mentioned earlier, easier will be to
achieve contraction phase of the Universe. We also clearly observe that, even if the potential
satisfies the deceleration condition, just like as for standard cosmological scenario, it is not necessary
that it will also cause deceleration in DGP brane world model at turnaround as of now the deceleration is also the cross over length scale 
$r_{c}$ dependent in the present context.

\begin{figure*}[htb]
\centering
\subfigure[ An illustration of the turnaround condition for a universe with an equation of state $w=1, \  k=0,\ \mid r\mid=10$.]{
    \includegraphics[width=7.2cm,height=8.2cm] {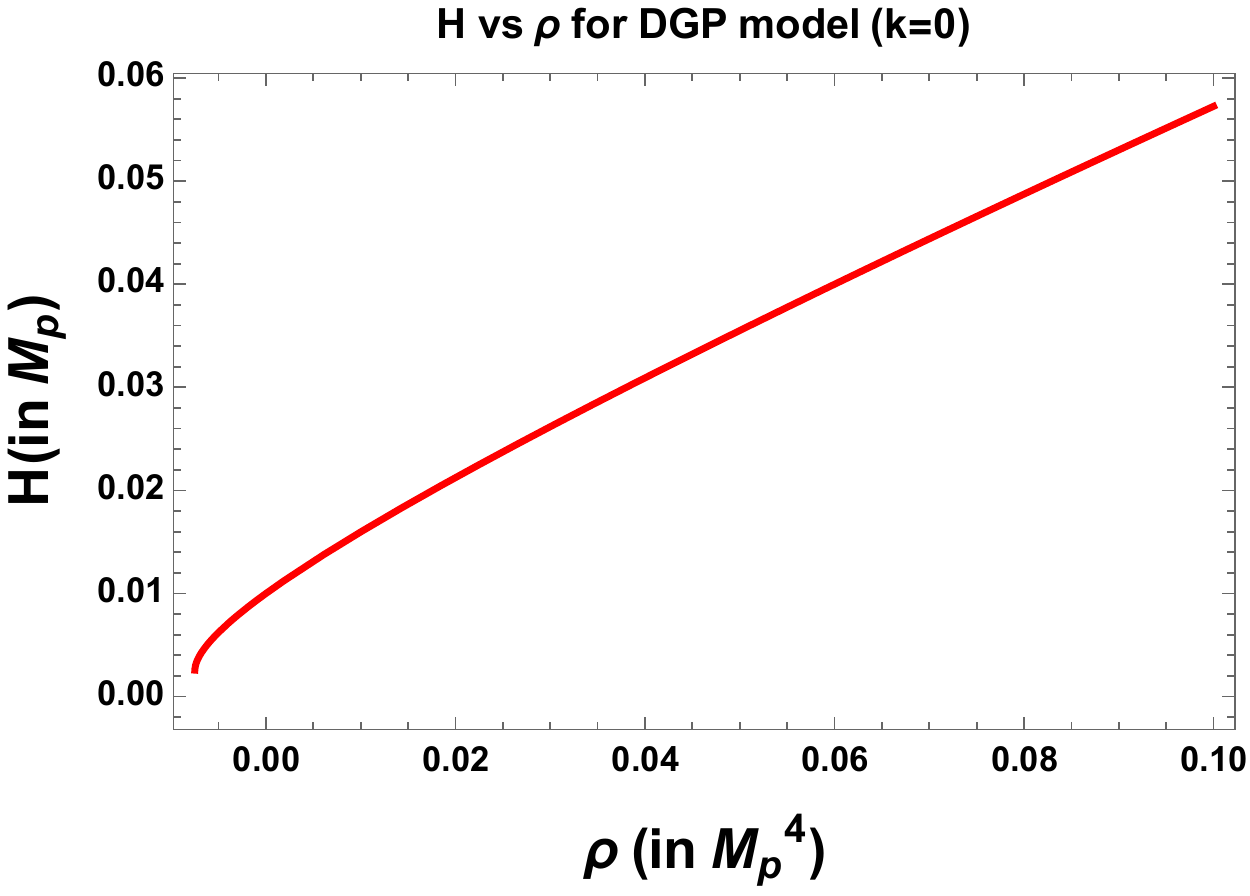}
    \label{dgp5}
}
\subfigure[ An illustration of the turnaround condition for a universe with an equation of state $w=1, \  k=1,\ \mid r_{c}\mid=1.44$.]{
    \includegraphics[width=7.2cm,height=8.2cm] {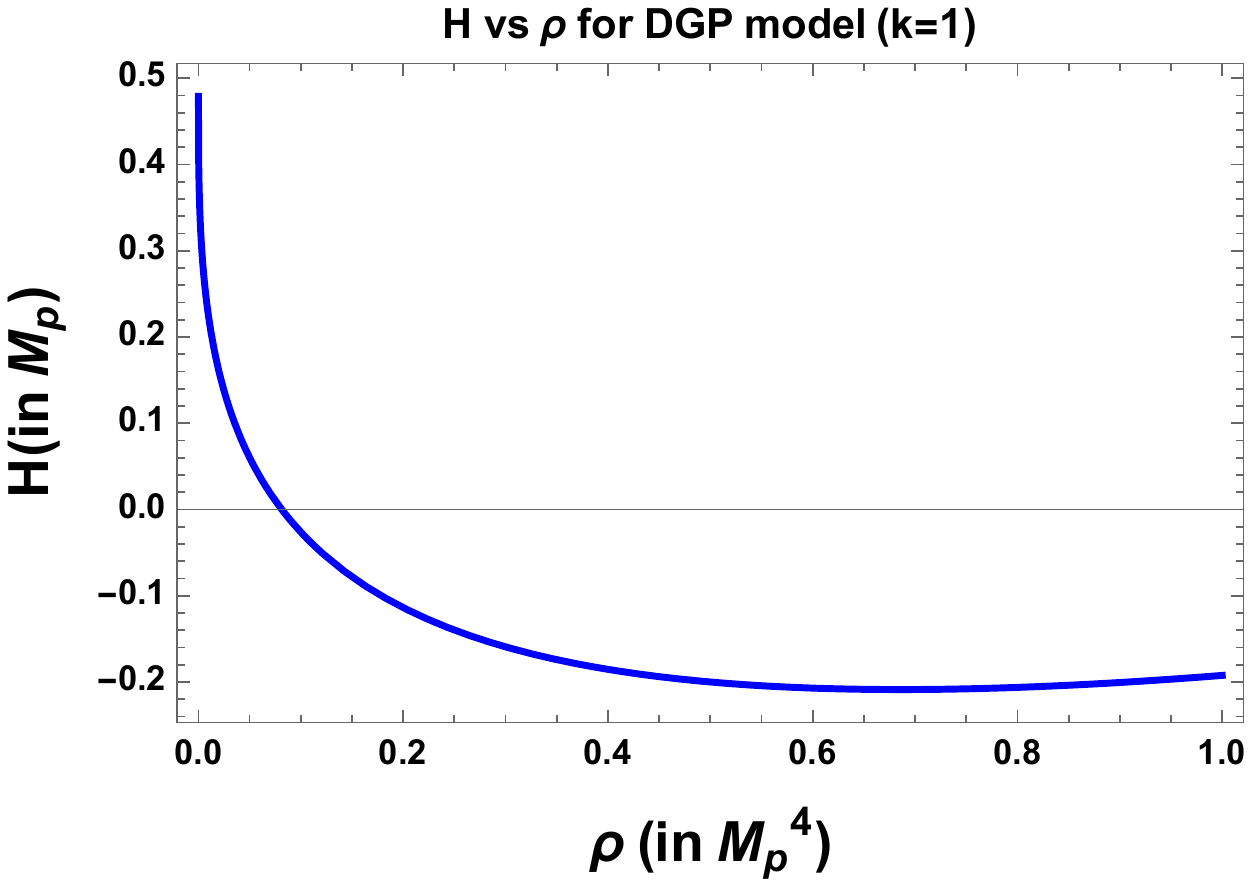}
    \label{dgp6}
}
\subfigure[An illustration of the deceleration condition at turnaround for a universe with an equation of state $w=1,\ k=1, \mid r_{c}\mid=10$.]{
    \includegraphics[width=7.2cm,height=8.2cm] {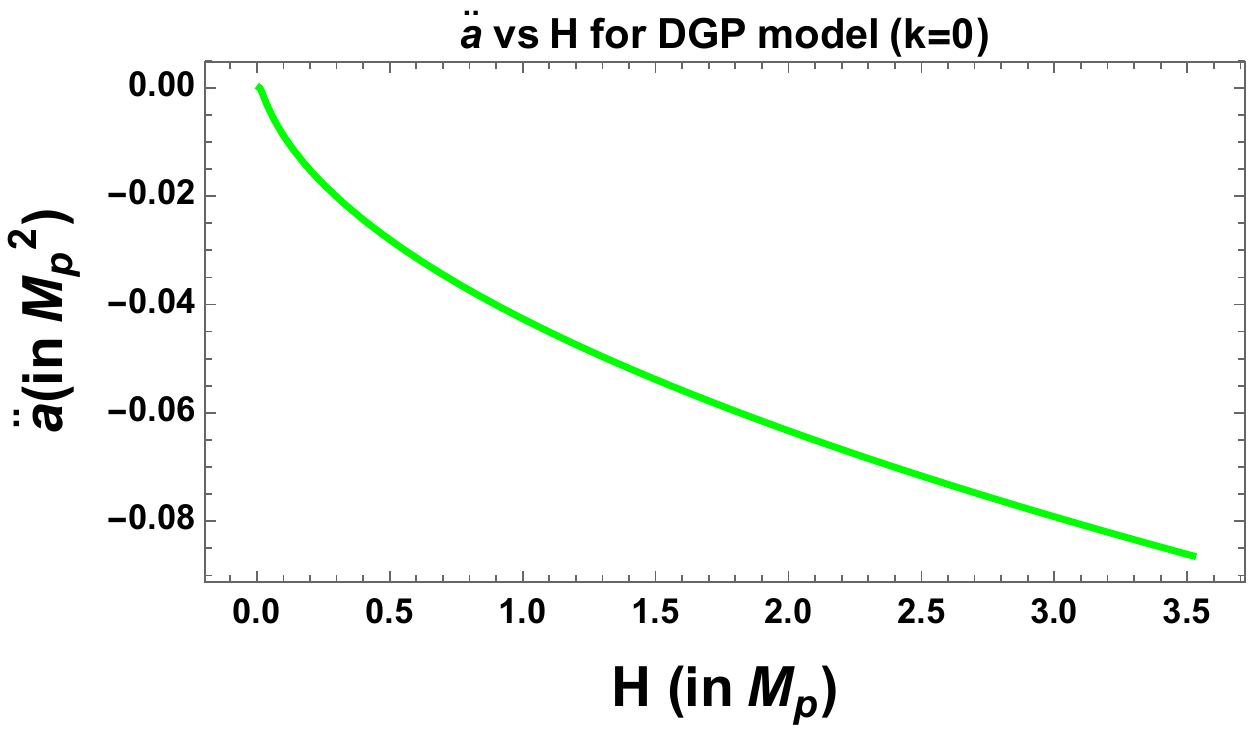}
    \label{dgp7}
}
\subfigure[An illustration of the deceleration condition at turnaround for a universe with an equation of state $w=1,\ k=1, \mid r_{c}\mid=1.44$.]{
    \includegraphics[width=7.2cm,height=8.2cm] {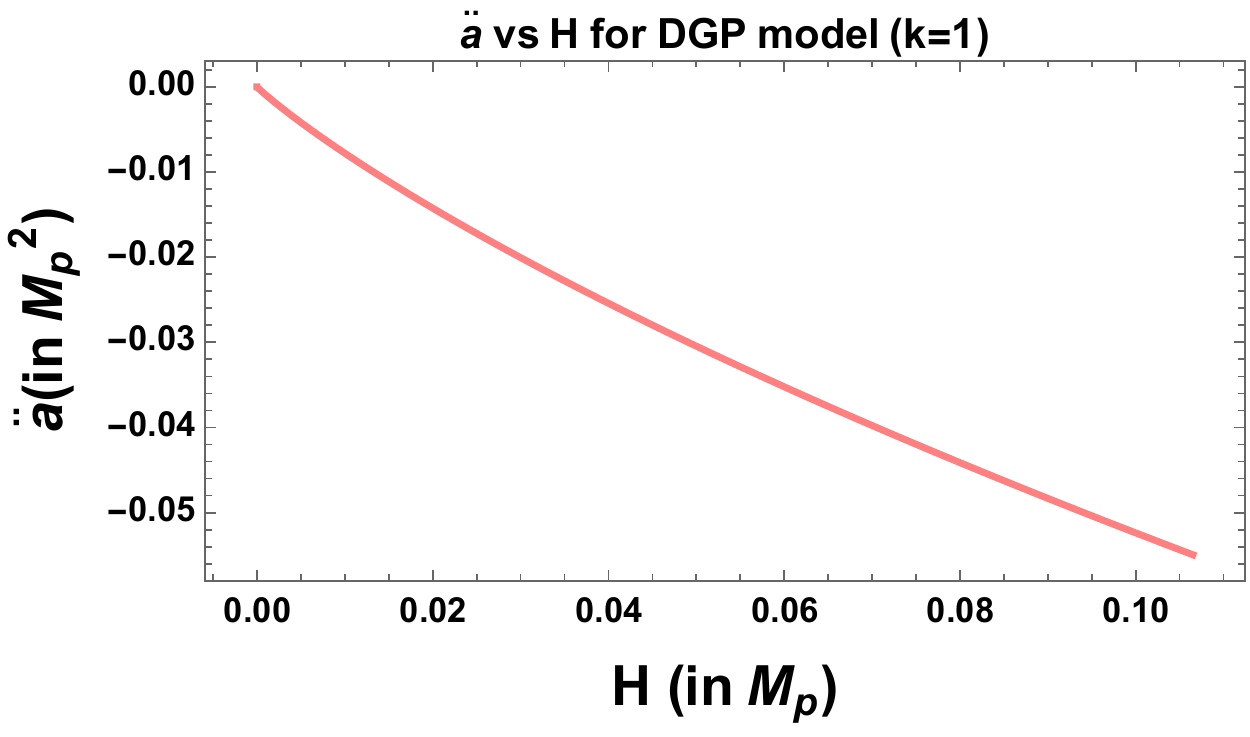}
    \label{dgp8}
}
\caption[Optional caption for list of figures]{ Graphical representation of the phenomena of turnaround and deceleration for DGP model.} 
\label{fig16}
\end{figure*}

In Fig. \ref{fig16}, we have shown the phenomena of turnaround and deceleration in dgp model. We can draw the following conclusions from the above figures:
\begin{itemize}
\item In Figs. \ref{dgp5} and \ref{dgp6}, we have plotted the r.h.s of Eqns. (\ref{dgp4}) and (\ref{DGPfr1}) respectively, using the relation $\rho=a^{-3(1+w)}$. But for this case we have used $w=1$, since we require a stiff equation of state for causing contraction and deceleration. 
\item From Figs. \ref{dgp5} and \ref{dgp6}, we get the turnaround at $\rho=\rho_{t}=0$ and at $\rho=\rho_{t}=0.08M_{p}^{4}$ respectively.
\item Figs. \ref{dgp7} and \ref{dgp8} show the necessary condition of deceleration ($\ddot{a}< 0$) at the time of turnaround. Here we have plotted the r.h.s of Eqns. (\ref{DGPfr2})and (\ref{dgp4}), (\ref{DGPfr1}) respectively. This plot has also been obtained for the same parameter values as the earlier graph.
\item Thus Fig. \ref{fig16} shows graphically that the phenomenon of turnaround is possible for the DGP model having $w=1$.
\end{itemize}

\subsection{Evaluation of the work done for one cycle}
Let us now compute the explicit contribution from the 
work done for aone complete cycle for DGP brane world cosmological setup. To serve this purpose we start with the following closed loop integral
: 
\begin{eqnarray}
\oint pdV &=& \int_{cont} pdV + \int_{exp} pdV  \nonumber \\
&=& \int_{a_{max}^{i-1}}^{a_{min}^{i-1}} pdV + \int_{a_{min}^{i-1}}^{a_{max}^i} pdV 
\label{hyst}
\end{eqnarray}
where $i$ and $i-1$ refer to the two successive cycle i.e. $i$th and $i-1$th cycle of expansion and contraction phase of the Universe. In the 
present context the work done corresponding to the expanding phase and the contracting phase can be expressed in terms of the work done 
between two successive cycle as:
\bea
\int_{cont} pdV &=& \int_{a_{max}^{i-1}}^{a_{min}^{i-1}} pdV,\\
\int_{exp} pdV &=& \int_{a_{min}^{i-1}}^{a_{max}^i} pdV\eea
where $(a_{max}^{i-1},a_{max}^i)$ are the maximum magnitude of the scale factor 
for $i-1$th and $i$th cycle of the Universe. Similarly, $(a_{min}^{i-1})$ represents the the minimum magnitude of the scale factor 
for $i-1$th cycle of the Universe for DGP brane world model.

Now the volume of the Universe can be written as, \be V = a^3 \ee  and an infinitesimal change in the volume can be written as, \be
dV = 3a^{2}da= 3a^{2}(t)\frac{da(t)}{dt}dt = 3a^{2}\dot{a}dt,\ee which is frequently used for further computation of the work done.
And we also know from Eq.~ (\ref{eq:scalar_rho}), in the presence of scalar field, the pressure for the 
scalar field can be written as: $p = \dot{\phi}^{2}/2 - V(\phi)$. Substituting these in Eq.~ (\ref{hyst}), we get
\begin{eqnarray}
\oint pdV &=& \int_{cont} 3\left(\frac{\dot{\phi}^{2}}{2} - V(\phi)\right)a^{2}\dot{a}dt + \int_{exp}  3\left(\frac{\dot{\phi}^{2}}{2} - V(\phi)\right)a^{2}\dot{a}dt \nonumber \\
&=& \int_{a_{max}^{i-1}}^{a_{min}^i-1} 3\left(\frac{\dot{\phi}^{2}}{2} - V(\phi)\right)a^{2}\dot{a}dt + \int_{a_{min}^{i-1}}^{a_{max}^i} 3\left(\frac{\dot{\phi}^{2}}{2} - V(\phi)\right)a^{2}\dot{a}dt \nonumber \\
\label{hyst1}
\end{eqnarray}
Further using the solution of scalar field $\phi$ and the scale factor $a$, one can solve the above equation to get the estimation of this
integral. Also it is important to mention here that the above integral can also be expressed in terms of scale
factor only using Eq~(\ref{DGPfr1}), Eq~(\ref{DGPfr2}) and Eq~(\ref{eq:scalar_rho}). But since
the Friedmann equations in case of DGP brane world model are
highly complicated, we can use the late time and early time approximations of the Friedmann equations in order to get an
physically relevant approximate analytical expression for the integral for the work done. Therefore,
the work done can be decomposed into four parts as follows:
\begin{equation}
\oint pdV = \underbrace{\int_{a_{max}^{i-1}}^{a'^{i-1}} pdV + \int_{a'^{i-1}}^{a_{min}^{i-1}} pdV }_{\bf Contraction}  +  \underbrace{\int_{a_{min}^{i-1}}^{a'^{i-1}} pdV + \int_{a'^{i-1}}^{a_{max}^{i}} pdV}_{\bf Expansion}
\label{hyst2}
\end{equation}
 where the first two terms corresponds to late and early times during the period of contraction respectively and the last two
 terms corresponds to early and late times during the period of expansion respectively. Here $ a'$ corresponds to the scale factor at the time of transition $t'$ from early to late time or vice-versa,
 $a_{max}$ and  $a_{min}$ corresponds to the values of the scale factor at the time of turnaround $t_{max}$ and bounce $t_{min}$ respectively.
 
At early time, from Eq.~(\ref{dgpexp}), keeping terms upto first order and neglecting the contribution from $1/r_{c}^{2}$ terms in the acceleration equation \cite{Gumjudpai:2003vv}, we get:
\begin{eqnarray}
\label{w11}\rho &=& 3M_{4}^{2}\left[\left\{\left(\frac{\dot{a}}{a}\right)^{2} + \frac{k}{a^{2}}\right\}^{1/2}-\frac{1}{2r_{c}}\right]^{2} \\
\label{w22}\frac{\ddot{a}}{a} &=& -\frac{(\rho + 3p)}{6M_{4}^{2}}
\end{eqnarray} 
Hence using Eq.~(\ref{w11}), Eq.~(\ref{w22}) and Eq.~(\ref{eq:scalar_rho}) for the pressure content of the Universe $p$, we get:
\begin{equation}
\frac{\dot{\phi}^{2}}{2} - V(\phi) = -M_{4}^{2}\left[\frac{2\ddot{a}}{a} + \left[\left\{\left(\frac{\dot{a}}{a}\right)^{2} + \frac{k}{a^{2}}\right\}^{1/2}-\frac{1}{2r_{c}}\right]^{2}\right]
\label{hyst3}
\end{equation}
Let this equation be valid upto to a cut-off time scale $t'$ where the value of the scale factor is $a'$. Therefore using Eq.~(\ref{hyst3}) into Eq.~(\ref{hyst1}), the second and third integral as appearing 
in Eq.~(\ref{hyst2}) can be expressed as: 
\begin{eqnarray}
\int_{a'^{i-1}}^{a_{min}^{i-1}} pdV &=& \int_{a_{min}^{i-1}}^{a'^{i-1}} pdV 
\\ &=& \int -3M_{4}^{2}\left[\frac{2\ddot{a}}{a} + \left[\left\{\left(\frac{\dot{a}}{a}\right)^{2} + \frac{k}{a^{2}}\right\}^{1/2}-\frac{1}{2r_{c}}\right]^{2}\right]a^{2}\dot{a} dt \nonumber \\
\label{hyst3.1} 
\end{eqnarray} 
At late time, using Eq.~(\ref{dgp5}) and Eq.~(\ref{dgpdecel}), we get: 
\begin{eqnarray}
\label{w1}\rho &=& \frac{3M_{4}^{2}}{2}\left[\left(\frac{\dot{a}}{a}\right)^{2} + \frac{k}{a^{2}} -\frac{1}{r_{c}^{2}}\right] \\
\label{w2}\frac{\ddot{a}}{a} &=& -\frac{(\rho + p)}{M_{4}^{2}} + \frac{1}{r_{c}^{2}} + \frac{2\rho}{3M_{4}^{2}}
\end{eqnarray}
Further using Eq.~(\ref{w1}), Eq.~(\ref{w2}) and Eq.~(\ref{eq:scalar_rho}) for the pressure content of the Universe $p$, we get:
\begin{equation}
\frac{\dot{\phi}^{2}}{2} - V(\phi) = M_{4}^{2}\left[\frac{3}{2r_{c}^{2}} - \frac{\ddot{a}}{a} - \frac{1}{2}\left(\frac{\dot{a}}{a}\right)^{2} - \frac{k}{2a^{2}}\right]
\label{hyst4}
\end{equation}
This equation is also valid upto to a cut-off time scale $t'$ where the value of the scale factor is $a'$. Therefore using Eq.~(\ref{hyst4}) into Eq.~(\ref{hyst1}), the first and last integral as appearing in Eq.~(\ref{hyst2}) can be expressed as: 
\begin{eqnarray}
\int_{a_{max}^{i-1}}^{a'^{i-1}} pdV &=& \int_{a'^{i-1}}^{a_{max}^{i}} pdV \\
&=& \int 3M_{4}^{2}\left[\frac{3}{2}a^{2}\dot{a} - \ddot{a}a\dot{a} - \frac{\dot{a}^{3}}{2} - \frac{k\dot{a}}{2}\right]dt 
\label{dgphyst5} 
\end{eqnarray}
Therefore by substituting Eq.~(\ref{hyst3.1}) and Eq.~(\ref{dgphyst5}) into Eq.~(\ref{hyst2}) the complete expression for the work done in a single expansion-contraction cycle is governed by the following expression:
\begin{eqnarray}
\oint pdV &=& \underbrace{\int_{a_{max}^{i-1}}^{a'^{i-1}} 3M_{4}^{2}\left[\frac{3}{2}a^{2}\dot{a} - \ddot{a}a\dot{a} 
- \frac{\dot{a}^{3}}{2} - \frac{k\dot{a}}{2}\right]dt}_{\bf I} \nonumber \\
&-& \underbrace{\int_{a'^{i-1}}^{a_{min}^{i-1}} 3M_{4}^{2}\left[\frac{2\ddot{a}}{a} + \left[\left(\left(\frac{\dot{a}}{a}\right)^{2}
+ \frac{k}{a^{2}}\right)^{1/2}-\frac{1}{2r_{c}}\right]^{2}\right]a^{2}\dot{a} dt}_{\bf II} \nonumber 
\\ &-& \underbrace{\int_{a_{min}^{i-1}}^{a'^{i-1}} 3M_{4}^{2}\left[\frac{2\ddot{a}}{a} +
\left[\left(\left(\frac{\dot{a}}{a}\right)^{2} + \frac{k}{a^{2}}\right)^{1/2}-\frac{1}{2r_{c}}\right[^{2}\right]a^{2}\dot{a} dt}_{\bf III} \nonumber \\ 
 &+& \underbrace{\int_{a'^{i-1}}^{a_{max}^{i}} 3M_{4}^{2}\left[\frac{3}{2}a^{2}\dot{a} - \ddot{a}a\dot{a} - \frac{\dot{a}^{3}}{2}
 - \frac{k\dot{a}}{2}\right]dt}_{\bf IV}
 \label{dgpwork}
\end{eqnarray}
In the present context, we clearly visualize from our analysis that, by knowing the solution of the scale factor in
the early Universe and as well as in the late Universe, we can get an idea of the nature of the hysteresis loop.
Additionally it is important to note that while the evaluation of work done is independent of the parameters of this model at early times,
parameter dependence enters through late time evaluation of the integral for work done.

\subsection{Semi-analytical analysis for cosmological potentials}
In this section, we try to find simple analytical expressions for the work done during one cycle
of expansion and contraction from various cosmological models in order to get some idea of the behavior of the cosmological hysteresis loop.
Though the analysis is independent of any particular functional form of potential, but in order to study the physical significance as well as the nature of
the derived results in the previous section, we need to specify the functional form of the cosmological potential. While doing the analysis for DGP brane world model and for
all the other subsequent cosmological models, we will consider three different potentials i.e. Hilltop, Natural and Coleman-Weinberg potential, which will be discussed in the next sections in detail.

Since from Eq.~(\ref{dgpwork}), we observe that the cosmological work done can be evaluated if we know the explicit form of the
variation of the scale factor with time, our main motivation is to find an explicit expression for the
scale factor in terms of time $`t'$. For this we need to solve Eq.~(\ref{DGPfr1}), Eq.~(\ref{eq:scalar_rho}) and Eq.~(\ref{eq:scalar field})
consistently. Therefore, we will basically substitute the expression for the Hubble parameter from the
Friedmann equation into Eq.~(\ref{eq:scalar field}) and replace $\rho$ by Eq.~(\ref{eq:scalar_rho})
with a specified form of the cosmological potential. Then we can solve for the scalar field and substitute it back to the
expression as appearing in the Friedmann equation to get an expression for scale factor in terms of $`t'$. For the sake of simplicity henceforth we will only concentrate for the  
case of $k=0$, which is also a valid and natural assumption in the present context, since we know that our present
observations predict a nearly flat universe. In order to simplify the analysis further, we will consider the case
when the contribution from the kinetic term is lesser than the potential energy i.e. \be\dot{\phi}^{2}<<V(\phi)\ee during the expansion and similarly in the physical situation
where the kinetic term is larger than the potential energy i.e.
\be \dot{\phi}^{2}>>V(\phi)\ee during the contraction phase of the Universe. Hence
the approximate forms of Eq.~(\ref{eq:scalar_rho}) and Eq.~(\ref{eq:scalar field}) during expansion and
contraction phase of the Universe will be given by:
\begin{eqnarray}
3H\dot{\phi}+\frac{dV}{d\phi}&\approx&0 ,\ \rho\approx V(\phi) \  \ {\rm (\bf during\ expansion)} \label{modeqn1} \\
\ddot{\phi}+3H\dot{\phi}&\approx&0, \  \rho\approx\frac{\dot\phi^{2}}{2} \  \ {\rm (\bf during\ contraction)}
\label{modeqn}
\end{eqnarray}
Next step is to specify the specific form of the potential, in order to get further informations and the constraints from the equations.
In the next subsections we have explicitly shown the analysis for three different potentials i.e. Hilltop, Natural and origanted Coleman-Weinberg potential.

\subsubsection{Case I: Hilltop potential }
In case of hilltop models the potential can be represented by the following functional form \cite{Choudhury:2015jaa}:
\be\label{m64}
V(\phi)=V_{0}\left[1+\beta\left(\frac{\phi}{M_4}\right)^{p}\right]
\ee
where $V_{0}=M^4$ is the tunable energy scale and $\beta$ is the index which characterizes the feature of the potential. In principle $\beta$ can be both positive and negative.
Additionally it is important to note that, in the present context, $V_{0}$ mimics the role of vacuum energy.
Since our present job is to substitute the expression for the scale factor into Eq.~ (\ref{dgpwork}), we need to find separate
expressions for the scale factor for both early and late times during the expansion and contraction phases of the Universe. 
\\ \\
\textbf{A. Expansion}
\\ \\
\underline{\bf i) Early time}:- 
\\ \\
When the scale factor is lying within the window, $a_{min}<a<\ a'$ or equivalently when $\rho r_{c}^{2}/M_{4}^{2}>>1$, we can use the approximated version of the Friedmann equation
as given by Eq.~ (\ref{dgpexp}), with $k=0$, where the energy density of the scalar field $\rho$ is for now described by the hilltop potentials. But instead of considering only the zeroth order term
(as we had done in order to find the condition for bounce), here we will consider upto the first order terms (as was done for finding an expression for cosmological work done).
Then substituting the resulting expression for the Hubble parameter $H$ into Eq.~(\ref{modeqn1}), we get the following integral equation in DGP brane world as:
\begin{equation}
\int d\left(\frac{\phi}{M_{4}}\right)\frac{\left\{\left[\frac{V_{0}}{3M_{4}^{2}}\left(1+\beta\left(\frac{\phi}{M_{4}}\right)^{p}\right)\right]^{1/2}+\frac{1}{2r_{c}}\right\}}{\beta p\left(\frac{\phi}{M_{4}}\right)^{p-1}}  = -\frac{V_{0}}{3M_{4}^{2}}\int dt
\label{dgpearly1}
\end{equation}
The exact solutions of the above integral equation is given in the Appendix, which we see has a very complicated form for Randall Sundrum (RSII) limiting situation. Hence to simplify the integrals,
we use the following redefinition of the field variables:
\be\label{opi} \frac{\phi(t)}{M_{4}}=e^{\lambda}\ee where now we will solve for $\lambda$. In order to further simplify the expressions, we solve for two limiting cases:
\\ \\
\underline{\bf a) $\phi/M_{4}<<1$:}
\\ \\
For this case we can expand the exponentials upto linear order and then using the result the integral on the left hand side of Eq.~(\ref{dgpearly1}) becomes:
\begin{equation}
\int\left[\left(\frac{V_{0}}{M_{4}^{2}}\right)^{1/2}\frac{(1+\beta)^{1/2}}{\beta p}\left(1+\frac{\beta p \lambda}{2(1+\beta)}\right)+\frac{1}{2r_{c}}\right](1+\lambda)[1-(p-1)\lambda] d\lambda = -\frac{V_{0}}{3M_{4}^{2}}t+c
\end{equation}
which is true under the assumption that the quantity \be\frac{\beta\lambda p}{(1+\beta)}<<1\ee is small. It is important to note that in the present context $c$ is an arbitrary integration constant.

Hence solving the above integral on the left hand side, we get the expression for $\lambda$, hence for $\phi/M_{4} = e^{\lambda}$ as: 
\begin{equation}
\lambda=\frac{\frac{V_{0}}{3M_{4}^{2}}}{\left(\frac{V_{0}}{3M_{4}^{2}}\right)^{1/2}\frac{(1+\beta)^{1/2}}{\beta p}+\frac{1}{2r_{c}}}(t_{i}-t)+\lambda_{i}
\label{potential15}
\end{equation}
Here $\lambda_{i}$ is the value of $\lambda$ at the initial time scale $t=t_{i}$. 

Substituting the expression for $\lambda$ back into the Friedmann equation, we get the expression for scale factor as
\begin{eqnarray}
a(t) &=& a_{i}\exp\left[\left(\left(\frac{V_{0}}{3M_{4}^{2}}\right)^{1/2}(1+\beta)^{1/2}+\frac{1}{2r_{c}}\right.\right.\nonumber\\&& \left.\left.~~~~~~~~~~~~ +\beta p\frac{\left(\frac{V_{0}}{3M_{4}^{2}}\right)^{1/2}}{2(1+\beta)^{1/2}}
\left\{\lambda_{i}+\frac{\frac{V_{0}}{3M_{4}^{2}}}{\left(\frac{V_{0}}{3M_{4}^{2}}\right)^{1/2}\frac{(1+\beta)^{1/2}}{\beta p}+\frac{1}{2r_{c}}}t_{i}\right\}\right)t\right.\nonumber\\&& \left.~~~~~~~~~~~~~~~~~~~~~~~~~~~~~~~~~~~~~~~~~-\left(\beta p\frac{\left(\frac{V_{0}}{3M_{4}^{2}}\right)^{1/2}}{2(1+\beta)^{1/2}}\right)\frac{\frac{V_{0}}{3M_{4}^{2}}}{2 \left(\frac{V_{0}}{3M_{4}^{2}}\right)^{1/2}\frac{(1+\beta)^{1/2}}{\beta p}+\frac{1}{2r_{c}}}t^{2}\right] 
\label{dgpearly2}
\end{eqnarray}
where $a_{i}$ is the value of the scale factor at time scale $t=t_{i}$ which is in our case fixed at the bouncing time scale $t_{i}=t_{b}$.
\\ 
\\
\underline{\bf b) $\phi/M_{4}>>1$:} 
\\ 
\\
For this physical situation Eq.~(\ref{dgpearly1}) simplifies to the following expression:
\begin{equation}
\int\frac{1}{\beta^{1/2} p}\left[\left(\frac{V_{0}}{3M_{4}^{2}}\right)^{1/2}e^{p\lambda/2}+\frac{1}{2r_{c}}\right]e^{(2-p)\lambda} d\lambda = -\frac{V_{0}}{3M_{4}^{2}}t + A_{0}
\end{equation}
where we use Eq.~(\ref{opi}). Here we have also assumed that $\beta e^{p\lambda}>>1$, which is a valid assumption since we are working in the limit where $\phi/M_{4}>>1$.

Hence solving the above integral we get the expression for $\lambda$ as
\begin{equation}
\lambda=\ln\left[\frac{\beta^{1/2}p}{\left(\frac{V_{0}}{3M_{4}^{2}}\right)^{1/2}}\left(A_{0}-\frac{V_{0}}{3M_{4}^{2}}t\right)\left(2-\frac{p}{2}\right)\right]
\end{equation}
Here $A_{0}$ is an arbitrary integration constant which can be expressed as: 
\begin{equation}
A_{0}=\frac{\left(\frac{V_{0}}{3M_{4}^{2}}\right)^{1/2}}{\beta^{1/2}p}\left(\frac{2}{4-p}\right)exp\left[\frac{(4-p)}{2}\lambda_{i}\right]+\frac{V_{0}}{3M_{4}^{2}}t_{i}
\end{equation}
Here we have neglected $1/2r_{c}$ as the following condition holds good:
\be 
\frac{1}{r_{c}}<<\left(\frac{V_{0}}{3M_{4}^{2}}\right)^{1/2}\beta^{1/2}e^{p \lambda/2},
\ee
which is a valid assumption since we are studying the cosmological consequences in the context of early universe in this limiting case.

Further substituting back the expression for $\lambda$ into the Friedmann equation and integrating, we get the following 
expression for the scale factor as:
\begin{equation}
a(t)=a_{i}\exp\left[\frac{\beta p(p-4)\left(A_{0}-\frac{V_{0}}{3M_{4}^{2}}t\right)\left(\left(2-\frac{p}{2}\right)
\left(A_{0}-\frac{V_{0}}{3M_{4}^{2}}t\right)\right)^{\frac{p}{4-p}}}{4\left(\frac{V_{0}}{3M_{4}^{2}}\right)}\right].
\end{equation} 
\\ \\
\underline{\bf ii) Late time:}
\\
\\
When the scale factor is lying within the window, $a' <a< a_{max}$ or equivalently
when $\rho r_{c}^{2}/M_{4}^{2}<<1$, we can use the approximate Friedmann equation
given by Eq.~(\ref{dgp5}) with $k=0$, where $\rho$ is now characterized by the hilltop potential. Then substituting
the resulting expression for the Hubble parameter $H$ into Eq.~(\ref{modeqn1}), we get an integral equation of the following form:
\begin{equation}
\int d\left(\frac{\phi}{M_{4}}\right)\frac{\left(\frac{2}{3}\frac{V_{0}}{M_{4}^{2}}+\frac{1}{r_{c}^{2}}+\frac{2}{3}\frac{\beta}{M_{4}^{2}}
\left(\frac{\phi}{M_{4}}\right)^{p}\right)^{1/2}}{\beta p\left(\frac{\phi}{M_{4}}\right)^{p-1}}  = -\frac{V_{0}}{3M_{4}^{2}}\int dt
\label{dgplate}
\end{equation}
The exact solution of the above integral is given in the Appendix for the Randall Sundrum (RSII) model. For the sake of simplicity, we follow the same
procedure as we have done in the previous case i.e. first use the field redefinition $\phi/M_{4} = e^{\lambda}$ and then study two limiting cases.
\\
\\
\underline{\bf $\phi/M_{4}<<1$:}
\\
\\
Once again expanding the exponentials upto linear order we get
\begin{equation}\label{edt}
\int d\lambda \frac{\left(\frac{2}{3}\frac{V_{0}+V_{0}\beta}{M_{4}^{2}}+\frac{1}{r_{c}^{2}}\right)^{1/2}}{\beta p}
\left(1+\frac{\frac{2}{3}\frac{V_{0}\beta}{M_{4}^{2}}p}{2\left(\frac{2}{3}\frac{V_{0}+V_{0}\beta}{M_{4}^{2}}+\frac{1}{r_{c}^{2}}\right)}\lambda\right)= - -\frac{V_{0}}{3M_{4}^{2}}t+A_{1}
\end{equation}
where $A_{1}$ is the arbitrary integration constant. The expression on the left hand side has been
obtained under the underlying assumption that the following condition:
\be \frac{\frac{2}{3}\frac{V_{0}\beta}{M_{4}^{2}}p}{2\left(\frac{2}{3}\frac{V_{0}+V_{0}\beta}{M_{4}^{2}}+\frac{1}{r_{c}^{2}}\right)}\lambda<<1\ee
is satisfied. As we performing the analysis in late time in the present context, hence $1/r_{c}^{2}$ is a large quantity, but to strictly satisfy this condition,
we need to choose $V_{0}$ in such a way that the following condition:
\be \frac{2}{3}\frac{V_{0}}{M_{4}^{2}}+\frac{1}{r_{c}^{2}}+\frac{2}{3}\frac{V_{0}\beta}{M_{4}^{2}}>>1\ee
is always satisfied during the late time.

Further solving the integral as stated in Eq.~(\ref{edt}), we get the following simplified expression for $\lambda$ as:
\begin{equation}
\lambda = \frac{\left(A_{1}-\frac{V_{0}}{3M_{4}^{2}}t\right)}{\left(\frac{2}{3}\frac{V_{0}+V_{0}\beta}{M_{4}^{2}}+\frac{1}{r_{c}^{2}}\right)^{1/2}}\beta p
\label{potential16}
\end{equation}
where $A_{1}$ is given by:
\begin{equation}
A_{1}=\frac{\lambda_{f}\left(\frac{2}{3}\frac{V_{0}+V_{0}\beta}{M_{4}^{2}}+\frac{1}{r_{c}^{2}}\right)^{1/2}}{\beta p}+\frac{V_{0}}{3M_{4}^{2}}t_{f}
\end{equation}
where $\lambda_{f}$ is the value of $\lambda$ at turnaround corresponding to time scale $t=t_{f}$.

In order to get the expression of the scale factor, we substitute this expressions back into the Friedmann equation and integrate the equation to get the following result:
\begin{equation}
a(t)=A'_{1}\exp\left[\left(\left(\frac{2}{3}\frac{V_{0}+V_{0}\beta}{M_{4}^{2}}+\frac{1}{r_{c}^{2}}\right)^{1/2}
+A_{1}\frac{\frac{2}{3}\frac{V_{0}\beta}{M_{4}^{2}}p^2}{4\left(\frac{2}{3}\frac{V_{0}+V_{0}\beta}{M_{4}^{2}}
+\frac{1}{r_{c}^{2}}\right)}\right)t-\frac{\frac{2}{9}\frac{V^2_{0}\beta}{M_{4}^{4}}p^2}{8\left(\frac{2}{3}\frac{V_{0}+V_{0}\beta}{M_{4}^{2}}+\frac{1}{r_{c}^{2}}\right)}t^{2}\right]
\label{scalefactor16}
\end{equation}
where
\begin{equation}
A'_{1}=a_{f}\exp\left[-\left(\left(\frac{2}{3}\frac{V_{0}+V_{0}\beta}{M_{4}^{2}}+\frac{1}{r_{c}^{2}}\right)^{1/2}
+A_{1}\frac{\frac{2}{3}\frac{V_{0}\beta}{M_{4}^{2}}p^2}{4\left(\frac{2}{3}\frac{V_{0}+V_{0}\beta}{M_{4}^{2}}
+\frac{1}{r_{c}^{2}}\right)}
\right)t_{f}+\frac{\frac{2}{9}\frac{V^2_{0}\beta}{M_{4}^{4}}p^2}{8\left(\frac{2}{3}\frac{V_{0}+V_{0}\beta}{M_{4}^{2}}+\frac{1}{r_{c}^{2}}\right)}t_{f}^{2}\right]
\end{equation}
Here $a_{f}$ is the value of the scale factor at the time of turnaround.
\\ \\ \\
\textbf{B. Contraction}
\\ \\
From Eq.~(\ref{modeqn}), we observe that for the contraction phase of the Universe, under the approximation which
we have assumed, the analysis becomes independent of the specific form of the cosmological potential. Hence, the results
for the DGP brane world model hold for any form of the potential.
 \\ \\ \\ 
\underline{\bf i) Early time:}
\\ \\ 
When the scale factor is lying within the window, $a_{min}<a< a'$, following the same analysis as we did for the case of expansion, with the new expression for density, the integral equation is now expressed in the following form:
\begin{equation}
\int\frac{1}{\left(\frac{\dot{\phi}}{\sqrt{6}M_{4}}+\frac{1}{2r_{c}}\right)\dot{\phi}} d\dot{\phi} = -\int dt
\end{equation}
Solving the above integral equation we get the following expression for $\dot{\phi}$ as:
\begin{equation}
\dot{\phi}=\frac{\sqrt{6}M_{4}}{2r_{c}\left[exp\left(\frac{(t-A_{2})}{2r_{c}}\right)-1\right]}
\label{dgpcont}
\end{equation}
where $A_{2}$ is the arbitrary integration constant given by: 
\begin{equation}
A_{2}=t_{i}-2r_{c}\ln\left[\frac{\sqrt{6}M_{4}}{2r_{c}\dot{\phi}_{i}}+1\right].
\end{equation}
Here $\dot{\phi}_{i}$ is the value of the derivative of the scalar field at time of bounce.

Further integrating the above equation once again, we get the expression for the field $\phi$ as:
\begin{equation}\label{eqsol}
\phi(t)=\frac{\sqrt{6}M_{4}}{r_c}\left[\ln\left(1-2e^{-\frac{A'_{2}}{r_{c}}+\frac{t}{r_{c}}}\right)r_{c}-t\right]
\end{equation}
where $A^{'}_{2}$ is the arbitrary integration constant given by: 
\begin{equation}
A'_{2}=t_{i}-r_{c}~\ln\left[\frac{1}{2}\left\{1-\exp\left(\left(\frac{\phi_{i}r_{c}}{\sqrt{6}M_{4}}+t_{i}\right)\frac{1}{r_{c}}\right)\right\}\right]
\end{equation}
where $\phi_{i}$ is the value of the field at the time of bounce $t=t_{i}$. Hence
substituting Eq.~(\ref{eqsol}) in the Friedmann equation, we get the following solution for the scale factor as:
\begin{equation}
a(t)=A''_{2}\left(e^{\frac{t}{2r_{c}}}-e^{\frac{A_{2}}{2r_{c}}}\right)
\label{scalefactor18}
\end{equation}
where $A^{''}_{2}$ is the arbitrary integration constant given by: 
\begin{equation}
A''_{2}=\frac{a_{i}}{\left(e^{\frac{t}{2r_{c}}}-e^{\frac{A_{2}}{2r_{c}}}\right)}.
\end{equation}
\\ \\
\underline{\bf ii) Late time:}
\\
\\
When the scale factor is lying within the window, $a'< a< a_{max}$,
using the Friedmann equation given by Eq.~ (\ref{dgp5}) and the late time condition, we get the following solutions from DGP brane world model:
\begin{eqnarray}
\dot\phi(t) &=& \frac{2}{r_{c}^{2}}exp\left[\frac{3}{r_{c}}(A_{3}-t)\right]\\
\phi(t) &=& -\frac{2}{3r_{c}}exp\left[\frac{3}{r_{c}}(A_{4}-t)\right]
\label{potential19}
\end{eqnarray}
where $A_{3}$ and $A_{4}$ are the arbitrary integration constants given by: 
\begin{eqnarray}
A_{3}&=&\frac{r_{c}}{3}\ln\left[\frac{r_{c}^{2}}{2}\dot{\phi}_{f}\right]+t_{f}\\
A_{4}&=&\frac{r_{c}}{3}\ln\left[-\frac{3r_{c}}{2}\phi_{f}\right]+t_{f}
\end{eqnarray}
Here $\phi_{f}$ and $\dot{\phi}_{f}$ are the values of the scalar field and its derivative at turnaround time scale $t=t_{f}$.

Further substituting the expression for $\dot{\phi}$ in the Friedmann equation, finally we get the expression for the scale factor as:
\begin{equation}
a(t)=A_{5}\exp\left[-\frac{r_{c}}{3}\left\{\sqrt{\frac{1}{r_{c}^{2}}+\frac{4}{3M_{4}^{2}r_{c}^{4}}e^{\frac{6}{r_{c}}(A_{4}-t)}}
-\frac{1}{r_{c}}\tanh^{-1}\left(r_{c}\sqrt{\frac{1}{r_{c}^{2}}+\frac{4}{3M_{4}^{2}r_{c}^{4}}e^{\frac{6}{r_{c}}(A_{4}-t)}}\right)\right\}\right]
\label{scalefactor19}
\end{equation}
where $A_5$ is the arbitrary integration constant given by:
\begin{equation}
A_{5}=a_{f}\exp\left[\frac{r_{c}}{3}\left\{\sqrt{\frac{1}{r_{c}^{2}}+\frac{4}{3M_{4}^{2}r_{c}^{4}}e^{\frac{6}{r_{c}}(A_{4}-t_f)}}
-\frac{1}{r_{c}}\tanh^{-1}\left(r_{c}\sqrt{\frac{1}{r_{c}^{2}}+\frac{4}{3M_{4}^{2}r_{c}^{4}}e^{\frac{6}{r_{c}}(A_{4}-t_f)}}\right)\right\}\right].
\end{equation}
\\ \\
\textbf{C. Expression for work done}
\\ \\
Using the solutions for the scale factor, one can further compute the expression for the integrals as appearing in Eq.~(\ref{dgpwork}), giving
the work done in one cycle for the cases for which we have the final expressions for the scale factor.

For the case $\phi/M_{4}<<1$, using the solutions of the scale factor as computed in the earlier section for hilltop potential,
we find that the expressions for work done for spatially flat case $k=0$ given by the following expression:
\bea
\underbrace{\int_{a_{max}^{i-1}}^{a'^{i-1}} 3M_{4}^{2}\left[\frac{3}{2}a^{2}\dot{a} - \ddot{a}a\dot{a} 
- \frac{\dot{a}^{3}}{2}\right]dt}_{\bf I} &=&0,\\
\underbrace{\int_{a'^{i-1}}^{a_{min}^{i-1}} 3M_{4}^{2}\left[\frac{2\ddot{a}}{a} + \left[\frac{\dot{a}}{a}-\frac{1}{2r_{c}}\right]^{2}\right]a^{2}
\dot{a} dt}_{\bf II} &=& 3M_{4}^{2}a'_{2}\left[-2e^{3a'_{1}t_{min}}+2e^{3a'_{1}t'}-18a_{1}(e^{a'_{1}t_{min}}-e^{a'_{1}t'})\nonumber\right.\\&&\left.~~~~
+9a_{2}(e^{2a'_{1}t_{min}}-e^{2a'_{1}t'})+6a_{3}(t_{min}-t')\right],~~~~~~~~~~~
\\  \underbrace{\int_{a_{min}^{i-1}}^{a'^{i-1}} 3M_{4}^{2}\left[\frac{2\ddot{a}}{a} +
\left[\frac{\dot{a}}{a}-\frac{1}{2r_{c}}\right]^{2}\right]a^{2}\dot{a} dt}_{\bf III}&=& 3M_{4}^{2} a_{4}\left\{
{\rm erf}[a_{5}(2a_{7}t_{min}-a_{6})]\nonumber\right.\\&&\left.~~~~~~-{\rm erf}[a_{5}(2a_{7}t^{'}-a_{6})]\right\}, \\ 
 \underbrace{\int_{a'^{i-1}}^{a_{max}^{i}} 3M_{4}^{2}\left[\frac{3}{2}a^{2}\dot{a} - \ddot{a}a\dot{a} - \frac{\dot{a}^{3}}{2}
 \right]dt}_{\bf IV}&=& 0,
\eea
and consequently the total work done in a one cycle can be expressed as:
\begin{eqnarray}
\oint pdV &=& -3M_{4}^{2}a'_{2}\left[-2e^{3a'_{1}t_{min}}+2e^{3a'_{1}t'}-18a_{1}(e^{a'_{1}t_{min}}-e^{a'_{1}t'})
\nonumber\right.\\&& \left. ~~~+9a_{2}(e^{2a'_{1}t_{min}}-e^{2a'_{1}t'})+6a_{3}(t_{min}-t')\right]\nonumber \\ &&- 
3M_{4}^{2} a_{4}({\rm Erf}[a_{5}(2a_{7}t_{min}-a_{6})]-{\rm Erf}[a_{5}(2a_{7}t^{'}-a_{6})])
\end{eqnarray}
where $a_{1}...a_{7}$ are constants that depends on the model 
parameters present in the expressions for the scale factor whose explicit expressions have been given in the appendix. Here we have quoted the results corresponding
to each integral of the work done as given in Eq.~(\ref{dgpwork}). Thus, we see that we get \be \oint pdV\neq 0,\ee
whose signature depends on the numerical values of the constants. Thus, one can conclude that
the phenomenon of hysteresis is fruitfully achieved for small field hilltop potentials.
\\ \\

 \subsubsection{Case II: Natural potential }
 In case of natural models the potential can be represented by the following functional
form \cite{Freese:2004un}:
\be V(\phi)=V_{0}\left[1+\cos\left(\frac{\phi}{f}\right)\right]\ee
where $V_{0}=M^4$ is the tunable energy scale and $f$ plays the mass scale of the pot. It is important to note that, the 
potential is periodic or equivalently shift symmetric 
for the shift in the field coordinate: \be \phi\rightarrow \phi +2\pi f,\ee
where $2\pi f$ mimics the role of angular shift in the field coordinate. Since we need to substitute the expression
for the scale factor into Eq.~(\ref{dgpwork}), we need to find separate expressions
for the scale factor for both early and late times during expansion and contraction phases. 
\\ \\
\textbf{A. Expansion}
\\ \\
\underline{\bf i) Early time:}
\\ \\
When the scale factor is lying within the window, $a_{min}<a< a'$ or equivalently 
when $\rho r_{c}^{2}/M_{4}^{2}>>1$, we can use the approximate Friedmann equation given by Eq.~(\ref{dgpexp}) with 
spatially flat case $k=0$, where $\rho$ is now given by the potential for natural potential. 
But instead of considering only the zeroth order term (as we had done in order to find the condition for bounce), 
here we will consider upto first order terms to compute the expression for work done.
Then substituting the resulting expression f or Hubble parameter $H$ into Eq.~(\ref{modeqn1}), we get an integral equation of the following form:
\begin{equation}
\int d\left(\frac{\phi}{f}\right)\sqrt{\frac{V_{0}}{3M_{4}^{2}}}\frac{\sqrt{1+\cos\left(\frac{\phi}{f}\right)}}{\sin\left(\frac{\phi}{f}\right)}= \frac{V_{0}}{3f^{2}}\int dt
\label{dgpearly6}
\end{equation}
The exact solutions of the above integrals are given in the Appendix for Radall Sundrum single brane world model (RSII).
Simplified analytical expressions could be obtained only for the small field case ($\phi/f<<1$) which has been discussed below. 

 For $\phi/f<<1$ case we take small argument approximations of the trigonometric functions after which we get the following 
 solution for the sclar field $\phi$ as:
\begin{equation}
\frac{\phi(t)}{f}=A_{6}exp\left[\frac{V_{0}}{3\sqrt{\frac{V_{0}}{3M_{4}^{2}}}\sqrt{2}f^{2}}t\right]
\label{potential20}
\end{equation}
where $A_{6}$ is the arbitrary integration constant is given by:
\begin{equation}
A_{6}=\frac{\phi_{i}}{f}\exp\left[-\frac{V_{0}}{3\sqrt{\frac{V_{0}}{3M_{4}^{2}}}\sqrt{2}f^{2}}t_{i}\right]
\end{equation}
Here $\phi_{i}$ is the value of $\phi$ at the initial time of bouncing time scale $t=t_{i}$. 

Further substituting the expression for $\phi$ back into the Friedmann equation, we get the following simplified expression for scale factor as:
\begin{equation}
a(t)=A_{7}\exp\left[\left(\sqrt{\frac{V_{0}}{3M_{4}^{2}}}\sqrt{2}+\frac{1}{2r_{c}}\right)t\right]
\label{dgpearly7}
\end{equation}
where $A_{7}$ is the arbitrary integration constant is given by:
\begin{equation}
A_{7}=a_{i}\exp\left[-\left(\sqrt{\frac{V_{0}}{3M_{4}^{2}}}\sqrt{2}+\frac{1}{2r_{c}}\right)t_{i}\right].
\label{dgpearly8}
\end{equation}
\\ \\ \\ \\
\underline{\bf ii) Late time:}\\ \\
When the scale factor is lying within the window, $a'<a< a_{max}$ or equivalently
when $\rho r_{c}^{2}/M_{4}^{2}<<1$, we can use the approximated form of the
Friedmann equation given by Eq.~ (\ref{dgp5}) with spatially flat case $k=0$. Hence
substituting the resulting expression for the Hubble parameter $H$ into Eq.~(\ref{modeqn1}), we get the following integral equation as given by:
\begin{equation}
\int d\left(\frac{\phi}{f}\right) \frac{\sqrt{\frac{2V_{0}}{3M_{4}^{2}}+\frac{1}{r_{c}^{2}}+\frac{2V_{0}}{3M_{4}^{2}}\cos\left(\frac{\phi}{f}\right)}}{\sin\left(\frac{\phi}{f}
\right)} = \frac{V_{0}}{3f^{2}}\int dt
\label{dgplate1}
\end{equation}
The exact solution of the above integral is given in the Appendix for Radall Sundrum single brane world model (RSII). Therefore,
in order to simplify the analysis, we again study small field limiting case as mentioned earlier.
For $\phi/f<<1$ case we take small argument approximations of the trigonometric functions and finally we get the following simplified 
expression for the scalar field $\phi$ as:
\begin{equation}
\frac{\phi(t)}{f} = A_{8}exp\left[\frac{1}{\sqrt{\frac{2V_{0}}{3M_{4}^{2}}+\frac{1}{r_{c}^{2}}+\frac{2V_{0}}{3M_{4}^{2}}}}\frac{V_{0}}{3f^{2}}t\right]
\label{potential21}
\end{equation}
where $A_{8}$ is the arbitrary integration constant is given by:
\begin{equation}
A_{8} = \frac{\phi_{F}}{f}exp\left[-\frac{1}{\sqrt{\frac{2V_{0}}{3M_{4}^{2}}+\frac{1}{r_{c}^{2}}
+\frac{2V_{0}}{3M_{4}^{2}}}}\frac{V_{0}}{3f^{2}}t_{F}\right].
\end{equation}
Here $\phi_{F}$ is the value of the scalar field at turnaround time scale $t=t_{F}$.

In order to get the expression of the scale factor, we substitute this expression back into
the Friedmann equation and integrate the equation to get following simplified expression for the scale factor as:
\begin{equation}
a(t)=A_{9}exp\left[\sqrt{\left(\frac{2V_{0}}{3M_{4}^{2}}+\frac{1}{r_{c}^{2}}+\frac{2V_{0}}{3M_{4}^{2}}\right)}t\right]
\label{scalefactor21}
\end{equation}
where $A_{9}$ is the arbitrary integration constant is given by:
\begin{equation}
A_{9}=a_{F}exp\left[-\sqrt{\left(\frac{2V_{0}}{3M_{4}^{2}}+\frac{1}{r_{c}^{2}}+\frac{2V_{0}}{3M_{4}^{2}}\right)}t_{F}\right]
\end{equation}
where $a_{F}$ is the value of the scale factor at turnaround time scale $t=t_{F}$.
\\ \\
\textbf{B. Contraction}
\\ \\
As has already been mentioned while performing the analysis for expansion, the conclusions for contraction phase is independent of
any potential, hence the analysis remains same for natural potential also.
\\ \\ \\
\textbf{C. Expression for work done}
\\ \\
Here we also study the work done for all the cases for which we have expressions for the scale factor for all the integrals appearing 
in Eq.~ (\ref{dgpwork}). We consider the case $\phi/f<<1$ for which the expression for work done for spatially flat case $k=0$ is given by:
\bea
\underbrace{\int_{a_{max}^{i-1}}^{a'^{i-1}} 3M_{4}^{2}\left[\frac{3}{2}a^{2}\dot{a} - \ddot{a}a\dot{a} 
- \frac{\dot{a}^{3}}{2}\right]dt}_{\bf I} &=&0,\\
\underbrace{\int_{a'^{i-1}}^{a_{min}^{i-1}} 3M_{4}^{2}\left[\frac{2\ddot{a}}{a} + \left[\frac{\dot{a}}{a}-\frac{1}{2r_{c}}\right]^{2}\right]a^{2}
\dot{a} dt}_{\bf II} &=& 3M_{4}^{2}a'_{2}\left[2e^{3a'_{1}t'}-2e^{3a'_{1}t_{min}}-18a_{1}(e^{a'_{1}t_{min}}-e^{a'_{1}t'})\nonumber\right.\\&& \left. 
+9a_{2}(e^{2a'_{1}t_{min}}-e^{2a'_{1}t'})+6a_{3}(t_{min}-t')\right],~~~~~~~~~~~
\\  \underbrace{\int_{a_{min}^{i-1}}^{a'^{i-1}} 3M_{4}^{2}\left[\frac{2\ddot{a}}{a} +
\left[\frac{\dot{a}}{a}-\frac{1}{2r_{c}}\right]^{2}\right]a^{2}\dot{a} dt}_{\bf III}&=& 3M_{4}^{2}b_{1}\left[e^{3b_{2}t'}-e^{3b_{2}t_{min}}\right], \\ 
 \underbrace{\int_{a'^{i-1}}^{a_{max}^{i}} 3M_{4}^{2}\left[\frac{3}{2}a^{2}\dot{a} - \ddot{a}a\dot{a} - \frac{\dot{a}^{3}}{2}
 \right]dt}_{\bf IV}&=& 0,
\eea
and consequently the total work done in a one cycle can be expressed as:
\begin{eqnarray}
\oint pdV &=& 0-3M_{4}^{2}a'_{2}\left[-2e^{3a'_{1}t_{min}}+2e^{3a'_{1}t'}-18a_{1}(e^{a'_{1}t_{min}}-e^{a'_{1}t'})
\nonumber\right.\\&& \left. ~~~+9a_{2}(e^{2a'_{1}t_{min}}-e^{2a'_{1}t'})+6a_{3}(t_{min}-t')\right]-3M_{4}^{2}b_{1}(e^{3b_{2}t'}-e^{3b_{2}t_{min}})+0\nonumber \\
\end{eqnarray}

Here $a_{1}..b_{2}$ are all constants dependent on the model 
parameters that appearing in the expressions for the scale factor. Their explicit forms are in the appendix. 
The results obtained in this section implies that, we get \be \oint pdV\neq 0\ee
for small field limit $\phi/f<<1$ and the signature of the integral is completely determined by 
the numerical values of the model parameters and the arbitrary integration constants. Most importantly,
the analysis shows that the phenomenon of hysteresis is fruitfully achieved for natural potential.
\\ \\
\subsubsection{Case III: Coleman-Weinberg potential}
Let us start with a theory of five dimensional ${\cal N}=2$ bulk supergravity in which by compactifying the extra fifth dimension 
it is possible to derive an four dimensional effective theory described by ${\cal N}=1$ supergravity theory in brane world. Within this prescription, 
 the four dimensional one-loop effective Coleman-Wienberg potential embedded in the brane world can be expressed as \cite{Choudhury:2011sq,Choudhury:2011rz,Choudhury:2012ib}:
\be V(\phi)=V_{0}\left[1+\left\{\alpha+\beta\ln\left(\frac{\phi}{M_{4}}\right)\right\}\left(\frac{\phi}{M_{4}}\right)^{4}\right]\ee
where $V_{0}$ sets the energy scale of supergravity theory. Additionally the model parameter 
$\alpha$ signifies the tree level effect and the parameter $\beta$ characterizes the effect of one-loop correction to the leading order result.
Here $M_{4}$ represents the background mass-scale of theory. For sake of simplicity one can consider $M_{4}$ to be the UV cut-off i.e. the Planck 
scale of the gravity theory.  
\\ \\
\textbf{A. Expansion}
\\ \\
\underline{\bf i) Early time:}
\\
\\
When the scale factor is lying within the window, $a_{min}<a<a'$ or equivalently
when $\rho r_{c}^{2}/M_{4}^{2}>>1$, we can use the approximated form of Friedmann equation given by
Eq.~ (\ref{dgpexp}) with spatially flat case $k=0$, where in the present context $\rho$ is given by the supergravity motivated 
potential. Instead of considering only the zeroth order term,
here we will consider upto first order terms. Then substituting the resulting expression for the Hubble parameter $H$
into Eq.~(\ref{modeqn1}), we get an integral equation of the following form:
\begin{equation}
\int d\left(\frac{\phi}{M_{4}}\right)\frac{\sqrt{\frac{V_{0}}{3M_{4}^{2}}\left[1+\left\{\alpha+\beta\ln
\left(\frac{\phi}{M_{4}}\right)\right\}\left(\frac{\phi}{M_{4}}\right)^{4}\right]}+\frac{1}{2r_{c}}}{\left(\frac{\phi}{M_{4}}\right)^{3}
\left[4\left\{\alpha+\beta\ln
\left(\frac{\phi}{M_{4}}\right)\right\}+\beta\right]}\   = -\frac{V_{0}}{3M_{4}^{2}}\int dt
\label{dgpearly10}
\end{equation}
For the sake of simplicity let us consider the following transformation or redefinition in the field:
\be\frac{\phi}{M_{4}}=e^{\lambda},\ee
as we had done for the previous cases, where now we will solve for the redefined field $\lambda$. In order to
further simplify the expressions, we solve for two limiting cases:
\\ \\
\underline{\bf a) $\phi/M_{4}<<1$:}\\ \\
For this case we can expand the exponentials upto linear order after which the integral on the left hand side of Eq.~ (\ref{dgpearly10}) becomes
\begin{equation}
\int d\lambda\frac{\left[\frac{V_{0}}{3M_{4}^{2}}\left(1+\left(\alpha+\beta\lambda\right)e^{4\lambda}\right)\right]^{1/2}+\frac{1}{2r_{c}}}{e^{3\lambda}
\left[4\left(\alpha+\beta\lambda\right)+\beta\right]}e^{\lambda} \   = -\frac{V_{0}}{3M_{4}^{2}}\int dt
\end{equation}
Next we compute the above integral equation on the left hand side under the assumption that the values of 
model parameters, $\alpha$ and $\beta$ satisfy the constraint:
\bea \alpha+4\alpha\lambda+\beta\lambda<<1,\\
\frac{4\beta\lambda}{(4\alpha+\beta)}<<1.\eea Consequently
we get the expression for $\lambda$, hence $\phi/M_{4} = e^{\lambda}$ as: 
\begin{equation}
\lambda=\frac{\left(A_{10}-\left(\frac{V_{0}}{3M_{4}^{2}}\right)t\right)}{\left[
\frac{\left(\frac{V_{0}}{3M_{4}^{2}}\right)^{1/2}\left(1+\frac{\alpha}{2}\right)+\frac{1}{2r_{c}}}{(4\alpha+\beta)}
\right]}
\label{potential22}
\end{equation}
where $A_{10}$ is the arbitrary integration constant given by:
\begin{equation}
A_{10}=\frac{\left(\frac{V_{0}}{3M_{4}^{2}}\right)^{1/2}\left(1+\frac{\alpha}{2}\right)+\frac{1}{2r_{c}}}{(4\alpha+\beta)}\lambda_{i}+\left(\frac{V_{0}}{3M_{4}^{2}}\right)t_{i}
\end{equation}
Here $\lambda_{i}$ is the value of $\lambda$ at the initial time $t=t_{i}$ of bounce. 

Substituting the expression for $\lambda$ back into the Friedmann equation, we get the expression for scale factor as:
\begin{equation}
a(t)=A_{11}exp\left[\left(\frac{V_{0}}{3M_{4}^{2}}\right)^{1/2}\left(1+\frac{\alpha}{2}\right)t+\frac{1}{2r_{c}}t\right]
\label{dgpearly11}
\end{equation}
where $A_{11}$ is the arbitrary integration constant given by:
\begin{equation}
A_{11}=a_{i}exp\left[-\left(\frac{V_{0}}{3M_{4}^{2}}\right)^{1/2}\left(1+\frac{\alpha}{2}\right)t_{i}-\frac{1}{2r_{c}}t_{i}\right]
\label{dgpearly11}
\end{equation}
Here $a_{i}$ is the value of the scale factor at the time of bounce.
\\ \\
\underline{\bf b) $\phi/M_{4}>>1$:}\\ \\ For this case Eq.~ (\ref{dgpearly10}) simplifies to
\begin{equation}
\int \left(\frac{V_{0}}{3M_{4}^{2}}\right)^{1/2}\beta^{1/2}\lambda^{-1/2}  d\lambda = -\frac{V_{0}}{3M_{4}^{2}}t + A_{12}
\end{equation}
The above expression is obtained provided that the following constraints are satisfied:
\bea \frac{\beta\lambda}{(1+\alpha)}>>1, \\
\frac{4\beta\lambda}{(4\alpha+\beta)}>>1.\eea We have also used the condition
$\rho r_{c}^{2}/M_{4}^{2}>>1$, which is true in the case of early universe.
Here $A_{12}$ is an arbitrary integration constant.

Solving the above integral equation we get the following expression for $\lambda$ as:
\begin{equation}
\lambda=\left[\left(-\frac{V_{0}}{3M_{4}^{2}}t+A_{12}\right)\frac{1}{2\left(\frac{V_{0}}{3M_{4}^{2}}\right)^{1/2}\beta^{1/2}}\right]^{2}
\end{equation}
where the explicit form of $A_{12}$ is given by:
\begin{equation}
A_{12}=\frac{V_{0}t_{i}}{3M_{4}^{2}}\pm 2\left(\frac{V_{0}}{3M_{4}^{2}}\right)^{1/2}\beta^{1/2}\lambda_{i}
\end{equation}
Here $\lambda_{i}$ is the value at the time of bounce $t=t_{i}$.

Further substituting back the expression for $\lambda$ into the Friedmann equation, we get the following expression for the scale factor as:

\begin{equation}
a(t)=A_{13}\exp\left[-\left(\frac{V_{0}}{3M_{4}^{2}}\right)^{1/2}\beta^{1/2}e^{\frac{2\left(A_{12}-\frac{V_{0}}{3M_{4}^{2}}t\right)^{2}}{\left(2\left(\frac{V_{0}}{3M_{4}^{2}}\right)^{1/2}\beta^{1/2}\right)^{2}}}\frac{\left(A_{12}-\frac{V_{0}}{3M_{4}^{2}}t\right)}{4\frac{V_{0}}{3M_{4}^{2}}\sqrt{\frac{\left(A_{12}-\frac{V_{0}}{3M_{4}^{2}}t\right)^{2}}{\left(2\left(\frac{V_{0}}{3M_{4}^{2}}\right)^{1/2}\beta^{1/2}\right)^{2}}}}\right]
\end{equation}
where $A_{13}$ is the arbitrary integration constant given by:
\begin{equation}
A_{13}=a_{i}\exp\left[\left(\frac{V_{0}}{3M_{4}^{2}}\right)^{1/2}\beta^{1/2}e^{\frac{2\left(A_{12}
-\frac{V_{0}}{3M_{4}^{2}}t_{i}\right)^{2}}{\left(2\left(\frac{V_{0}}{3M_{4}^{2}}\right)^{1/2}
\beta^{1/2}\right)^{2}}}\frac{\left(A_{12}-\frac{V_{0}}{3M_{4}^{2}}t_{i}\right)}{4
\frac{V_{0}}{3M_{4}^{2}}\sqrt{\frac{\left(A_{12}-\frac{V_{0}}{3M_{4}^{2}}t_{i}\right)^{2}}{\left(2\left(\frac{V_{0}}{3M_{4}^{2}}
\right)^{1/2}\beta^{1/2}\right)^{2}}}}\right].
\end{equation} 
\\ \\
\underline{\bf ii) Late time:}
\\ \\ 
When the scale factor is lying within the window, $a'<a< a_{max}$ or equivalently
when $\rho r_{c}^{2}/M_{4}^{2}<<1$, we can use the approximated version of the
Friedmann equation given by Eq.~ (\ref{dgp5}) with spatially flat case $k=0$. 
Hence substituting the resulting expression for the Hubble parameter $H$ into Eq.~(\ref{modeqn1}), we get the following integral equation given by:

\begin{equation}
\int d\left(\frac{\phi}{M_{4}}\right)\frac{\sqrt{\left[\frac{2}{3}\frac{V_{0}}{M_{4}^{2}}\left(1+\left\{\alpha+\beta\ln\left(\frac{\phi}{M_{4}}\right)\right\}\left(\frac{\phi}{M_{4}}
\right)^{4}\right)+\frac{1}{r_{c}^{2}}\right]}}{\left(\frac{\phi}{M_{4}}
\right)^{3}\left[4\alpha+\beta+4\beta\ln\left(\frac{\phi}{M_{4}}\right)\right]} = -\frac{V_{0}}{3M_{4}^{2}}\int dt
\label{dgpearly12}
\end{equation}
For the sake of simplicity during the analysis, we follow the same procedure as mentioned 
before and use the field redefinition: \be \frac{\phi}{M_{4}} = e^{\lambda}\ee and then study two limiting physical situations.
\\ \\ \\ \\
\underline{\bf a) $\phi/M_{4}<<1$:}\\ \\
For this small field limit expanding the exponentials upto linear order we get the following integral equation:

\begin{equation}
\int \frac{1}{r_{c}(4\alpha+\beta)}\left[1+\frac{1}{3}\frac{V_{0}r_{c}^{2}}{M_{4}^{2}}(1+\alpha+4\alpha\lambda+\beta\lambda)\right](1-2\lambda)\left(1-\frac{4\beta}{4\alpha+\beta}\lambda\right) d\lambda= -\int \frac{V_{0}}{3M_{4}^{2}}dt
\end{equation}
where the integral in the left hand side has been obtained under assumption that:
\be \frac{1}{3}\frac{V_{0}r_{c}^{2}}{M_{4}^{2}}(1+\alpha+4\alpha\lambda+\beta\lambda)<<1,\ee
which is a valid assumption since we are in the late universe and in the limit of small $\lambda$.

Solving the above integral, we get the expression for $\lambda$ as
\begin{equation}
\lambda = \frac{\left(-\frac{V_{0}}{3M_{4}^{2}}t+A_{14}\right)}{\left[\frac{1}{r_{c}(4\alpha+\beta)}
+\frac{V_{0}r_{c}}{3M_{4}^{2}(4\alpha+\beta)}+\frac{V_{0}\alpha r_{c}}{3M_{4}^{2}(4\alpha+\beta)}\right]}
\label{potential24}
\end{equation}
where $A_{14}$ is the arbitrary integration constant given by:
\begin{equation}
A_{14}=\lambda_{f}\left(\frac{1}{r_{c}(4\alpha+\beta)}+\frac{V_{0}r_{c}}{3M_{4}^{2}(4\alpha+\beta)}+\frac{V_{0}\alpha r_{c}}{3M_{4}^{2}(4\alpha+\beta)}\right)+\frac{V_{0}}{3M_{4}^{2}}t_{f}
\end{equation}
where $\lambda_{f}$ is the value at turnaround corresponding to time $t=t_{f}$.

In order to get the expression of the scale factor, we substitute this
expression back into the Friedmann equation and integrate the equation to get:
\begin{eqnarray}
a(t) &=& A_{15}\exp\left[\left(\frac{1}{r_{c}}+\frac{V_{0}r_{c}}{3M_{4}^{2}}(1+\alpha)+\frac{\frac{4V_{0}r_{c}}{3M_{4}^{2}}(4\alpha+\beta)A_{14}}{\frac{1}{r_{c}(4\alpha+\beta)}+\frac{V_{0}r_{c}}{3M_{4}^{2}(4\alpha+\beta)}+\frac{V_{0}\alpha r_{c}}{3M_{4}^{2}(4\alpha+\beta)}}\right)t\right] \nonumber \\
 && ~~~~~~~~~~~~~~~~~~\times \exp\left[-\frac{V_{0}}{3M_{4}^{2}}\frac{\frac{4V_{0}r_{c}}{3M_{4}^{2}}(4\alpha+\beta)}{\frac{1}{r_{c}(4\alpha+\beta)}+\frac{V_{0}r_{c}}{3M_{4}^{2}(4\alpha+\beta)}+\frac{V_{0}\alpha r_{c}}{3M_{4}^{2}(4\alpha+\beta)}}\frac{t^{2}}{2}\right]
 \label{scalefactor24}
\end{eqnarray}
where $A_{15}$ is the arbitrary integration constant given by:
\begin{eqnarray}
A_{15} &=& a_{f}\exp\left[-\left(\frac{1}{r_{c}}+\frac{V_{0}r_{c}}{3M_{4}^{2}}(1+\alpha)+\frac{\frac{4V_{0}r_{c}}{3M_{4}^{2}}(4\alpha+\beta)A_{14}}{\frac{1}{r_{c}(4\alpha+\beta)}+\frac{V_{0}r_{c}}{3M_{4}^{2}(4\alpha+\beta)}+\frac{V_{0}\alpha r_{c}}{3M_{4}^{2}(4\alpha+\beta)}}\right)t_{f}\right] \nonumber \\
 && ~~~~~~~~~~~~~~~~~\times \exp\left[\left(\frac{V_{0}}{3M_{4}^{2}}\frac{\frac{4V_{0}r_{c}}{
 3M_{4}^{2}}(4\alpha+\beta)}{\frac{1}{r_{c}(4\alpha+\beta)}+\frac{V_{0}r_{c}}{3M_{4}^{2}(4\alpha+\beta)}+\frac{V_{0}
 \alpha r_{c}}{3M_{4}^{2}(4\alpha+\beta)}}\frac{t_{f}^{2}}{2}\right)\right].
\end{eqnarray}
\\ \\
\underline{\bf b) $\phi>>M_{4}$:}\\ \\ For this case the integral in Eq.~ (\ref{dgpearly12}), writing $\phi/M_{4}=e^{\lambda}$, simplifies to:
\begin{equation}
\int \frac{\sqrt{\left(\frac{2}{3}\frac{V_{0}}{M_{4}^{2}}\beta\lambda e^{4\lambda}+\frac{1}{r_{c}^{2}}\right)}}{4\beta\lambda}e^{-2\lambda} d\lambda = -\frac{V_{0}}{3M_{4}^{2}}\int dt
\end{equation}
But in order to get an solution for $\lambda$, we need to further simplify the integral,
which is possible if we assume that the values of the parameters of the model satisfy
the condition: \be \frac{V_{0}\beta\lambda e^{4\lambda}}{M_{4}^{2}}>>\frac{1}{r_{c}^{2}}\ee is satisfied. This is possible because
though we are in the late time i.e. \be \frac{\rho}{M_{4}^{2}}=\frac{V_{0}}{M_{4}^{2}}+\frac{V_{0}\alpha e^{4\lambda}}{M_{4}^{2}}+\frac{V_{0}\beta\lambda e^{4\lambda}}{M_{4}^{2}}<<\frac{1}{r_{c}^{2}},\ee  
we have both the above conditions being satisfied by $1/r_{c}^{2}$ simultaneously. Under this assumption, we get the solution for $\lambda$ as:
\begin{equation}
\lambda=\left(-\frac{V_{0}}{3M_{4}^{2}}t+A_{16}\right)^{2}\frac{(4\beta)^{2}}{\frac{2}{3}\frac{V_{0}}{M_{4}^{2}}\beta}
\end{equation}
where $A_{16}$ is the arbitrary integration constant given by:
\begin{equation}
A_{16}=\frac{V_{0}}{3M_{4}^{2}}t_{f}\pm \frac{\left(\frac{2}{3}\frac{V_{0}}{M_{4}^{2}}\beta\right)^{1/2}\lambda_{f}^{1/2}}{4\beta}
\end{equation}
Substituting the expression for $\lambda$ back into the Friedmann equation, we get the expression for the scale factor as:
\begin{equation}
a(t)=A_{17}\exp\left[-\frac{3e^{\left(\frac{32\beta^{2}\left(A_{16}-\frac{tV_{0}}{3M_{4}^{2}}\right)^{2}}{\frac{2}{3}\frac{V_{0}}{M_{4}^{2}}\beta}\right)}\left(tV_{0}-3A_{16}M_{4}^{2}\right)}{16\frac{V_{0}}{M_{4}}\sqrt{\frac{\beta^{2}(-3A_{16}M_{4}^{2}+tV_{0})^{2}}{\frac{2}{3}\frac{V_{0}}{M_{4}^{2}}\beta M_{4}^{4}}}}\right]
\end{equation}
where $A_{17}$ is the arbitrary integration consant given by:
\begin{equation}
A_{17}=a_{f}\exp\left[\frac{3e^{\left(\frac{32\beta^{2}\left(A_{16}-\frac{t_{f}V_{0}}{3M_{4}^{2}}\right)^{2}}{
\frac{2}{3}\frac{V_{0}}{M_{4}^{2}}\beta}\right)}\left(t_{f}V_{0}-3A_{16}M_{4}^{2}\right)}{
16\frac{V_{0}}{M_{4}}\sqrt{\frac{\beta^{2}(-3A_{16}M_{4}^{2}+t_{f}V_{0})^{2}}{\frac{2}{3}\frac{V_{0}}{M_{4}^{2}}\beta M_{4}^{4}}}}\right].
\end{equation}
Here, $a_{f}$ is the value of the scale factor at turnaround $t=t_{f}$.
\\ \\
\textbf{B. Contraction}
\\ \\
As has already been mentioned that while performing the analysis for expansion, the conclusions for
contraction phase is independent of any potential, hence rest of the analysis remains same as mentioned earlier.
\\ \\
\textbf{C. Expression for work done}
\\ \\
The expression for work done in this case is same as we get for the case of natural potential
with the expressions for constants now given by the parameters of this supergravity motivated 
model. Hence we see that in this case also, applying certain approximations and physical limits, 
we can get an analytical expression for work done which is non zero, thus leading to the phenomenon of cosmological hysteresis in the present context.
\\ \\
\subsection{Graphical Analysis}
\subsubsection{Case I: Hilltop potential}

\underline{\textbf{{Graphical Analysis:}}}
\\ \\
All the graphs in this section and in the following sections have been plotted in units of $M_{p}=1,\ H_{0}=1,\ c=1$, where $M_{p}$ is the Planck mass, $H_{0}$ is the present value of the Hubble parameter and $c$ is the speed of light. Throughout the analysis $r_{c}$ will be expressed in units of $H_{0}^{-1}$, hence only its magnitude will be written explicitly
\\ \\
\begin{figure*}[htb]
\centering
\subfigure[ An illustration of the behavior of the scale factor with time during the early  expansion phase for $\phi<<M_{p}$ with $V_{0}=10^{-8}M_{p}^{4},\ p=2,\ \beta=0.001,\ \lambda_{i}=10^{-8},\ t_{i}=0.1,\ r_{c}=1$.]{
    \includegraphics[width=7.2cm,height=7.5cm] {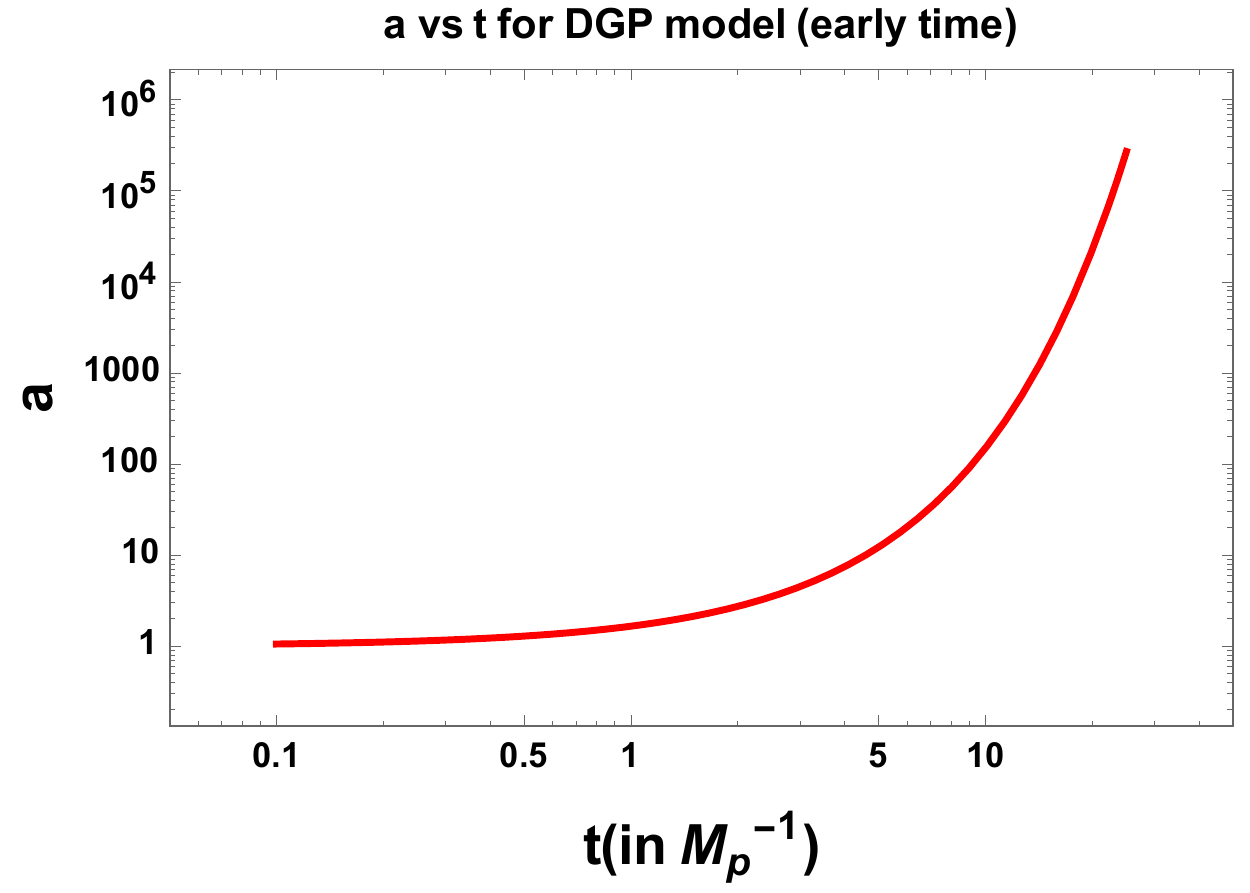}
    \label{dgp9}
}
\subfigure[An illustration of the behavior of the potential during early expansion phase for $\phi<<M_{p}$ with $V_{0}=4.1{\rm x}10^{-3}M_{p}^{4},\ \beta=0.72,\ p=2,\ \lambda_{i}=6.6,\ t_{i}=95M_{pl}^{-1},\ a_{i}=1,\ r_{c}=1$ .]{
    \includegraphics[width=7.2cm,height=7.5cm] {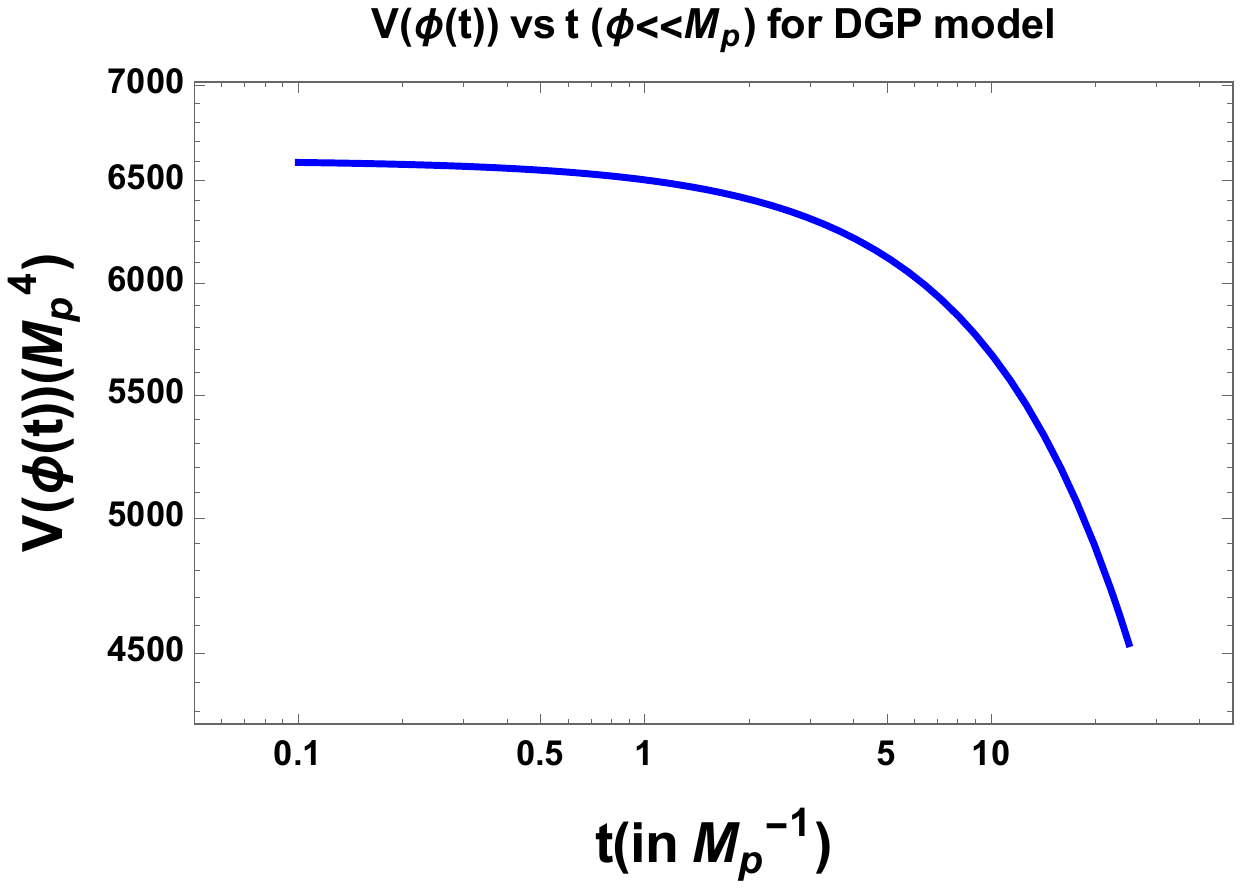}
    \label{dgp10}
}
\subfigure[An illustration of the behavior of the scale factor with time during late time expansion phase for $\phi<<M_{p}$ with $V_{0}=10^{-8}M_{p}^{4},\ p=3,\ \beta=0.001,\ A_{1}=10^{-8}M_{p},\ A_{1}'=10^{-8},\ r_{c}=1.0$.]{
    \includegraphics[width=7.2cm,height=7.5cm] {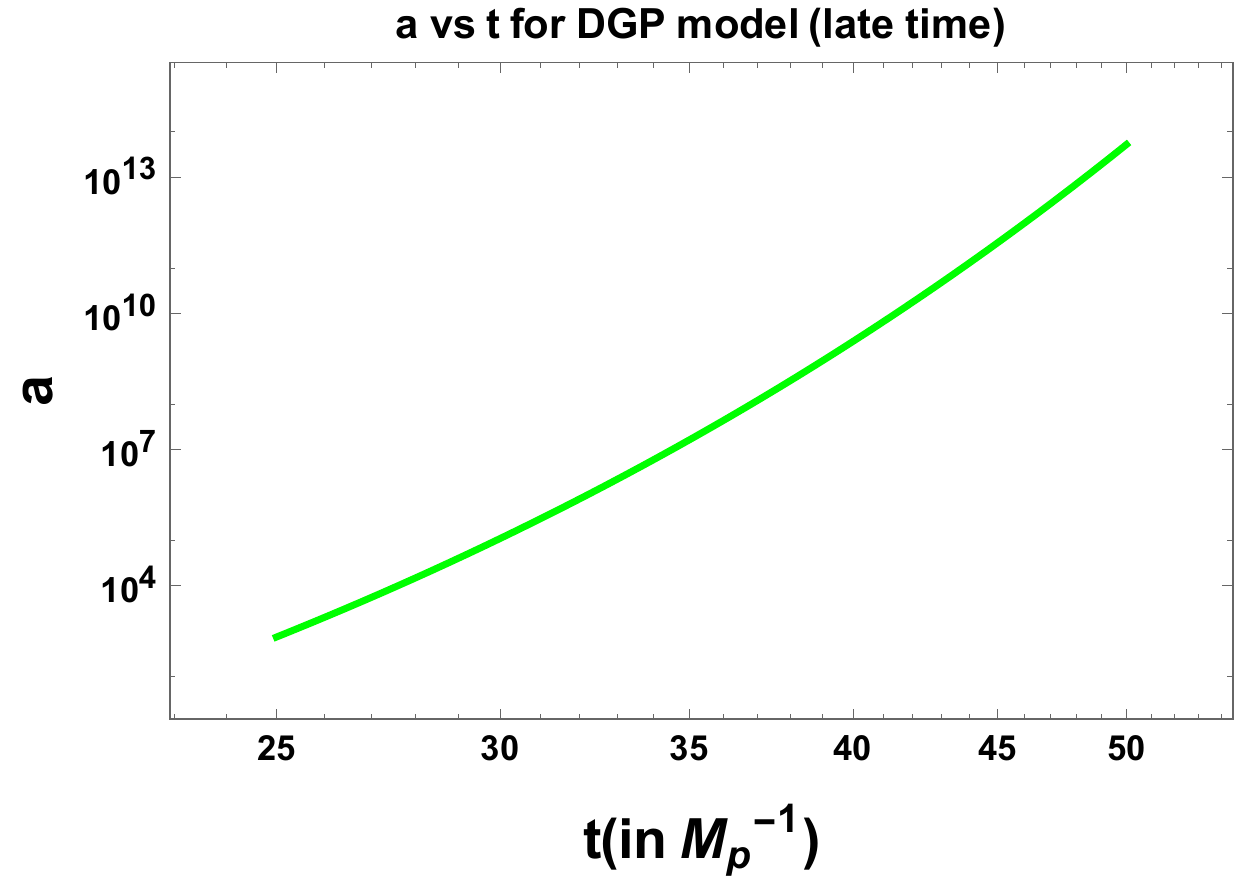}
    \label{dgp11}
}
\subfigure[An illustration of the behavior of the potential during late time expansion phase for $\phi<<M_{p}$ with $V_{0}=3.7{\rm x}10^{-3}M_{p}^{4},\ p=3,\ \beta=0.14,\ A_{1}=20.8M_{p},\ r_{c}=1.0$ .]{
    \includegraphics[width=7.2cm,height=7.5cm] {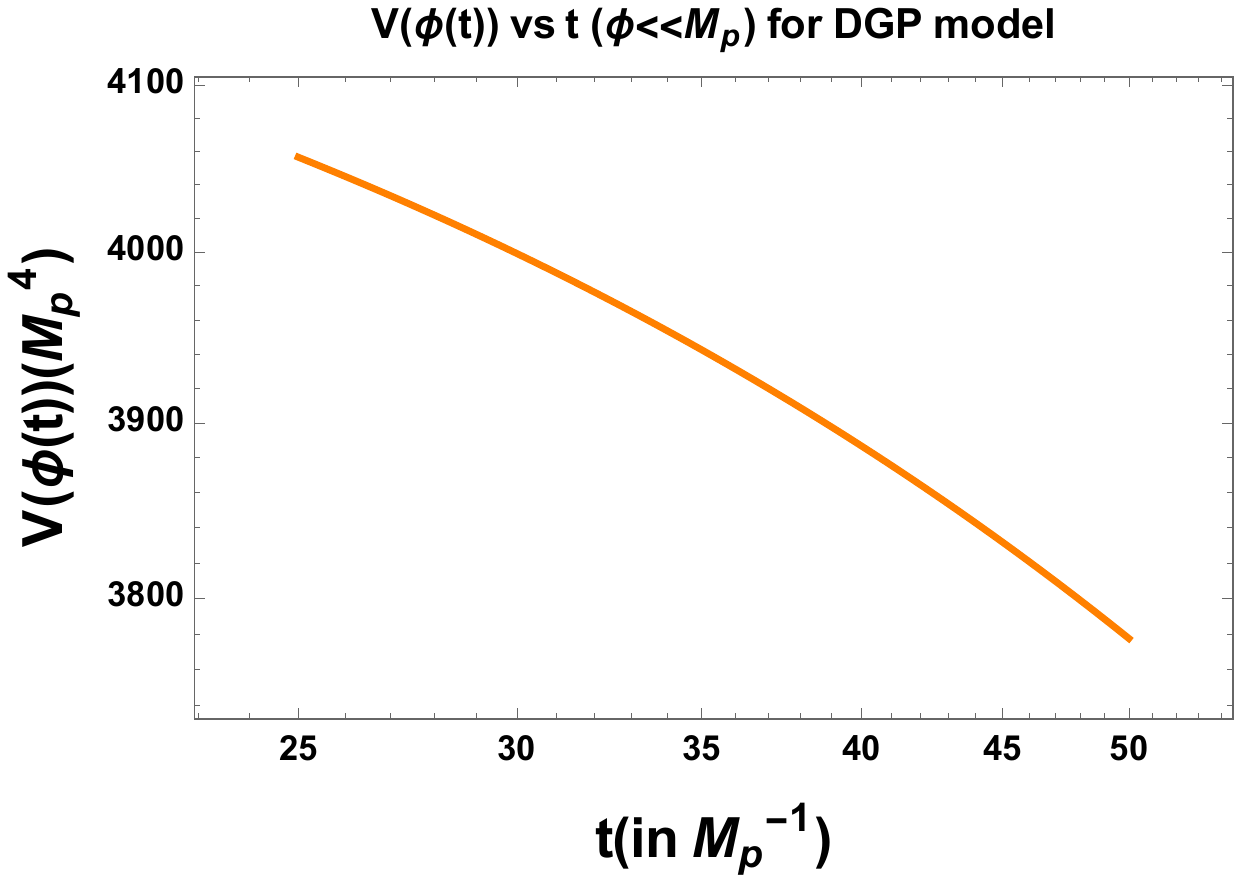}
    \label{dgp12}
}
\caption[Optional caption for list of figures]{ Graphical representation of the evolution of the scale factor and the potential during the expansion phase for DGP model.} 
\label{fig18}
\end{figure*}
In Fig. \ref{fig18} we have shown the evolution of the scale factor and the potential for early and late time expansion phase. From the above plots, we can draw the following conclusions:
\begin{itemize}
\item Fig. \ref{dgp9}, shows the plot of the scale factor in the small field limit for hilltop potential given by Eqn. (\ref{dgpearly2}) with the parameter values $V_{0}=10^{-8}M_{p}^{4},\ p=2,\ \beta=0.001,\ \lambda_{i}=10^{-8},\ t_{i}=0.1,\ a_{i}=1,\ r_{c}=1$. 
\item Higher values of the integration constants only results in an increase in the amplitude of the scale factor, not affecting the nature of the graph.
\item From Eqn. (\ref{dgpearly2}), we can conclude that solution will only be possible for $\beta\geq 0$. Changing the value of $\beta$ mildly modifies the amplitude of expansion, but the nature of the graph remains same. The plot is almost independent of any variation in $p$. Larger values of $r_{c}$ increases the amplitude of the graph. 
\item Fig. \ref{dgp10} shows the plot of the behavior of the potential with time for small field hilltop potential. This graph has been obtained with the help of Eqn. (\ref{potential15}) with parameter values $V_{0}=4.1{\rm x}10^{-3}M_{p}^{4},\ \beta=0.72,\ p=2,\ \lambda_{i}=6.6,\ t_{i}=95M_{p}^{-1},\ r_{c}=1$. The evolution of the potential is not much affected by the value of $p$. Large values of $r_{c}$  increases the height of the potential and makes the potential fall less steeply.
\item Fig. \ref{dgp11} shows the plot of the scale factor in the late time expansion phase for hilltop potential given by Eqn. (\ref{scalefactor16}). This plot has been obtained for $V_{0}=10^{-8}M_{p}^{4},\ p=3,\ \beta=0.001,\ A_{1}=10^{-8}M_{p},\ A_{1}'=10^{-8},\ r_{c}=1.0$. Detail graphical analysis have shown that the expansion of the universe is almost independent of the values that $\beta,\ p,\ A_{1}$ takes i.e in this case both negative and positive values of $\beta$ can give rise to an expanding universe. We do not get the required expansion for a value of $r_{c}<0.04$.
\item Fig. \ref{dgp12} shows the evolution of the potential in the late time expansion phase for hilltop potential. Fig. \ref{dgp12} has been obtained by with the help of Eqn. (\ref{potential16} with the parameter values $V_{0}=3.7{\rm x}10^{-3}M_{p}^{4},\ p=3,\ \beta=0.14,\ A_{1}=20.8M_{p},\ r_{c}=1.0$. Detail graphical analysis show that the larger the value of $\beta$, smaller is the the range of $r_{c}$ for which we get potential giving rise to expansion. The graph is almost independent of the values that $p,\ A_{1}$ takes. 
\item If we compare Fig. \ref{dgp10} and \ref{dgp12}, we find that the potential falls more steeply during the late time than in early time. This is expected because the potential is near its end of expansion phase, hence kinetic term starts dominating resulting in steeper fall of the potential. 
\end{itemize}
\begin{figure*}[htb]
\centering
\subfigure[ An illustration of the behavior of the scale factor with time during the late contraction phase with $A_{4}=100M_{p}^{-1},\ r_{c}=6$.]{
    \includegraphics[width=7.2cm,height=8cm] {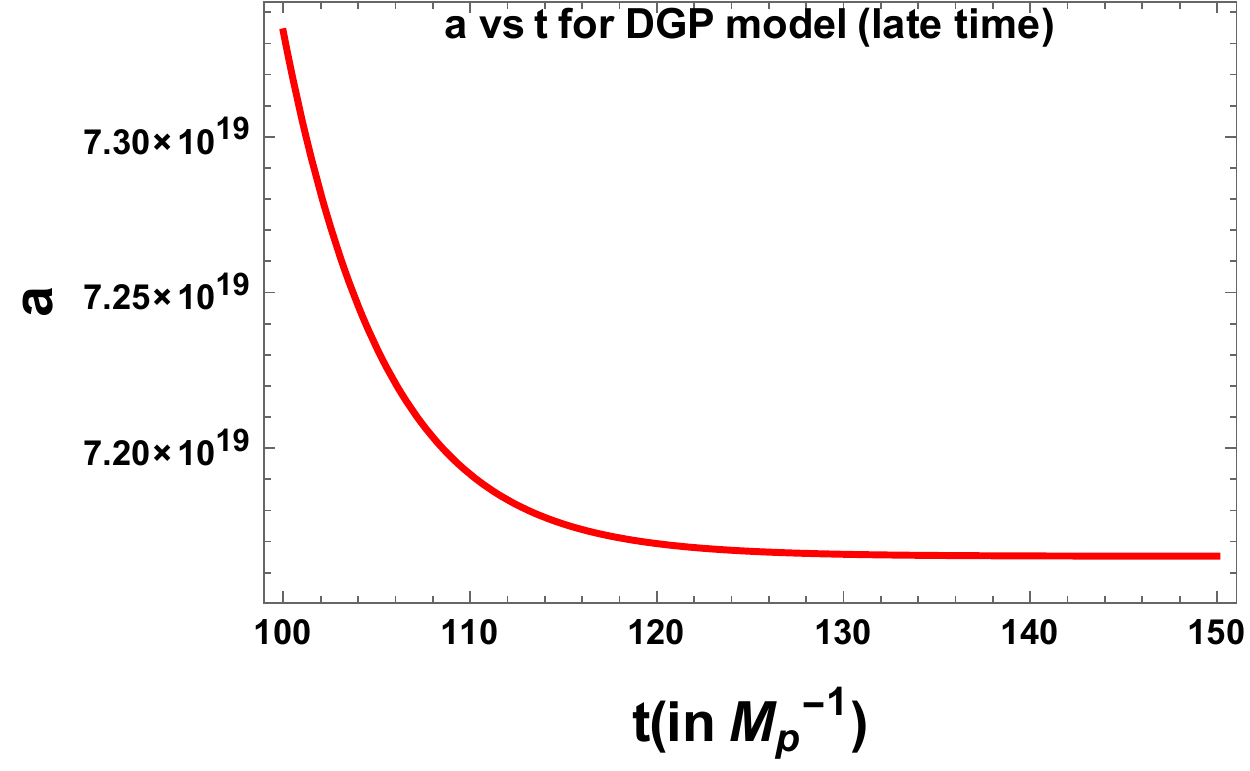}
    \label{dgp13}
}
\subfigure[An illustration of the behavior of the potential during late contraction phase with $V_{0}=6.1{\rm x}10^{-3}M_{p}^{4},\ A_{3}=152M_{p}^{-1},\ A_{4}=730M_{p},\ \beta=0.52,\ r_{c}=1.62,\ p=1$ .]{
    \includegraphics[width=7.2cm,height=8cm] {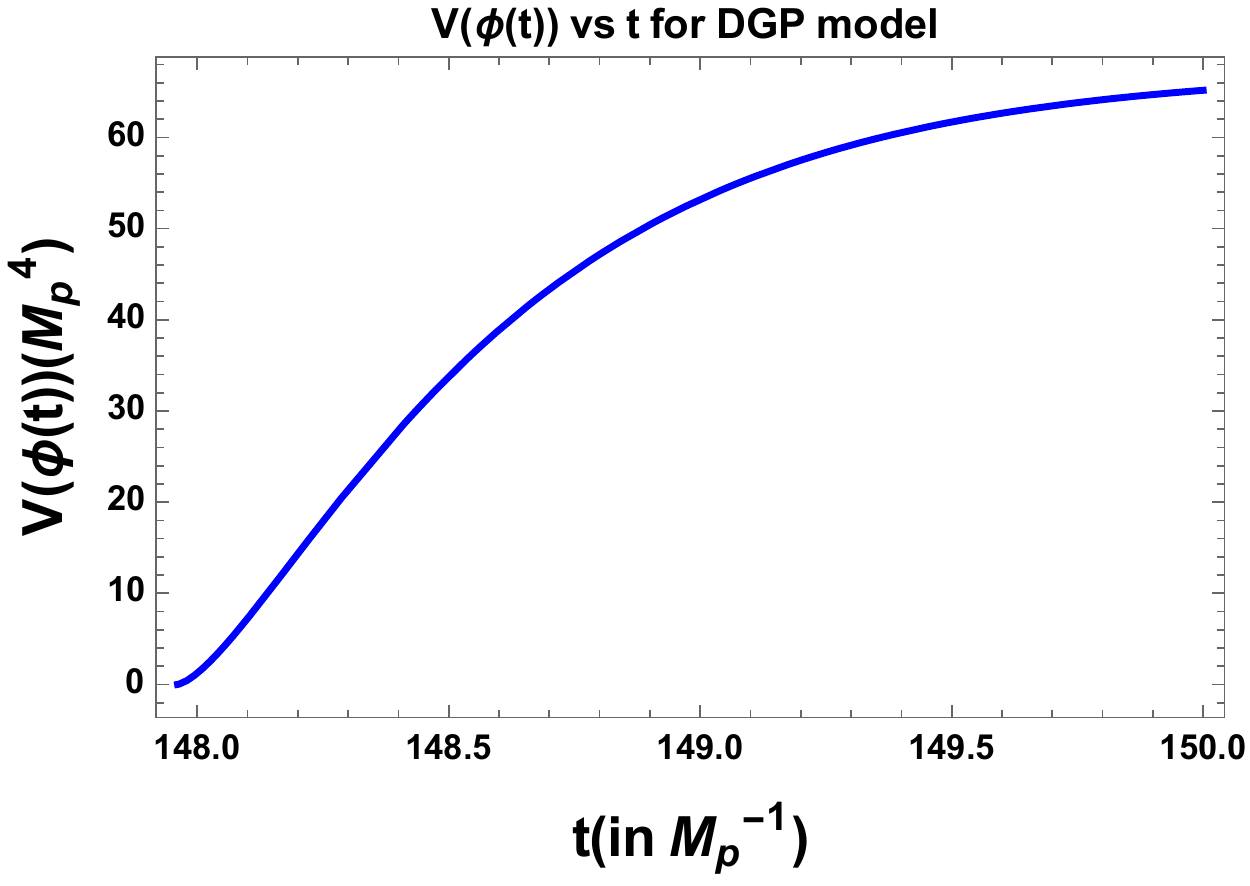}
    \label{dgp14}
}
\subfigure[An illustration of the behavior of the scale factor with time during early time contraction phase with $A_{2}=100M_{p}^{-1},\ A_{2}''=1,\ \mid r_{c}\mid=2.0$.]{
    \includegraphics[width=7.2cm,height=8cm] {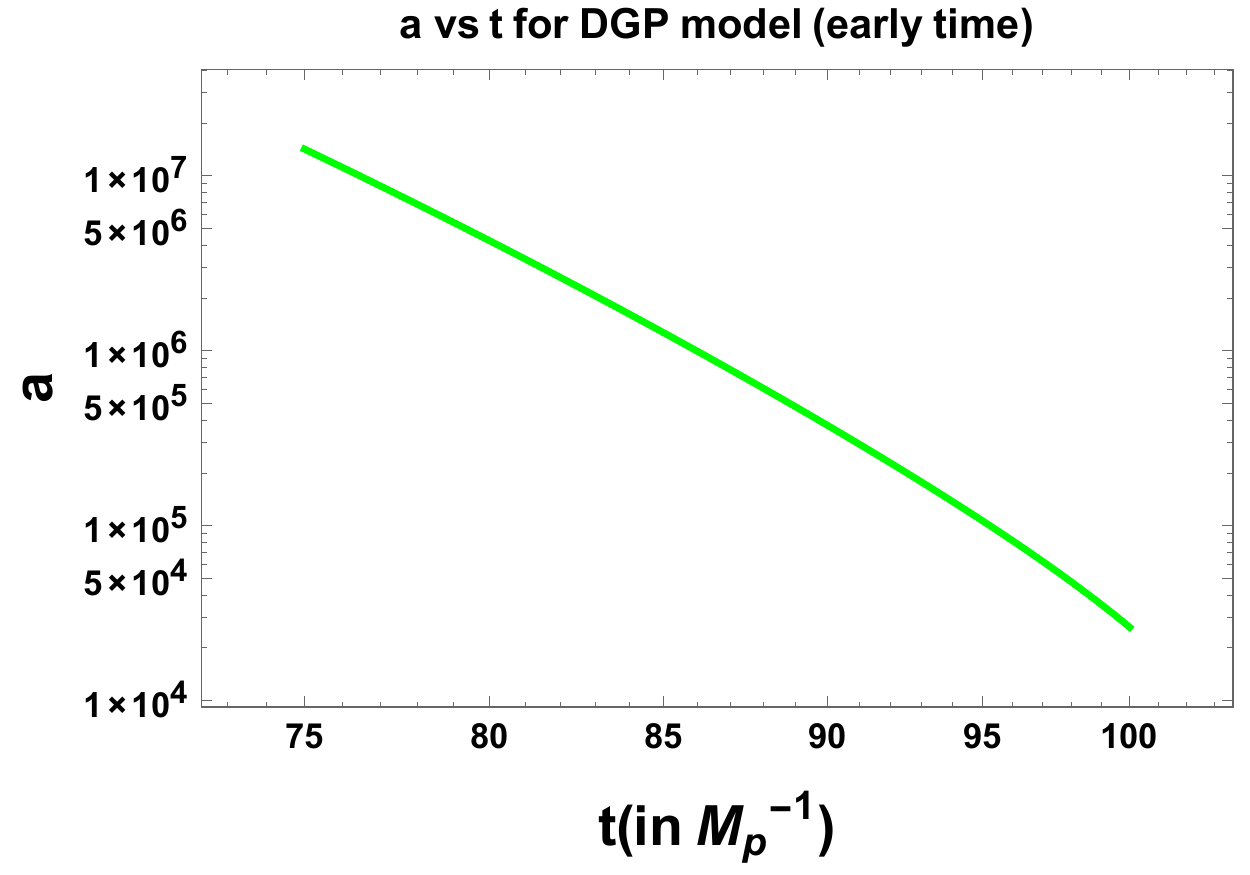}
    \label{dgp15}
}
\subfigure[An illustration of the behavior of the potential during early time contraction phase with $V_{0}=10^{-2}M_{p}^{4},\ p=1,\ \beta=0.392,\ A_{2}'=63M_{p}^{-1},\ \mid r_{c}\mid=1.4$ .]{
    \includegraphics[width=7.2cm,height=8cm] {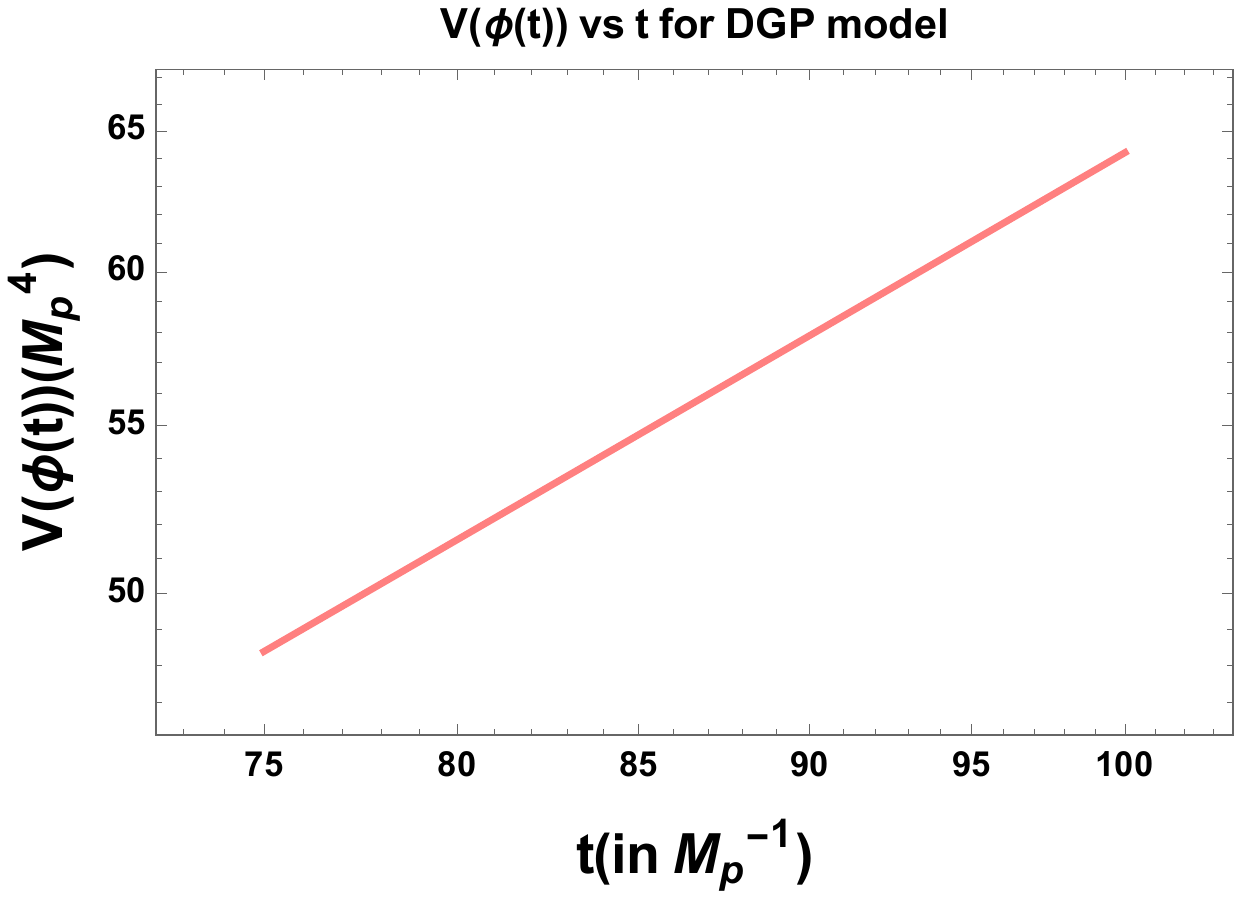}
    \label{dgp16}
}
\caption[Optional caption for list of figures]{ Graphical representation of the evolution of the scale factor and the potential during the expansion phase for DGP model.} 
\label{fig19}
\end{figure*}
In Fig.~\ref{fig19} we have shown the evolution of the scale factor and the potential for early and late time contraction phase. From the above plots, we can draw the following conclusions:
\begin{itemize}
\item Fig.~\ref{dgp13}, shows the plot of the scale factor during late time contraction phase for hilltop potential given by Eq.~(\ref{scalefactor19}) with the parameter values $A_{4}=100M_{p}^{-1},\ r_c=6$. 
\item The nature of the graph (i.e the decrease in amplitude of the scale factor with time, which is expected for contraction phase) remains almost unchanged if we change the parameter values. Larger values of $r$ decreases the amplitude of expansion (but the change is very small), but smaller values make the potential fall more steeply. 
\item As we can see from Fig.~\ref{dgp13}, the change in amplitude of the scale factor is very small. This change becomes negligibly small for very small values of $A_{4}$, and very large values of $A_{4}$ are also not allowed.
\item Fig.~\ref{dgp14} shows the plot of the behavior of the potential with time for late time contraction of hilltop potential. This graph has been obtained with the help of Eq.~(\ref{potential19}) with parameter values $V_{0}=6.1{\rm x}10^{-3}M_{p}^{4},\ A_{3}=152M_{p}^{-1},\ A_{4}=730M_{p},\ \beta=0.52,\ r_{c}=1.62,\ p=1$.
\item The potential in this case rises with time,which is expected since we are in the contraction phase and kinetic energy is dominating initially. Detail graphical analysis show that for very small value of $r_{c}$, the correct nature of the potential is obtained only if we take smaller values of the other parameters. Larger values of the parameters make the potential rise more steeply and attained the nearly flat region faster.  
\item Fig.~\ref{dgp15} shows the plot of the scale factor in the early time contraction phase for hilltop potential given by Eq.~(\ref{scalefactor18}). This plot has been obtained for $A_{2}=100M_{p}^{-1},\ A_{2}''=1,\ \mid r_{c}\mid=2.0$. Detail graphical analysis have shown that as we decrease the value of $\mid r_c \mid$, the fall of the scale factor becomes more linear. Larger values of $\mid r_{c}\mid$ increases the amplitude of the scale factor but makes contraction possible only for smaller values of $A_{2}$.
\item Fig.~\ref{dgp16} shows the evolution of the potential in the early time contraction phase for hilltop potential. Fig.~\ref{dgp16} has been obtained by with the help of Eq.~(\ref{eqsol}) with the parameter values $V_{0}=10^{-2}M_{p}^{4},\ p=1,\ \beta=0.392,\ A_{2}'=63M_{p}^{-1},\ \mid r_{c}\mid=1.4$. Detail graphical analysis show that the nature of the graph do not depend on the parameter values. Larger values of the parameters only increases or decrease the height of the potential. But in order to get the correct nature of the potential which will result in contraction, we need $\beta>0$. 
\item Though fig.~\ref{dgp13} and Fig.~\ref{dgp14} are for late time and Fig.~\ref{dgp15} and Fig.~\ref{dgp16} are for early time, in one complete cycle, the late time phase appears before the early time phase in contraction under the convention which we follow. But in this case, the time range has been chosen not according to the convention, but in order to get some suitable output due to the complicated nature of the equations.
\item If we compare Fig.~\ref{dgp13} and Fig.~\ref{dgp15}, we find that the decrease in the amplitude of the scale factor is much less during the late time contraction phase, as compared to early time phase.
\item If we compare Fig.~\ref{dgp16} with Fig.~\ref{dgp9}, we find that there is a net increase in the amplitude of the scale factor after one complete cycle.
\end{itemize}

\subsubsection{Case II: Natural potential}
\begin{figure*}[htb]
\centering
\subfigure[ An illustration of the behavior of the scale factor with time during the early  expansion phase for $\phi<<f$ with $V_{0}=10^{-8}M_{p}^{4},\ r_{c}=1,\ A_{7}=1$.]{
    \includegraphics[width=7.2cm,height=8cm] {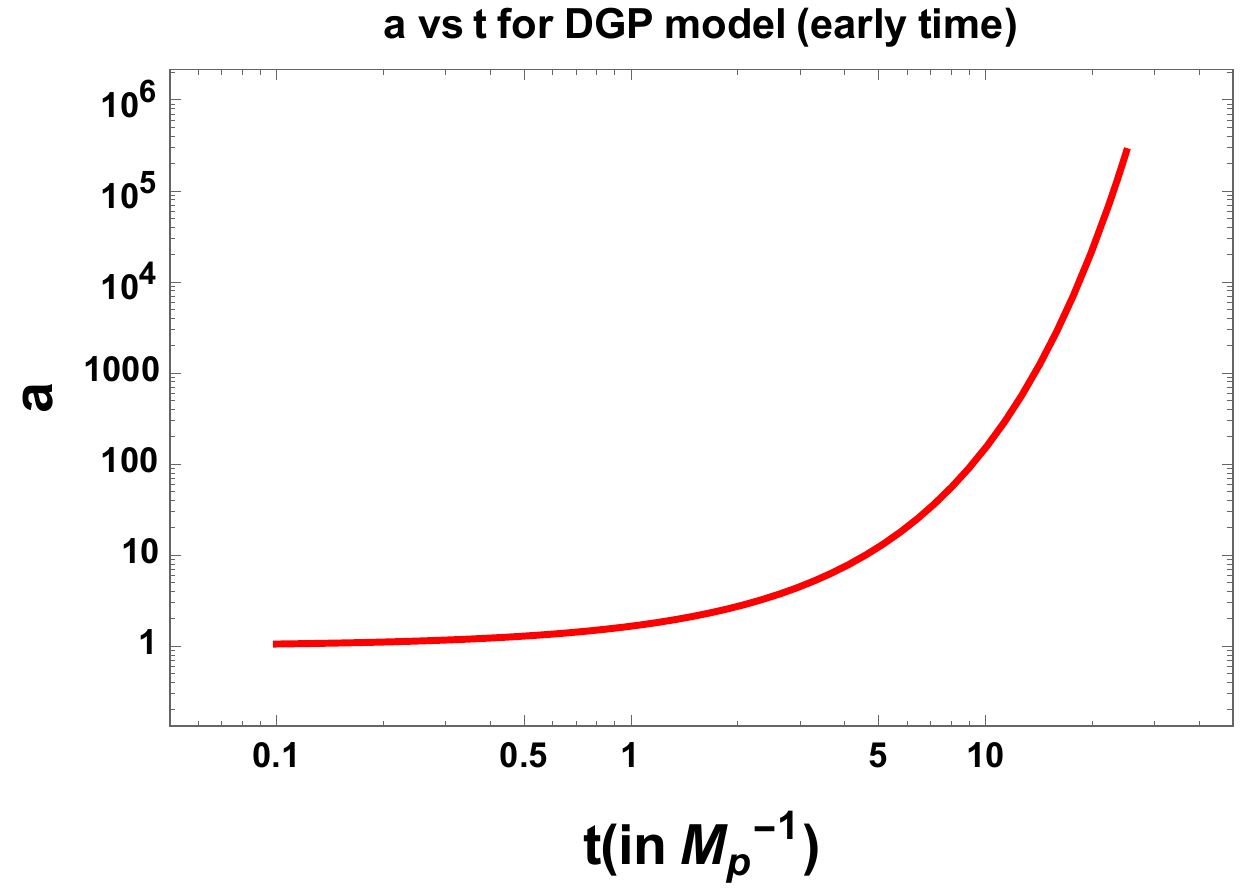}
    \label{dgp17}
}
\subfigure[An illustration of the behavior of the potential during early expansion phase for $\phi<<f$ with $V_{0}=4.7{\rm x}10^{-3}M_{p}^{4},\ r_{c}=1.67,\ A_{6}=32,\ f=6.7M_{p}$ .]{
    \includegraphics[width=7.2cm,height=8cm] {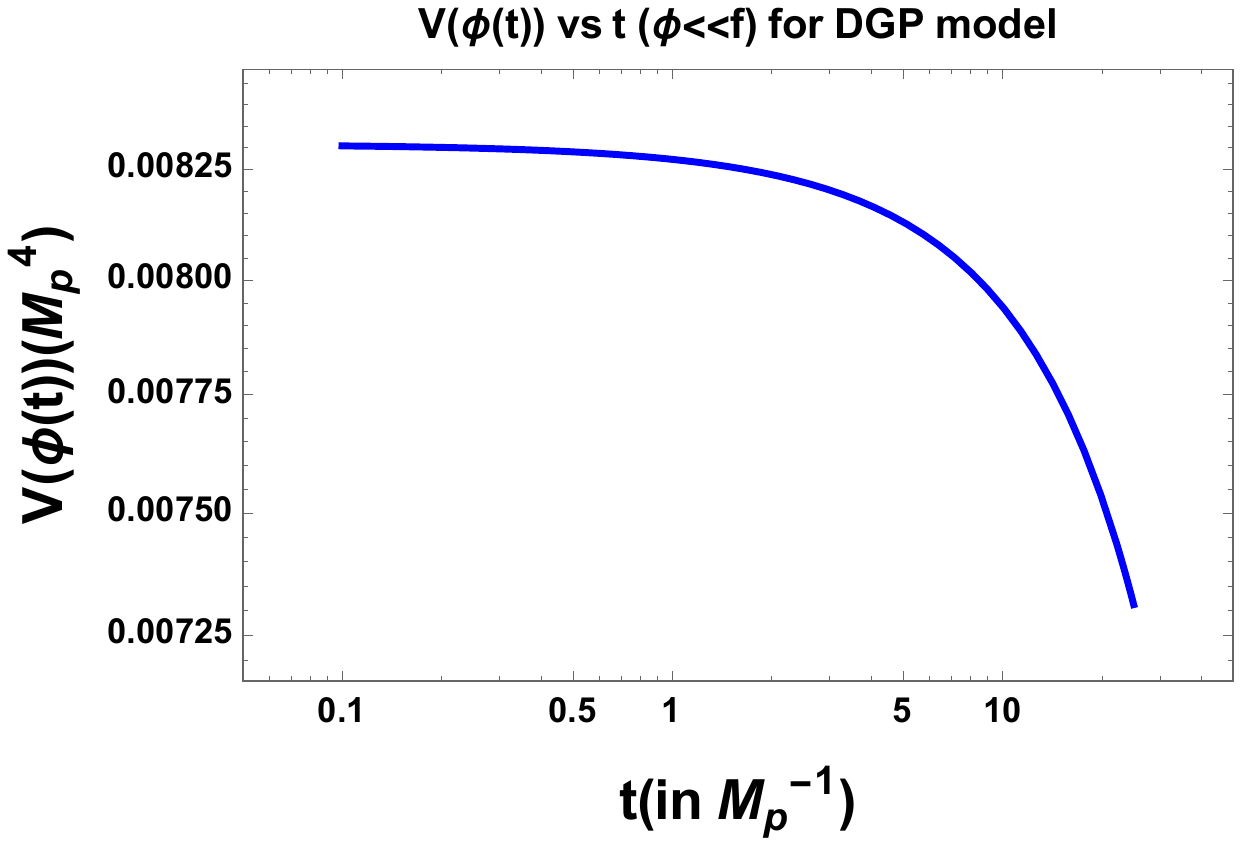}
    \label{dgp18}
}
\subfigure[An illustration of the behavior of the scale factor with time during late time expansion phase for $\phi<<f$ with $V_{0}=10^{-8}M_{p}^{4},\ r_{c}=1.17,\ A_{9}=1$.]{
    \includegraphics[width=7.2cm,height=8cm] {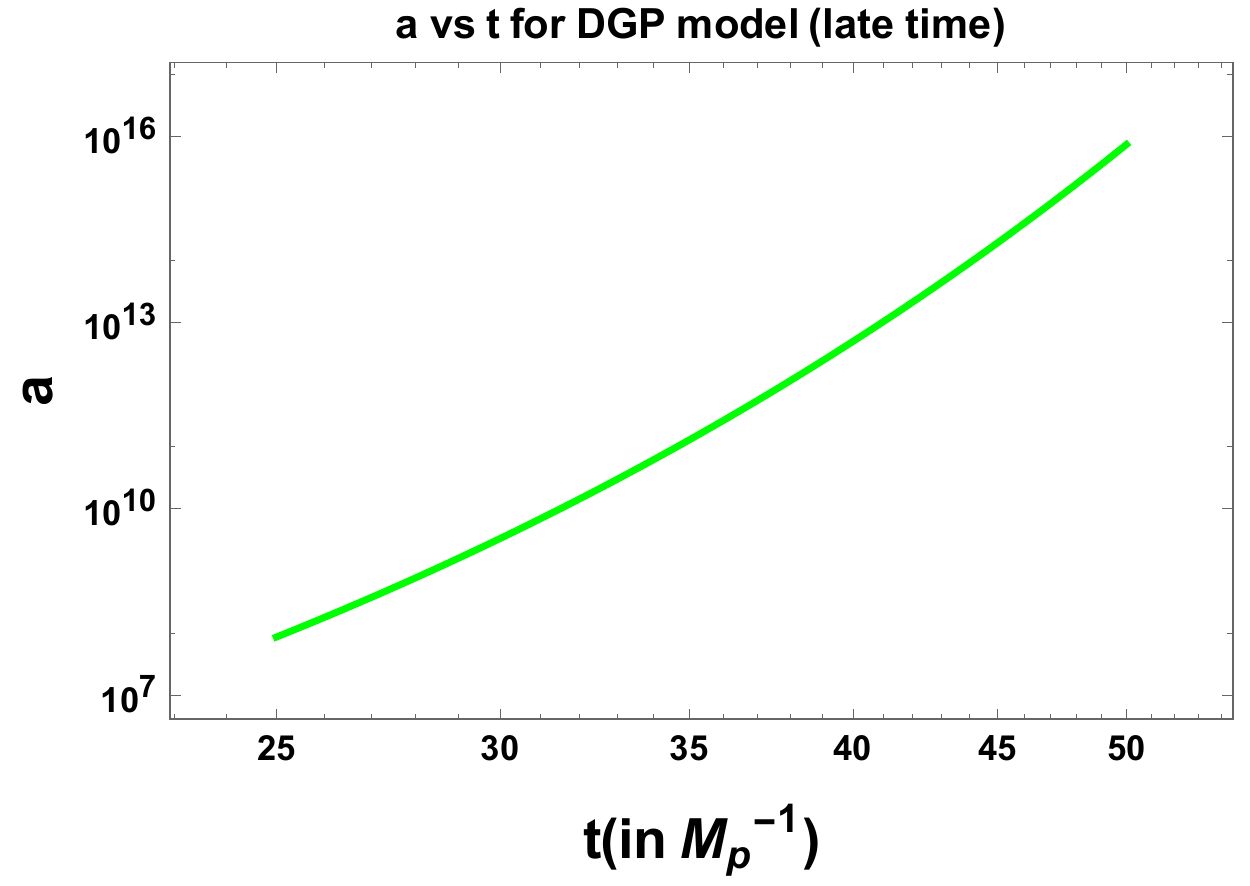}
    \label{dgp19}
}
\subfigure[An illustration of the behavior of the potential during late time expansion phase for $\phi<<f$ with $V_{0}=4.7{\rm x}10^{-3}M_{p}^{4},\ r_{c}=1.67,\ A_{8}=1,\ f=6.7M_{p}$ .]{
    \includegraphics[width=7.2cm,height=8cm] {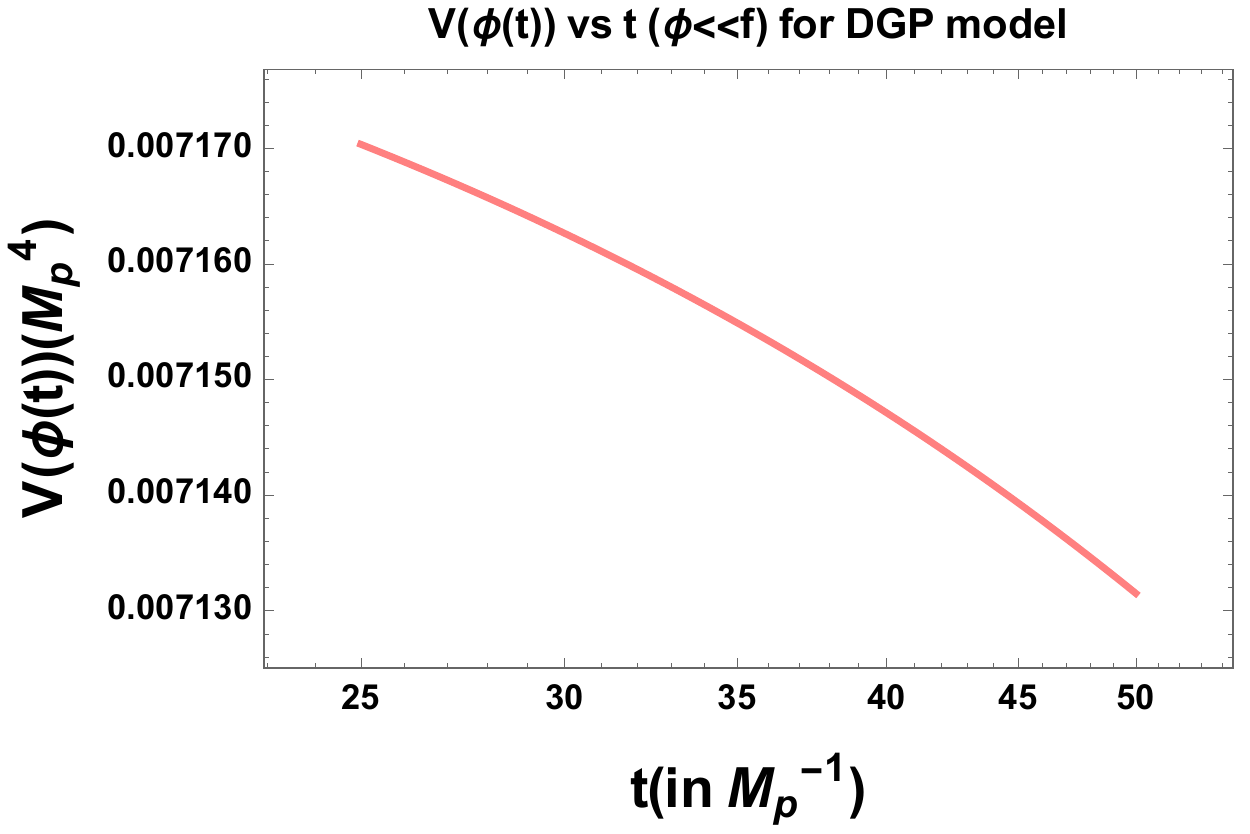}
    \label{dgp20}
}
\caption[Optional caption for list of figures]{ Graphical representation of the evolution of the scale factor and the potential during the expansion phase for the DGP model.} 
\label{fig20}
\end{figure*}
In Fig. \ref{fig20} we have shown the evolution of the scale factor and the potential for early and late time expansion phase. From the above plots, we can draw the following conclusions:
\begin{itemize}
\item Fig. \ref{dgp17}, shows the plot of the scale factor in the small field limit for natural  potential given by Eq. (\ref{dgpearly7}) with the parameter values $V_{0}=10^{-8}M_{p}^{4},\ r_{c}=1,\ A_{7}=1$ during the early phase of expansion. 
\item Higher values of the parameters only increase the amplitude of expansion, keeping the nature of the plot unchanged.
\item Fig. \ref{dgp18} shows the plot of the behavior of the potential with time for small field natural potential during the early phase of expansion. This graph has been obtained with the help of Eq. (\ref{potential20}) with parameter values $V_{0}=4.7{\rm x}10^{-3}M_{p}^{4},\ r_{c}=1.67,\ A_{6}=32,\ f=6.7M_{p}$. Large values of $r_{c}$ and $f$ increases the height of the potential. Only for certain range of values of $A_{6}$, we get the required evolution of the potential (such as expansion is possible if $A_{6}$ lies within ($1$ to $3$) or ($6$ to $9$) or ($13$ to $15$) etc.).
\item Fig. \ref{dgp19}, shows the plot of the scale factor in the small field limit for natural  potential given by Eq. (\ref{scalefactor21}) with the parameter values $V_{0}=10^{-8}M_{p}^{4},\ r_{c}=1.17,\ A_{9}=1$ during the late phase of expansion. 
\item Higher values of the parameters only increase the amplitude of expansion, keeping the nature of the plot unchanged.
\item Fig. \ref{dgp20} shows the plot of the behavior of the potential with time for small field natural potential. This graph has been obtained with the help of Eq. (\ref{potential21}) with parameter values $V_{0}=4.7{\rm x}10^{-3}M_{p}^{4},\ r_{c}=1.67,\ A_{8}=1,\ f=6.7M_{p}$. The conclusions regarding the variation of the nature of the potential with parameters remains same as for the early time expansion.
\item If we compare the amplitudes of the scale factor in Fig. \ref{dgp13} with Fig. \ref{dgp17}, we find that after one cycle of expansion and contraction, we can get a net increase in amplitude of the scale factor provided the parameters are chosen properly. 
\end{itemize}

\subsubsection{Case III: Coleman-Weinberg potential}
\begin{figure*}[htb]
\centering
\subfigure[ An illustration of the behavior of the scale factor with time during the early  expansion phase for $\phi<<M_{p}$ with $V_{0}=10^{-8}M_{p}^{4},\ r_{c}=0.86,\ A_{11}=1,\ \alpha=0.1$.]{
    \includegraphics[width=7.2cm,height=8cm] {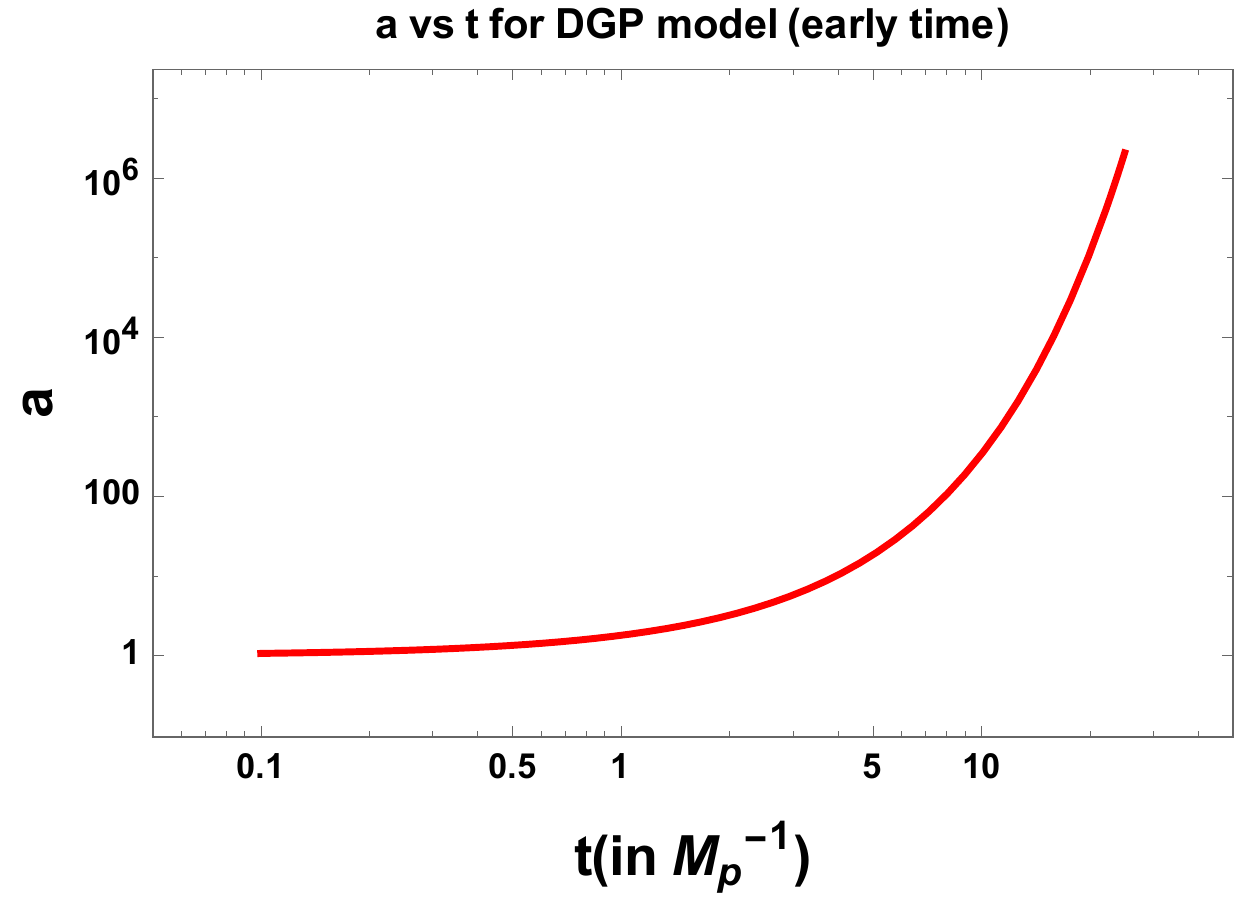}
    \label{dgp21}
}
\subfigure[An illustration of the behavior of the potential during early expansion phase for $\phi<<M_{p}$ with $V_{0}=10^{-5}M_{p}^{4},\ r_{c}=0.86,\ A_{10}=5M_{p},\ \alpha=0.02,\ \beta=0.03$ .]{
    \includegraphics[width=7.2cm,height=8cm] {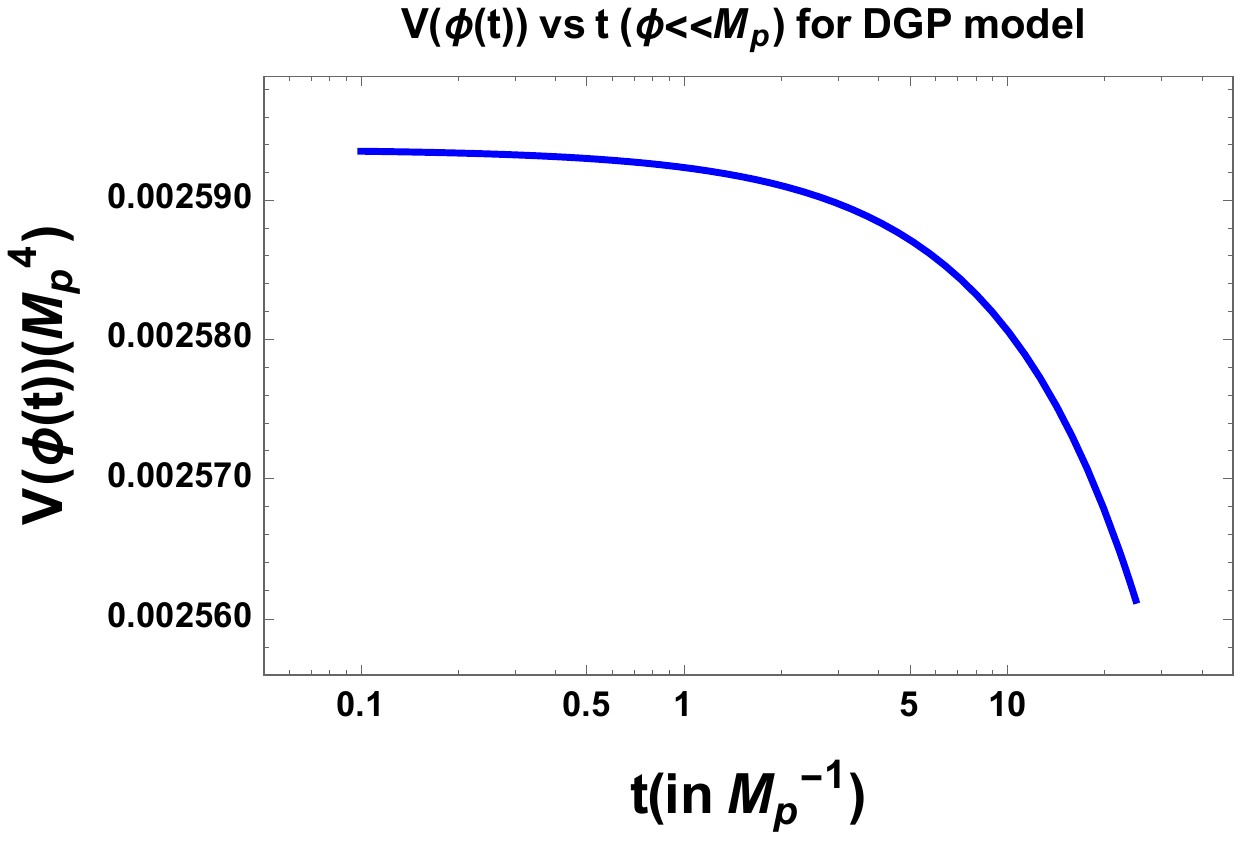}
    \label{dgp22}
}
\subfigure[An illustration of the behavior of the scale factor with time during late time expansion phase for $\phi<<M_{p}$ with $V_{0}=6{\rm x}10^{-4}M_{p}^{4},\ r_{c}=1.66,\ A_{14}=10^{-8}M_{p},\ A_{15}=1,\ \alpha=0.145$.]{
    \includegraphics[width=7.2cm,height=8cm] {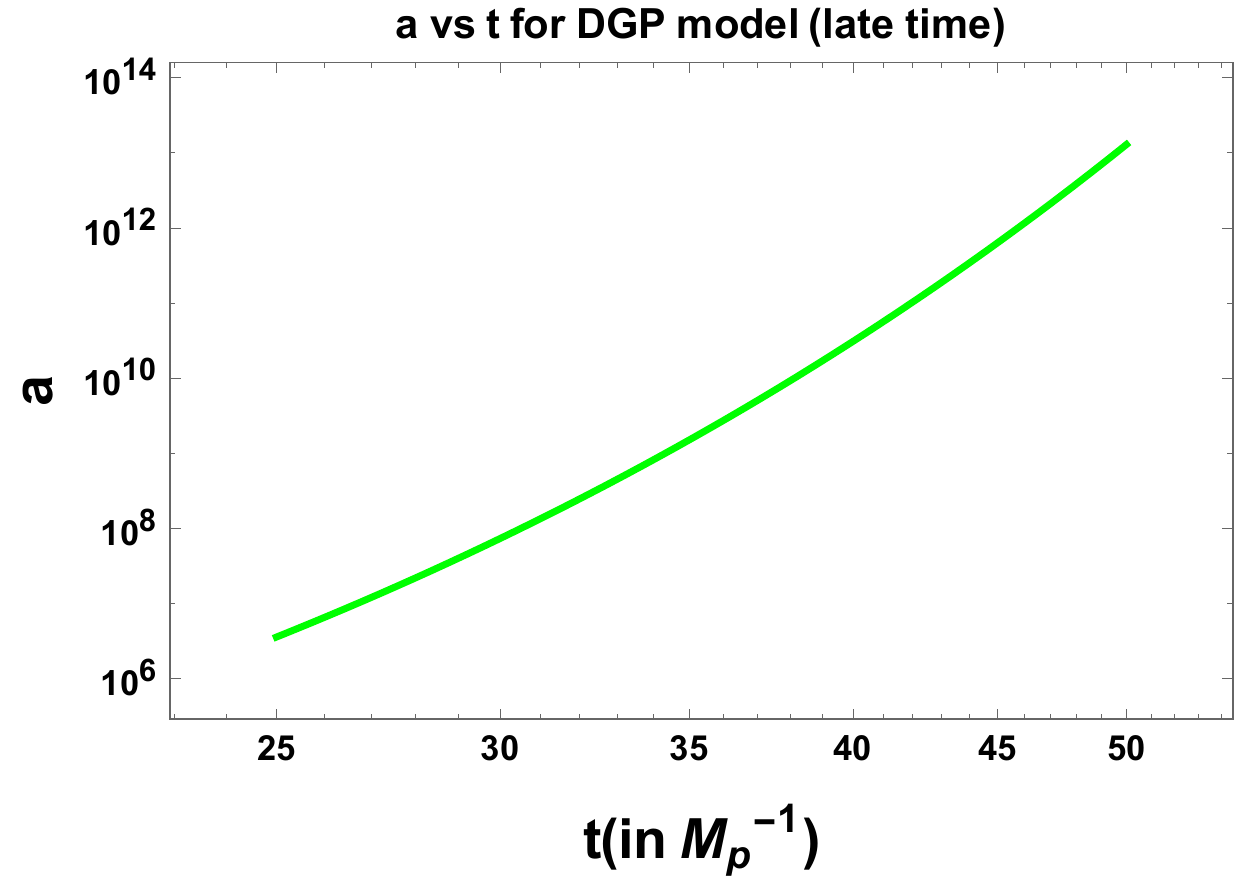}
    \label{dgp23}
}
\subfigure[An illustration of the behavior of the potential during late time expansion phase for $\phi<<M_{p}$ with $V_{0}=10^{-5}M_{p}^{4},\ r_{c}=1.32,\ A_{14}=10^{-8}M_{p},\ \alpha=0.032,\ \beta=-1.9$ .]{
    \includegraphics[width=7.2cm,height=8cm] {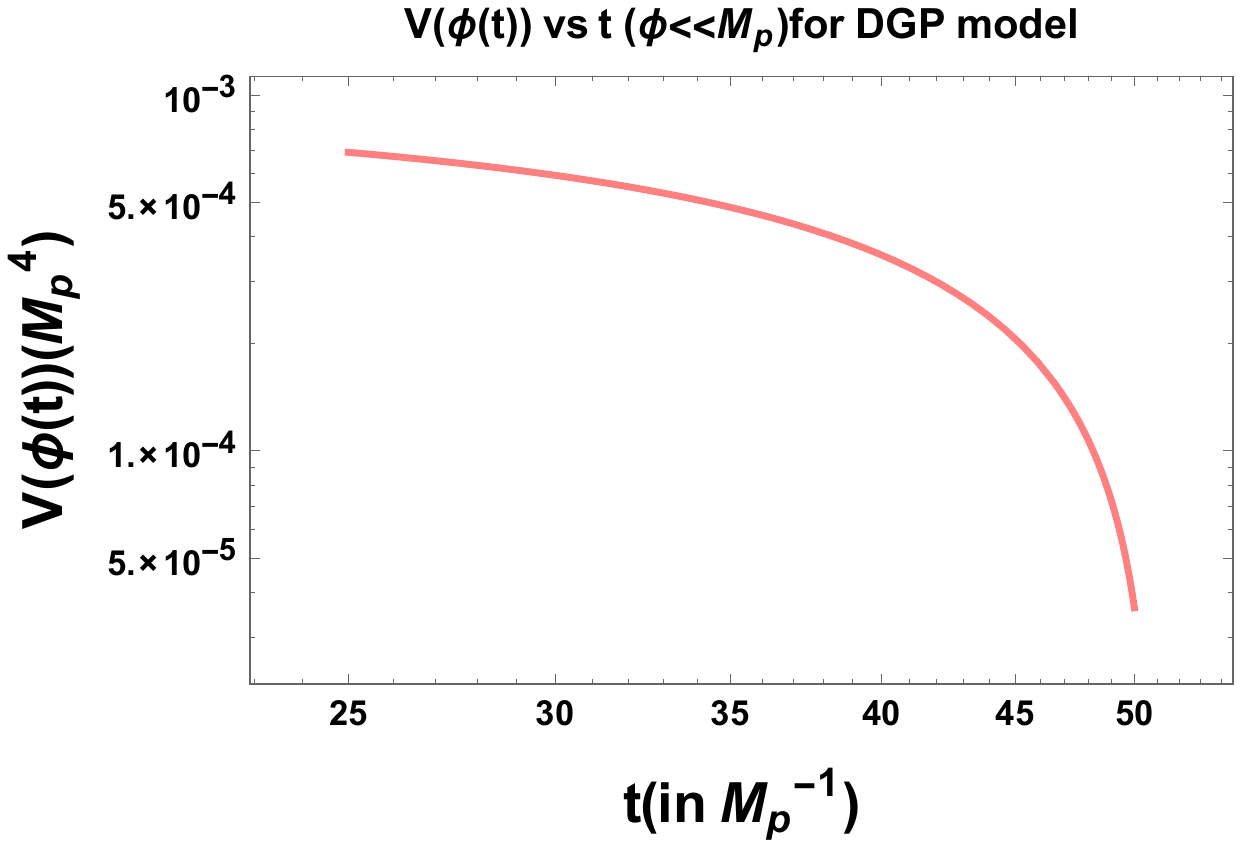}
    \label{dgp24}
}
\caption[Optional caption for list of figures]{ Graphical representation of the evolution of the scale factor and the potential during the expansion phase for the DGP model.} 
\label{fig21}
\end{figure*}
In Fig. \ref{fig21} we have shown the evolution of the scale factor and the potential for early and late time small field expansion phase for supergravity potential . From the above plots, we can draw the following conclusions:
\begin{itemize}
\item Fig. \ref{dgp21}, shows the plot of the scale factor in the small field limit for supergravity  potential given by Eqn. (\ref{dgpearly11}) with the parameter values $$V_{0}=10^{-8}M_{p}^{4},\ r_{c}=0.86,\ A_{11}=1,\ \alpha=0.1$$ during the early phase of expansion. 
\item Higher values of the parameters only increase the amplitude of expansion, keeping the nature of the plot unchanged. The variation in the nature and amplitude of the plot is almost independent of the value of $\alpha$ parameter.
\item Fig. \ref{dgp22} shows the plot of the behavior of the potential with time for small field supergravity potential during the early phase of expansion. This graph has been obtained with the help of Eqn. (\ref{potential22}) with parameter values $V_{0}=10^{-5}M_{p}^{4},\ r_{c}=0.86,\ A_{10}=5M_{p},\ \alpha=0.02,\ \beta=0.03$. Large values of the parameters causes an increase in the amplitude of the expansion, keeping the nature of the plot unchanged. Largest increase in amplitude occurs for any change in value of the parameter $\alpha$. Detail graphical analysis have shown that we do  not get the required nature of the plot for $\beta<0$. Thus only positive values of $\beta$ are allowed in this case.
\item Fig. \ref{dgp23}, shows the plot of the scale factor in the small field limit for supergravity  potential given by Eqn. (\ref{scalefactor24}) with the parameter values $V_{0}=6{\rm x}10^{-4}M_{p}^{4},\ r_{c}=1.66,\ A_{14}=10^{-8}M_{p},\ A_{15}=1,\ \alpha=0.145$ during the late phase of expansion. 
\item For $V_{0}>6{\rm x}10^{-4}M_{p}^{4}$, expansion is possible only if the values of $\beta,\ \alpha$ and $r_{c}$ lie close to or lesser than unity. If we make the values of any of the parameters $\beta, \alpha,\ r_{c}$ large, then expansion is possible only for the case when both $A_{14}$ and $V_{0}$ is O($10^{-4}$) or less.
\item Large and positive values of $\beta$ make the expansion more linear. Expansion is possible for $\beta<0$, provided $V_{0}$ takes large values and $\alpha$ and $A_{14}$ are smaller than unity.
\item Fig. \ref{dgp24} shows the plot of the behavior of the potential with time for small field supergravity potential during late time expansion. This graph has been obtained with the help of Eqn. (\ref{potential24}) with parameter values $V_{0}=10^{-5}M_{p}^{4},\ r_{c}=1.32,\ A_{14}=10^{-8}M_{p},\ \alpha=0.032,\ \beta=-1.9$. Expansion is possible only if $\beta<2$. Again for expansion to happen for positive value of $\beta$, all the other parameters must be much less than unity. Larger and positive values of $\beta$ makes the potential fall more linearly. For expansion to occur, $A_{14}$ must be $<<1$.
\item If we compare the amplitudes of the scale factor in Fig. \ref{dgp13} with Fig. \ref{dgp21}, we find that after one cycle of expansion and contraction, we can get a net increase in amplitude of the scale factor provided the parameters are chosen properly. 
\end{itemize}

For large field case ($\phi>>M_{p}$) during expansion,
the conclusions and the nature of the graphs are almost
similar, hence have not been shown explicitly. Thus, a 
increase in the amplitude of the scale factor after one
complete cycle of expansion and contraction, is possible for this case also.

\section{Hysteresis from cosmological constant dominated Einstein gravity model}
\label{sq4}
In this section we will consider three different cases by taking different functional forms of
the cosmological constant like term in the effective action. One will be the case of standard model
of cosmology i.e when the cosmological constant is a constant. Then we will
further extend our analysis for the case when the cosmological constant
is field dependent. This dependence will be included either in the form
of a power series form of the scalar field or by including a dilaton field. 
\\ \\
\underline{\textbf{Case 1: Cosmological constant of $\Lambda$CDM model}}
\\ \\
Let us start our discussion with $\Lambda$CDM model, which 
is the standard model of Big Bang cosmology, where the Einstein's field equations are modified by the additional
cosmological constant term denoted as $\Lambda$.
The modified Friedmann equations in this model can be expressed as:
\begin{equation}
H^{2} =\left(\frac{\dot{a}}{a}\right)^2 = \frac{\rho}{3M^{2}} - \frac{k}{a^{2}} + \frac{\Lambda}{3},
\label{constant1}
\end{equation}
\begin{equation}
H^{2}+\dot{H}=\frac{\ddot a}{a} = -\frac{(\rho + 3p)}{6M^{2}} + \frac{\Lambda}{3},
\label{constant2}
\end{equation}
where we are denoting the Planck mass by M. 
\\ \\
\underline{\textbf{Case 2: Scalar field dependent cosmological constant}}
\\ \\
Let us consider a situation where we make the cosmological constant term 
scalar field dependent i.e we replace $\Lambda$ by $\Lambda(\phi)$, where \be \Lambda(\phi) = \sum\limits_{i=0}^4 \Lambda_{i}\phi^{i},\ee
then the representative four dimensional effective action for this case is given by:
\begin{eqnarray}
S &=& \int d^{4}x \sqrt{-g}\left(R + \sum\limits_{i=0}^4 \Lambda_{i}\phi^{i} + \frac{1}{2}g^{\mu\nu}\partial_{\mu}\phi\partial_{\nu}\phi - V(\phi)\right) \nonumber\\ 
&=& \underbrace{\int d^{4}x \sqrt{-g}\left(R + \Lambda_{0}\right)}_{S_{\Lambda EH}} + \underbrace{
\int d^{4}x \sqrt{-g}\left(\sum\limits_{i=1}^4 \Lambda_{i}\phi^{i}
+ \frac{1}{2}g^{\mu\nu}\partial_{\mu}\phi\partial_{\nu}\phi - V(\phi)\right)}_{S_{\Lambda\phi}},
\end{eqnarray}
where $\Lambda_{0}$ is the general cosmological constant ({\bf denoted by $\Lambda$ in case 1}) and
$\Lambda_{1},\  \Lambda_{2},\ \Lambda_{3} \& \ \Lambda_{4}$ are the new constants.

 The first action $S_{\Lambda EH}$ is the normal action for $\Lambda$CDM model, hence the Friedmann equations
 for this part are same as Eqns. (\ref{constant1}) and (\ref{constant2}). The scalar field dynamics
 get modified due to the action $S_{\Lambda\phi}$. The equation of motion for the scalar field is given by:
 \begin{equation}
 \frac{\partial^{\mu}\delta(\sqrt{-g}{\cal L}_{\Lambda\phi})}{\delta\partial^{\mu}\phi} - \frac{\delta(\sqrt{-g}{\cal L}_{\Lambda\phi})}{\delta\phi} = 0
 \label{philambda}
 \end{equation}
where ${\cal L}_{\Lambda\phi}$ is the Lagrangian density for the scalar field and for FLRW metric $\sqrt{-g} = a^{3}$. Solving
the above equation we get the equation of motion for $\phi$ as:
\begin{eqnarray}
\ddot{\phi} + 3H\dot{\phi} - \sum\limits_{i=1}^{4} i\Lambda_{i}\phi^{i-1} + V_{,\phi} &=& 0
\label{philambda1}
\end{eqnarray}
where the first two terms come from the first term in Eq.~ (\ref{philambda}) and the rest of the terms come from the last term in Eq.~ (\ref{philambda}).

The energy momentum tensor for scalar field is given by 
\begin{eqnarray}
T_{\mu\nu}^{\phi} &=& \partial_{\mu}\phi\partial_{\nu}\phi - g_{\mu\nu}{\cal L}_{\Lambda\phi} \\ \nonumber
&=& \partial_{\mu}\phi\partial_{\nu}\phi - g_{\mu\nu}\left(\frac{1}{2}\partial^{\sigma}\phi\partial_{\sigma}\phi - V(\phi) + \sum\limits_{i=1}^4 \Lambda_{i}\phi^{i}\right) 
\end{eqnarray} 
For perfect fluid the expressions for the energy density $\rho$ and pressure $p$ can be expressed as:
\begin{eqnarray}
\rho &=& \frac{1}{2}\dot{\phi}^{2} + V(\phi) - \sum\limits_{i=1}^4 \Lambda_{i}\phi^{i} \\
p &=& \frac{1}{2}\dot{\phi}^{2} - V(\phi) + \sum\limits_{i=1}^4 \Lambda_{i}\phi^{i}
\label{philambda2}
\end{eqnarray}
Solving Eq.~(\ref{philambda1}), Eq.~(\ref{philambda2}) and Eq.~(\ref{constant1}) simultaneously, we get the solutions for the scalar field 
$\phi$ and scale factor $a$. Thus we see that both the density, pressure as well as the equation of motion of the scalar field gets modified from the standard case, which was expected. 
\\ \\
\underline{\textbf{Case 3: Cosmological constant along with a dilaton field}}
\\ \\\
In this case we add a dilaton field along with the general cosmological constant i.e we replace $\Lambda$ by $\Lambda(\phi)$, 
where \be \Lambda(\phi) = \Lambda_{0}+\Lambda e^{\phi/M_{p}},\ee Then the representative action for this case is then given by:
\begin{eqnarray}
S &=& \int d^{4}x \sqrt{-g}\left(R + \Lambda_{0} + \Lambda e^{\phi/M_{p}} + \frac{1}{2}g^{\mu\nu}\partial_{\mu}\phi\partial_{\nu}\phi - V(\phi)\right)\\ 
&=& \underbrace{\int d^{4}x \sqrt{-g}\left(R + \Lambda_{0}\right)}_{S_{\Lambda EH}} + \underbrace{\int d^{4}x \sqrt{-g}\left(\Lambda e^{\phi/M_{p}}+ 
\frac{1}{2}g^{\mu\nu}\partial_{\mu}\phi\partial_{\nu}\phi - V(\phi)\right)}_{S_{\Lambda\phi}},\nonumber 
\end{eqnarray}
where $\Lambda_{0}$ is the general cosmological constant and $\Lambda$ is another constant.

Repeating the analysis as we did for the previous case 2, we get the following equation of motion for the scalar field $\phi$ as:
\begin{equation}
\ddot{\phi} + 3H\dot{\phi} - \frac{\Lambda}{M_p} e^{\phi/M_{p}} + V_{,\phi} = 0
\label{dilaton1}
\end{equation}

Similarly the expressions for density and pressure for a perfect fluid in this case can be expressed as:
\begin{eqnarray}
\rho &=& \frac{1}{2}\dot{\phi}^{2} + V(\phi) - \Lambda e^{\phi/M_{p}} \\
p &=& \frac{1}{2}\dot{\phi}^{2} - V(\phi) + \Lambda e^{\phi/M_{p}} 
\label{dilaton2}
\end{eqnarray}
Solving Eq.~(\ref{dilaton1}), Eq.~(\ref{dilaton2}) and Eq.~(\ref{constant1}) simultaneously, we get the solutions for the scalar field 
$\phi$ and scalae factor $a$.
\\
\subsection{Condition for bounce}

\underline{\textbf{For case 1, 2, 3}}
\\ \\
In the analysis mentioned below, we
have first shown the results for open and closed universe only. The analysis for flat universe has been shown separately.

Let us start the discussion for the case 1 where at bounce, setting the Hubble parameter $H = 0$ in Eq.~(\ref{constant1}) we get:
\begin{equation}
\rho_{b} = 3M^{2}\left(\frac{k}{a_{b}^2} -\frac{\Lambda}{3}\right)
\label{condbounce}
\end{equation}
where $\rho_{b}$ and $\ a_{b}$ are the representative density and scale factor at bounce.

The mass content at bounce (neglecting the overall constant factor) is given by: \be M_{b} = \rho_{b} a_{b}^{3} = 3M^{2}\left(ka_{b} -\frac{\Lambda a_{b}^{3}}{3}\right).\ee
Therefore the infinitesimal change in the mass content can be expressed as: \be\delta M =  3M^{2}\left(k - \Lambda a_{b}^{2}\right)\delta a_{b}.\ee 
Now using energy conservation one can write:  \be \delta M + \delta W = 0,\ee
where $\delta W$ is the work done during each expansion-contraction cycle.

Further setting \be \delta M = -\delta W = -\oint pdV,\ee we get the expression for change
in amplitude of the scale factor at each successive cycle in terms of the work done as: 
\begin{equation}
\delta a_{min} = -\frac{1}{3M^{2}\left(k - \Lambda a_{min}^{2}\right)}\oint pdV
\label{constant3}
\end{equation}
Thus we clearly observe that just like for DGP model as discussed in the earlier section,
here also the increase in amplitude of the scale factor depends on the parameter of the model i.e the variants of 
cosmological constant as mentioned in case 1, case 2 and case 3.
\be\begin{array}{lll}\label{rk9xcxc}
 \displaystyle\delta a_{min} =\left\{\begin{array}{ll}
                    \displaystyle   -\frac{1}{3M^{2}\left(1 - \Lambda a_{min}^{2}\right)}\oint pdV &
 \mbox{\small {\bf for {$k=+1$}}}  \\ \\
         \displaystyle   \frac{1}{3M^{2}\left(1 + \Lambda a_{min}^{2}\right)}\oint pdV & \mbox{\small {\bf for {$k=-1$}}}.
          \end{array}
\right.
\end{array}\ee
Let us now briefly mention the characteristic feature of the results for cosmological bounce for the previously mentioned 
variants of cosmological constant model 
in the following: \\ \\
\begin{enumerate}
 \item For a closed universe, with $k=+1$:
 \begin{itemize}
  \item The hysteresis loop integral \be \oint pdV < 0\ee or equivalently \be p_{exp} < p_{cont}\ee if $(1 - \Lambda a_{min}^{2}) > 0$ or $\Lambda a_{min}^{2} < 1$ and $\delta a_{min}>0$,
  \item The hysteresis loop integral \be \oint pdV > 0\ee or equivalently \be p_{exp} > p_{cont}\ee if $(1 - \Lambda a_{min}^{2}) < 0$ or $\Lambda a_{min}^{2} > 1$ and $\delta a_{min}>0$. 
 \end{itemize}
\item On the other hand, for the open universe, with $k=-1$:
\begin{itemize}
  \item The hysteresis loop integral \be\oint pdV < 0\ee or equivalently \be p_{exp} < p_{cont}\ee if $(1 + \Lambda a_{min}^{2}) < 0$ or $\Lambda a_{min}^{2} <-1$ and $\delta a_{min}>0$,
  \item The hysteresis loop integral \be\oint pdV > 0\ee or equivalently \be p_{exp} > p_{cont}\ee if $(1 + \Lambda a_{min}^{2}) > 0$ or $\Lambda a_{min}^{2} >-1$ and $\delta a_{min}>0$. 
 \end{itemize}
\end{enumerate}

For the other two cases i.e case 2 and case 3, since the Friedmann equations
remain the same, the condition for bounce in both the cases will be given
by Eq.~ (\ref{constant3}). Hence the above analysis holds true for case 2 and case 3 also.
\\ 
\subsection{Condition for acceleration}

\underline{\textbf{For case 1}}
\\ \\
From Eq.~ (\ref{constant2}), at bounce the condition for acceleration is given by
\begin{equation}
\rho_{b} + 3p_{b} < 2\Lambda M^{2}
\label{constantaccel}
\end{equation}

Substituting the expression for $\rho_{b}$ for different values of k we get the conditions for acceleration at bounce as:
\be\begin{array}{lll}\label{rk9xcxcz}
 \displaystyle p_{b} =\left\{\begin{array}{ll}
                    \displaystyle   \Lambda M^{2} - \frac{M^{2}}{a_{b}^{2}} &
 \mbox{\small {\bf for {$k=+1$}}}  \\ \\
         \displaystyle  \Lambda M^{2} + \frac{M^{2}}{a_{b}^{2}} & \mbox{\small {\bf for {$k=-1$}}}.
          \end{array}
\right.
\end{array}\ee
Thus we see that the condition for acceleration at bounce now depends not only on the scale factor but also on the cosmological constant.

Substituting the expressions of $\rho$ and p from Eqns. (\ref{eq:scalar_rho}) into Eq.~ (\ref{constantaccel}), we get the condition for acceleration in terms of the scalar field as
\begin{equation}
\dot\phi^{2} < V(\phi) + \Lambda M^{2}
\label{constantphi}
\end{equation}
Thus we see that even if the contribution from the potential is lesser than the
standard canonical kinetic term i.e. even if $\dot\phi^{2} > V(\phi)$, the
above condition will still be satisfied if $\dot\phi^{2}$ is lesser than the sum of $V(\phi) +  \Lambda M^{2}$. Thus
we can use potentials of smaller values than standard case and yet produce bounce. Thus the presence of cosmological constant modifies the effective potential present at bounce.

\begin{figure*}[htb]
\centering
\subfigure[ An illustration of the bouncing condition for a universe with an equation of state w=1/3 and k=1.]{
    \includegraphics[width=7.2cm,height=8cm] {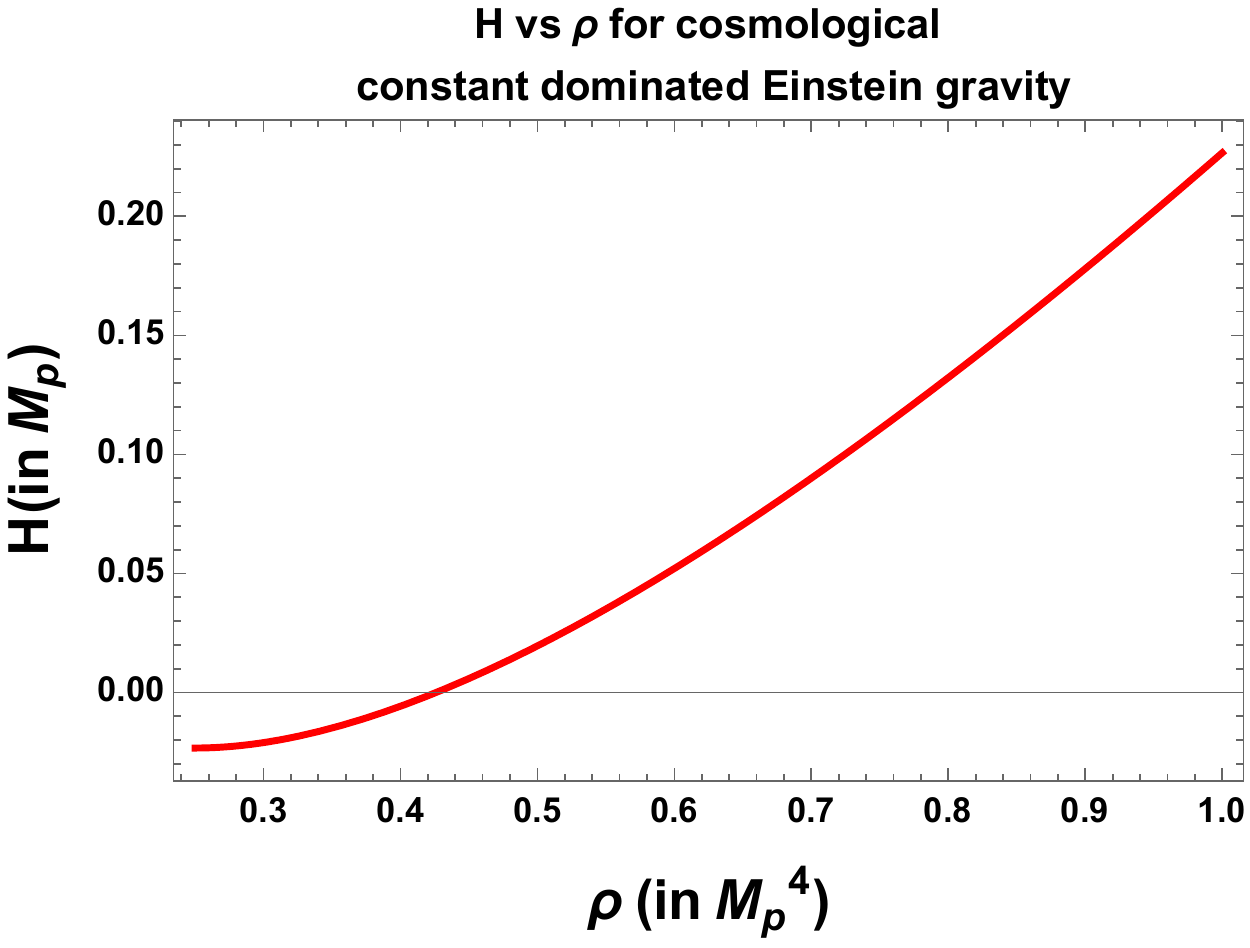}
    \label{lambda1}
}
\subfigure[An illustration of the acceleration condition at the time of bounce for a universe with an equation of state w=1/3, k=1.]{
    \includegraphics[width=7.2cm,height=8cm] {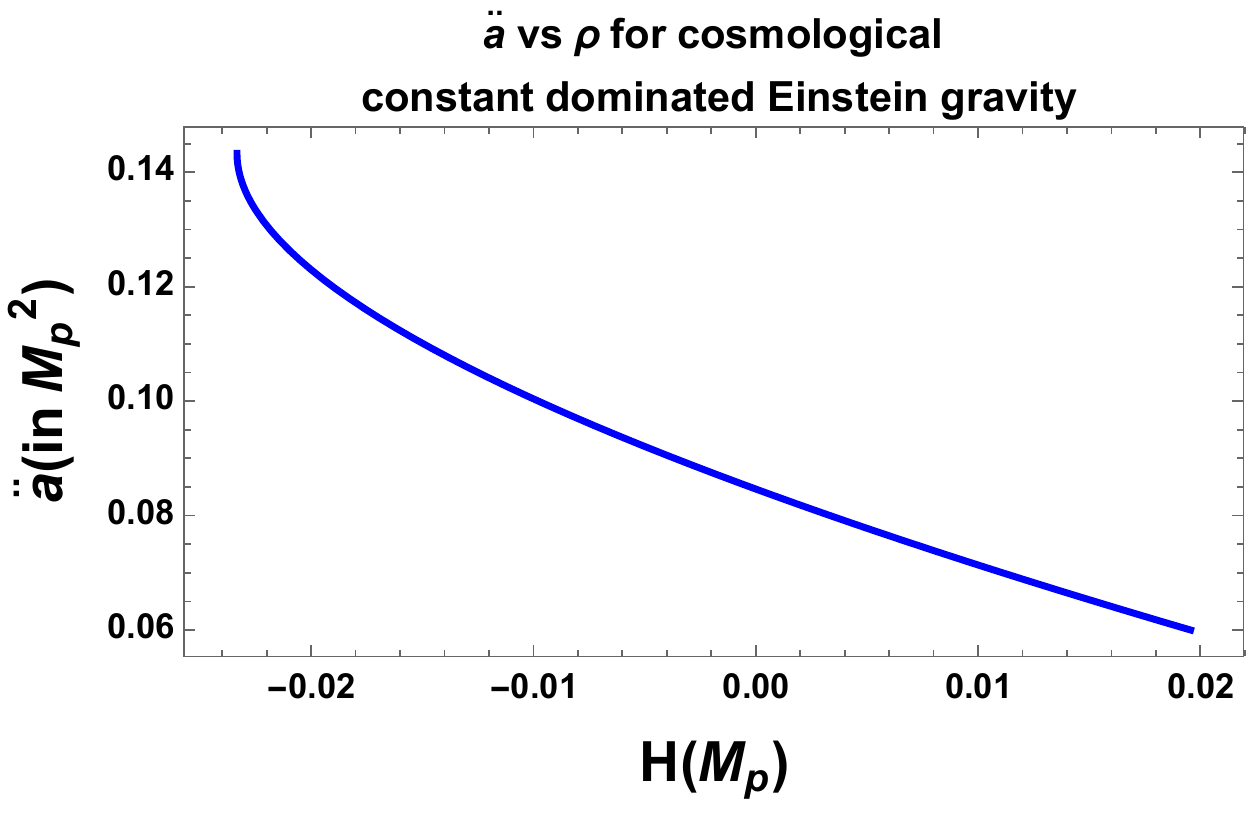}
    \label{lambda3}
}
\caption[Optional caption for list of figures]{ Graphical representation of the phenomena of bounce and acceleration for cosmological constant dominated Einstein gravity.} 
\label{fig3}
\end{figure*}

In Figs. \ref{lambda1} and \ref{lambda3}, we have shown the phenomena of bounce and acceleration in this cosmological constant dominated model. We can draw the following conclusions from the above figures:
\begin{itemize}
\item In Fig. \ref{lambda1}, we have plotted the r.h.s of Eqn. (\ref{constant1}) using the relation $\rho=a^{-3(1+w)}$. The graph has been plotted only for the case when $k=1$. This is because, Eq. (\ref{condbounce}) shows that density becomes negative for the cases when $k=0,-1$, which is unphysical. 
\item We have used $w=1/3$, since we require a soft equation of state for causing the acceleration and expansion. The case $w=0$ made $H$ only approximately zero.
\item From Fig. \ref{lambda1}, using the value of $\Lambda$ from Planck 2015 data \cite{Ade:2015xua}, we get the bounce at $\rho=\rho_{b}=0.43 M_{pl}^{4}$. This value increases if we take higher values of $\Lambda$.
\item Fig. \ref{lambda3} shows the necessary condition of acceleration ($\ddot{a}> 0$) at the time of bounce. Here we have plotted the r.h.s of Eqns. (\ref{constant1}) and (\ref{constant2}). This plot has also been obtained for the same parameter values as the earlier graph.
\item Thus Fig. \ref{fig3} shows graphically that the phenomenon of bounce is possible for a closed universe dominated by cosmological constant like term and having $w=1/3$.
\item Since the condition of bounce is true for all the three cases being studied in this section, we can say that bouncing universe is possible for all the three cases. 
\end{itemize}
\underline{\textbf{For case 2}}
\\ \\
Condition for acceleration in terms of the pressure and density is again given by Eq.~(\ref{constantaccel}). Then
substituting the expressions for $\rho$ and p from Eq.~(\ref{philambda2}), we get the condition
for acceleration in terms of the scalar field as: 
\begin{equation}
\dot{\phi}^{2} < 2\Lambda_{0}M^{2} + V(\phi) - \sum\limits_{i=1}^{4} \Lambda_{i}\phi^{i}
\end{equation}
Thus depending on the signs of constants ($\Lambda_{1},\ \Lambda_{2},\ \Lambda_{3},\ \Lambda_{4}$), 
we observe that less or more contribution from the potential than the canonical kinetic term as appearing in the standard
case may be required. To realize the essence of this statement let us consider an example, in which all the new constants are positive or the resultant sign of the summation is positive, then
this decreases the right hand side from standard case given by Eq.~(\ref{constantphi}) and
this condition for acceleration can be achieved only if the contribution from the potential is more than compared to the standard case.
Again if all the constants are negative or the resultant sign of the summation terms is negative, then we need
lesser contribution from the potential than the standard case to attain the condition of acceleration. Thus we see
that apart from the general cosmological constant term, the presence of different powers of scalar fields along with different
constants increases or decreases the effective potential at the time of bounce.
\\ \\
\underline{\textbf{For case 3}}
\\ \\
Condition for acceleration in terms of the pressure and density
is again given by Eq.~ (\ref{constantaccel}). Then substituting
the expressions for $\rho$ and p from Eqns. (\ref{dilaton2}),
we get the condiition for acceleration in terms of the scalar field as: 
\begin{equation}
\dot{\phi}^{2} < 2\Lambda_{0}M^{2} + V(\phi) - \Lambda e^{\phi/M_{p}}
\end{equation}
Thus if $\Lambda$ is positive,  then this decreases the right hand side from standard
case and this condition for acceleration can be achieved only if more
potential than the standard case is present. Again if $\Lambda$ is negative,
even potential lesser than the standard case can satisfy the condition of acceleration.

Therefore from the above analysis we can conclude that, whether the
standard potential will be able to cause acceleration at bounce
now depends on the extra terms present in the action.
\\
\subsection{Condition for turnaround}   
\underline{\textbf{For case 1, 2 ,3}}
\\ \\
The condition for turnaround is exactly same as the condition for bounce appearing in case 1 i.e. one can write:
\begin{equation}
\delta a_{max} = -\frac{1}{3M^{2}\left(k - \Lambda a_{max}^{2}\right)}\oint pdV
\label{constant5}
\end{equation}
Rest all the conclusions remain same as what we had got for the bounce case.
It is important to mention here that the conclusions for case 2 and case 3 also remains the same as appearing in the case 1.
\\

\subsection{Condition for deceleration}

\underline{\textbf{For case 1}}
\\ \\
From Eq.~ (\ref{constant2}), at turnaround the condition for deceleration is given by
\begin{equation}
\rho_{t} + 3p_{t} > 2\Lambda M^{2}
\label{constantdecel}
\end{equation}
i.e turnaround can be obtained without violating the strong energy condition.

In place of Eq.~(\ref{rk9xcxcz}), we get the conditions for deceleration at turnaround as:
\be\begin{array}{lll}\label{turnaround}
 \displaystyle p_{t} >\left\{\begin{array}{ll}
                    \displaystyle   \Lambda M^{2} - \frac{M^{2}}{a_{t}^{2}} &
 \mbox{\small {\bf for {$k=+1$}}}  \\ \\
         \displaystyle  \Lambda M^{2} + \frac{M^{2}}{a_{t}^{2}} & \mbox{\small {\bf for {$k=-1$}}}.
          \end{array}
\right.
\end{array}\ee
Substituting the expressions of $\rho$ and p from Eqns. (\ref{eq:scalar_rho}) into Eq.~ (\ref{constantdecel}), we get the condition for expansion in terms of the scalar field as
\begin{equation}
\dot\phi^{2} > V(\phi) + \Lambda M^{2}
\label{constantphi1}
\end{equation}
Thus we see that a potential lesser than the standard case is required to satisfy this condition because even if the condition $\phi > V(\phi)$ is satisfied, we will get acceleration only if $\phi >  V(\phi) + \Lambda M^{2}$. Thus, now the cosmological constant increases the effective potential relative to the standard case. 

\begin{figure*}[htb]
\centering
\subfigure[ An illustration of the turnaround condition for a universe with an equation of state w=1 and k=1.]{
    \includegraphics[width=7.2cm,height=8cm] {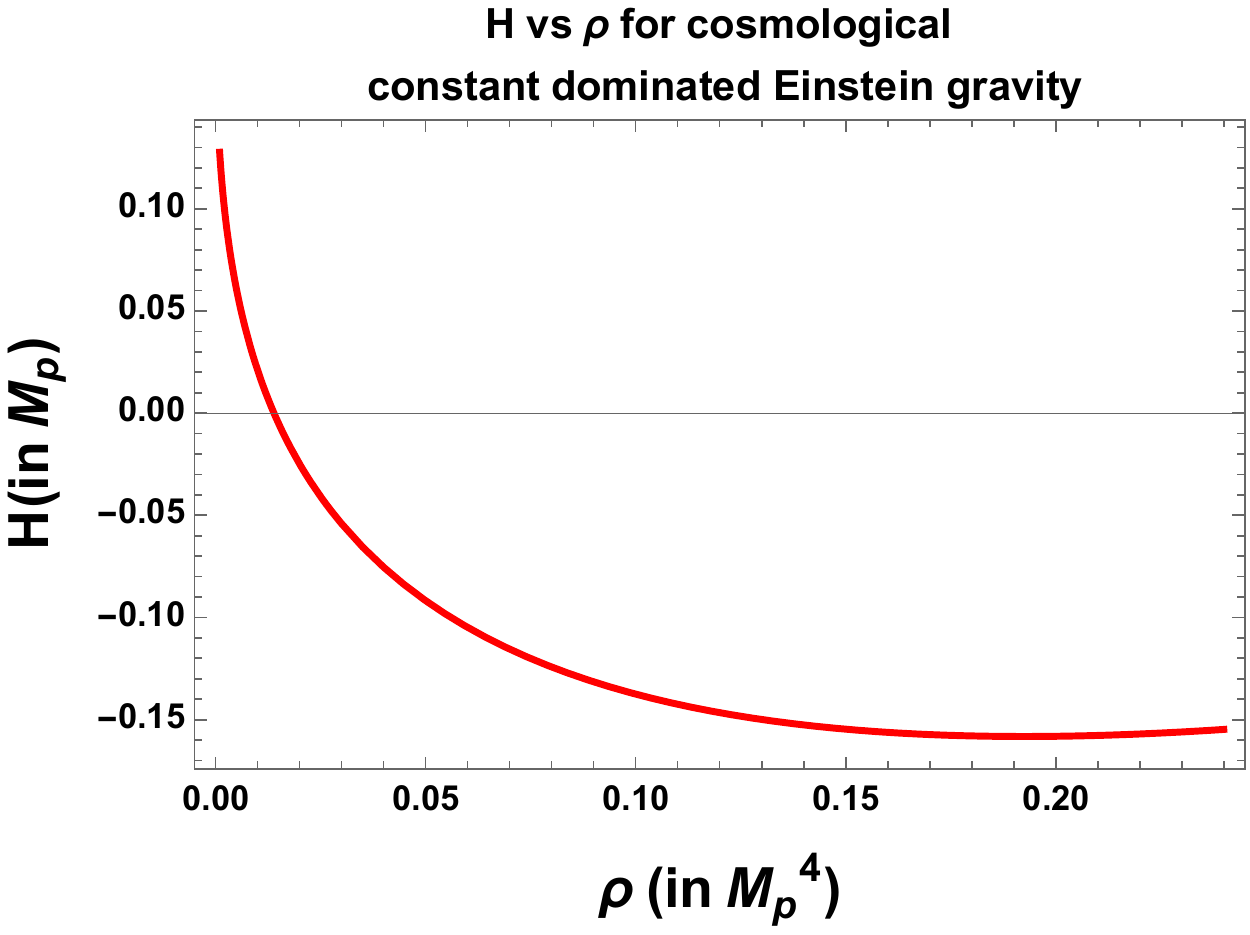}
    \label{lambda2}
}
\subfigure[An illustration of the deceleration condition at turnaround for a universe with an equation of state w=1, k=1.]{
    \includegraphics[width=7.2cm,height=8cm] {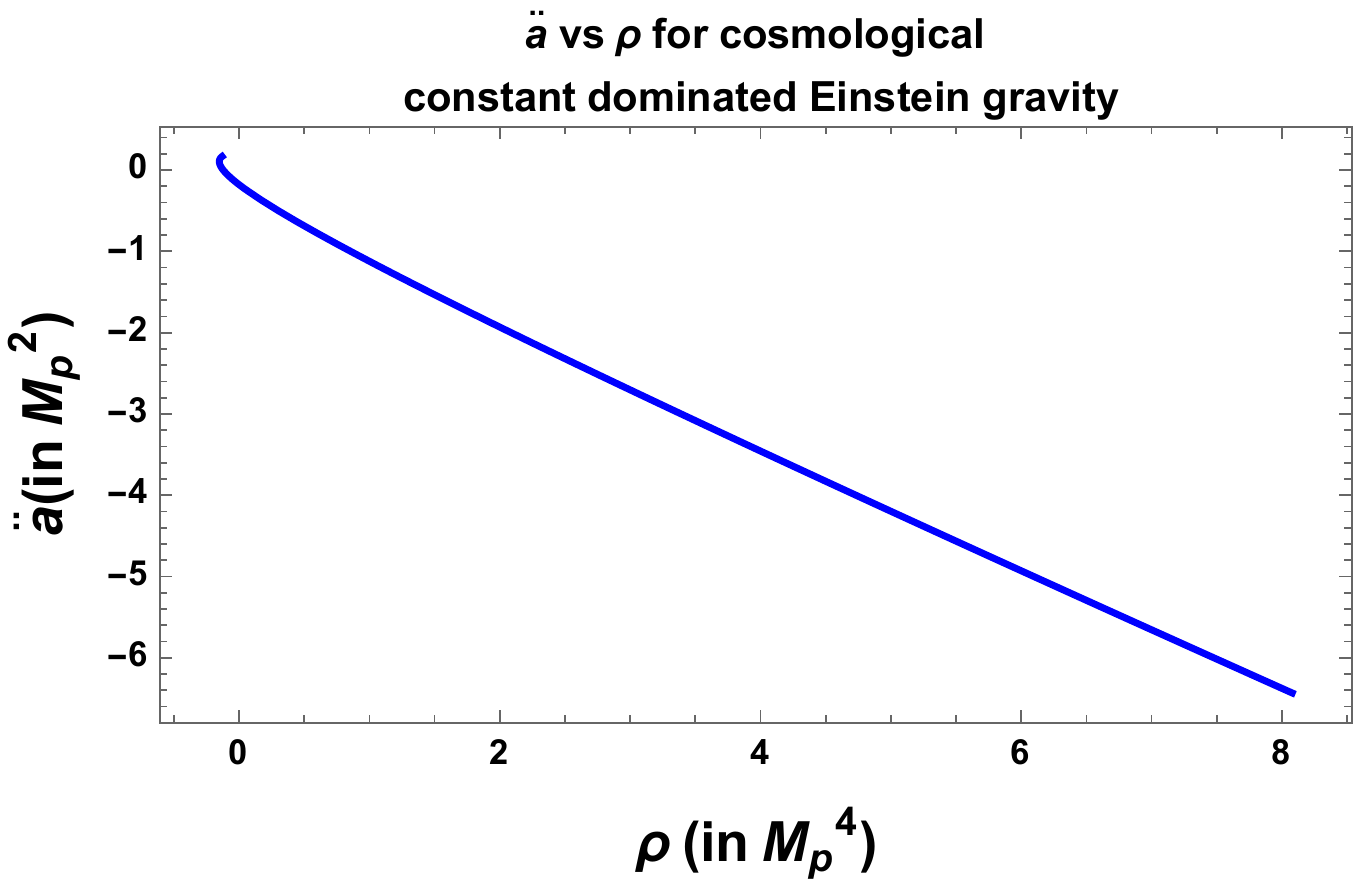}
    \label{lambda4}
}
\caption[Optional caption for list of figures]{ Graphical representation of the phenomena of turnaround and deceleration for cosmological constant dominated Einstein gravity.} 
\label{fig4}
\end{figure*}

In Figs. \ref{lambda2} and \ref{lambda4}, we have shown the phenomena of turnaround and deceleration in this cosmological constant dominated model. We can draw the following conclusions from the above figures:
\begin{itemize}
\item In Fig. \ref{lambda2}, we have once again plotted the r.h.s of Eq. (\ref{constant1}) using the relation $\rho=a^{-3(1+w)}$ for a closed universe. But for this case we have used $w=1$, since we require a stiff equation of state for causing contraction and deceleration.
\item In Fig. \ref{lambda2}, $H$ going to negative values may be interpreted as the universe changing its direction of motion at turnaround. As had been discussed in \cite{Kanekar:2001qd}, the condition of bounce/turnaround can be imposed by either the condition of making the scale factor changing sign with other quantities remaining same, or, $\dot{a}$ going to negative values with other quantities remaining same. 
\item From Fig. \ref{lambda2}, using the value of $\Lambda$ from Planck 2015 data \cite{Ade:2015xua}, we get the turnaround at $\rho=\rho_{t}=0.014 M_{pl}^{4}$. This value increases if we take higher values of $\Lambda$.
\item Fig. \ref{lambda3} shows the necessary condition of acceleration ($\ddot{a}< 0$) at the time of turnaround. Here we have plotted the r.h.s of Eqns. (\ref{constant1})and (\ref{constant2}). This plot has also been obtained for the same parameter values as the earlier graph.
\item Thus Fig. \ref{fig4} shows graphically that the phenomenon of turnaround is possible for a closed universe dominated by cosmological constant like term and having $w=1$.
\item Since the condition of turnaround is true for all the three cases being studied in this section, we can say that bouncing universe is possible for all the three cases. 
\end{itemize}
\underline{\textbf{For case 2}}
\\ \\
Condition for deceleration in terms of the pressure and density is 
given by Eq.~ (\ref{constantdecel}) with $\Lambda$ replaced by $\Lambda_{0}$ in this specific case.
Then substituting the expressions for density $\rho$ and pressure $p$ from Eq.~(\ref{philambda2}), we get the following 
condition for deceleration in terms of the scalar field as:
\begin{equation}\label{we}
\dot{\phi}^{2} > 2\Lambda_{0}M^{2} + V(\phi) - \sum\limits_{i=1}^{4} \Lambda_{i}\phi^{i}
\end{equation}
Thus depending on the signs of constants ($\Lambda_{1},\ \Lambda_{2},\ \Lambda_{3},\ \Lambda_{4}$), we see that
less or more contribution from the potential than the standard case may be required. 
To justify the validity of this statement let us consider an example, where
all the new constants are independently positive or the cumulative effect appearing through 
their summation is positive, then this decreases the right hand side of Eq.~(\ref{we})
from standard case and the condition for deceleration can be achieved even if more contribution from the 
potential plays crucial role compared to the standard case.  
On the other hand, if all the constants are independently negative or the cumulative effect appearing through 
their summation is negative, deceleration can be achieved even if the contribution from the potential is lesser than the standard case.
\\ \\
\underline{\textbf{For case 3}}
\\ \\
Condition for deceleration in terms of the pressure and density is again given by Eq.~(\ref{constantdecel}) with
$\Lambda$ replaced by $\Lambda_{0}$. Then substituting the expressions for density $\rho$ and pressure $p$
from Eq.~(\ref{dilaton2}), we get the condition for deceleration in terms of the scalar field as: 
\begin{equation}
\dot{\phi}^{2} > 2\Lambda_{0}M^{2} + V(\phi) - \Lambda e^{\phi/M_{p}}
\end{equation}
Thus if the signature of $\Lambda$ is positive, then this decreases the right hand side
from the standard case and this condition for acceleration can be achieved even if more contribution from the 
potential compared to the standard case is present. On the other hand, if $\Lambda$ is negative, we need lesser potential than the standard case to attain the condition of deceleration.
\\ \\
\underline{\textbf{For $k = 0$ case}}
 \\ \\
The results below have been shown only for case 1. But they perfectly hold good for case 2 and case 3 also.
In the present context the Friedmann equations for this case are given by:
\bea
H^{2} = \left(\frac{\dot{a}}{a}\right)^2 =\frac{\rho}{3M^{2}}  + \frac{\Lambda}{3},\\
\label{constant1new}
\dot{H}+H^2=\frac{\ddot a}{a} = -\frac{(\rho + 3p)}{6M^{2}} + \frac{\Lambda}{3}.
\label{constant2new}
\eea
From Eq.~(\ref{constant1new}), we can infer that the same Friedmann equation cannot give rise to both
bounce and turnaround condition for $k=0$ case. This is because $\Lambda$ is constant in the present setup, hence it is not possible for it to be equal to both the high density at the time
of bounce and low density at the time of turnaround. 

For causing bounce/turnaround the following condition holds good:
\begin{equation}
\rho_{b/t} = -\Lambda M^{2}
\end{equation}
Therefore in place of Eq.~(\ref{constant3}), in the present context we get:
\begin{equation}
\delta(a_{min/max})^{3} = \frac{1}{\Lambda M^{2}}\oint pdV
\end{equation}
Thus, the amplitude of the scale factor increases:
\begin{itemize}
 \item if $\oint pdV > 0$ for $\Lambda>0$. 
 
 \item if $\oint pdV < 0$ for $\Lambda<0$. 
\end{itemize}
Similarly the condition for acceleration/deceleration for $k=0$ case can be written as:
\begin{eqnarray}
p_{b} &<& \Lambda M^{2} \ {\rm (\bf for\ acceleration\ at\ bounce)}\\
p_{b} &>& \Lambda M^{2} \ {\rm (\bf for\ deceleration\ at\ turnaround)}
\end{eqnarray}
Thus we observe that the pressure at bounce/turnaround required to cause acceleration/deceleration becomes
independent of the scale factor at the time of bounce/turnaround. Also the results shows that it depends only on the cosmological constant.
\subsection{Evaluation of work done in one cycle} 
\underline{\textbf{For case 1}}
\\ \\
Following the same procedure as we have done for DGP model, here also the general expression for the hysteresis loop is given by Eq.~(\ref{hyst1}). In order to
express the loop in terms of the scale factor only, using Eq.~(\ref{constant1}) and Eq.~(\ref{constant2}), we get:
\begin{eqnarray}
\rho=\frac{\dot{\phi}^2}{2}+V(\phi) &=& 3M^{2}\left[\left(\frac{\dot{a}}{a}\right)^{2} + \frac{k}{a^{2}} - \frac{\Lambda}{3}\right], \\
H^2+\dot{H}=\frac{\ddot{a}}{a} &=& -\frac{(\rho + 3p)}{6M^{2}} + \frac{\Lambda}{3}
\end{eqnarray}
Further using the above equations and Eq.~(\ref{eq:scalar_rho}) for pressure $p$ we get:
\begin{equation}
\frac{\dot{\phi}^{2}}{2} - V(\phi) = M^{2}\left[\Lambda - \frac{2\ddot{a}}{a} - \left(\frac{\dot{a}}{a}\right)^{2} - \frac{k}{a^{2}}\right].
\label{hyst6}
\end{equation}
Therefore the expression for work done is given by:
\begin{equation}
\oint pdV = 3M^{2}\oint \left(\Lambda a^{2}\dot{a} - 2\ddot{a}\dot{a}a - \dot{a}^{3} - k\dot{a}\right)dt.
\label{constant4}
\end{equation}
It is important to note that the total work done in one cycle now depends on the cosmological constant also.
\\ \\
\underline{\textbf{For case 2}}
\\ \\
For case 2 using the expression of p given by Eq.~(\ref{philambda2}) into Eq.~(\ref{hyst1}), we get the expression for the total work done in one expansion-contraction cycle in terms of the $\phi$ and a as
\begin{eqnarray}
\oint pdV &=& \int_{cont} 3\left(\frac{1}{2}\dot{\phi}^{2} - V(\phi) + \sum\limits_{i=1}^{4} \Lambda_{i}\phi^{i}\right)a^{2}\dot{a}dt \nonumber \\ &&~~~~~~~~~~~~~~~~~+ \int_{exp}  3\left(\frac{1}{2}\dot{\phi}^{2} - V(\phi) + \sum\limits_{i=1}^{4} \Lambda_{i}\phi^{i}\right) a^{2}\dot{a}dt \nonumber \\ \\
 &=& \int_{a_{max}^{i-1}}^{a_{min}^{i-1}} 3\left(\frac{1}{2}\dot{\phi}^{2} - V(\phi) + \sum\limits_{i=1}^{4} \Lambda_{i}\phi^{i}\right)a^{2}\dot{a}dt  \nonumber \\ &&~~~~~~~~~~~~~~~~~+ \int_{a_{min}^{i-1}}^{a_{max}^{i}} 3\left(\frac{1}{2}\dot{\phi}^{2} - V(\phi) + \sum\limits_{i=1}^{4} \Lambda_{i}\phi^{i}\right) a^{2}\dot{a}dt \nonumber \\
\label{philambda3}
\end{eqnarray}
Since the Friedmann equations are same as that appearing in the earlier section, the expression for the above equation in terms
of the scale factor is again given by Eq.~(\ref{constant4}) in which $\Lambda$ is replaced by a new constant $\Lambda_{0}$. Thus from Eq.~(\ref{philambda3}) we find that the work 
done now depends on the value of all the new constants ($\Lambda_{1},\ \Lambda_{2},\ \Lambda_{3},\ \Lambda_{4}$) of the model. Thus by allowing sufficient amount of tuning in these constants
we can make the signature of the representative integral positive or negative.
\\ \\
\underline{\textbf{For case 3}}
\\ \\
For case 3 using the expression for pressure $p$ as given by Eq.~(\ref{dilaton2}), into Eq.~(\ref{hyst1}), we get the expression for the total work done in one expansion-contraction cycle in terms of the scalar field 
$\phi$ and scale factor $a$ as:
\begin{eqnarray}
\oint pdV &=& \int_{cont} 3\left(\frac{1}{2}\dot{\phi}^{2} - V(\phi) + \Lambda e^{\phi/M_{p}}\right)a^{2}\dot{a}dt \nonumber \\ &&~~~~~~~~~~~~~~~~~+ \int_{exp}  3\left(\frac{1}{2}\dot{\phi}^{2} - V(\phi) + \Lambda e^{\phi/M_{p}}\right) a^{2}\dot{a}dt \nonumber \\ \\
 &=& \int_{a_{max}^{i-1}}^{a_{min}^{i-1}} 3\left(\frac{1}{2}\dot{\phi}^{2} - V(\phi) + \Lambda e^{\phi/M_{p}}\right)a^{2}\dot{a}dt  \nonumber \\ &&~~~~~~~~~~~~~~~~~+ \int_{a_{min}^{i-1}}^{a_{max}^{i}} 3\left(\frac{1}{2}\dot{\phi}^{2} - V(\phi) + \Lambda e^{\phi/M_{p}}\right) a^{2}\dot{a}dt \nonumber \\
\label{dilaton3}
\end{eqnarray}
In this case also, the expression for the above equation in terms of the scale factor is again given by Eq.~(\ref{constant4}) where $\Lambda$ is replaced by the constant 
$\Lambda_{0}$. Thus adjusting the values and signatures of the constants $(\Lambda_{0},\Lambda)$ we can change the signature of the representative integral.

\subsection{Semi-analytical analysis for cosmological potentials}
Here also we will perform the analysis for the case when $k=0$. Though, the exact and complete treatment would have been for the case $k=1$, simple analytical solutions were obtained only for flat universe which has been discussed below.  Also, since we know that for $k=0$, this model can
cause either bounce or turnaround, hence to get an expression for the complete work done in one cycle,, we
will consider that cosmological constant is present at late times (i.e causing turnaround) whereas the early
universe is governed by brane world cosmology model for which the analysis has been shown in the Appendix in the context of Randall-Sundrum single brane world (RSII). Now,
if we compare Eq.~(\ref{bouncenew}) with Eq.~(\ref{lqcnew}), we see that they are both same. Hence the results
for early universe which will be calculated for loop quantum model, can be used here. Below, we show the late
time analysis for $\Lambda$CDM. In this context, we have denoted the Planck mass by $M_{p}$.

\subsubsection{Case I: Hilltop potential}
\underline{\textbf{{For Case 1:}}}
\\ \\ \\
\textbf{A. Expansion}
\\ \\ \\
At late times within the window $a'<a< a_{max}$ using Eq.~(\ref{constant1}) and Eq.~(\ref{modeqn1}) with the expression for density $\rho$ now given by the hilltop potential, we get an 
integral equation of the following form:
\begin{equation}
\int d\left(\frac{\phi}{M_{p}}\right) \frac{\sqrt{\left[\frac{V_{0}\left(1+\beta\left(\frac{\phi}{M_{p}}\right)^{p}\right)}{3M_{p}^{2}}+\frac{\Lambda}{3}\right]}}{\beta p\left(\frac{\phi}{M_{p}}\right)^{p-1}}
= -\frac{V_{0}}{3M_{p}^{2}}\int dt.
\label{lambdaexpand}
\end{equation} 
The exact solution of this integral is given in the Appendix for RSII model. To compute the left hand side of the above integral equation we again follow the same procedure as we done for DGP model i.e.
we redefine the field as: \be \frac{\phi}{M_{p}}=e^{\lambda}\ee and hence using this new definition we study two limiting cases.
\\ \\
\underline{\bf a)$\phi/M_{p}<<1$:} \\ \\
In this limit, we can expand the exponentials upto linear order and upon further simplification we get the following expression for the redefined field $\lambda$ as:
\begin{equation}
\lambda=\left(-\frac{V_{0}}{3M_{p}^{2}}t+B_{0}\right)\frac{\beta p}{\sqrt{\left(\frac{V_{0}}{3M_{p}^{2}}(1+\beta)+\frac{\Lambda}{3}\right)}}
\label{potential1}
\end{equation}
where $B_{0}$ is the arbitrary integration constant given by:
\begin{equation}
B_{0}=\frac{\lambda_{f}}{\beta p}\sqrt{\left(\frac{V_{0}}{3M_{p}^{2}}(1+\beta)+\frac{\Lambda}{3}\right)}+\frac{V_{0}t_{f}}{3M_{p}^{2}}.
\end{equation}
Here $\lambda_{f}$ is the value at the time of turnaround i.e at $t=t_{f}$.

The above solution has been obtained under the assumption that the quantity \be \frac{\frac{V_{0}}{3M_{p}^{2}}}{\left(\frac{V_{0}}{3M_{p}^{2}}(1+\beta)+\frac{\Lambda}{3}\right)}\beta p\lambda <<1,\ee which 
is possible if we choose the value of $V_{0}$ and $\Lambda$ accordingly.

Substituting this expression back into the Friedmann equation, we get the expression for the scale factor as
\bea
a(t)=B_{1}\exp\left[\left(\sqrt{\left(\frac{V_{0}}{3M_{p}^{2}}(1+\beta)+\frac{\Lambda}{3}\right)}+\frac{\frac{V_{0}}{3M_{p}^{2}}\beta^{2}p^{2}}{2\left(\frac{V_{0}}{3M_{p}^{2}}(1+\beta)+\frac{\Lambda}{3}\right)}B_{0}\right)t\nonumber
\right.\\ \left.~~~~~~~~~~~~~~~~~~~~~~~~~~~~~~~-\frac{\left(\frac{V_{0}}{3M_{p}^{2}}\right)^{2}\beta^{2}p^{2}}{4\left(\frac{V_{0}}{3M_{p}^{2}}(1+\beta)+\frac{\Lambda}{3}\right)}\frac{t^{2}}{2}\right]
\label{scalefactor1}
\eea
where
\bea
B_{1}=a_{f}\exp\left[-\left(\sqrt{\left(\frac{V_{0}}{3M_{p}^{2}}(1+\beta)+\frac{\Lambda}{3}\right)}+\frac{\frac{V_{0}}{3M_{p}^{2}}\beta^{2}p^{2}}{2\left(\frac{V_{0}}{3M_{p}^{2}}(1+\beta)+\frac{\Lambda}{3}\right)}B_{0}\right)t_{f}\nonumber
\right.\\ \left.~~~~~~~~~~~~~~~~~~~~~~~~~~~~~~~+\frac{\left(\frac{V_{0}}{3M_{p}^{2}}\right)^{2}\beta^{2}p^{2}}{4\left(\frac{V_{0}}{3M_{p}^{2}}(1+\beta)+\frac{\Lambda}{3}\right)}\frac{t_{f}^{2}}{2}\right].
\eea
Here $a_{f}$ is the value of scale factor at turnaround time scale $t=t_{f}$.
 \\ \\ \\ \\
\underline{\bf ii) $\phi/M_{p}>>1$:}\\ \\ 
In this limit, the solution for the redefined field $\lambda$ is given by:
\begin{equation}
\lambda=\frac{2}{p}\left[2-\ln\left(\left(\frac{V_{0}t}{3M_{p}^{2}}+B_{2}\right)\frac{\beta^{1/2}p^{2}}{2\left(\frac{V_{0}t}{3M_{p}^{2}}\right)^{1/2}}\right)\right]
\label{potential2}
\end{equation} 
where $B_{2}$ is the arbitrary integration constant given by:
\begin{equation}
B_{2}=\frac{2\left(\frac{V_{0}t}{3M_{p}^{2}}\right)^{1/2}e^{(2-\frac{p}{2})\lambda_{f}}}{\beta^{1/2}p^{2}}-\frac{V_{0}t_{f}}{3M_{p}^{2}}.
\end{equation}
The above solution has been obtained under the assumption that the conditions: \bea \beta e^{p\lambda}&>>&1,\\  \frac{V_{0}\beta e^{p\lambda}}{3M_{p}^{2}}&>>&\frac{\Lambda}{3}, \eea are satisfied.
These conditions can be satisfied since we are in the large field limit and we can choose the values of the parameters of this model $(\beta, p, V_{0})$ accordingly.

Substituting the above expression into the Friedmann equation, we get the expression for the scale factor as:
\begin{equation}
a(t)=B_{3}exp\left[\left(\frac{V_{0}}{3M_{p}^{2}}\right)^{1/2}\beta^{1/2}e^{2}\frac{\ln\left(\frac{\beta^{1/2}p^{2}B_{2}}{2\left(\frac{V_{0}}{3M_{p}^{2}}\right)^{1/2}}
+\frac{\beta^{1/2}p^{2}\left(\frac{V_{0}}{3M_{p}^{2}}\right)^{1/2}t}{2}\right)}{\left(\frac{\beta^{1/2}p^{2}\left(\frac{V_{0}}{3M_{p}^{2}}\right)^{1/2}}{2}\right)}\right],
\label{scalefactor2}
\end{equation}
where $B_{3}$ is the arbitrary integration constant given by:
\begin{equation}
B_{3}=a_{f}exp\left[-\left(\frac{V_{0}}{3M_{p}^{2}}\right)^{1/2}\beta^{1/2}e^{2}\frac{\ln\left(\frac{\beta^{1/2}p^{2}B_{2}}{2\left(\frac{V_{0}}{3M_{p}^{2}}\right)^{1/2}}
+\frac{\beta^{1/2}p^{2}\left(\frac{V_{0}}{3M_{p}^{2}}\right)^{1/2}t_{f}}{2}\right)}{\left(\frac{\beta^{1/2}p^{2}\left(\frac{V_{0}}{3M_{p}^{2}}\right)^{1/2}}{2}\right)}\right].
\end{equation}
\\ \\ \\ \\ \\ \\
\textbf{B. Contraction}
\\ \\
This phase is independent of any choice of the potential provided the condition $\dot{\phi}^{2}>>V(\phi)$ holds good in this phase.
Using Eq.~(\ref{modeqn}) and Eq.~(\ref{constant1}), we get the following solutions:
\begin{equation}
\dot{\phi}^{2}=\left[\frac{\Lambda B_{4}}{3}\left(\frac{e^{\sqrt{2\Lambda}t}+1}{e^{\sqrt{2\Lambda}t}-1}\right)^{2}-\frac{\Lambda}{3}\right],
\label{kinetic}
\end{equation}
where $B_{4}$ is the arbitrary integration constant given by:
\begin{equation}
B_{4}=\left(\dot{\phi}_{f}^{2}+\frac{\Lambda}{3}\right)\frac{3}{\Lambda}\left(\frac{e^{\sqrt{2\Lambda}t_{f}}-1}{e^{\sqrt{2\Lambda}t_{f}}+1}\right)^{2}.
\end{equation}
From the above expressions, we get the solution of the scale factor as:
\begin{equation}
a(t)=B_{5}\left[\frac{(e^{\sqrt{2\Lambda}t}-1)^{2}}{e^{\sqrt{2\Lambda}t}}\right]^{\frac{1}{3M_{p}}\sqrt{\frac{\Lambda B_{4}}{2}}},
\label{scalefactor3}
\end{equation}
where $B_{5}$ is the arbitrary integration constant given by:
\begin{equation}
B_{5}=a_{f}\left[\frac{(e^{\sqrt{2\Lambda}t_{f}}-1)^{2}}{e^{\sqrt{2\Lambda}t_{f}}}\right]^{-\frac{1}{3M_{p}}\sqrt{\frac{\Lambda B_{4}}{2}}}.
\end{equation}
\\ \\
\textbf{C. Expression for work done}
\\ \\
As will be discussed in the next section, that the expression for integral for work done in the early universe for loop quantum model contains a large no. of terms, hence will not be shown here. But the results of the integral corresponding to late universe have been discussed below.
\\ \\
\underline{\bf i) $\phi/M_{p}<<1$:}
\\ \\
\begin{eqnarray}
\oint pdV &=& b_{1}\left({\rm erf}[-b_{2}+b_{3}t_{max}]-{\rm erf}[-b_{2}+b_{3}t']\right)+3M_{p}^{2}(45\sinh(\sqrt{\Lambda}t_{max})\nonumber \\ &&~~~~~~~~~~~~-9\sinh(2\sqrt{\Lambda}t_{max})+\sinh(3\sqrt{\Lambda}t_{max})
-45\sinh(\sqrt{\Lambda}t')\nonumber\\ &&~~~~~~~~~~~~+9\sinh(2\sqrt{\Lambda}t')-\sinh(3\sqrt{\Lambda}t')-30t_{max}+30t')
\end{eqnarray}
\\ \\
\underline{\bf ii) $\phi/M_{p}>>1$:}
\\ \\
\begin{eqnarray}
\oint pdV &=& B_{3}^{3}\frac{\sqrt{\frac{\pi}{2}}(-3+\Lambda)({\rm Erfi}(\sqrt{3b_{5}}(b_{6}+b_{7}t_{max}))-{\rm Erfi}(\sqrt{3b_{5}}(b_{6}+b_{7}t_{min}))}{2b_{7}}\nonumber \\&&~~~~~~~+3M_{p}^{2}(45\sinh(\sqrt{\Lambda}t_{max})\nonumber \\ &&~~~~~~~~~~~~-9\sinh(2\sqrt{\Lambda}t_{max})+\sinh(3\sqrt{\Lambda}t_{max})-45\sinh(\sqrt{\Lambda}t')
\nonumber \\ &&~~~~~~~~~~~~+9\sinh(2\sqrt{\Lambda}t')-\sinh(3\sqrt{\Lambda}t')-30t_{max}+30t')
\end{eqnarray}
Here $b_{1}...b_{7}$ are constants that depend on the parameters present in the expression for the scale factor. Their explicit forms have been given in the appendix.

Including the contributions from the other integrals also we see that we get a non-zero work done for both large and small field cases
for hilltop potential when we have pure cosmological constant in the background.
\\ \\
\underline{\textbf{{For Case 2}}}\\ \\
\textbf{A. Expansion}\\ \\
Within the window $a_{min}<a<a_{max}$ at late times using Eq.~(\ref{philambda}) and Eq.~(\ref{philambda2}) along 
with energy density of scalar field, $\rho$, now given by the hilltop potential, we get an integral equation of the form (neglecting $\ddot{\phi}$):
\begin{equation}
\int d\left(\frac{\phi}{M_{p}}\right) \frac{\sqrt{\left[\frac{V_{0}\left(1+\beta\left(\frac{\phi}{M_{p}}\right)^{p}\right)-\sum_{i=1}^{4}
\Lambda_{i}(M_{p})^{i}(\frac{\phi}{M_{p}})^{i}}{3M_{p}^{2}}
+\frac{\Lambda_{0}}{3}\right]}}{\left[-\beta p\left(\frac{\phi}{M_{p}}\right)^{p-1}+\sum_{i=1}^{4}\Lambda_{i}
M_{p}^{i-1}(\frac{\phi}{M_{p}})^{i-1}\right]} = \frac{1}{3M_{p}}\int dt.
\label{scalarhilltop}
\end{equation} 
For the sake of clarity, we again follow the same procedure as we have done for DGP model i.e. redefine the field variable,
as: \be \frac{\phi}{M_{p}}=e^{\lambda}.\ee
Simplified analytical expressions can only be obtained for the small field limit given by $\phi/M_{p}<<1$, so that we will concentrate in 
this specific physical situation for this case. In this limit, we can expand the exponentials upto linear order and upon further
simplification we get the following expression for $\lambda$ as:
\begin{equation}
\lambda=\frac{\left(\frac{t}{3M_{p}}+B_{14}\right)}{\sqrt{J}/K}
\label{potential4}
\end{equation}
where the arbitrary integration constants $J$, $K$ and $B_{14}$ is given by:
\bea
J&=&\frac{V_{0}}{3M_{p}^{2}}+\frac{\Lambda_{0}}{3}+\frac{V_{0}\beta}{3M_{p}^{2}}
-\left(\frac{\Lambda_{1}}{3M_{p}}+\frac{\Lambda_{2}}{3}+\frac{\Lambda_{3}M_{p}}{3}+\frac{\Lambda_{4}M_{p}^{2}}{3}\right),\\
K&=&-\frac{V_{0}}{M_{p}}p\beta +\Lambda_{1}+2\Lambda_{2}M_{p}+3\Lambda_{3}M_{p}^{2}+4\Lambda_{4}M_{p}^{3},\\
B_{14}&=&\lambda_{f}(\sqrt{J}/K)-\frac{t_{f}}{3M_{p}}.
\eea
Here $\lambda_{f}$ is the value at the time of turnaround i.e. at the time scale $t=t_{f}$.

The above solution has been obtained under the assumption that the conditions: \bea \frac{J\lambda}{M}&<<&1,\\ \frac{N\lambda}{K}&<<&1,\eea 
are satisfied. This is possible because we are in the small $\lambda$ limit and we can choose the values of the other constants accordingly. Here
we introduce two new constants $M$ and $K$ defined as:
\begin{equation}
M=\frac{V_{0}p\beta}{3M_{p}^{2}}-\left(\frac{\Lambda_{1}}{3M_{p}}+\frac{2\Lambda_{2}}{3}+\frac{3\Lambda_{3}M_{p}}{3}+\frac{4\Lambda_{4}M_{p}^{2}}{3}\right)
\end{equation} 
\begin{equation}
K=-\frac{V_{0}}{M_{p}}\beta p(p-1)+2\Lambda_{2}M_{p}+9\Lambda_{3}M_{p}^{2}+12\Lambda_{4}M_{p}^{3}
\end{equation}

Further substituting this expression back into the Friedmann equations, we get the following expression for the scale factor as:
\begin{equation}
a(t)=B_{15}\exp\left[\sqrt{J}\left\{t\left(1+\frac{B_{14}KM}{6M_{p}J^{3/2}}\right)+\frac{t^{2}}{2}\left(\frac{KM}{6M_{p}J^{3/2}}\right)\right\}\right],
\label{scalefactor4}
\end{equation}
where $B_{15}$ is the arbitrary integration constant given by:
\begin{equation}
B_{15}=a_{f}\exp\left[-\sqrt{J}\left\{t_{f}\left(1+\frac{B_{14}KM}{6M_{p}J^{3/2}}\right)+\frac{t_{f}^{2}}{2}\left(\frac{KM}{6M_{p}J^{3/2}}\right)\right\}\right]
\end{equation} 
Here $a_{f}$ is the value of scale factor at turnaround time scale $t=t_{f}$.\\ \\
\textbf{B. Contraction}
\\ \\
The conclusion remains same as for $\Lambda$CDM model, because in this case all the extra terms containing $\phi$, will be neglected at leading order approximation.
\\ \\
\textbf{C. Expression for work done}
\\ \\
The solution of the scale factor (i.e its time dependence) is same as that for hilltop potential for constant $\Lambda$ case. Hence the expression for work done is same except that the constant will now depend on the parameters of this model. Thus the phenomenon of hysteresis also holds true for this model.
\\ \\
\underline{\textbf{{For Case 3}}}
\\ \\
\textbf{A. Expansion}
\\ \\
At late times within the window $a'<a<a_{max}$, using Eq.~(\ref{dilaton1}) and Eq.~(\ref{dilaton2})
with scalr field density $\rho$ now given by the hilltop potential, we get an integral equation of the following form (neglecting $\ddot{\phi}$):
\begin{equation}
\int d\left(\frac{\phi}{M_p}\right)~\frac{\sqrt{\left(\frac{V_{0}\left(1+\beta\left(\frac{\phi}{M_{p}}\right)^{p}\right)-\Lambda e^{\phi/M_{p}}}{3M_{p}^{2}}
+\frac{\Lambda_{0}}{3}\right)}}{\left[-\frac{V_{0}}{M_{p}}\beta p\left(\frac{\phi}{M_{p}}\right)^{p-1}
+\Lambda e^{\phi/M_{p}}\right]} = \frac{1}{3M_{p}}\int dt.
\label{scalardilaton}
\end{equation} 
In order to simplify the above integral for computational purpose,
we follow the same procedure as we have already done for DGP model i.e. redefine the field by using:
\be \frac{\phi}{M_{p}}=e^{\lambda}.\ee
Simplified solutions were possible only for small field case i.e. for $\phi/M_{p}<<1$, which has been discussed below.

In this limit, we can expand the exponentials upto linear order and upon further
simplification we get the expression for $\lambda$ as
\begin{equation}
\lambda=\left(\frac{t}{3M_{p}}+B_{18}\right)\frac{(\frac{-V_{0}\beta p}{M_{p}}+\frac{\Lambda e}{M_{p}})}{\sqrt{\frac{V_{0}}{3M_{p}^{2}}+\frac{V_{0}\beta}{3M_{p}^{2}}-\frac{\Lambda e}{3M_{p}^{2}}+\frac{\Lambda_{0}}{3}}}
\end{equation}
where $B_{18}$ is the arbitrary integration constant given by:
\begin{equation}
B_{18}=\frac{\lambda_{f}\sqrt{\frac{V_{0}}{3M_{p}^{2}}+\frac{V_{0}\beta}{3M_{p}^{2}}-\frac{\Lambda e}{3M_{p}^{2}}+\frac{\Lambda_{0}}{3}}}{(\frac{-V_{0}\beta p}{M_{p}}+\frac{\Lambda e}{M_{p}})}-\frac{t_{f}}{3M_{p}}
\end{equation}
Here $\lambda_{f}$ is the value at the time of turnaround i.e at the time scale $t=t_{f}$.

The above solutions have been obtained under the assumption that the conditions:
\bea \frac{(V_{0}\beta p-\Lambda e)\lambda}{(V_{0}+V_{0}\beta-\Lambda e+\Lambda_{0}M_{p}^{2})}&<<&1,\\
\frac{(\left(\Lambda e-V_{0}\beta p(p-1)\right)/M_{p})\lambda}{((-V_{0}\beta p+\Lambda e)/M_{p})}&<<&1\eea
are satisfied, which is possible if we choose the values of the constants accordingly.

Further substituting this expression back into the Friedmann equation, we get the following expression for the scale factor as:
\begin{equation}
a(t)=B_{19}\exp\left[\sqrt{O'}\left(\left(1+\frac{P'B_{18}}{2O'^{3/2}}\right)t+\frac{P't^{2}}{12M_{p}O'^{3/2}}\right)\right]
\label{scalefactor5}
\end{equation}
where $B_{19}$ is the arbitrary integration constant given by:
\begin{equation}
B_{19}=a_{f}exp\left[-\sqrt{O'}\left(\left(1+\frac{P'B_{18}}{2O'^{3/2}}\right)t_{f}+\frac{P't_{f}^{2}}{12M_{p}O'^{3/2}}\right)\right]
\end{equation}
in which we introduce two new symbols given by:
\bea
O'&=& \frac{V_{0}}{3M_{p}^{2}}+\frac{V_{0}\beta}{3M_{p}^{2}}-\frac{\Lambda e}{3M_{p}^{2}}+\frac{\Lambda_{0}}{3},\\
P'&=&\left(\frac{V_{0}\beta p-\Lambda e}{3M_{p}^{2}}\right)\left(-\frac{V_{0}\beta p}{M_{p}}+\frac{\Lambda e}{M_{p}}\right).
\eea
Here $a_{f}$ is the value of scale factor at turnaround.
\\ \\ \\ \\
\textbf{B. Contraction}
\\ \\
Solutions remain same as obtained before. So the conclusion also remain unchanged for contraction phase of the Universe.
 \\ \\
 \textbf{C. Expression for work done}
 \\ \\
The solution of the scale factor (i.e its time dependence) is
same as that obtained for hilltop potential. Hence the expression for work
done is exactly same except that the constant will now depend on the
parameters of this model. Thus the phenomenon of hysteresis also holds good for this model.
\\ \\

\subsubsection{Case II: Natural potential}

\underline{\textbf{{For Case 1:}}}
\\ \\
\textbf{A. Expansion}
\\ \\
At late times within the window $a'< a< a_{max}$, using Eq.~(\ref{modeqn1}) and Eq.~(\ref{constant1}), we get the following 
integral equation of the form:
\begin{equation}
\int d\left(\frac{\phi}{f}\right)~\frac{\sqrt{\left(\frac{V_{0}}{3M_{p}^{2}}+\frac{\Lambda}{3}+\frac{V_{0}}{3M_{p}^{2}}\cos\left(\frac{\phi}{f}\right)\right)}}{
\sin\left(\frac{\phi}{f}\right)}=\frac{V_{0}}{3f^{2}}\int dt.
\label{lambdanatural}
\end{equation}
The exact solution of the left hand side of the above equation is given in the Appendix for RSII model. 
For the sake of simplicity, here we study solutions for two limiting physical situations.
\\ \\
\underline{\bf i) $\phi/f<<1$:}
\\ \\
Using the small argument approximations for trigonometric functions, we get the expression for scalar field $\phi$ as:
\begin{equation}
\frac{\phi(t)}{f}=B_{6}\exp\left[\frac{V_{0}t}{3f^{2}\left(1+\frac{\frac{V_{0}}{3M_{p}^{2}}}{\frac{V_{0}}{3M_{p}^{2}}
+\frac{\Lambda}{3}}\right)^{1/2}\left(\frac{V_{0}}{3M_{p}^{2}}+\frac{\Lambda}{3}\right)^{1/2}}\right],
\label{potential6}
\end{equation}
where $B_{6}$ is the arbitrary integration constant given by:
\begin{equation}
B_{6}=\frac{\phi_{F}}{f}exp\left[-\frac{V_{0}t_{F}}{3f^{2}\left(1+\frac{\frac{V_{0}}
{3M_{p}^{2}}}{\frac{V_{0}}{3M_{p}^{2}}+\frac{\Lambda}{3}}\right)^{1/2}\left(\frac{V_{0}}{3M_{p}^{2}}+\frac{\Lambda}{3}\right)^{1/2}}\right].
\end{equation}
Here $\phi_{F}$ is the value of the scalar field at turnaround corresponding to $t=t_{F}$.

Further substituting the expression for $\phi$ into the Friedmann equation, we get the following expression for the scale factor as:
\begin{equation}
a(t)=B_{7}\exp\left[\left(1+\frac{\frac{V_{0}}{3M_{p}^{2}}}{\frac{V_{0}}{3M_{p}^{2}}+\frac{\Lambda}{3}}\right)^{1/2}\left(\frac{V_{0}}{3M_{p}^{2}}+\frac{\Lambda}{3}\right)^{1/2}t\right]
\label{scalefactor6}
\end{equation}
where $B_{7}$ is the arbitrary integration constant given by:
\begin{equation}
B_{7}=a_{F}\exp\left[-\left(1+\frac{\frac{V_{0}}{3M_{p}^{2}}}{\frac{V_{0}}{3M_{p}^{2}}+\frac{\Lambda}{3}}\right)^{1/2}\left(\frac{V_{0}}{3M_{p}^{2}}+\frac{\Lambda}{3}\right)^{1/2}t_{F}\right]
\end{equation}
Here $a_{F}$ is the value of the scale factor at turnaround.
\\ \\
\underline{\bf ii) $\phi/f>>1$:}\\ \\
Since we know that for large argument $\phi/f>>1$, value of cosine function is very small, hence using this concept and the following constraint condition:
\be {\rm cosec}^{2}\left(\frac{\phi}{f}\right)>>\frac{\frac{V_{0}}{3M_{p}^{2}}\cos\left(\frac{\phi}{f}\right)}{2\left(\frac{V_{0}}{3M_{p}^{2}}
+\frac{\Lambda}{3}\right)\sin^{2}\left(\frac{\phi}{f}\right)},\ee
we get the solution for the scalar field $\phi$ as:
\begin{equation}
\frac{\phi(t)}{f}=2\tan^{-1}\left[\left(\frac{V_{0}t}{3M_{p}^{2}}+B_{8}\right)\frac{1}{\sqrt{\frac{V_{0}}{3M_{p}^{2}}+\frac{\Lambda}{3}}}\right],
\label{potential7}
\end{equation}
where $B_{8}$ is the arbitrary integration constant given by:
\begin{equation}
B_{8}=\tan\left(\frac{\phi_{F}}{2f}\right)\left(\sqrt{\frac{V_{0}}{3M_{p}^{2}}+\frac{\Lambda}{3}}-\frac{V_{0}t_{F}}{3M_{p}^{2}}\right).
\end{equation}
Here $\phi_{F}$ is the value of the scalar field at turnaround.

Further using the expression for $\phi$ in the Friedmann equation, we get the following expression for the scale factor as:
\begin{eqnarray}
a(t)&=& B_{9}\exp\left[\left(\frac{V_{0}}{3M_{p}^{2}}+\frac{\Lambda}{3}\right)^{1/2}G(t)
\right],\nonumber\\
\label{scalefactor7}
\end{eqnarray}
where $G(t)$ is defined as:
\bea
G(t)&=&\left(t-\frac{\frac{\frac{V_{0}}{3M_{p}^{2}}}{2\left(\frac{V_{0}}{3M_{p}^{2}}+\frac{\Lambda}{3}\right)}
\left(B_{8}+\frac{V_{0}t}{3M_{p}^{2}}\right)-\frac{\frac{V_{0}}{3M_{p}^{2}}}{\left(\frac{V_{0}}{3M_{p}^{2}}+\frac{\Lambda}{3}\right)^{1/2}}
\tan^{-1}\left(\frac{B_{8}+\frac{V_{0}t}{3M_{p}^{2}}}{\left(\frac{V_{0}}{3M_{p}^{2}}+\frac{\Lambda}{3}\right)^{1/2}}\right)}{\left(\frac{V_{0}}{3M_{p}^{2}}\right)}\right),~~~~~
\eea
and $B_{8}$ is the arbitrary integration constant given by:
\begin{eqnarray}
B_{9}&=& a_{F}\exp\left[-\left(\frac{V_{0}}{3M_{p}^{2}}+\frac{\Lambda}{3}\right)^{1/2}G(t_{F})\right].\nonumber\\
\end{eqnarray}
Here $a_{F}$ is the value of the scale factor at turnaround.
\\ \\
\textbf{B. Contraction}
\\ \\
This phase is independent of any choice of the potential provided the condition $\dot{\phi}^2>>V(\phi)$ holds good in this phase.
This implies that the final conclusion remains same as obtained for previous hilltop potential in the background of $\Lambda$CDM model.
\\ \\
\textbf{C. Expression for work done}
\\ \\
As will be discussed in the next section, that the expression for representative integral for work done in
the early universe for loop quantum model contains a large number of terms, hence will
not be explicitly shown in the present context. But the results of the integral corresponding to the late universe have been discussed below.
\\ \\
\underline{\bf i) $\phi/f<<1$:}\\ \\
In this limit, the total work done is given by:
\begin{eqnarray}
\oint pdV &=& b_{8}(e^{3b_{9}t'}-e^{3b_{9}t_{max}})+3M_{p}^{2}(45\sinh(\sqrt{\Lambda}t_{max})-9\sinh(2\sqrt{\Lambda}t_{max})\nonumber \\&&~~~~+\sinh(3\sqrt{\Lambda}t_{max})
-45\sinh(\sqrt{\Lambda}t')+9\sinh(2\sqrt{\Lambda}t')-\sinh(3\sqrt{\Lambda}t')\nonumber \\&&~~~~~~~~~~~~~~~~~~~~~~~~~~~~~~~~~~~~~~~~~~~~~~~~~~~~~~~~~~~~~-30t_{max}+30t').
\end{eqnarray}
Here $b_{4}...b_{9}$ are constants that depend on the parameters present in the expression for the scale factor whose explicit expressions have been given in the appendix.

Including the contributions from the other integrals also we see that we get a non-zero work done for small field case for hilltop potential when we have cosmological constant in the background.
\\ \\ 
\underline{\bf ii) $\phi/f>>1$:}\\ \\
In this limit, the total work done is given by:
\begin{eqnarray}
\oint pdV &=& \frac{1}{3 b_{10}(-1+b_{11}b_{12}-b_{12}b_{13})}\times\nonumber\\&&~~~~~~~\left[-3b_{14}^{3}\left(e^{3b_{10}(t_{max}-b_{11}+b_{13})(b_{15}+b_{12}t_{max})}-e^{3b_{10}(t_{min}-b_{11}+b_{13})
(b_{15}+b_{12}t_{min})}\right)\nonumber \right.\\&&\left.~~~~~~~~~ +b_{14}^{4}(e^{3b_{10}(t_{max}-b_{11}+b_{13})(b_{15}+b_{12}t_{max})}-e^{3b_{10}(t_{min}-b_{11}+b_{13})
(b_{15}+b_{12}t_{min})})\nonumber \right.\\&&\left. ~~~~~~~~~ -3b_{14}(e^{b_{10}(t_{max}-b_{11}+b_{13})(b_{15}+b_{12}t_{max})}
-e^{b_{10}(t_{min}-b_{11}+b_{13})(b_{15}+b_{12}t_{min})})\nonumber \right.\\&&\left. ~~~~~~~~~+3M_{p}^{2}(45\sinh(\sqrt{\Lambda}t_{max})-9\sinh(2\sqrt{\Lambda}t_{max})+\sinh(3\sqrt{\Lambda}t_{max})
\nonumber \right.\\&&\left.~~~~~~~~~ -45\sinh(\sqrt{\Lambda}t')+9\sinh(2\sqrt{\Lambda}t')-\sinh(3\sqrt{\Lambda}t')\nonumber -30t_{max}+30t'\right].\\
\end{eqnarray}

Here $b_{4}...b_{14}$ are constants that depend on the parameters present in the expression for the scale factor.
Thus we get a non-zero work done for large field case for hilltop potential when we have cosmological constant.
\\ \\
\underline{\textbf{{For Case 2}}}
\\ \\
\textbf{A. Expansion}
\\ \\
Analytical solutions in this case can also be obtained for small field limit i.e. when $\phi/f<<1$. Hence taking the
small angle approximations of the trigonometric functions and redefining the field using following transformation equation: \be \frac{\phi}{f}=e^{\lambda},\ee
we get the following solution for the transformed field $\lambda$ as:
\begin{equation}
\lambda=\left(\frac{t}{3f}+B_{16}\right)\frac{K'}{\sqrt{M'}}
\label{potential8}
\end{equation}
where $B_{16}$ is the arbitrary integration constant given by:
\begin{equation}
B_{16}=\lambda_{F}\frac{\sqrt{M'}}{K'}-\frac{t_{F}}{3f}
\end{equation}
where we introduce two new constants $K'$ and $M'$ given by:
\bea
K'&=&\frac{V_{0}}{f}+\Lambda_{1}+2\Lambda_{2}M_{p}+3\Lambda_{3}M_{p}^{2}+4\Lambda_{4}M_{p}^{3},\\
M'&=&\frac{2V_{0}}{3M_{p}^{2}}+\frac{\Lambda_{0}}{3}-\left(\frac{\Lambda_{1}}{3M_{p}}+\frac{\Lambda_{2}}{3}+\frac{\Lambda_{3}M_{p}}{3}+\frac{\Lambda_{4}M_{p}^{2}}{3}\right).
\eea
Here $\lambda_{F}$ is the value at turnaround.

The above expressions have been obtained under the assumption that the following constraint conditions are satisfied:
\bea \frac{L'\lambda}{M'}&<<&1,\\ \frac{N'\lambda}{K'}&<<&1.\eea 
Since we are in the small field limit, these conditions can be satisfied if we choose the numerical values of the model parameters accordingly.
Here we additionally introduce two new constants $L'$ and $N'$ given by: 
\begin{equation}
L'=\frac{\Lambda_{1}}{3M_{p}}+\frac{2\Lambda_{2}}{3}+\frac{3\Lambda_{3}M_{p}}{3}+\frac{4\Lambda_{4}M_{p}^{2}}{3},
\end{equation}
\begin{equation}
N'=\frac{V_{0}}{f}+2\Lambda_{2}M_{p}+9\Lambda_{3}M_{p}^{2}+12\Lambda_{4}M_{p}^{3}.
\end{equation}
The expression for the scale factor that follows from the above solution is given by:
\begin{equation}
a(t)=B_{17}\exp\left[\sqrt{M'}\left(\left(1-\frac{L'B_{16}K'}{2M'^{3/2}}\right)t-\frac{L'K't^{2}}{12M'^{3/2}f}\right)\right],
\label{scalefactor8}
\end{equation}
where $B_{17}$ is the arbitrary integration constant given by:
\begin{equation}
B_{17}=a_{F}\exp\left[-\sqrt{M'}\left(\left(1-\frac{L'B_{16}K'}{2M'^{3/2}}\right)t_{F}-\frac{L'K't_{F}^{2}}{12M'^{3/2}f}\right)\right].
\end{equation}
Here $a_{F}$ is the value of the scale factor at turnaround.
\\ \\
\textbf{B. Contraction}
\\ \\
The conclusion remains same as for $\Lambda$CDM model, because in this case all the extra terms containing $\phi$, will be neglected at leading order approximation.
\\ \\
\textbf{C. Expression for work done}
\\ \\
The solution of the scale factor (i.e its time dependence) is same as that for hilltop potential.
Hence the expression for work done is same except that the constant will now depend on
the parameters of this model. Thus the phenomenon of hysteresis also holds good for this model.
\\ \\ \\ 
\underline{\textbf{{For Case 3}}}
\\ \\
At late times within the interval $a'<a<a_{max}$, using Eq.~(\ref{dilaton1}) and Eq.~(\ref{dilaton2}), we get the integral
equation of the form similar to Eq.~(\ref{scalardilaton}) with $V(\phi)$ given by the natural potential.
Simple analytical expressions for the scalar field and scale factor could be obtained only for the  case when $\phi/f<<1$ which has been discussed below.
In this limiting situation using the small argument approximations for trigonometric functions, we get the following expression for the scalar field $\phi$ as:
\begin{equation}
\frac{\phi(t)}{f}=\left(\frac{t}{3f}+B_{20}\right)\frac{\Lambda}{f\sqrt{\frac{2V_{0}}{3M_{p}^{2}}-\frac{\Lambda}{3M_{p}^{2}}+\frac{\Lambda_{0}}{3}}},
\label{potential9}
\end{equation}
where $B_{20}$ is the arbitrary integration constant given by:
\begin{equation}
B_{20}=\frac{f\phi_{F}\sqrt{\frac{2V_{0}}{3M_{p}^{2}}-\frac{\Lambda}{3M_{p}^{2}}+\frac{\Lambda_{0}}{3}}}{\Lambda}-\frac{t_{F}}{3f}.
\end{equation}
Here $\phi_{F}$ is the value of the scalar field at turnaround corresponding to $t=t_{F}$.

Finally, substituting the expression for $\phi$ into the Friedmann equation, we get the following expression for the scale factor as:
\begin{equation}
a(t)=B_{21}\exp\left[\sqrt{Q'}\left(\left(1-\frac{B_{20}\Lambda^{2}}{2Q'^{3/2}fM_{p}^{2}}\right)t-\frac{\Lambda^{2}t^{2}}{12M_{p}^{2}f^{2}Q'^{3/2}}\right)\right],
\label{scalefactor9}
\end{equation}
where $B_{21}$ is the arbitrary integration constant given by:
\begin{equation}
B_{21}=a_{F}\exp\left[-\sqrt{Q'}\left(\left(1-\frac{B_{20}\Lambda^{2}}{2Q'^{3/2}fM_{p}^{2}}\right)t_{F}-\frac{\Lambda^{2}t_{F}^{2}}{12M_{p}^{2}f^{2}Q'^{3/2}}\right)\right].
\end{equation}
Here we introduce a new constant $Q'$ defined as:
\begin{equation}
Q'=\frac{2V_{0}}{3M_{p}^{2}}-\frac{\Lambda}{3M_{p}^{2}}+\frac{\Lambda_{0}}{3}
\end{equation}
Here $a_{F}$ is the value of the scale factor at turnaround.
\\ \\ \\
\textbf{B. Contraction}
\\ \\
Solutions remain same as obtained before. So the conclusion also remain unchanged for contraction phase of the Universe.
 \\ \\
 \textbf{C. Expression for work done}
 \\ \\
The solution of the scale factor (i.e its time dependence) is same as that for hilltop potential. Hence the expression for work done is same except that the constant will now depend on the parameters of this model.
Thus the phenomenon of hysteresis also holds good for this model.
\\ \\ 

\subsubsection{Case III:  Coleman-Weinberg potential}

\underline{\textbf{{For Case 1:}}}
\\ \\
\textbf{A. Expansion}
\\ \\
At late times within the interval $a'<a<a_{max}$, using Eq.~(\ref{modeqn1}) and Eq.~(\ref{constant1}), we get the following integral equation of the form:
\begin{equation}
\int d\left(\frac{\phi}{M_{p}}\right)~\frac{\sqrt{\left[\frac{V_{0}}{3M_{p}^{2}}\left(1+\left(\alpha+\beta\ln\left(\frac{\phi}{M_{p}}\right)\right)\left(\frac{\phi}{M_{p}}
\right)^{4}\right)+\frac{\Lambda}{3}\right]}}{\left(\frac{\phi}{M_{p}}\right)^{3}\left[4\alpha+\beta+4\beta\ln\left(\frac{\phi}{M_{p}}\right)\right]}=-\frac{V_{0}}{3M_{p}^{2}}\int dt.
\end{equation}
To compute the left hand side of the above integral equation, we once again use the field redefinition via the following transformation:
\be \frac{\phi}{M_{p}}=e^{\lambda},\ee and study the two limiting physical situations as mentioned below.
\\ \\
\underline{\bf i) $\phi/M_{p}<<1$:}
\\ \\
In this limit, we expand the exponentials upto linear order and get the resulting solution for the transformed field $\lambda$ as:
\begin{equation}
\lambda=\left(-\frac{V_{0}}{3M_{p}^{2}}t+B_{10}\right)\frac{(4\alpha+\beta)}{\left(\frac{V_{0}}{3M_{p}^{2}}+\frac{V_{0}\alpha}{3M_{p}^{2}}+\frac{\Lambda}{3}\right)^{1/2}},
\label{potential10}
\end{equation}
where $B_{10}$ is the arbitrary integration constant given by:
\begin{equation}
B_{10}=\frac{\left(\frac{V_{0}}{3M_{p}^{2}}+\frac{V_{0}\alpha}{3M_{p}^{2}}+\frac{\Lambda}{3}\right)^{1/2}\lambda_{f}}{4\alpha+\beta}+\frac{V_{0}}{3M_{p}^{2}}t_{f},
\end{equation}
The above expressions have been obtained using the assumptions that the constraint conditions:
\bea \frac{4\beta\lambda}{(4\alpha+\beta)}&<<&1,\\ \frac{\frac{V_{0}}{3M_{p}^{2}}(4\alpha+\beta)\lambda}{\frac{V_{0}}{3M_{p}^{2}}+\frac{V_{0}\alpha}{3M_{p}^{2}}+\frac{\Lambda}{3}}&<<&1,\eea  are satisfied.
Since we are already in the limit in which the value of $\lambda$ is very small, hence by choosing the other parameters of the model properly, we can easily satisfy the above constraint conditions.

Further substituting the above expression into the Friedmann equation, we get the following expression for the scale factor as:
\begin{eqnarray}
a(t)&=& B_{11}\exp\left[\left(\left(\frac{V_{0}}{3M_{p}^{2}}+\frac{V_{0}\alpha}{3M_{p}^{2}}+\frac{\Lambda}{3}\right)^{1/2}+\frac{\frac{V_{0}}{3M_{p}^{2}}(4\alpha+\beta)}{2\left(\frac{V_{0}}
{3M_{p}^{2}}+\frac{V_{0}\alpha}{3M_{p}^{2}}+\frac{\Lambda}{3}\right)}B_{10}(4\alpha+\beta)\right)t\nonumber\right.\\ &&\left.~~~~~~~~~~~~~~~~~~~~~~~~~~
~~~~~~~~~~~-\frac{V_{0}\frac{V_{0}}{3M_{p}^{2}}(4\alpha+\beta)}{12\left(\frac{V_{0}}{3M_{p}^{2}}+\frac{V_{0}\alpha}{3M_{p}^{2}}+\frac{\Lambda}{3}\right)M_{p}^{2}}(4\alpha+\beta)t^{2}\right],
\label{scalefactor10}
\end{eqnarray}
where $B_{11}$ is the arbitrary integration constant given by:
\begin{eqnarray}
B_{11} &=&a_{f}\exp\left[-\left(\left(\frac{V_{0}}{3M_{p}^{2}}+\frac{V_{0}\alpha}{3M_{p}^{2}}+\frac{\Lambda}{3}\right)^{1/2}+\frac{\frac{V_{0}}{3M_{p}^{2}}(4\alpha+\beta)}
{2\left(\frac{V_{0}}{3M_{p}^{2}}+\frac{V_{0}\alpha}{3M_{p}^{2}}+\frac{\Lambda}{3}\right)}B_{10}(4\alpha+\beta)\right)t_{f}\nonumber\right.\\ &&\left.~~~~~~~~~~~~~~~~~~~~~~~~~~
~~~~~~~~~~~+\frac{V_{0}\frac{V_{0}}{3M_{p}^{2}}(4\alpha+\beta)}{12\left(\frac{V_{0}}{3M_{p}^{2}}+\frac{V_{0}\alpha}{3M_{p}^{2}}+\frac{\Lambda}{3}\right)M_{p}^{2}}(4\alpha+\beta)t_{f}^{2}\right].
\end{eqnarray}
Here $a_{f}$ is the value of the scale factor at turnaround.
\\ \\
\underline{\bf ii) $\phi/M_{p}>>1$:}
\\ \\
In this case we get an integral of the form
\begin{equation}
\int  d\lambda~\frac{\sqrt{\left[\frac{V_{0}}{3M_{p}^{2}}+\frac{\Lambda}{3}+\frac{V_{0}}{3M_{p}^{2}}(\alpha+\beta\lambda)e^{4\lambda}\right]}}{(4\alpha+\beta+4\beta\lambda)}e^{-2\lambda} = -\frac{V_{0}}{3M_{p}^{2}}\int dt.
\end{equation}
To compute the left hand side of the above integral equation and hence to get an analytical expression for $\lambda$, we assume the following three constraint conditions 
i.e. 
\bea \left(\frac{V_{0}}{3M_{p}^{2}}+\frac{\Lambda}{3}\right)&<<&\frac{V_{0}}{3M_{p}^{2}}(\alpha+\beta\lambda)e^{4\lambda},\\ 
4\alpha+\beta&<<&4\beta\lambda,\\ 
\alpha&<<&\beta\lambda.\eea 
Since we are in large $\lambda$ limit, by choosing the values of the parameters of the model,
the above conditions can be easily satisfied. Under these assumptions, we get the expression for $\lambda$ as:
\begin{equation}
\lambda=\left(-\frac{V_{0}}{3M_{p}^{2}}t+B_{12}\right)^{2}\frac{4\beta}{\frac{V_{0}}{3M_{p}^{2}}},
\label{potential11}
\end{equation}
where $B_{12}$ is the arbitrary integration constant given by:
\begin{equation}
B_{12}=\frac{V_{0}}{3M_{p}^{2}}t_{f}\pm \left(\frac{V_{0}\lambda_{f}}{12M_{p}^{2}\beta}\right)^{1/2}
\end{equation}
Here $\lambda_{f}$ is the value at turnaround.

Further substituting the expression for $\lambda$, hence $\phi$ into the Friedmann equation and applying the above mentioned constraint conditions, we get the following expression for the scale factor as:
\begin{equation}
a(t)=B_{13}\exp\left[2e^{\left(\frac{8\left(B_{12}-\frac{V_{0}t}{3M_{p}^{2}}\right)^{2}\beta}{\frac{V_{0}}{3M_{p}^{2}}}
\right)}t\left(\frac{\left(-\frac{V_{0}}{3M_{p}^{2}}t+B_{12}\right)^{2}\beta^{2}}{\frac{V_{0}}{3M_{p}^{2}}}\right)^{1/2}\right],
\label{scalefactor11}
\end{equation}
where $B_{13}$ is the arbitrary integration constant given by:
\begin{equation}
B_{13}=a_{f}\exp\left[-2e^{\left(\frac{8\left(B_{12}-\frac{V_{0}t_{f}}{3M_{p}^{2}}\right)^{2}\beta}{\frac{V_{0}}{3M_{p}^{2}}}
\right)}t_{f}\left(\frac{\left(-\frac{V_{0}}{3M_{p}^{2}}t_{f}+B_{12}\right)^{2}\beta^{2}}{\frac{V_{0}}{3M_{p}^{2}}}\right)^{1/2}\right].
\end{equation}
Here $a_{f}$ is the value of the scale factor at turnaround. 
\\ \\
\textbf{B. Contraction}
\\ \\
This phase is independent of any choice of the potential provided the condition $\dot{\phi}^2>>V(\phi)$ holds good in this phase.
This implies that the final conclusion remains same as obtained for previous hilltop and natural potential in the background of $\Lambda$CDM model.
\\ \\
\textbf{C. Expression for work done}
\\ \\
The solution of the scale factor (i.e its time dependence) is same as that for hilltop and natural potential.
Hence the expression for work done is same except that the constant will now depend on the parameters of this model. Thus the phenomenon of hysteresis also holds perfectly for this model.
\\ \\
\underline{\textbf{{For Case 2}}}
\\ \\
\textbf{A. Expansion}
\\ \\
At late times within the window $a'<a<a_{max}$, using Eq.~(\ref{philambda}) and Eq.~(\ref{philambda1}), we get an integral equation similar to Eq.~(\ref{scalarhilltop}), 
with the potential $V(\phi)$ now given by the Supregravity motivated Coleman-Weinberg potential. For the case 2 the mathematical form of the governing integral equation is exacly same as 
that obtained in case 2, only in the present case the constant $\Lambda$ is replaced by the field dependent $\Lambda(\phi)$, which is defined earlier.
To compute the left hand side of the master integral equation we will once again use the previously mentioned field redefinition: \be \frac{\phi}{M_{p}}=e^{\lambda}.\ee
One can find the analytical solutions only for the case when the small field limiting approximation is valid i.e. $\phi/M_{p}<<1$ is satisfied. 

In this limit, we expand the exponentials upto linear order and get the resulting solution for the redefined field $\lambda$ as:
\begin{equation}
\lambda=\frac{\left(\frac{t}{3M_{p}}+B_{16}\right)O}{\sqrt{P}}
\label{potential12}
\end{equation}
where $B_{16}$ is the arbitrary integration constant given by:
\begin{equation}
B_{16}=\frac{\lambda_{f}\sqrt{P}}{O}-\frac{t_{f}}{3M_{p}}.
\end{equation}
Here we introduce two new constants $O$ and $P$ given by the following expressions:
\begin{equation}
O=-(4\alpha+\beta)+\Lambda_{1}+2\Lambda_{2}M_{p}+3\Lambda_{3}M_{p}^{2}+4\Lambda_{4}M_{p}^{3},
\end{equation}
\begin{equation}
P=\frac{V_{0}}{3M_{p}^{2}}+\frac{\Lambda_{0}}{3}-\left(\frac{\Lambda_{1}}{3M_{p}}+\frac{\Lambda_{2}}{3}+\frac{\Lambda_{3}M_{p}}{3}+\frac{\Lambda_{4}M_{p}^{2}}{3}\right).
\end{equation}
Here $\lambda_{f}$ is the value at turnaround when $t=t_{f}$.

The above expressions have been obtained using the assumptions that the conditions, \bea \frac{Q\lambda}{P}&<<&1,\\ \frac{L\lambda}{O}&<<&1,\eea are satisfied.
Since we are already in the limit in which the value of $\lambda$ is very small, hence by choosing the other parameters of the model properly, we can easily satisfy the above conditions.
Here we introduce two new constants $Q$ and $L$ given by the following expressions:
\begin{equation}
Q=\frac{V_{0}\beta}{3M_{p}^{2}}+\frac{4V_{0}\alpha}{3M_{p}^{2}}-\left(\frac{\Lambda_{1}}{3M_{p}}+\frac{2\Lambda_{2}}{3}+\frac{3\Lambda_{3}M_{p}}{3}+\frac{4\Lambda_{4}M_{p}^{2}}{3}\right),
\end{equation}
\begin{equation}
L=-\left(\frac{V_{0}}{M_{p}}\right)(7\beta+12\alpha)+2\Lambda_{2}M_{p}+6\Lambda_{3}M_{p}^{2}+12\Lambda_{4}M_{p}^{3}.
\end{equation}
Further substituting the above expression into the Friedmann equation, we get the following expression for the scale factor as:
\begin{equation}
a(t)= B_{17}\exp\left[\sqrt{P}\left(\left(1+\frac{OQ B_{16}}{2P^{3/2}}\right)t+\frac{OQ}{12P^{3/2}M_{p}}t^{2}\right)\right],
\label{scalefactor12}
\end{equation}
where $B_{17}$ is the arbitrary integration constant given by:
\begin{equation}
B_{17}=a_{f}\exp\left[-\sqrt{P}\left(\left(1+\frac{OQB_{16}}{2P^{3/2}}\right)t_{f}+\frac{OQ}{12P^{3/2}M_{p}}t_{f}^{2}\right)\right].
\end{equation}
Here $a_{f}$ is the value of the scale factor at turnaround.
\\ \\
\textbf{B. Contraction}
\\ \\
The conclusion remains same as for $\Lambda$CDM model, because in this case all the extra terms containing $\phi$, will be neglected at leading order approximation.
\\ \\
\textbf{C. Expression for work done}
\\ \\
The solution of the scale factor (i.e its time dependence) is same as that for hilltop and natural potential. Hence the expression for work done is same except that the constant will now depend on the parameters of this model. Thus the phenomenon of hysteresis also holds true for this model.
\\ \\
\underline{\textbf{{For Case 3}}}
\\ \\
At late times within the window $a'<a<a_{max}$, using Eq.~(\ref{dilaton1}) and Eq.~(\ref{dilaton2}), we get the integral equation of the form similar to Eq.~ (\ref{scalardilaton})
with the potential now given by the supergravity motivated Coleman-Weinberg potential. Simplified analytical solutions can be obtained if we use the following field redefinition: 
\be \frac{\phi}{M_{p}}=e^{\lambda},\ee which we have used earlier and take small field limiting approximation $\phi/M_{p}<<1$ in the present context.

In this limit, we expand the exponentials upto linear order and get the resulting solution for the redefined field variable $\lambda$ as:
\begin{equation}
\lambda=\left(\frac{t}{3M_{p}}+B_{22}\right)\frac{M_{p}^{2}}{\frac{1+\frac{R'}{2S'}}{\frac{U'}{V'}}+\frac{(\frac{U'}{V'}-(1+\frac{R'}{2S'}))}{\frac{U'^{2}}{V'^{2}}}},
\label{potential13}
\end{equation}
where $B_{22}$ is the arbitrary integration constant given by:
\begin{equation}
B_{22}=\frac{\lambda_{f}}{M_{p}^{2}}\left(\frac{1+\frac{R'}{2S'}}{\frac{U'}{V'}}+\frac{(\frac{U'}{V'}-(1+\frac{R'}{2S'}))}{\frac{U'^{2}}{V'^{2}}}\right)-\frac{t_{f}}{3M_{p}}.
\end{equation}
Here for the sake of clarity we introduce four model dependent constants $R', S', U', V'$ given by:
\bea
R'&=&\frac{V_{0}}{3M_{p}^{2}}(4\alpha+\beta-\frac{\Lambda e}{V_{0}}),\\
S'&=&\frac{V_{0}}{3M_{p}^{2}}+\frac{\Lambda_{0}}{3}+\frac{V_{0}\alpha}{3M_{p}^{2}}-\frac{\Lambda e}{3M_{p}^{2}},\\
U'&=&-\frac{4V_{0}\beta}{M_{p}}+\frac{3V_{0}(4\alpha+\beta)}{M_{p}}+
\frac{\Lambda e}{M_{p}},\\
V'&=&-\frac{V_{0}}{M_{p}}(4\alpha+\beta)+\frac{\Lambda e}{M_{p}}.
\eea
The above expressions have been obtained using the assumptions that the conditions, \bea \frac{R'\lambda}{S'}&<<&1,\\ \frac{U'\lambda}{V'}&<<&1, \eea are satisfied.
Since we are already in the limit in which the value of $\lambda$ is very small, hence by choosing the other parameters of the model properly, we can easily satisfy the above conditions.

Further substituting the above expression into the Friedmann equation, we get the following expression for the scale factor as:
\begin{equation}
a(t)= B_{23}\exp\left[\sqrt{S'}\left(\left(1+\frac{R'M_{p}^{2}B_{22}}{2S'\left(\frac{1+\frac{R'}{2S'}}{\frac{U'}{V'}}+\frac{(\frac{U'}{V'}-(1+\frac{R'}{2S'}))}{\frac{U'^{2}}{V'^{2}}}\right)}\right)t
+\frac{R't^{2}M_{p}^{2}}{12S'\left(\frac{1+\frac{R'}{2S'}}{\frac{U'}{V'}}+\frac{(\frac{U'}{V'}-(1+\frac{R'}{2S'}))}{\frac{U'^{2}}{V'^{2}}}\right)}\right)\right],
\label{scalefactor13}
\end{equation}
where
\begin{equation}
B_{23}=a_{f}\exp\left[-\sqrt{S'}\left(\left((1+\frac{R'M_{p}^{2}B_{22}}{2S'\left(\frac{1+\frac{R'}{2S'}}{\frac{U'}{V'}}+\frac{(\frac{U'}{V'}
-(1+\frac{R'}{2S'}))}{\frac{U'^{2}}{V'^{2}}}\right)}\right)t_{f}+\frac{R't_{f}^{2}M_{p}}{12S'\left(\frac{1+\frac{R'}{2S'}}{\frac{U'}{V'}}+\frac{(\frac{U'}{V'}
-(1+\frac{R'}{2S'}))}{\frac{U'^{2}}{V'^{2}}}\right)}\right)\right].
\end{equation}
Here $a_{f}$ is the value of the scale factor at turnaround.
\\ \\
\textbf{B. Contraction}
\\ \\
 Solutions remain same as obtained before. So the conclusion also remain unchanged for contraction phase of the Universe.
\\ \\
\textbf{C. Expression for work done}
\\ \\
The solution of the scale factor (i.e. its time dependence) is same as that for hilltop potential. Hence the expression for work done is same except that
the constant will now depend on the parameters of this model. Thus the phenomenon of hysteresis also holds true for this model.
\\ \\ 
\\ \\
\subsection{Graphical Analysis}
\subsubsection{Case I: Hilltop potential}
All the graphs in this section and in the following sections have been plotted in units of $M_{p}=1,\ H_{0}=1,\ c=1$, where $M_{p}$ is the Planck mass, $H_{0}$ is the present value of the Hubble parameter and $c$ is the speed of light. While performing the analysis, for the value of the standard cosmological constant, we have used Planck  2015 data \cite{Ade:2015xua}. The analysis for all the three potentials has been done for late times since, in the early times the solutions are those that we get for Randall-Sundrum single brane world (RSII).
\\ \\
\textbf{For Case 1:}
\\ \\
\begin{figure*}[htb]
\centering
\subfigure[ An illustration of the behavior of the scale factor with time  during expansion phase for $\phi<<M_{p}$ with $V_{0}=10^{-8}M_{p}^{4},\ p=2,\ \beta=1,\ B_{0}=M_{p},\ B_{1}=1$.]{
    \includegraphics[width=7.2cm,height=7.8cm] {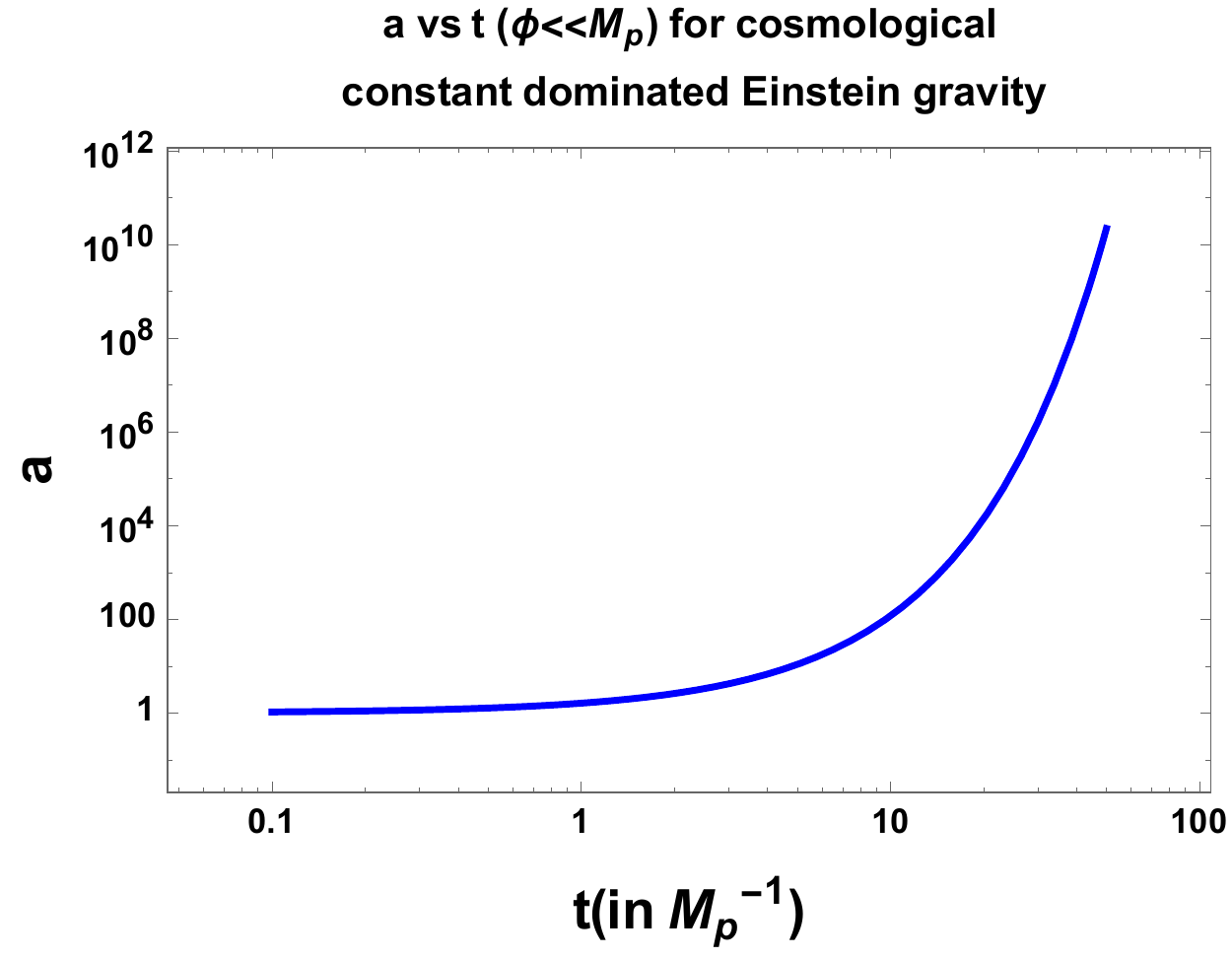}
    \label{lambda5}
}
\subfigure[An illustration of the behavior of the scale factor with time during expansion phase for $\phi<<M_{p}$ with $V_{0}=10^{-2}M_{p}^{4},\ p=3,\ \beta=-8,\ B_{0}=M_{p},\ B_{1}=1$.]{
    \includegraphics[width=7.2cm,height=7.8cm] {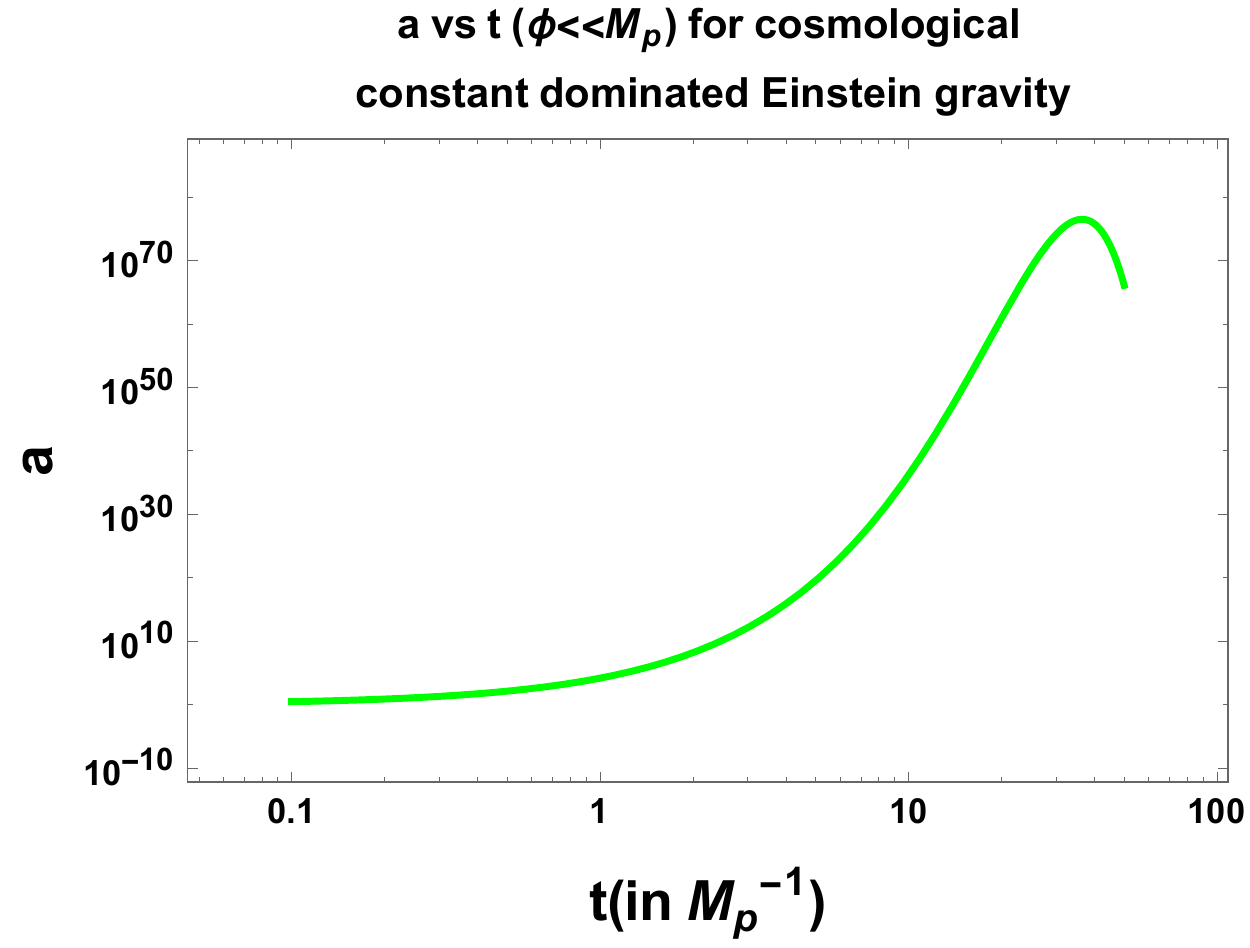}
    \label{lambda6}
}
\subfigure[An illustration of the behavior of the potential during expansion phase for $\phi<<M_{p}$ with $V_{0}=10^{-4}M_{p}^{4},\ \beta=2,\ p=2,\ B_{0}=M_{p}$ .]{
    \includegraphics[width=10.2cm,height=7.8cm] {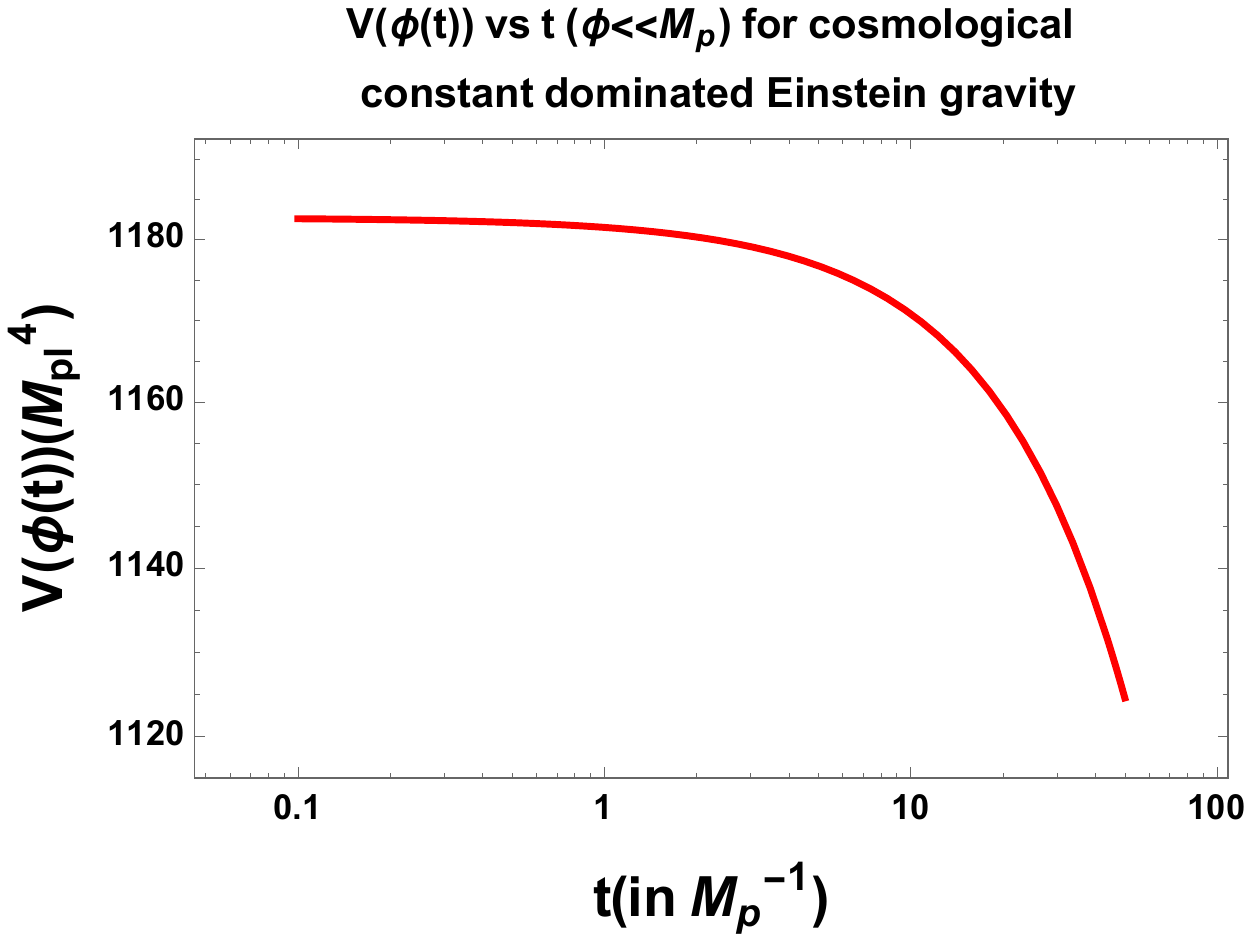}
    \label{lambda7}
}
\caption[Optional caption for list of figures]{ Graphical representation of the evolution of the scale factor and the potential during the expansion and contraction phase for the cosmological constant dominated Einstein gravity for case 1.} 
\label{fig4}
\end{figure*}
\begin{figure*}[htb]
\centering
\subfigure[An illustration of the behavior of the scale factor with time  during contraction phase with $B_{4}=10^{-3}M_{p}^{2},\ B_{5}=10^{4}$.]{
    \includegraphics[width=7.2cm,height=8cm] {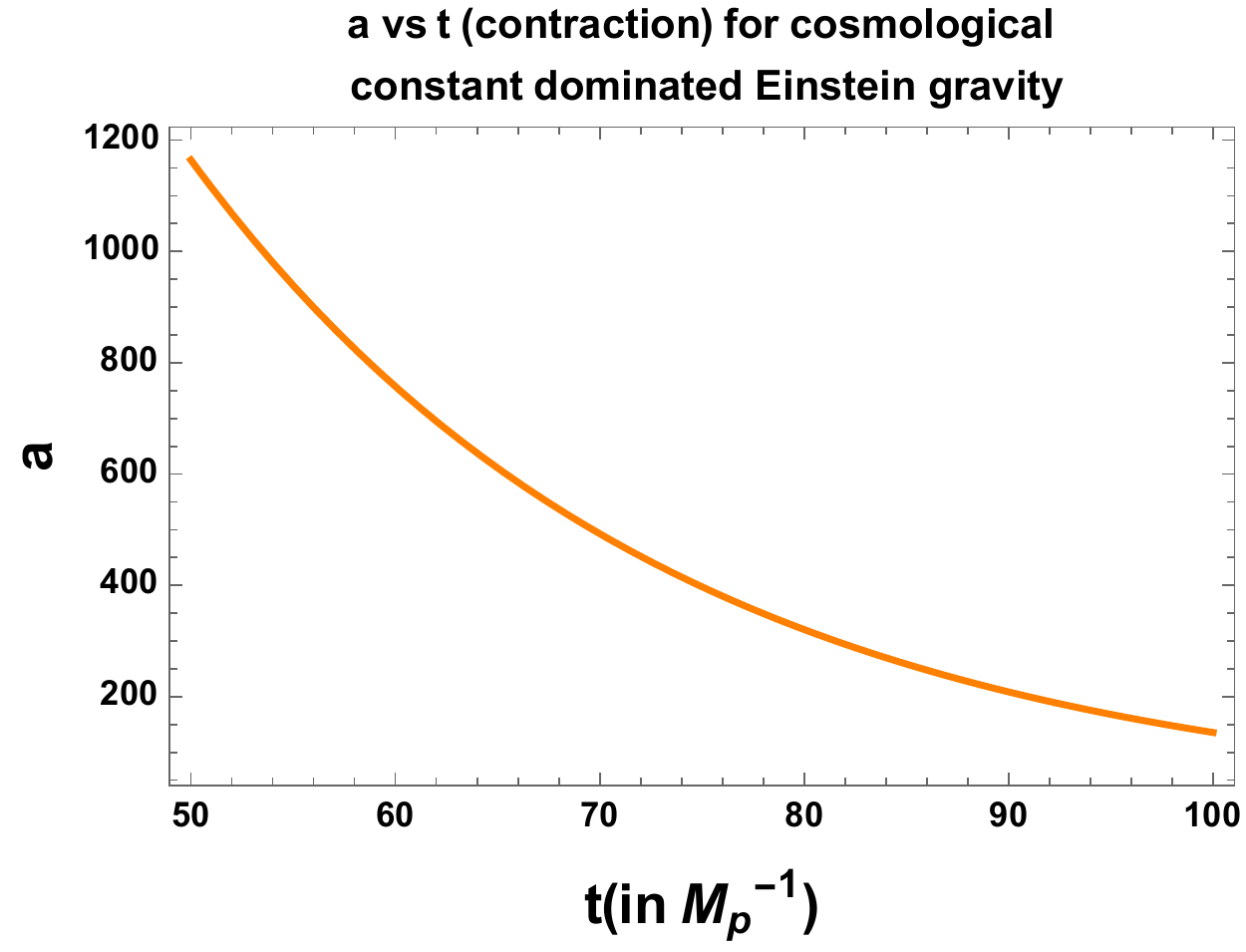}
    \label{lambda8}
}
\subfigure[An illustration of the behavior of the kinetic term of the potential with time  during contraction phase with $B_{4}=5.4M_{p}^{2}$.]{
    \includegraphics[width=7.2cm,height=8cm] {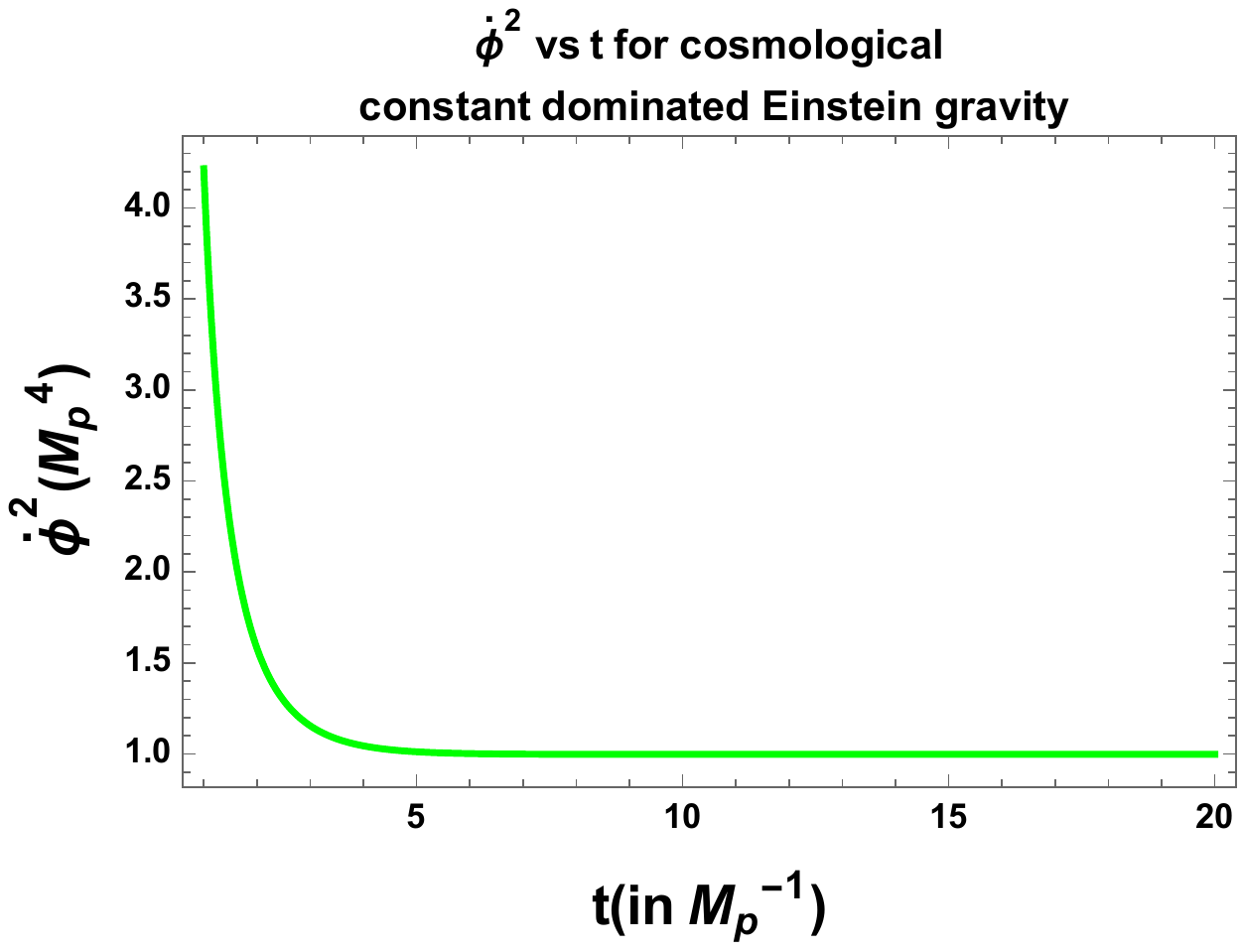}
    \label{lambda88}
}
\caption[Optional caption for list of figures]{ Graphical representation of the evolution of the scale factor and the kinetic term of the potential during the contraction phase for the cosmological constant dominated Einstein gravity for case 1.} 
\label{fig4.5}
\end{figure*}
\begin{figure*}[htb]
\centering
\subfigure[An illustration of the behavior of the scale factor with time  during expansion phase for $\phi>>M_{p}$ with $V_{0}=3{\rm x} 10^{-1}M_{p}^{4},\ p=3,\ \beta=1,\ B_{2}=M_{p},\ B_{3}=1$.]{
    \includegraphics[width=7.2cm,height=8cm] {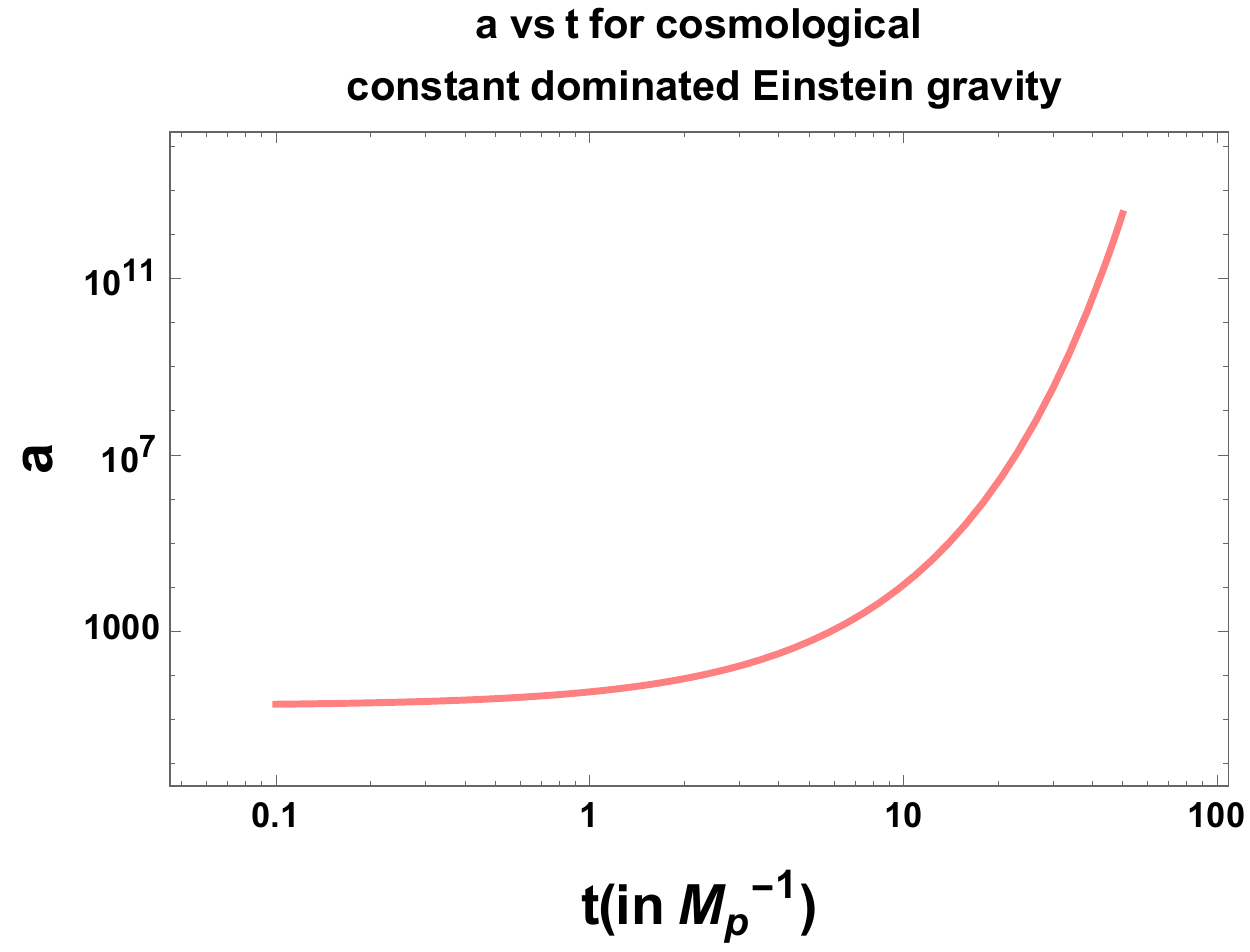}
    \label{lambda9}
}
\subfigure[An illustration of the behavior of the potential with time  during expansion phase for $\phi>>M_{p}$ with $V_{0}=3{\rm x} 10^{-1}M_{p}^{4},\ p=3,\ \beta=1,\ B_{2}=M_{p},$.]{
    \includegraphics[width=7.2cm,height=8cm] {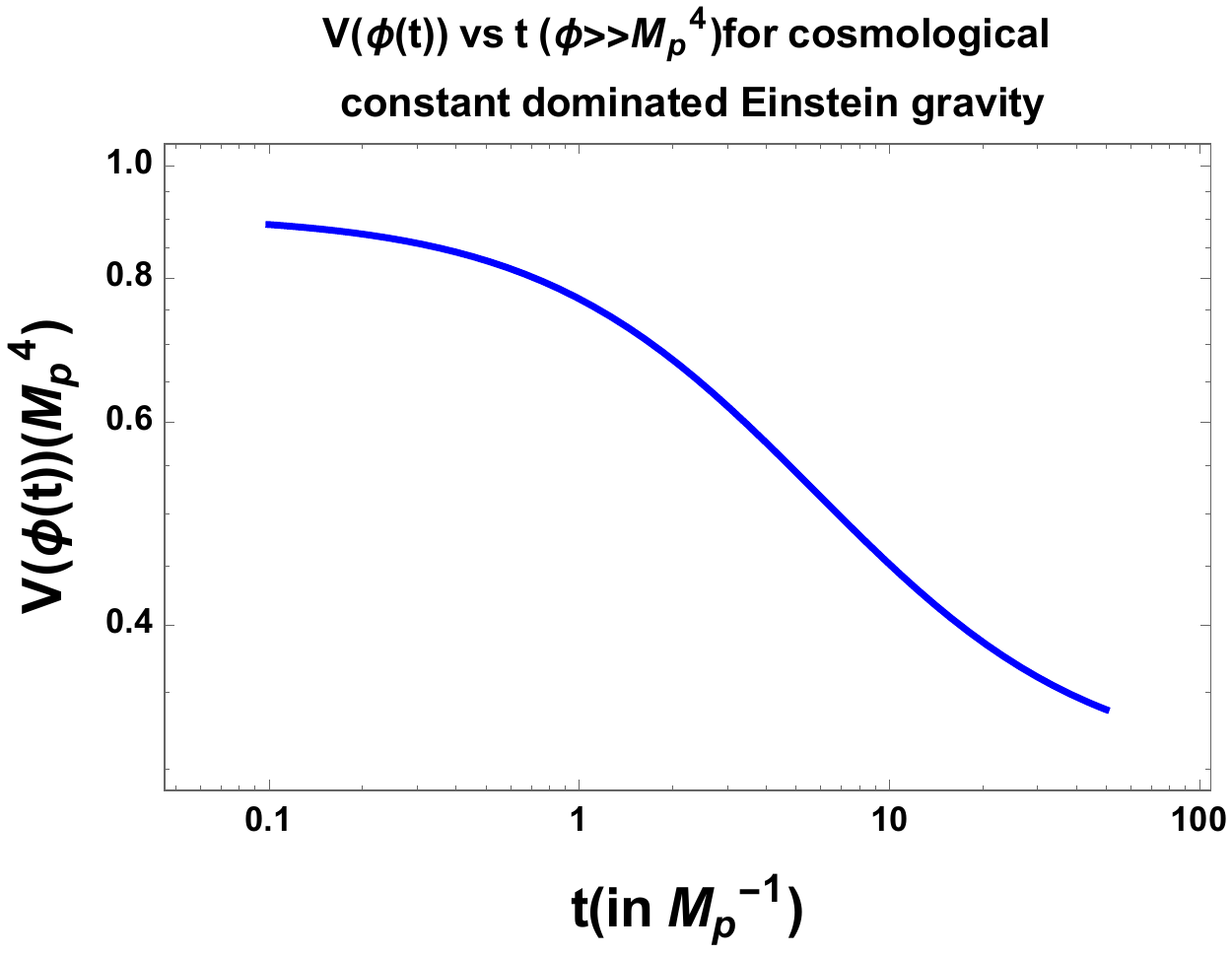}
    \label{lambda10}
}
\caption[Optional caption for list of figures]{ Graphical representation of the evolution of the scale factor and the potential during the expansion phase for the cosmological constant dominated Einstein gravity for case 1.} 
\label{fig5}
\end{figure*}
Figs. \ref{fig4} and \ref{fig4.5} show the evolution of the scale factor and kinetic term of the potential during contraction phase. We can draw the following conclusions from the plots:
\begin{itemize}
\item Fig. \ref{lambda5}, shows the plot of the scale factor in the small field limit for hilltop potential given by Eqn. (\ref{scalefactor1}). 
\item While obtaining the above plot, we kept the value of the integration constants $B_{0}$ and $B_{1}$ as 1. Higher values of the integration constants only results in an increase in the amplitude of the scale factor, not affecting the nature of the graph.
\item In Figs. \ref{lambda5} and \ref{lambda6}, the plot has been shown for two particular cases which gives interesting results. The red curve has been obtained for the case when $V_{0}=10^{-8}M_{p}^{4},\ p=2,\ \beta=1,\ B_{0}=1,\ B_{1}=1$. While the blue curve has been obtained for the case when $V_{0}=10^{-2}M_{p}^{4},\ p=3,\ \beta=-8,\ B_{0}=1,\ B_{1}=1$. 
\item While analyzing the results graphically, we have found that when the value of $V_{0}$ is low (for example $\approx 10^{-8}M_{p}^{4}$), the universe undergoes expansion for any values of the parameters $p,\ \beta$, etc. But if we make $V_{0}>10^{-3}M_{p}^{4}$, then expansion is possible for any values of the parameters provided $B_{0}$ takes larger value. One such example has been shown in Fig. \ref{lambda6}, where the universe undergoes starts contracting instead of expanding at later  times. Thus the values of parameters for which Fig. \ref{lambda6} has been obtained, are not favourable for the present scenario.
\item Fig. \ref{lambda7} shows the plot of the behavior of the potential with time for small field hilltop potential. This graph has been obtained with the help of Eqn. (\ref{potential1}) with parameter values $V_{0}=10^{-4}M_{p}^{4},\ \beta=2,\ p=2,\ B_{0}=1$. Negative values of $\beta$ and very large values of $B_{0}$ does not give the required variation of the potential for causing expansion. Thus, such values of the parameters are not favorable for our model.
\item In Fig. \ref{lambda8}, we have plotted Eqn. (\ref{scalefactor3}), which results in the contraction phase of the universe provided we choose the value of the constants accordingly. This plot has been obtained for $B_{4}=10^{-3},\ B_{5}=10^{4}$. Taking higher values of $B_{4}$ decreases the amplitude of the scale factor, whereas higher values of $B_{5}$ increases the amplitude of the scale factor. If we compare Fig. \ref{lambda1} with Fig. \ref{lambda8}, we find that there occurs a net increase in the amplitude of the scale factor after one expansion-contraction cycle.
\item In Fig. \ref{lambda88}, we have plotted the kinetic term of the potential during the contraction phase with the help of Eqn. (\ref{kinetic}) using $B_{4}=5.4M_{p}^{2}$. Negative values of $B_{4}$ do not give the correct nature of the kinetic term required for causing contraction. The graph shows a decreasing kinetic term with time because this will result in contraction phase.
\item Fig. \ref{lambda9} shows the plot of the scale factor in the large field limit for hilltop potential given by Eqn. (\ref{scalefactor2}). This plot has been obtained for $V_{0}=3{\rm x} 10^{-1}M_{p}^{4},\ \beta=1,\ p=3,\ B_{2}=M_{p},\ B_{3}=1$. From Eqn. (\ref{scalefactor2}), we can conclude that negative values of $\beta$ are not allowed in this case, since this will make the expression imaginary. Larger and positive values of $\beta$ results in a decrease in amplitude of the scale factor Also, if we make $p>3$, then the amplitude of the scale factor decreases to a very low value. If we make $p<2$, then the evolution of the scale factor becomes very uneven, i.e.it remains constant for most part of its evolution, with a sudden increase in amplitude at the end. Thus for smooth and large amplitude expansion for the values of the other parameters that we have chosen here, the allowed range of $p$ is $2\leq p\geq 3$.
\item Fig. \ref{lambda10} shows the plot of the behavior of the potential with time for large field hilltop potential. This graph has been obtained with the help of Eqn. (\ref{potential2}) with parameter values $V_{0}=3{\rm x}10^{-1}M_{p}^{4},\ \beta=1,\ p=3,\ B_{2}=M_{p}^{4}$. 
\end{itemize}

\textbf{For Case 2:}
\\ \\
\begin{figure*}[htb]
\centering
\subfigure[\scriptsize An illustration of the behavior of the scale factor with time  during expansion phase for $\phi<<M_{p}$ with $V_{0}=10^{-8}M_{p}^{4},\ p=2,\ \beta=1,\ B_{14}=M_{p}^{-2},\ B_{15}=1,\ \Lambda_{1}=10^{-8}M_{p}^{3},\ \Lambda_{2}=10^{-8}M_{p}^{2},\ \Lambda_{3}=10^{-8}M_{p},\ \Lambda_{4}=10^{-8}$ for red curve and $V_{0}=10^{-8}M_{p}^{4},\ p=3,\ \beta=2,\ B_{14}=M_{p}^{-2},\ B_{15}=1,\ \Lambda_{1}=-10^{-2}M_{p}^{3},\ \Lambda_{2}=-10^{-2}M_{p}^{2},\ \Lambda_{3}=-10^{-2}M_{p},\ \Lambda_{4}=-10^{-2}$ for blue curve .]{
    \includegraphics[width=7.0cm,height=6.8cm] {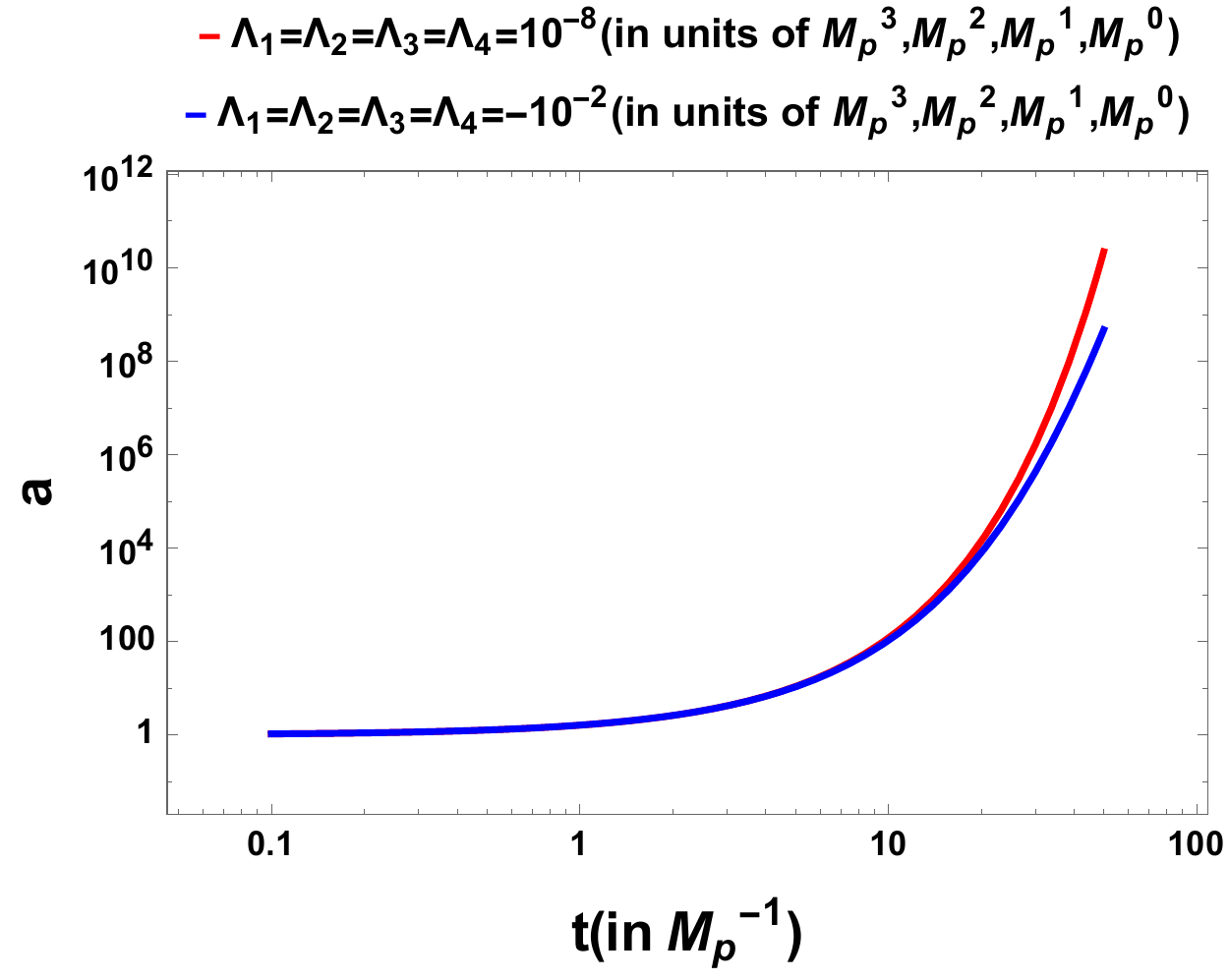}
    \label{lambda12}
}
\subfigure[\scriptsize An illustration of the behavior of the potential with time  during expansion phase for $\phi<<M_{p}$ with $V_{0}=10^{-6}M_{p}^{4},\ p=3,\ \beta=1,\ B_{14}=M_{p},\ \Lambda_{1}=-10^{-2}M_{p}^{3},\ \Lambda_{2}=-10^{-2}M_{p}^{2},\ \Lambda_{3}=-10^{-2}M_{p},\ \Lambda_{4}=-10^{-2}$.]{
    \includegraphics[width=7.0cm,height=6.8cm] {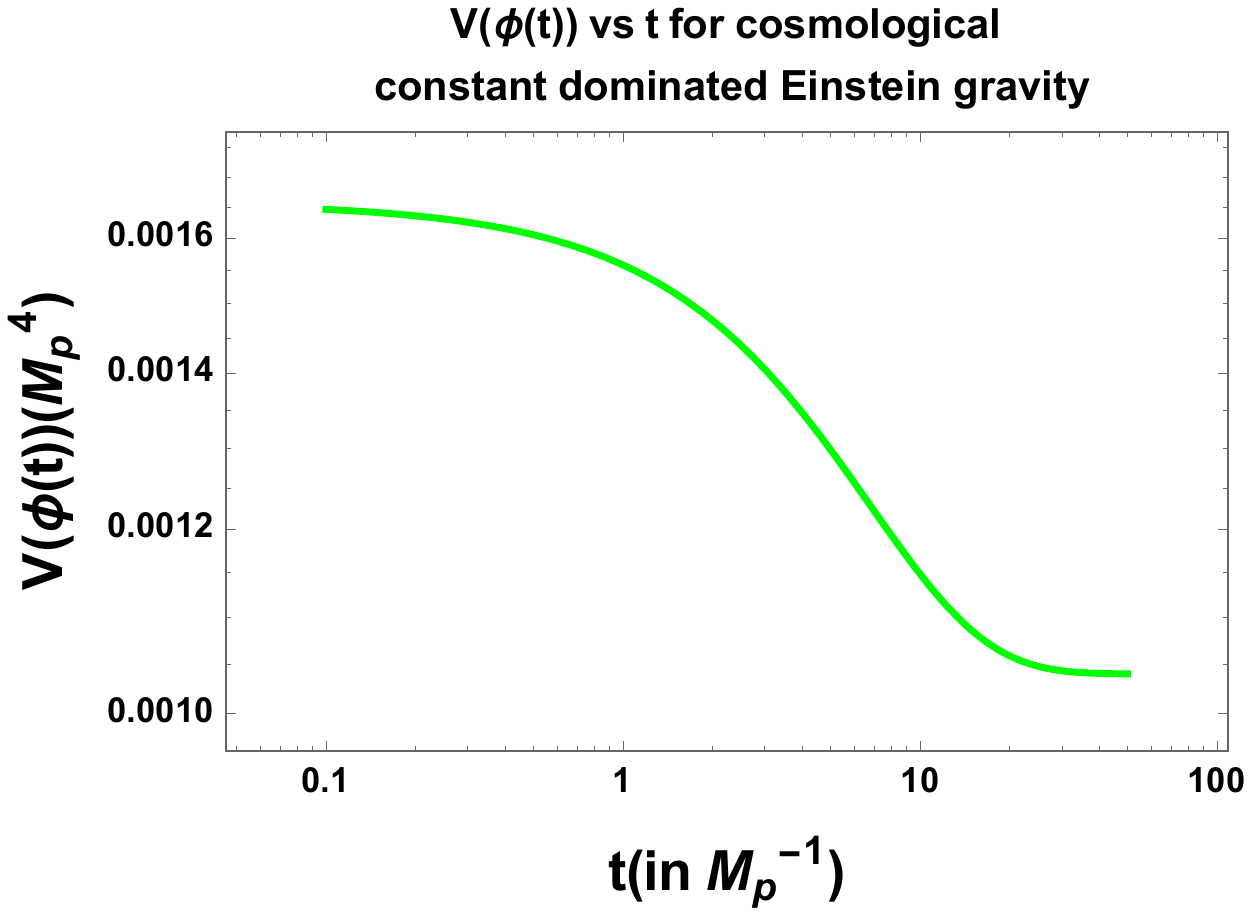}
    \label{lambda13}
}
\subfigure[\scriptsize An illustration of the behavior of the potential with time  during expansion phase for $\phi<<M_{p}$ with $V_{0}=4{\rm x}10^{-6}M_{p}^{4},\ p=3,\ \beta=1,\ B_{14}=M_{p},\ \Lambda_{1}=10^{-4}M_{p}^{3},\ \Lambda_{2}=10^{-3}M_{p}^{2},\ \Lambda_{3}=4{\rm x}10^{-3}M_{p},\ \Lambda_{4}=-3{\rm x}10^{-3}$.]{
    \includegraphics[width=11.0cm,height=6.8cm] {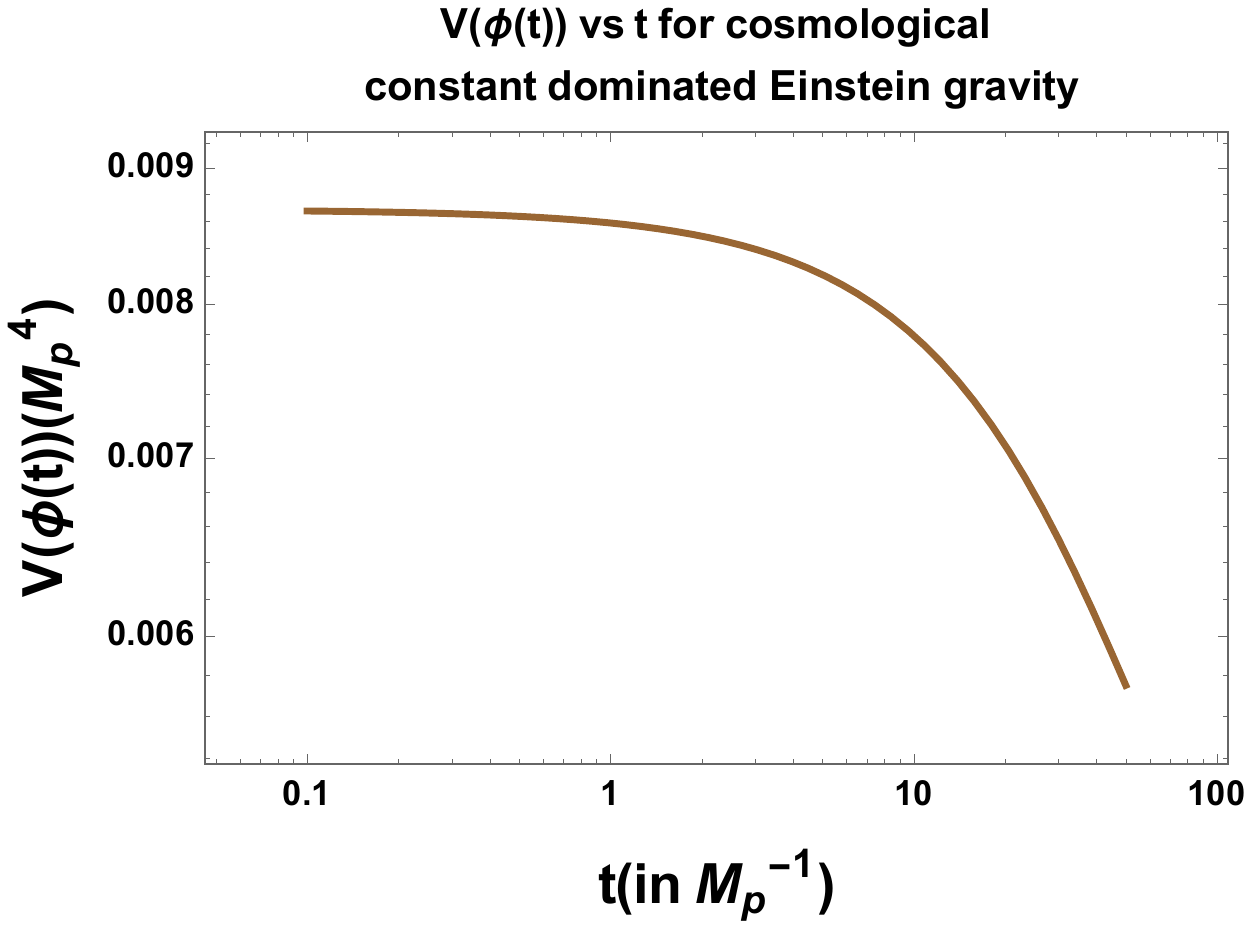}
    \label{lambda14}
}
\caption[Optional caption for list of figures]{ Graphical representation of the evolution of the scale factor and the potential during the expansion phase for the cosmological constant dominated Einstein gravity for case 2.} 
\label{fig6}
\end{figure*}
In Fig. \ref{fig6}, we have plotted the evolution of the scale factor and the potential during expansion for case 2. From the figures we can draw the following conclusions:
\begin{itemize}
\item Fig. \ref{lambda12} shows the variation of the scale factor with time for small field hilltop potential for two cases of parameter values for a scalar field dependent cosmological constant.
\item Fig. \ref{lambda12} has been obtained from Eq. (\ref{scalefactor4}). Detail graphical analysis show that an expanding universe in the present scenario is possible only if the values of $\Lambda_{1},\ \Lambda_{2},\ \Lambda_{3},\ \Lambda_{4}$ lies between $-10^{-2}$ to $10^{-2}$. They can take any sign provided their magnitudes remain within the specified range. The expansion in this case is almost independent of the values of $\beta,\ p$ and $V_{0}$.
\item In Fig. \ref{lambda12}, we have shown two such cases which give rise to an expanding universe.
\item Fig. \ref{lambda13} and Fig. \ref{lambda14} show the variation of the potential with time during expansion for case 2. Detail graphical analysis have shown that keeping the magnitude of the parameters within the above mentioned range, if we make $\Lambda_{1}$ negative, then expansion is possible only if $\Lambda_{4}$ is also negative or only slightly positive. This also requires a higher value of $V_{0}$. The other two parameters can be either positive or negative. Also, if the values of $\Lambda_{1}$, $\Lambda_{3}$ and $\Lambda_{4}$ are very close to the upper bound, then we won't get the correct nature of the potential required for evolution.
\item From Figs. \ref{lambda13} and \ref{lambda14}, we can conclude that higher and positive  parameter values increases the height of the potential. But smaller and negative values of the constants make the potential flatter. Also, from Fig. \ref{lambda14}, only a negative value of $\Lambda_{4}$ can produce the correct evolution of the potential if all the other constants are positive and small.
\end{itemize}

\textbf{For Case 3:}
\\ \\
\begin{figure*}[htb]
\centering
\subfigure[An illustration of the behavior of the potential with time  during expansion phase for $\phi<<M_{p}$ with $V_{0}=10^{-8}M_{p}^{4},\ B_{18}=M_{p}^{-2},\ \Lambda=10^{-2}M_{p}^{4},\ B_{19}=1$.]{
    \includegraphics[width=7.2cm,height=8cm] {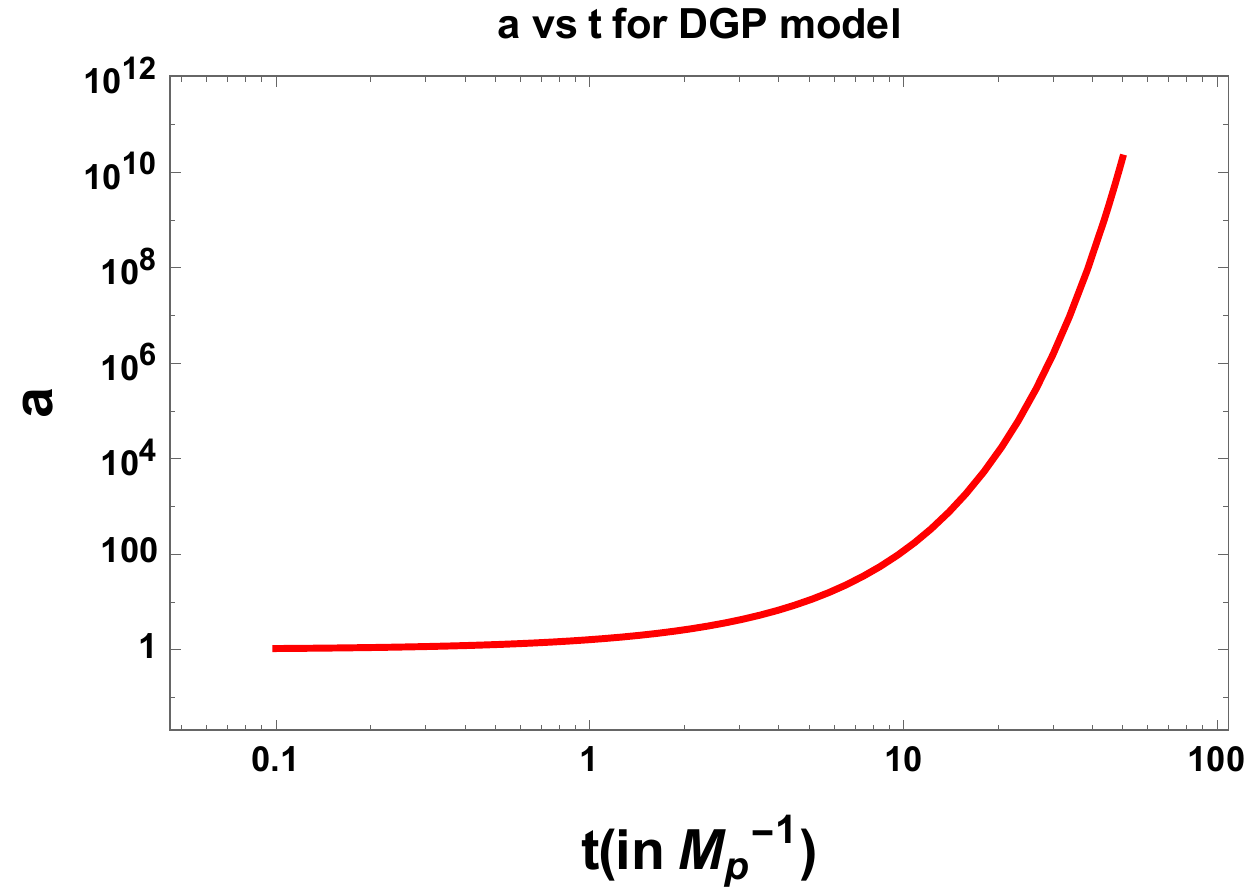}
    \label{lambda15}
}
\subfigure[An illustration of the behavior of the potential with time  during expansion phase for $\phi<<M_{p}$ with $V_{0}=2{\rm x}10^{-4}M_{p}^{4},\ p=3,\ \beta=1,\ B_{18}=M_{p}^{-2},\ \Lambda=-10^{-3}M_{p}^{4}$.]{
    \includegraphics[width=7.2cm,height=8cm] {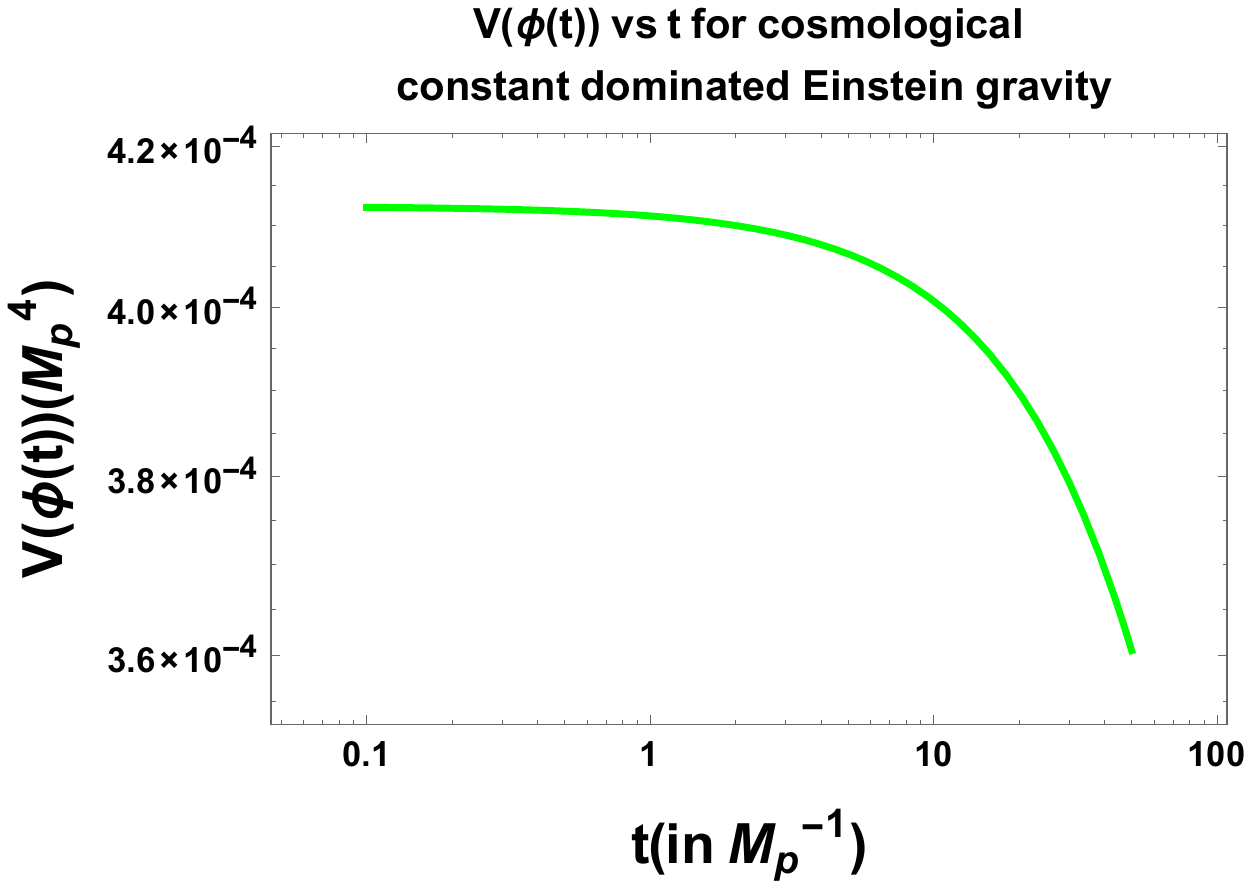}
    \label{lambda16}
}
\caption[Optional caption for list of figures]{ Graphical representation of the evolution of the scale factor and the potential during the expansion phase for the cosmological constant dominated Einstein gravity for case 3.} 
\label{fig7}
\end{figure*}
In Fig. \ref{fig7}, we have plotted the evolution of the scale factor and the potential during expansion for case 3. From the figures we can draw the following conclusions:
\begin{itemize}
\item Fig. \ref{lambda15} shows the variation of the scale factor with time for small field hilltop potential for $V_{0}=10^{-8}M_{p}^{4},\ B_{18}=M_{p}^{-2},\ \Lambda=10^{-2}M_{p}^{4},\ B_{19}=1$.
\item Fig. \ref{lambda15} has been obtained from Eqn. (\ref{scalefactor5}). Detail graphical analysis show that an expanding universe in the present scenario is possible only if the value of $\Lambda$ lies within the range $-10^{-2}$ to $10^{-2}$. But as we go nearer to the upper bound, expansion occurs only for very small values (very close to unity) of $B_{18}$. The expansion in this case is independent of the values of $\beta,\ p$ . Higher values of $V_{0}$ increases the amplitude of the expansion. But values of $V_{0}$ greater than $10^{-3}M_{p}^{4}$ do not give rise to an expanding universe.
\item Fig. \ref{lambda16} shows the variation of the hilltop potential for an expanding universe for case 3 with $\phi<<M_{p}$ with $V_{0}=2{\rm x}10^{-4}M_{p}^{4},\ p=3,\ \beta=1,\ B_{18}=M_{p}^{-2},\ \Lambda=-10^{-3}M_{p}^{4}$. Graphical analysis show that within the allowed range of the parameter values, as discussed before, only a negative $\Lambda$ can give the correct nature of the evolution of the potential when $V_{0}$ is very small (around $10^{-8}M_{p}^{4}$. For this case, $\beta$ and $p$ can take any values. But for larger values of $V_{0}$, we can increase the range of $\Lambda$ and make it positive by making $\beta$ positive. For example, if $V_{0}=10^{-3}M_{p}^{4}$ and $\Lambda=8{\rm x}10^{-3}M_{p}^{4}$, then the minimum value of $\beta$ required to give the correct evolution of the potential is $5$.
\end{itemize} 

\subsubsection{Case II: Natural potential}
\textbf{For Case 1:}
\\ \\
\begin{figure*}[htb]
\centering
\subfigure[ An illustration of the behavior of the scale factor with time  during expansion phase for $\phi<<f$ with $V_{0}=10^{-8}M_{p}^{4},\ B_{6}=1,\ B_{7}=1$.]{
    \includegraphics[width=7.2cm,height=8cm] {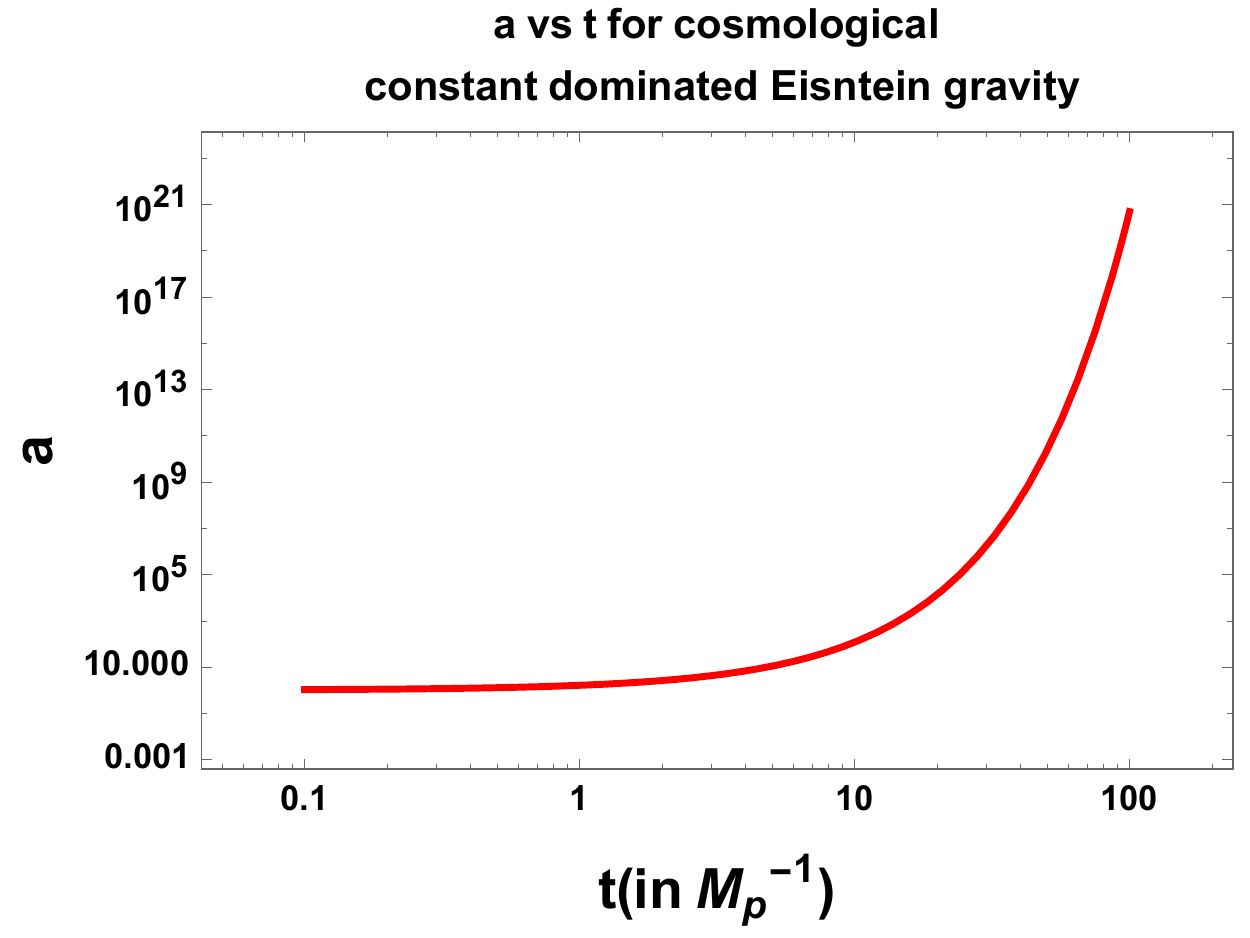}
    \label{lambda17}
}
\subfigure[An illustration of the behavior of the potential during expansion phase for $\phi<<f$ with $V_{0}=3.6{\rm x}10^{-3}M_{p}^{4},\  B_{6}=1,\ f=8M_{p}$ .]{
    \includegraphics[width=7.2cm,height=8cm] {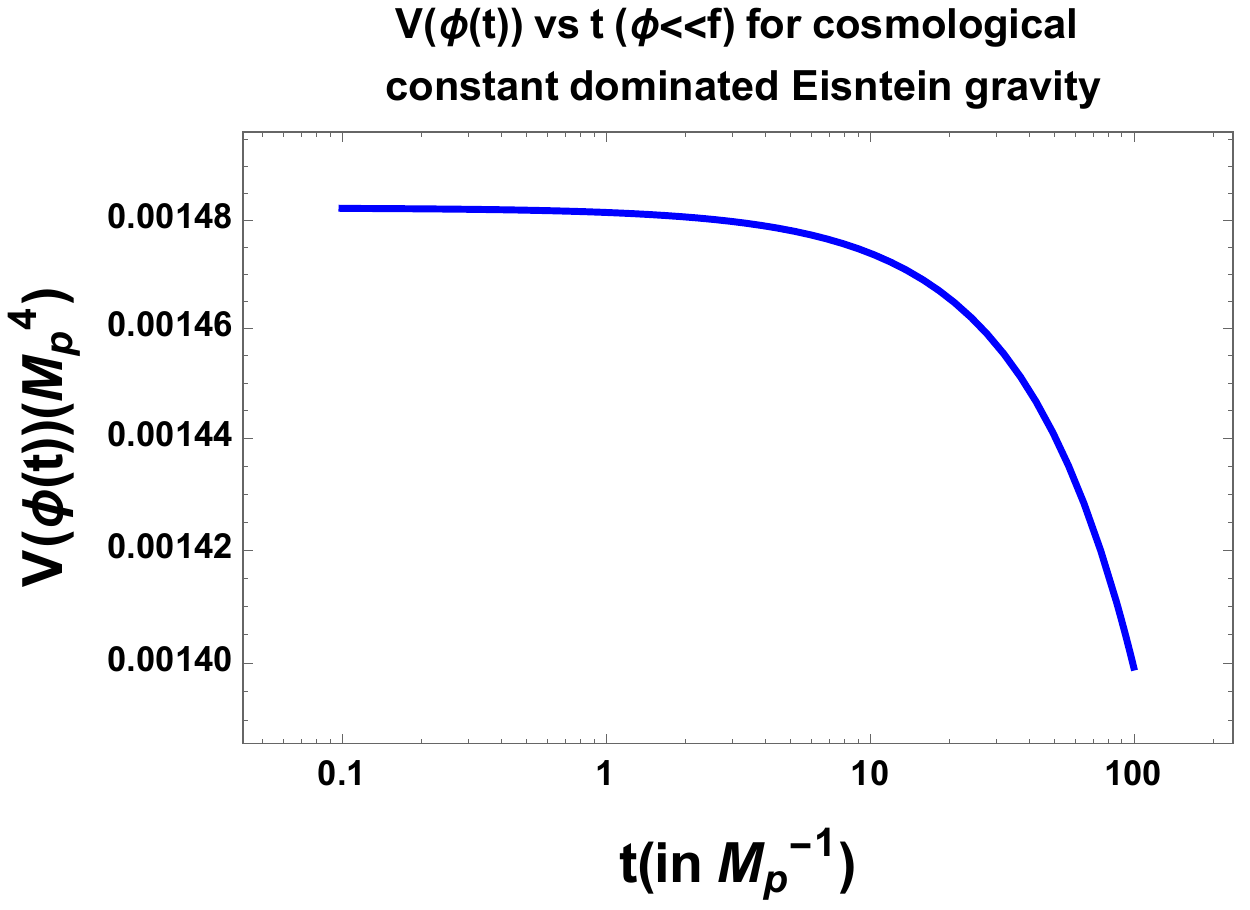}
    \label{lambda18}
}
\subfigure[An illustration of the behavior of the scale factor with time  during expansion phase for $\phi>>f$ with $V_{0}=10^{-2}M_{p}^{4},\ B_{8}=M_{p},\ B_{9}=1$.]{
    \includegraphics[width=7.2cm,height=8cm] {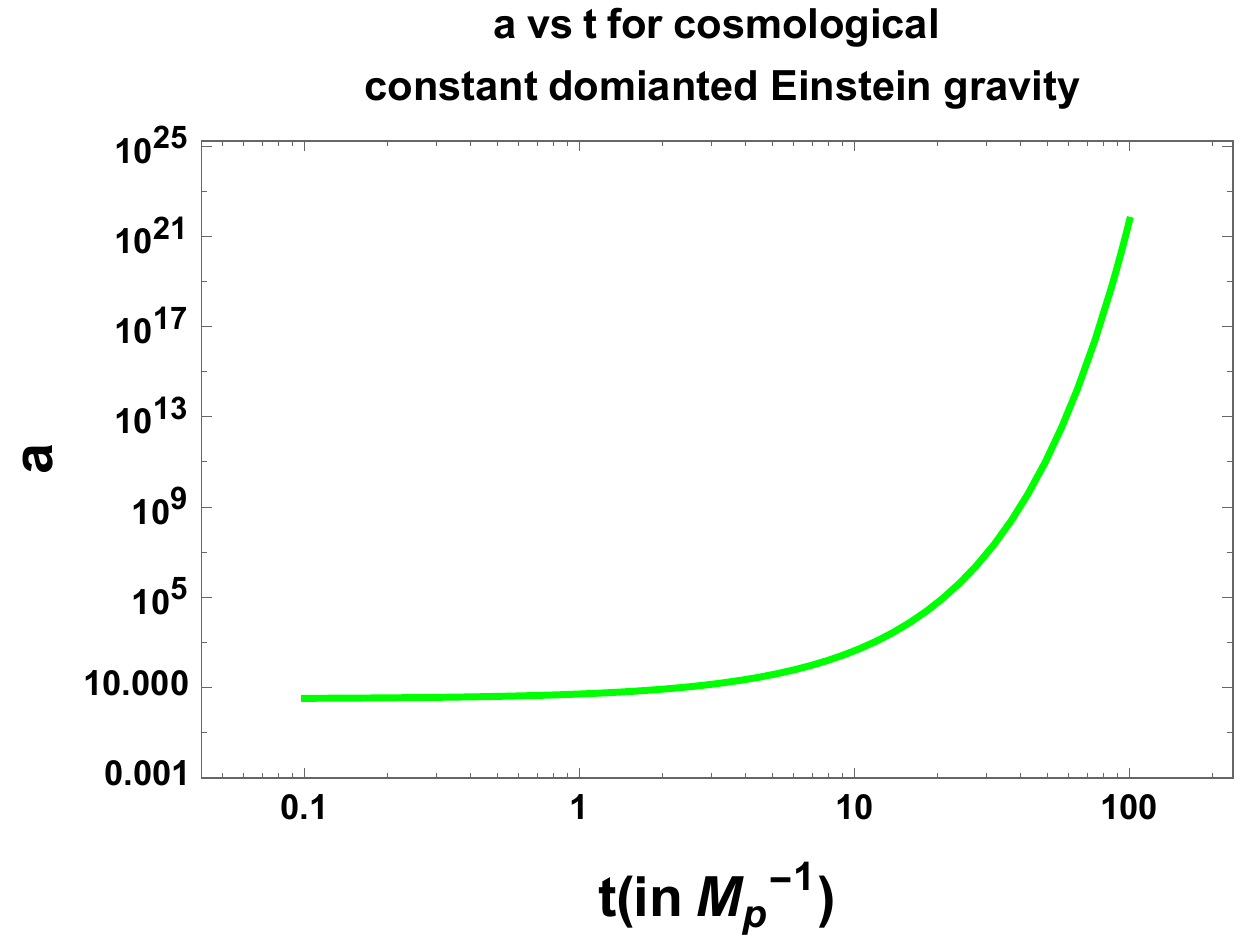}
    \label{lambda19}
}
\subfigure[An illustration of the behavior of the potential with time  during expansion phase for $\phi>>f$ with $V_{0}=10^{-2}M_{p}^{4},\ B_{8}=M_{p},$.]{
    \includegraphics[width=7.2cm,height=8cm] {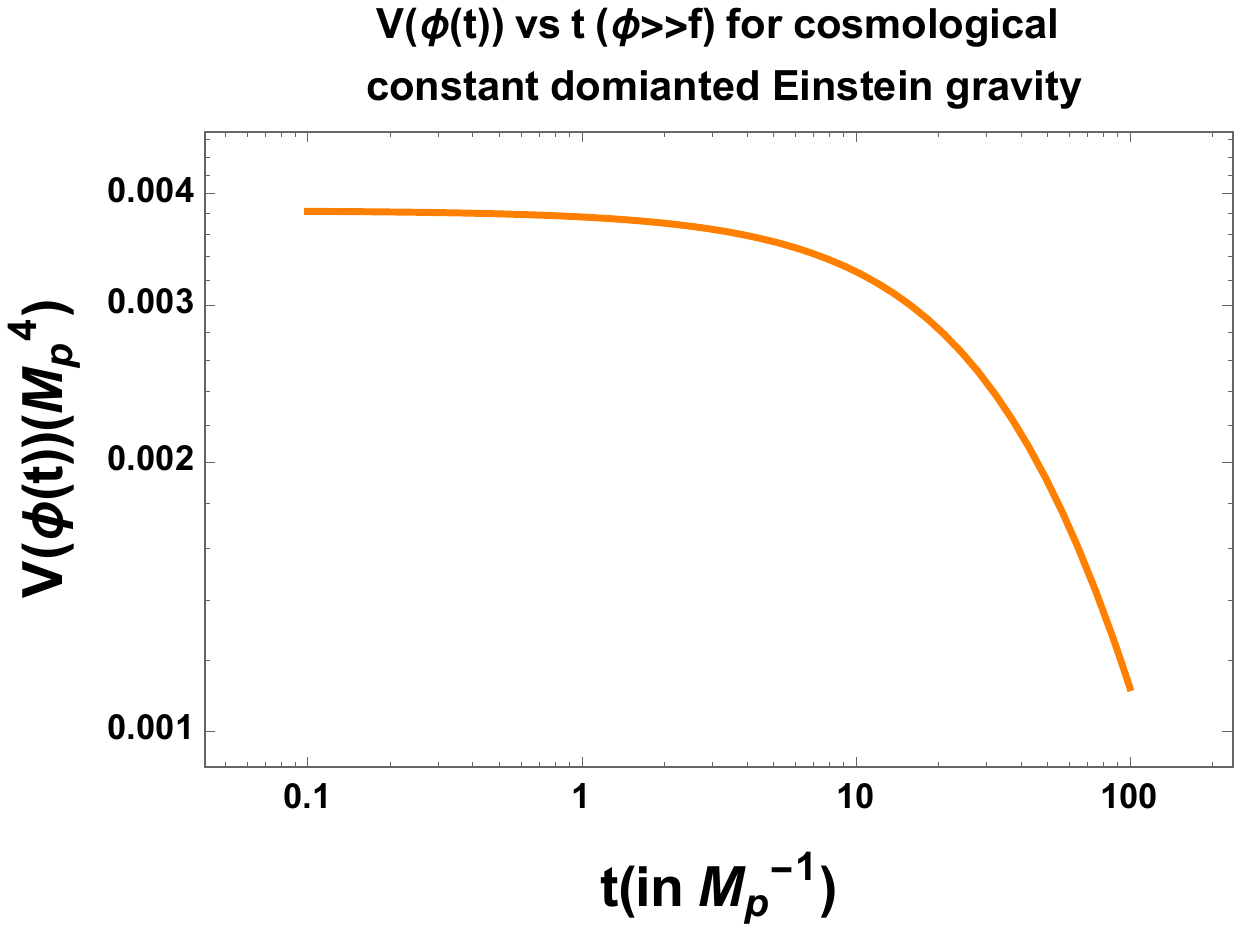}
    \label{lambda20}
}
\caption[Optional caption for list of figures]{ Graphical representation of the evolution of the scale factor and the potential during the expansion and contraction phase for the cosmological constant dominated Einstein gravity for case 1.} 
\label{fig9}
\end{figure*}

Fig. \ref{fig9} show the evolution of the scale factor and the potential for the case of natural potential for case 1 during expansion. The following conclusions can be drawn from the graphs:
\begin{itemize}
\item Fig. \ref{lambda17}, shows the plot of the scale factor in the small field limit for natural potential given by Eqn. (\ref{scalefactor6}). 
\item While obtaining the above plot, we kept the value of the integration constants $B_{6}$ and $B_{7}$ as 1. Higher values of the integration constants only results in an increase in the amplitude of the scale factor, not affecting the nature of the graph. Higher values of $V_{0}$ also only results in an increase in amplitude of the scale factor.
\item Fig. \ref{lambda18} shows the plot of the behavior of the potential with time for small field natural potential. This graph has been obtained with the help of Eqn. (\ref{potential6}) with parameter values $V_{0}=3.6{\rm x}10^{-3}M_{p}^{4},\  B_{6}=1, \ f=8 M_{p}$. Larger values of the parameters gives rise to larger amplitude of expansion, but the nature of the graph remains the same. If we compare Fig. \ref{lambda17} with Fig. \ref{lambda8}, we find that there occurs a net increase in the amplitude of the scale factor after one expansion-contraction cycle.
\item Fig. \ref{lambda19} shows the plot of the scale factor in the large field limit for natural potential given by Eqn. (\ref{scalefactor7}). This plot has been obtained for $V_{0}=10^{-2}M_{p}^{4},\ B_{8}=M_{p},\ B_{9}=1$. Expansion is obtained only if the value of $V_{0}$ is lesser than $2M_{p}^{4}$.
\item Fig. \ref{lambda20} shows the plot of the behavior of the potential with time for large field natural potential. This graph has been obtained with the help of Eqn. (\ref{potential7}) with parameter values $V_{0}=10^{-2}M_{p}^{4},\ B_{8}=M_{p}$. 
\end{itemize}

\textbf{For Case 2:}
\\ \\
\begin{figure*}[htb]
\centering
\subfigure[\scriptsize An illustration of the behavior of the scale factor with time  during expansion phase for $\phi<< f $ with $V_{0}=10^{-8}M_{p}^{4},\ f=2 M_{p},\ B_{16}=M_{p}^{-2},\ B_{17}=1,\ \Lambda_{1}=-10^{-2}M_{p}^{3},\ \Lambda_{2}=-10^{-2}M_{p}^{2},\ \Lambda_{3}=-10^{-2}M_{p},\ \Lambda_{4}=-10^{-2}$ for red curve and $V_{0}=10^{-8}M_{p}^{4},\ f=6 M_{p},\ B_{16}=M_{p}^{-2},\ B_{17}=1,\ \Lambda_{1}=3{\rm x}10^{-3}M_{p}^{3},\ \Lambda_{2}=3{\rm x}10^{-3}M_{p}^{2},\ \Lambda_{3}=3{\rm x}10^{-3}M_{p},\ \Lambda_{4}=3{\rm x}10^{-3}$ for blue dashed curve .]{
    \includegraphics[width=7.2cm,height=8cm] {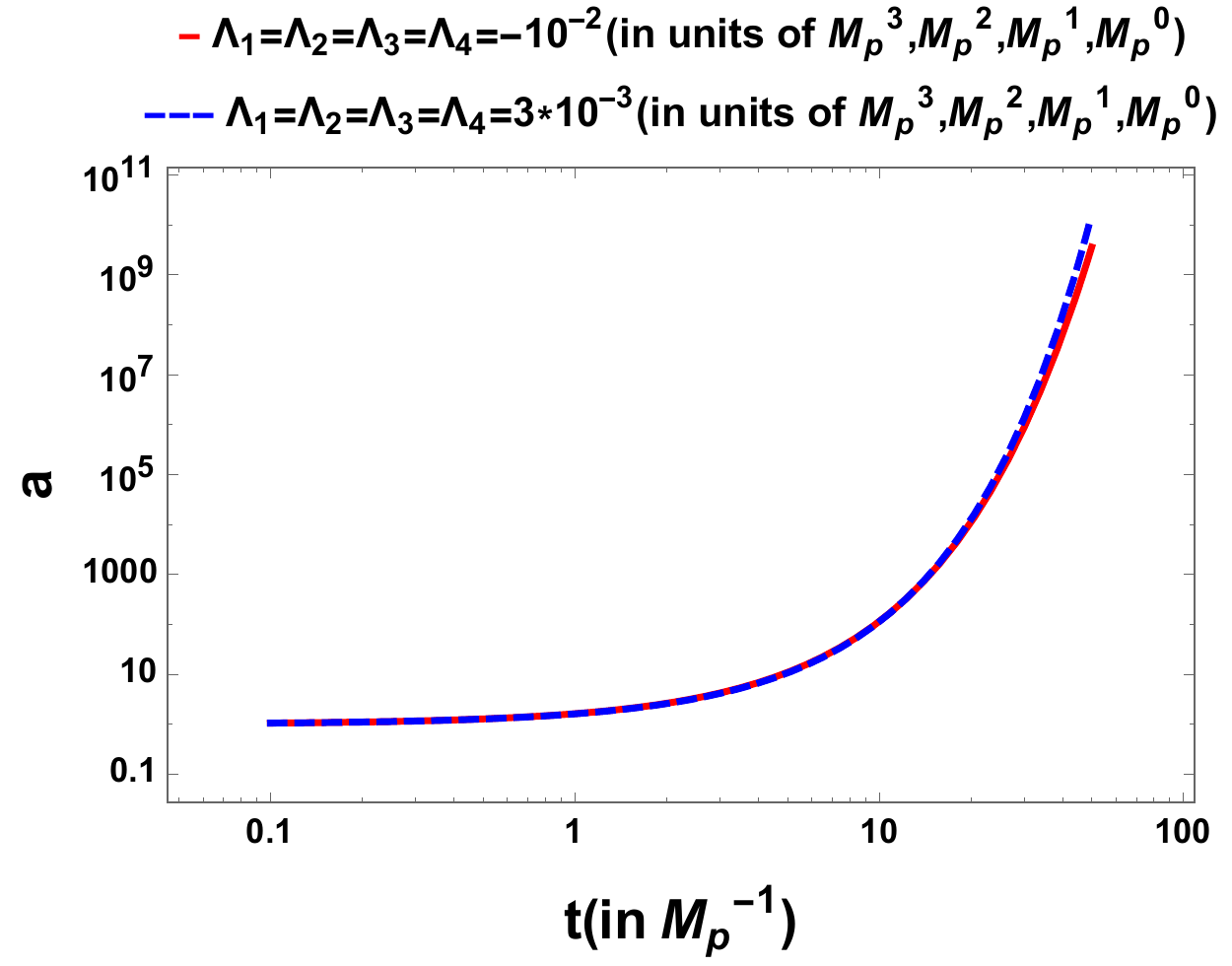}
    \label{lambda21}
}
\subfigure[\scriptsize An illustration of the behavior of the potential with time  during expansion phase for $\phi<<f$ with $V_{0}=4.4{\rm x}10^{-3}M_{p}^{4},\ f=4 M_{p},\ B_{16}=M_{p}^{-2},\  \Lambda_{1}=-10^{-2}M_{p}^{3},\ \Lambda_{2}=1.2{\rm x}10^{-3}M_{p}^{2},\ \Lambda_{3}=2.4{\rm x}10^{-3}M_{p},\ \Lambda_{4}=6.5{\rm x}10^{-3}$.]{
    \includegraphics[width=7.2cm,height=8cm] {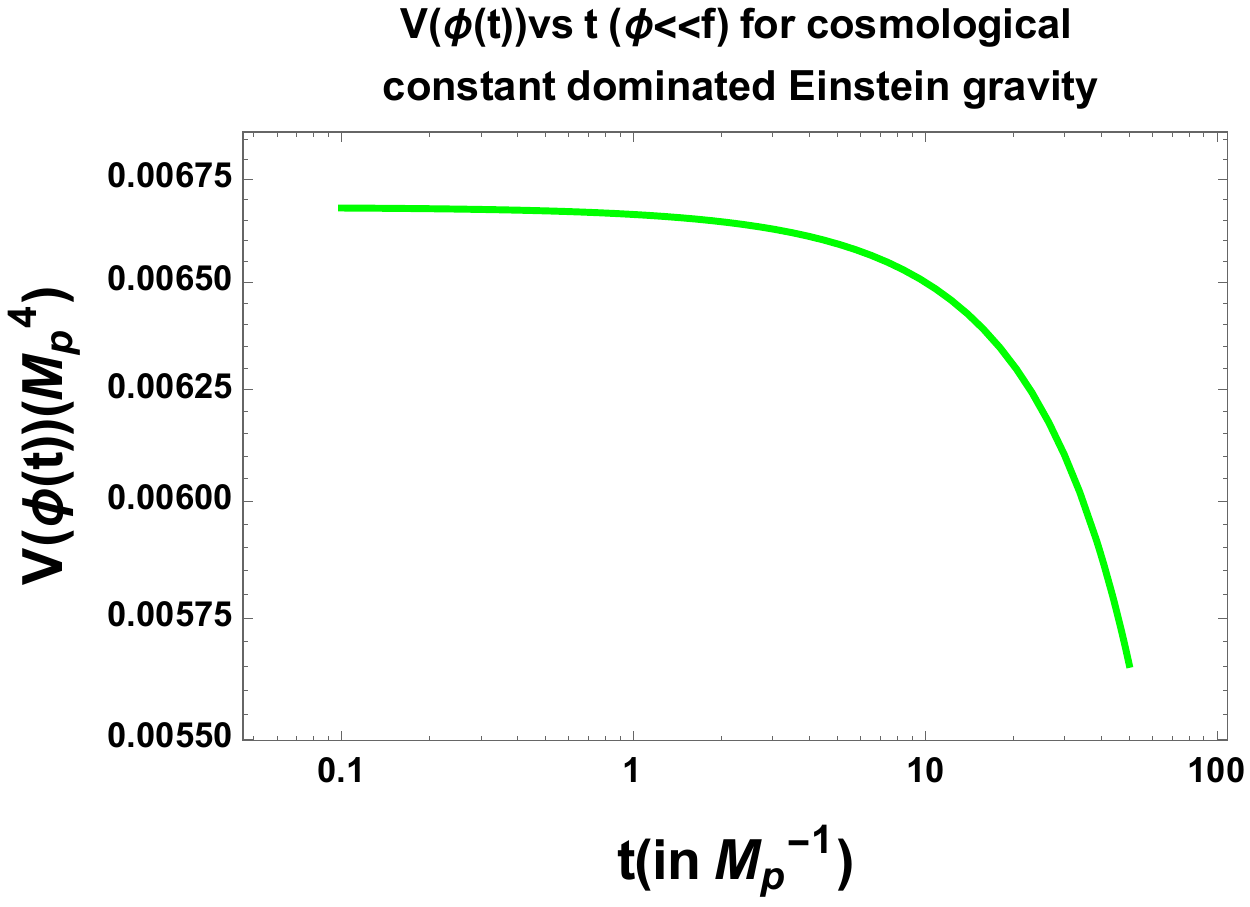}
    \label{lambda22}
}
\caption[Optional caption for list of figures]{ Graphical representation of the evolution of the scale factor and the potential during the expansion phase for the cosmological constant dominated Einstein gravity for case 2.} 
\label{fig10}
\end{figure*}
In Fig. \ref{fig10}, we have plotted the evolution of the scale factor and the potential during expansion for case 2. From the figures we can draw the following conclusions:
\begin{itemize}
\item Fig. \ref{lambda21} shows the variation of the scale factor with time for small field natural potential for two cases of parameter values for a scalar field dependent cosmological constant.
\item Fig. \ref{lambda21} has been obtained from Eq. (\ref{scalefactor8}). Detail graphical analysis show that an expanding universe in the present scenario is possible only if the values of $\Lambda_{1},\ \Lambda_{2},\ \Lambda_{3},\ \Lambda_{4}$ lies between $-10^{-2}$ to $10^{-2}$ similar to the case for hilltop potential. They can take any sign provided their magnitudes remain within the specified range. But expansion is possible only if $f>1 M_{p}$ and value of $B_{16}$ is very close to unity. If we make the value of $B_{16}$ too large, then expansion is possible eithe rif all the other constants are positive except $\Lambda_{2}$ or if only $\Lambda_{2}$ is positive and any two or all three other constants are negative. Increase in the value of $V_{0}$ only increases the amplitude of scale factor.
\item In Fig. \ref{lambda21}, we have shown two such cases which give rise to an expanding universe.
\item Fig. \ref{lambda22} shows the variation of the potential with time during expansion for case 2. Detail graphical analysis have shown that keeping the magnitude of the parameters within the above mentioned range, expansion is possible only if $\Lambda_{3}>3{\rm x}10^{-3}M_{p},\ \Lambda_{4}>5{\rm x}10^{-3}$ and $V_{0}>10^{-6}M_{p}^{4}$.
\item  Higher and positive  parameter values increases the height of the potential. 
\end{itemize}

\textbf{For Case 3:}
\\ \\
\begin{figure*}[htb]
\centering
\subfigure[An illustration of the behavior of the potential with time  during expansion phase for $\phi<<f$ with $V_{0}=10^{-8}M_{p}^{4},\ B_{20}=M_{p}^{-2},\ \Lambda=10^{-3}M_{p}^{4},\ B_{21}=1$.]{
    \includegraphics[width=7.2cm,height=8cm] {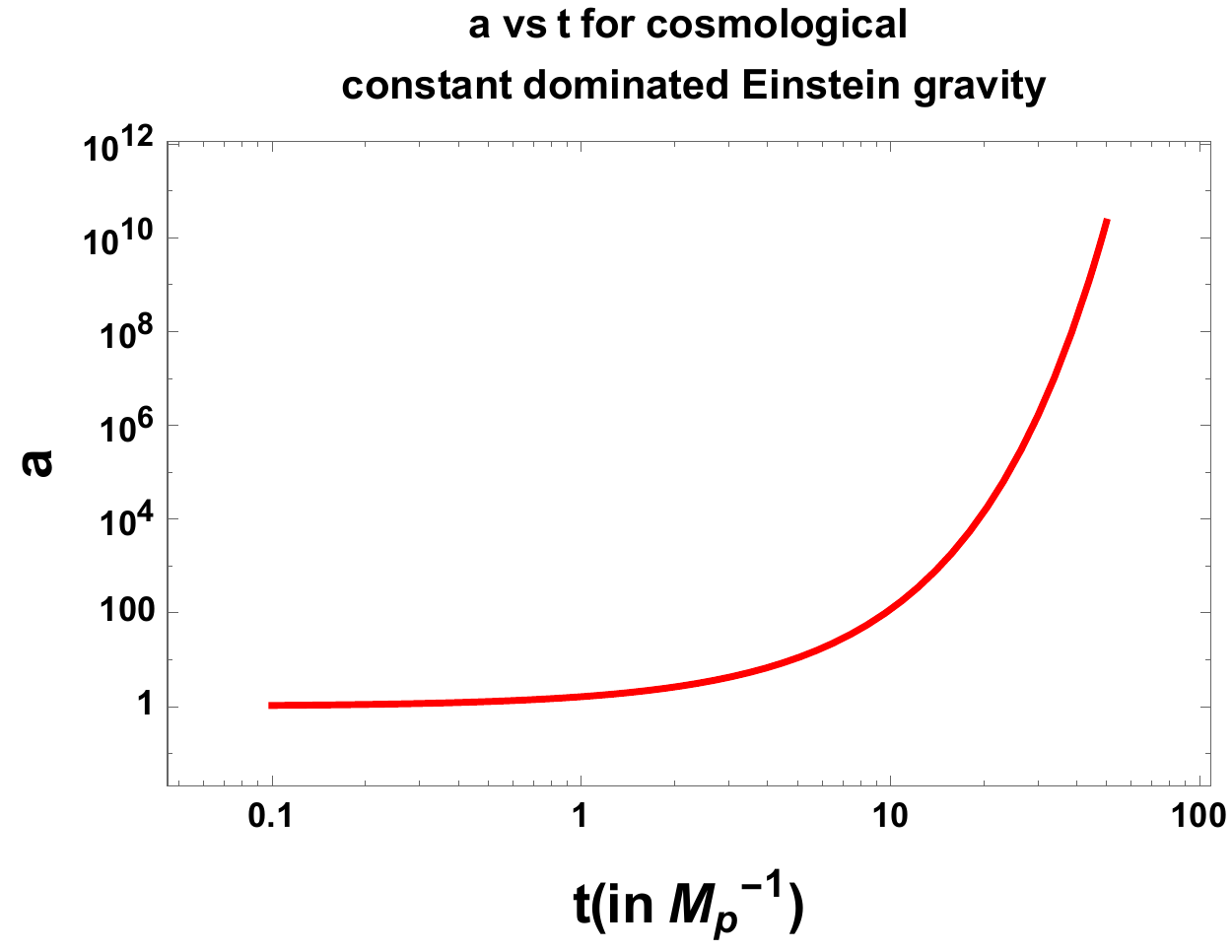}
    \label{lambda23}
}
\subfigure[An illustration of the behavior of the potential with time  during expansion phase for $\phi<<f$ with $V_{0}=4{\rm x}10^{-3}M_{p}^{4},\ f=1,\ B_{20}=M_{p}^{-2},\ \Lambda=-10^{-2}M_{p}^{4}$.]{
    \includegraphics[width=7.2cm,height=8cm] {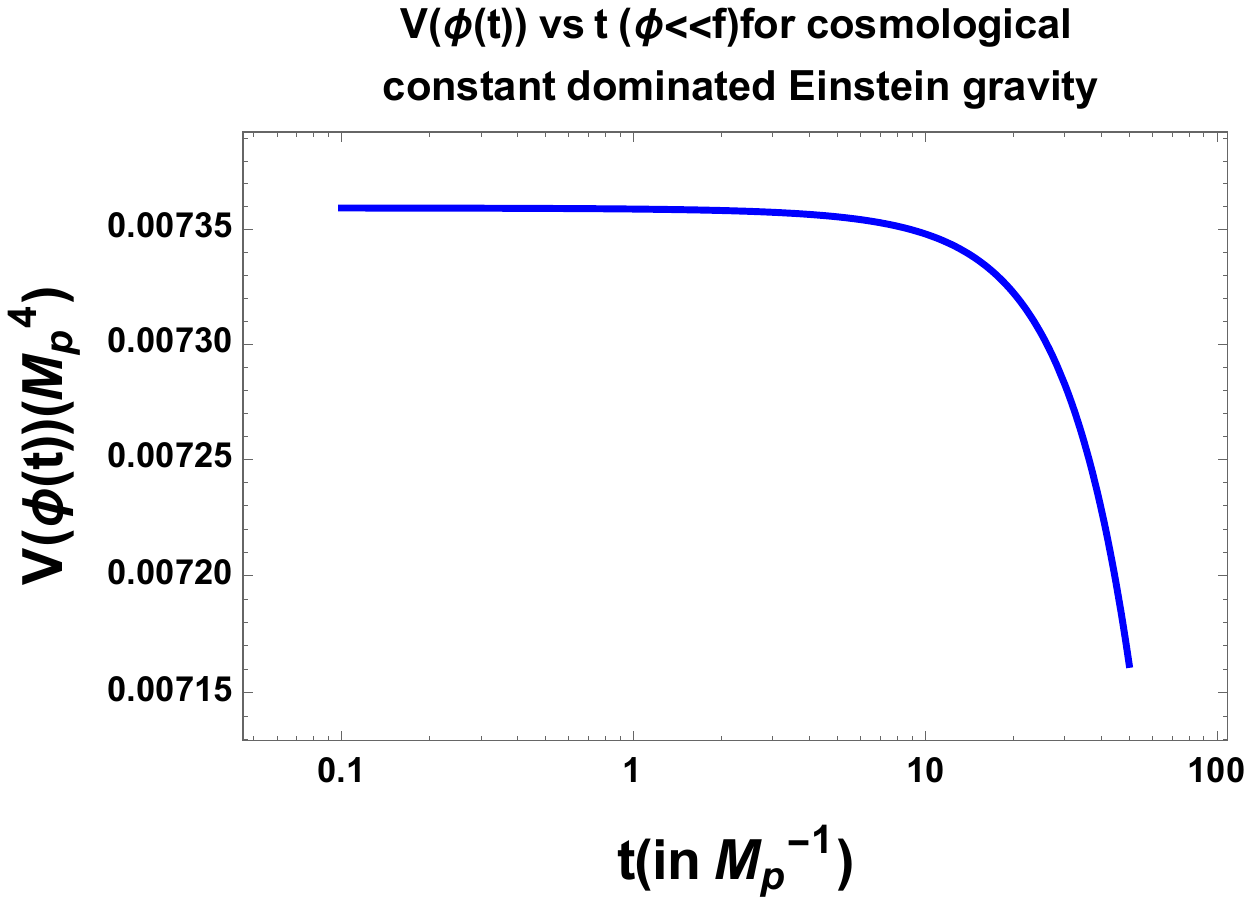}
    \label{lambda24}
}
\caption[Optional caption for list of figures]{ Graphical representation of the evolution of the scale factor and the potential during the expansion phase for the cosmological constant dominated Einstein gravity for case 3.} 
\label{fig11}
\end{figure*}
In Fig. \ref{fig11}, we have plotted the evolution of the scale factor and the potential during expansion for case 3. From the figures we can draw the following conclusions:
\begin{itemize}
\item Fig. \ref{lambda23} shows the variation of the scale factor with time for small field natural potential for $V_{0}=10^{-8}M_{p}^{4},\ B_{20}=M_{p}^{-2},\ \Lambda=10^{-3}M_{p}^{4},\ B_{21}=1$.
\item Fig. \ref{lambda23} has been obtained from Eq. (\ref{scalefactor9}). Detail graphical analysis show that an expanding universe in the present scenario is possible only if the value of $\Lambda$ lies within the range $-10^{-2}$ to $10^{-2}$. But if $f<1 M_{p}$, expansion is possible only if $B_{20}$ is very close to unity and $\Lambda > -2.8{\rm x}10^{-2}M_{p}^{4}$. For $f >1 M_{p}$, expansion is possible if $\Lambda$ lies within the range $-10^{-2}M_{p}^{4}$ to $10^{-2}M_{p}^{4}$
\item Fig. \ref{lambda24} shows the variation of the natural potential for an expanding universe for case 3 with $\phi<<f$ with $V_{0}=4{\rm x}10^{-3}M_{p}^{4},\ f=1,\ B_{20}=M_{p}^{-2},\ \Lambda=-10^{-2}M_{p}^{4}$. Fig. \ref{lambda24} has been obtained with the help of Eqn. (\ref{potential9}). Graphical analysis show that within the allowed range of the parameter values, as discussed before, larger values of $V_{0}$ and $\Lambda$ results in steeper fall of the potential and larger potential height.

\end{itemize}

\subsubsection{Case III:  Coleman-Weinberg potential}

\textbf{For Case 1:}
\\ \\
\begin{figure*}[htb]
\centering
\subfigure[ An illustration of the behavior of the scale factor with time  during expansion phase for $\phi<<M_{p}$ with $V_{0}=2{\rm x}10^{-3}M_{p}^{4},\ B_{10}=M_{p},\ B_{11}=1,\ \alpha=0.5,\ \beta=1$.]{
    \includegraphics[width=7.2cm,height=7.5cm] {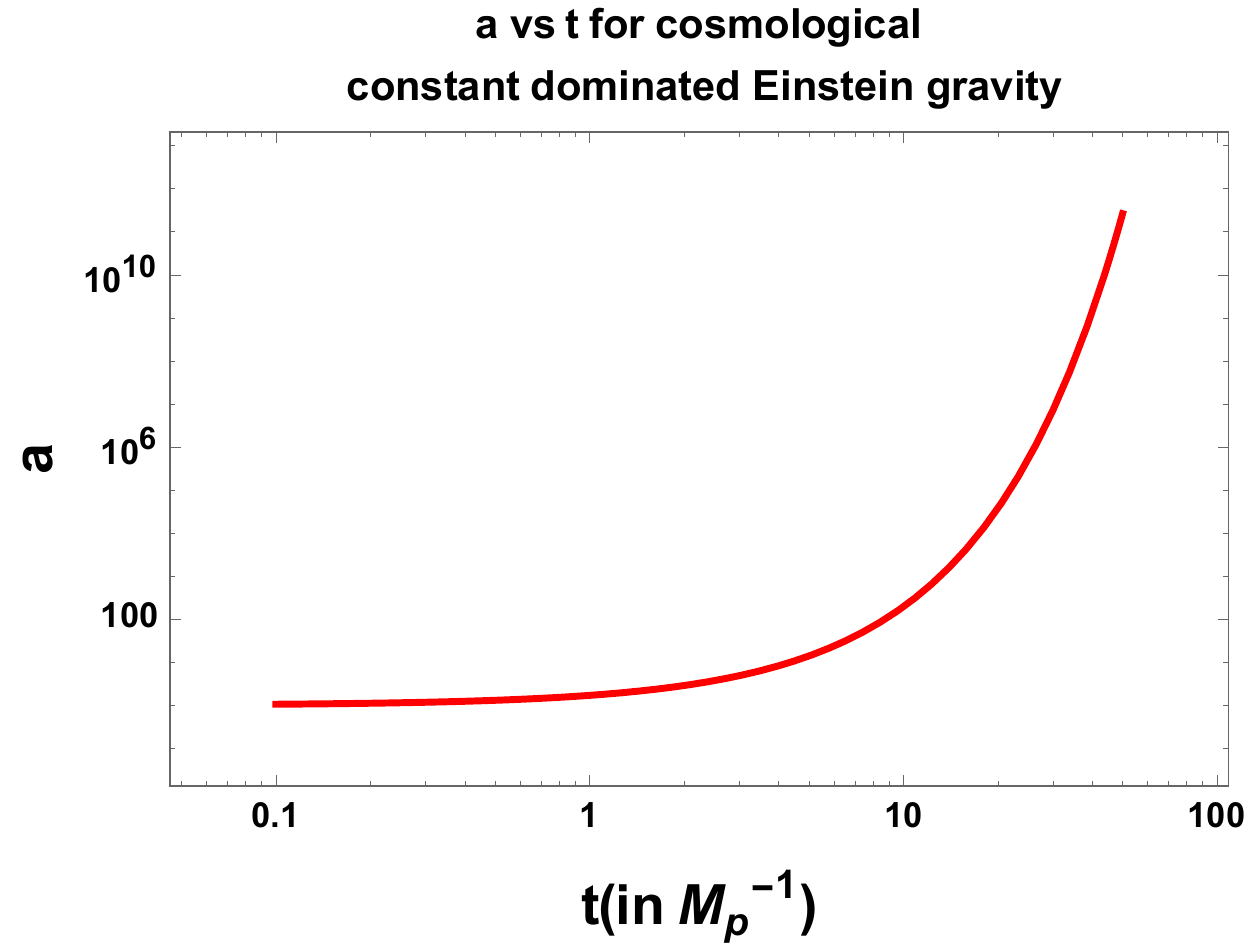}
    \label{lambda25}
}
\subfigure[An illustration of the behavior of the potential during expansion phase for $\phi<<M_{p}$ with $V_{0}=4.6{\rm x}10^{-3}M_{p}^{4},\  B_{10}=0.5,\ \alpha=0.18,\ \beta=2$ .]{
    \includegraphics[width=7.2cm,height=7.5cm] {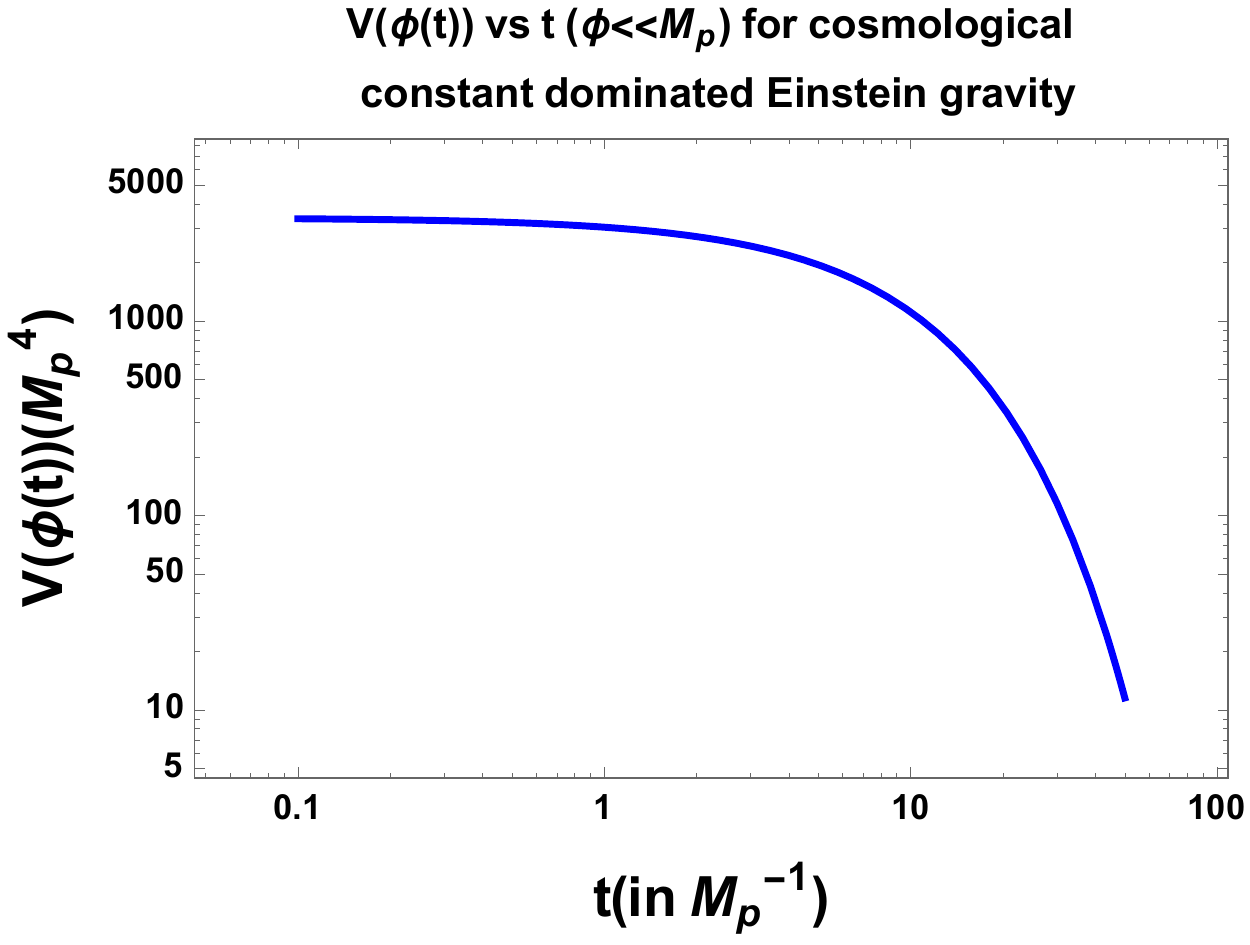}
    \label{lambda26}
}
\subfigure[An illustration of the behavior of the scale factor with time  during expansion phase for $\phi>>M_{p}$ with $V_{0}=10^{-8}M_{p}^{4},\ B_{12}=10^{-8}M_{p},\ B_{13}=1,\ \beta=7$.]{
    \includegraphics[width=7.2cm,height=7.5cm] {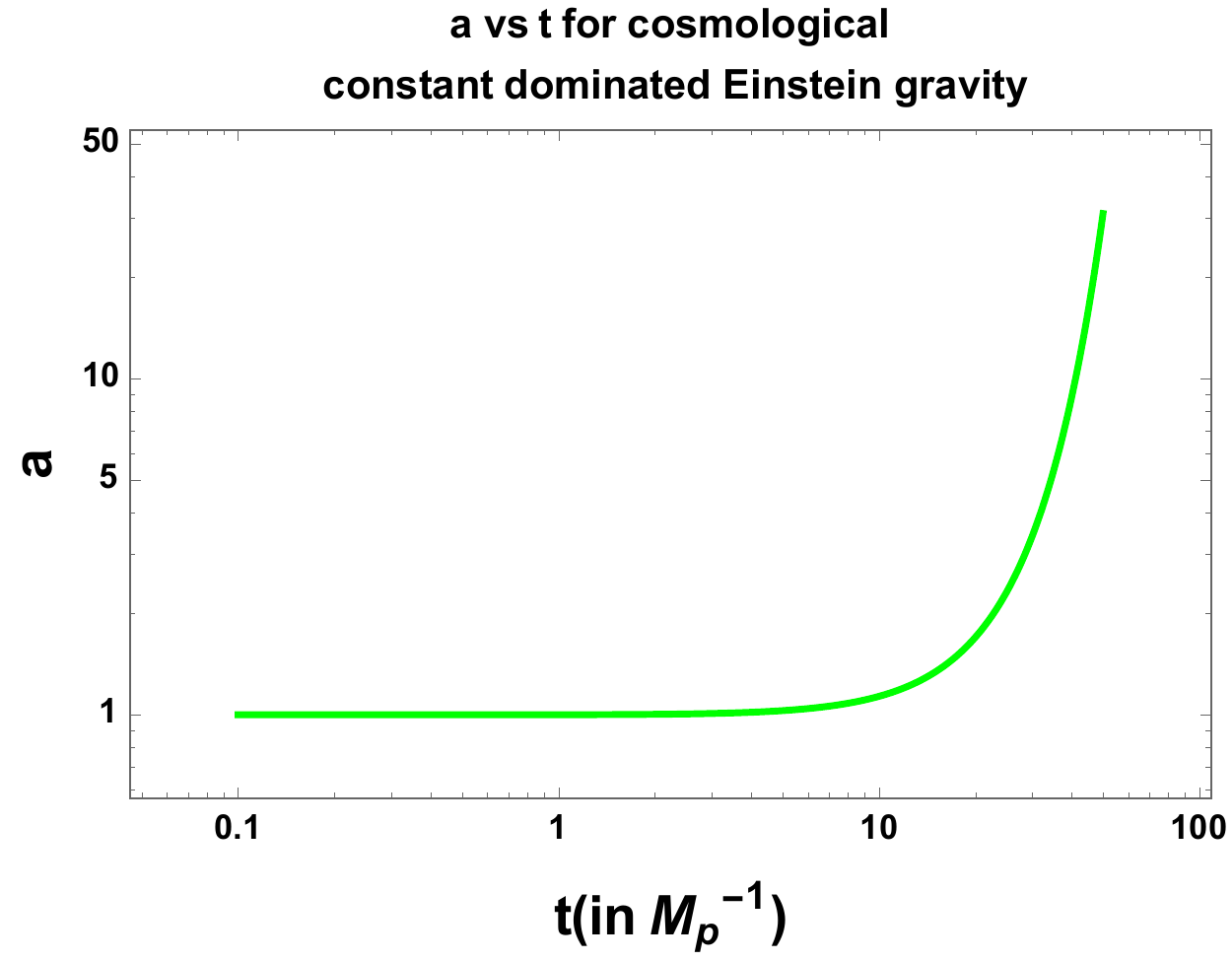}
    \label{lambda27}
}
\subfigure[An illustration of the behavior of the potential with time  during expansion phase for $\phi>>M_{p}$ with $V_{0}=8{\rm x}10^{-3}M_{p}^{4},\ B_{12}=10^{-8}M_{p},\ \alpha=1,\ \beta=-0.01$.]{
    \includegraphics[width=7.2cm,height=7.5cm] {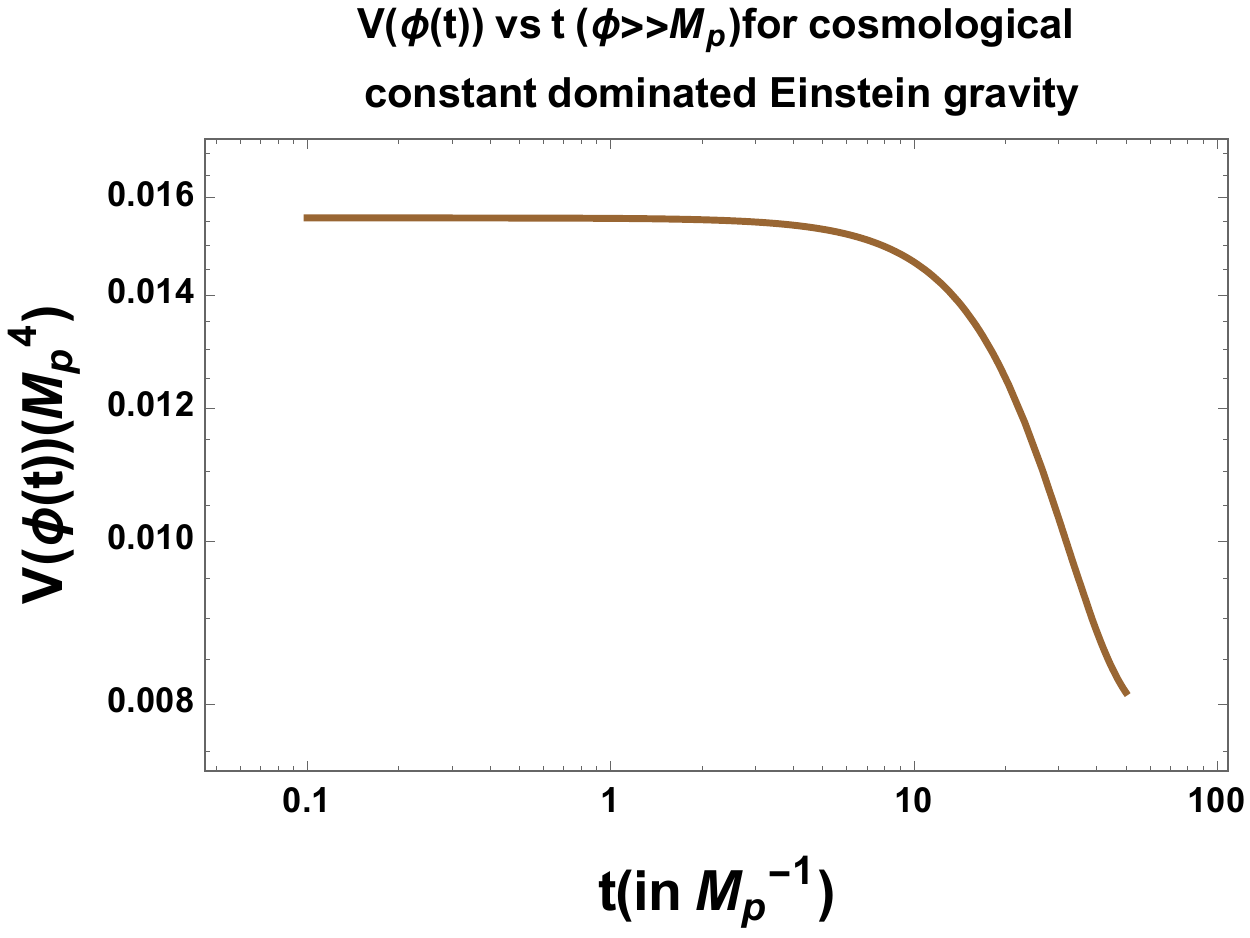}
    \label{lambda28}
}
\caption[Optional caption for list of figures]{ Graphical representation of the evolution of the scale factor and the potential during the expansion and contraction phase for the cosmological constant dominated Einstein gravity for case 1.} 
\label{fig12}
\end{figure*}

Fig. \ref{fig12} show the evolution of the scale factor and the potential for the case of supergravity potential for case 1 during expansion. The following conclusions can be drawn from the graphs:
\begin{itemize}
\item Fig. \ref{lambda25}, shows the plot of the scale factor in the small field limit for supergravity potential given by Eqn. (\ref{scalefactor10}) with the parameter values $V_{0}=2{\rm x}10^{-3}M_{p}^{4},\ B_{10}=M_{p},\ B_{11}=1,\ \alpha=0.5,\ \beta=1$. 
\item While obtaining the above plot, we kept the value of the integration constants $B_{10}$ and $B_{11}$ as 1. Higher values of the integration constants only results in an increase in the amplitude of the scale factor, not affecting the nature of the graph. Higher values of $V_{0}$ also only results in an increase in amplitude of the scale factor. The evolution of the scale factor is almost independent of $\alpha$. Large increase in amplitude of the scale factor occurs for large value of $\beta$ irrespective of its sign. 
\item Fig. \ref{lambda26} shows the plot of the behavior of the potential with time for small field supergravity potential. This graph has been obtained with the help of Eqn. (\ref{potential10}) with parameter values $V_{0}=4.6{\rm x}10^{-3}M_{p}^{4},\  B_{10}=0.5,\ \alpha=0.18,\ \beta=2$. Larger values of the parameters gives rise to larger amplitude of expansion, but the nature of the graph remains the same. If we compare Fig. \ref{lambda25} with Fig. \ref{lambda8}, we find that there occurs a net increase in the amplitude of the scale factor after one expansion-contraction cycle.
\item Fig. \ref{lambda27} shows the plot of the scale factor in the large field limit for supergravity potential given by Eqn. (\ref{scalefactor11}). This plot has been obtained for $V_{0}=10^{-8}M_{p}^{4},\ B_{12}=10^{-8}M_{p},\ B_{13}=1,\ \beta=7$. Proper expansion is obtained only if the value of $B_{12}$ is $<<1$ and the value of $\beta$ is large. Larger value of $V_{0}$ results in nearly constant expansion amplitude with a huge increase towards the later times. Expansion is possible for both negative and positive values of $\beta$.
\item Fig. \ref{lambda28} shows the plot of the behavior of the potential with time for large field supergravity potential. This graph has been obtained with the help of Eqn. (\ref{potential11}) with parameter values $V_{0}=8{\rm x}10^{-3}M_{p}^{4},\ B_{12}=10^{-8}M_{p},\ \alpha=1,\ \beta=-0.01$. Proper evolution of the potential which will give rise to expanding universe is possible only if $\beta<0$ and $B_{12}$ is $<<1$. Larger values of $\alpha$ makes the potential fall more steeply. Larger values of $V_{0}$ mildly increase the height of the potential. 
\end{itemize}

\textbf{For Case 2:}
\\ \\
\begin{figure*}[htb]
\centering
\subfigure[\scriptsize An illustration of the behavior of the scale factor with time  during expansion phase for $\phi<< M_{p}$ with $V_{0}=1.3{\rm x}10^{-3}M_{p}^{4},\ B_{16}=M_{p}^{-2},\ B_{17}=1,\ \Lambda_{1}=2.6{\rm x}10^{-2}M_{p}^{3},\ \Lambda_{2}=-10^{-1}M_{p}^{2},\ \Lambda_{3}=-10^{-1}M_{p},\ \Lambda_{4}=8.8{\rm x}10^{-2},\ \alpha=0.13,\ \beta=-0.4$ .]{
    \includegraphics[width=7.2cm,height=8cm] {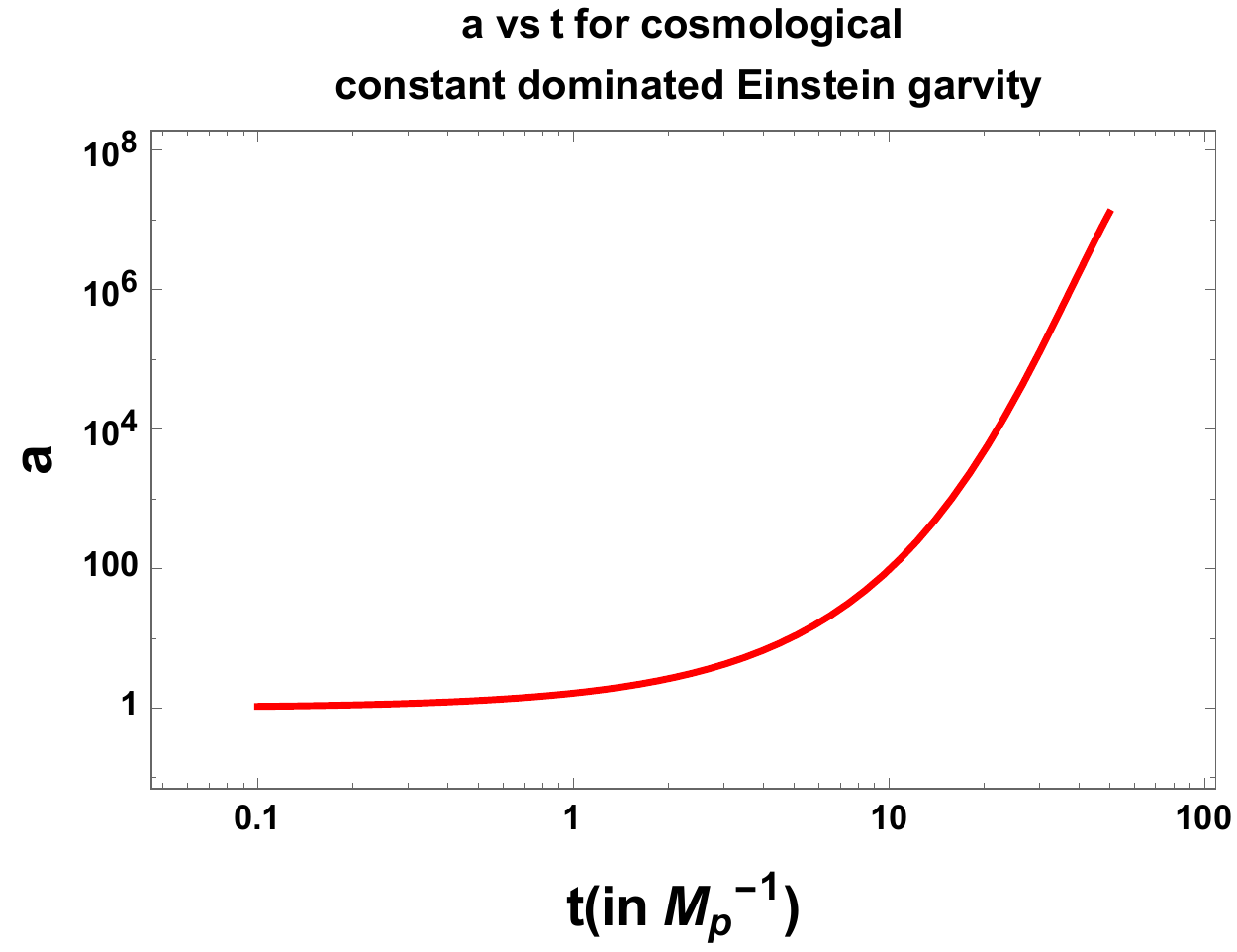}
    \label{lambda32}
}
\subfigure[\scriptsize An illustration of the behavior of the potential with time  during expansion phase for $\phi<< M_{p} $ with $V_{0}=2.1{\rm x}10^{-2}M_{p}^{4},
\  \alpha=10^{-4},\  B_{16}=10^{-4}M_{p}^{-2},\ \Lambda_{1}=-10^{-1}M_{p}^{3},\ \Lambda_{2}=7.9{\rm x}10^{-2}M_{p}^{2},\ \Lambda_{3}=7.9{\rm x}10^{-2}M_{p},\ \Lambda_{4}=8.4{\rm x}10^{-1},\ \beta=-1.85$.]{
    \includegraphics[width=7.2cm,height=8cm] {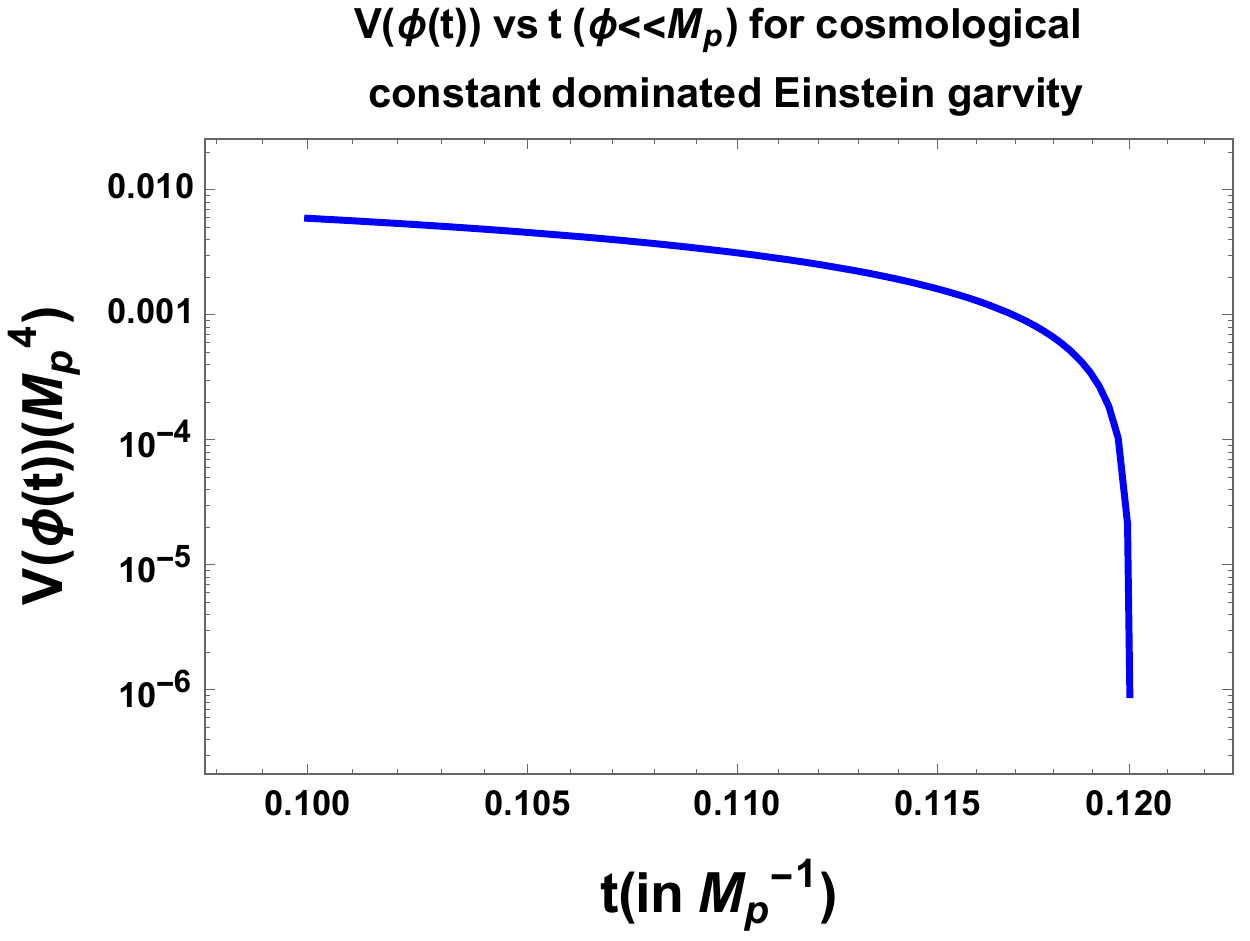}
    \label{lambda33}
}
\caption[Optional caption for list of figures]{ Graphical representation of the evolution of the scale factor and the potential during the expansion phase for the cosmological constant dominated Einstein gravity for case 2.} 
\label{fig13}
\end{figure*}
In Fig. \ref{fig13}, we have plotted the evolution of the scale factor and the potential during expansion for case 2. From the figures we can draw the following conclusions:
\begin{itemize}
\item Fig. \ref{lambda32} shows the variation of the scale factor with time for small field supergravity potential for case 2 with $V_{0}=1.3{\rm x}10^{-3}M_{p}^{4},\ B_{16}=M_{p}^{-2},\ B_{17}=1,\ \Lambda_{1}=2.6{\rm x}10^{-2}M_{p}^{3},\ \Lambda_{2}=-10^{-1}M_{p}^{2},\ \Lambda_{3}=-10^{-1}M_{p},\ \Lambda_{4}=8.8{\rm x}10^{-2},\ \alpha=0.13,\ \beta=-0.4$
\item Fig. \ref{lambda32} has been obtained from Eqn. (\ref{scalefactor12}). Detail graphical analysis show that an expanding universe in the present scenario is possible only if the values of $\Lambda_{1},\ \Lambda_{2},\ \Lambda_{3},\ \Lambda_{4}$ lies between $-10^{-1}$ to $10^{-1}$. They can take any sign provided their magnitudes remain within the specified range. But expansion is possible for $\beta>0$ only if any two of the constants ($\Lambda_{1},\ \Lambda_{2},\ \Lambda_{3},\ \Lambda_{4}$) are positive and the other two take negative values. For $\beta<0$, exoansion is possible for any possible combination of signs of the constants. Large values of $\alpha$ results in larger amplitude of expansion, but the nature remains the same.
\item Fig. \ref{lambda33} shows the variation of the potential with time during expansion for case 2 obtained with the help of Eqn. (\ref{potential12}) and with parameter values $V_{0}=2.1{\rm x}10^{-2}M_{p}^{4},\  \alpha=10^{-4},\  B_{16}=10^{-4}M_{p}^{-2},\ \Lambda_{1}=-10^{-1}M_{p}^{3},\ \Lambda_{2}=7.9{\rm x}10^{-2}M_{p}^{2},\ \Lambda_{3}=7.9{\rm x}10^{-2}M_{p},\ \Lambda_{4}=8.4{\rm x}10^{-1},\ \beta=-1.85$. Detail graphical analysis have shown that keeping the magnitude of the parameters within the above mentioned range, potential evolves with the required nature to cause expansion only if $\beta<0$ and $\alpha\leq 1$. 
\item  Higher and positive  parameter values increases the height of the potential. 
\end{itemize}

\textbf{For Case 3:}
\\ \\
\begin{figure*}[htb]
\centering
\subfigure[An illustration of the behavior of the potential with time  during expansion phase for $\phi<<M_{p}$ with $V_{0}=1.5{\rm x}10^{-3}M_{p}^{4},\ B_{22}=M_{p}^{-2},\ \Lambda=4{\rm x}10^{-3}M_{p}^{4},\ B_{23}=1,\ \alpha=0.96,\ \beta=1$.]{
    \includegraphics[width=7.2cm,height=7.7cm] {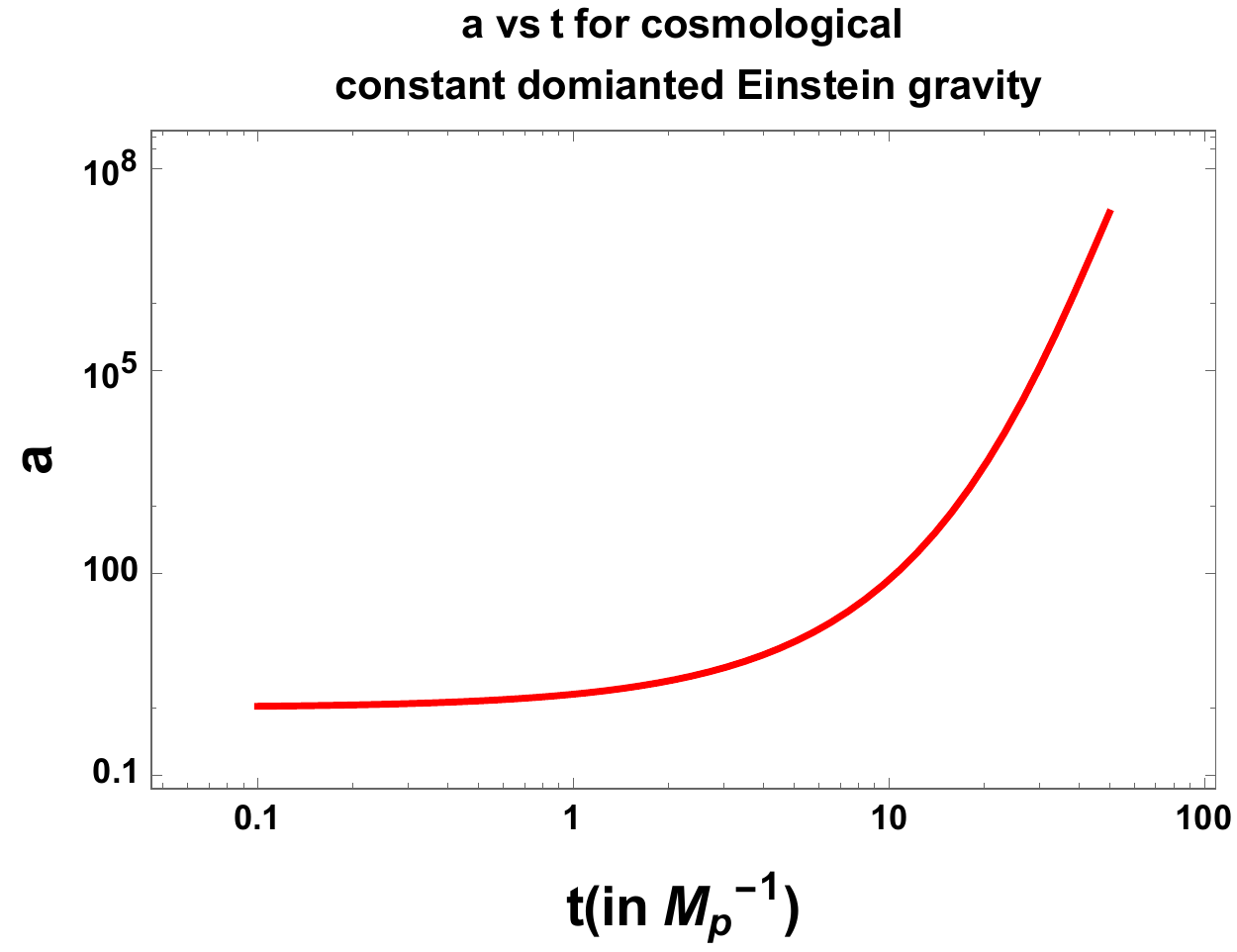}
    \label{lambda29}
}
\subfigure[An illustration of the behavior of the potential with time  during expansion phase for $\phi<<M_{p}$ with $V_{0}=8{\rm x}10^{-3}M_{p}^{4},\ \alpha=0.8,\ B_{22}=M_{p}^{-2},\ \beta=0.15,\ B_{23}=1,\ \Lambda=-3.6{\rm x}10^{-2}M_{p}^{4}$.]{
    \includegraphics[width=7.2cm,height=7.7cm] {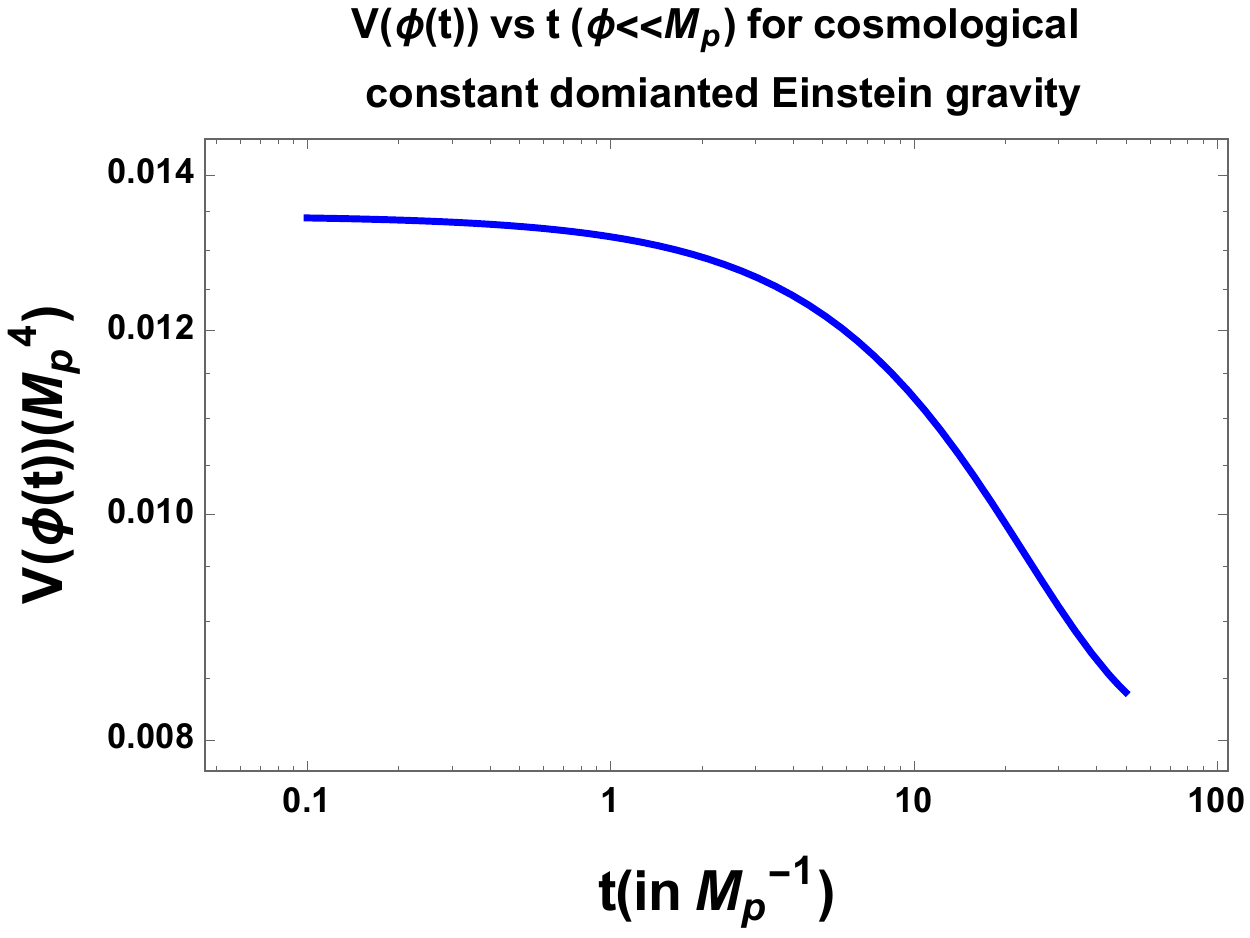}
    \label{lambda31}
}
\subfigure[An illustration of the behavior of the potential with time  during expansion phase for $\phi<<M_{p}$ with $V_{0}=1.2{\rm x}10^{-3}M_{p}^{4},\ \alpha=10^{-8},\ B_{22}=10^{-8}M_{p}^{-2},\ \beta=-0.7,\ B_{23}=1,\ \Lambda=4.3{\rm x}10^{-3}M_{p}^{4}$.]{
    \includegraphics[width=11.2cm,height=7.7cm] {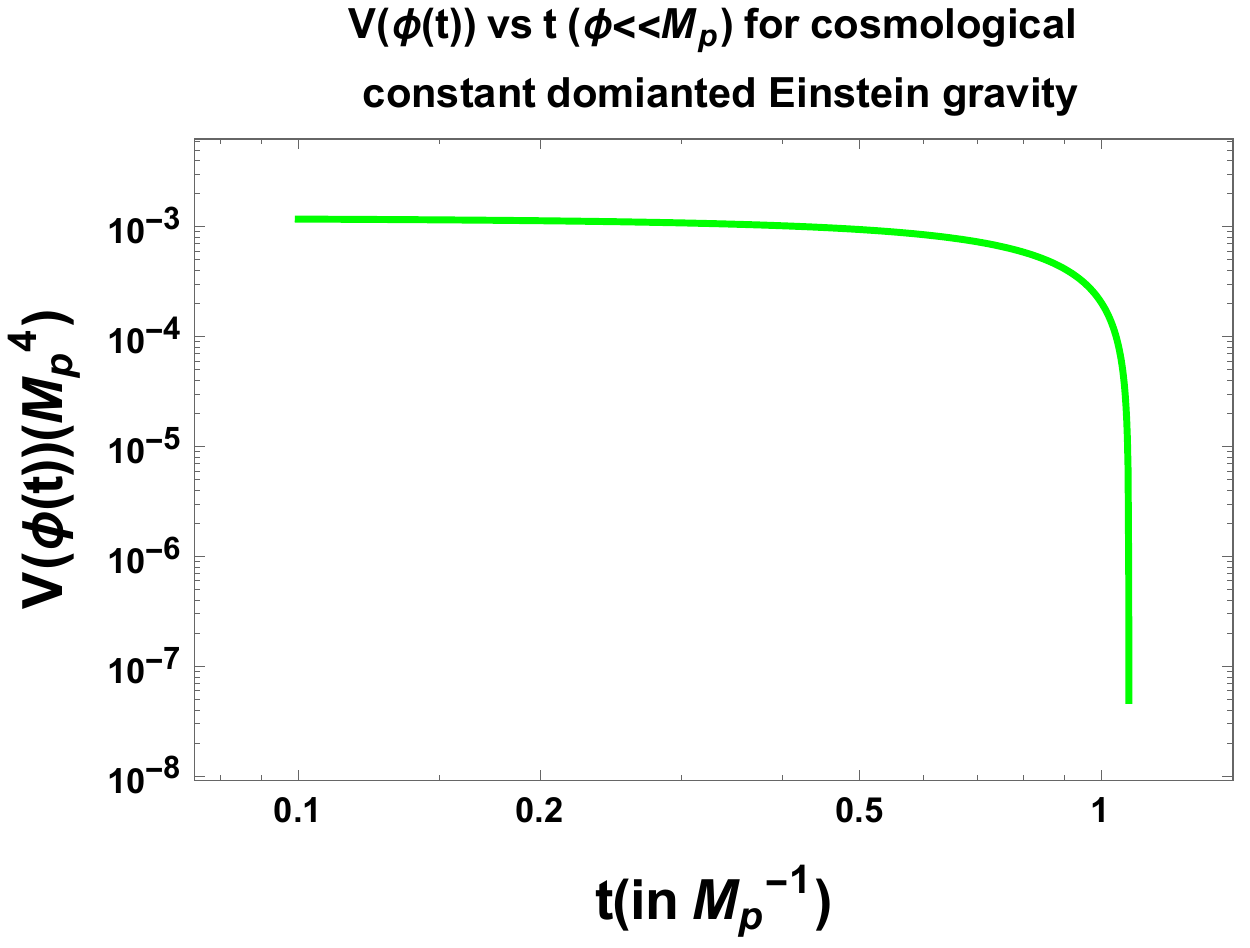}
    \label{lambda30}
}
\caption[Optional caption for list of figures]{ Graphical representation of the evolution of the scale factor and the potential during the expansion phase for the cosmological constant dominated Einstein gravity for case 3.} 
\label{fig14}
\end{figure*}
In Fig. \ref{fig14}, we have plotted the evolution of the scale factor and the potential during expansion for case 3. From the figures we can draw the following conclusions:
\begin{itemize}
\item Fig. \ref{lambda29} shows the variation of the scale factor with time for small field supergravity potential for $V_{0}=1.5{\rm x}10^{-3}M_{p}^{4},\ B_{22}=M_{p}^{-2},\ \Lambda=4{\rm x}10^{-3}M_{p}^{4},\ B_{23}=1,\ \alpha=0.96,\ \beta=1$.
\item Fig. \ref{lambda29} has been obtained from Eqn. (\ref{scalefactor13}). Detail graphical analysis show that an expanding universe in the present scenario is possible only if the value of $\Lambda$ lies within the range $-10^{-1}$ to $10^{-2}$. If we decreas the value of $V_{0}$ to $10^{-7}M_{p}^{4}$(say), then expansion is possible only for values of $B_{22} <<M_{p}^{-2}$. Expansion is possible for both positive and negative values of $\beta$, however small positive values of $\beta$ are more favourable for expansion. 
\item Figs. \ref{lambda31} and \ref{lambda30} show the variation of the supergravity potential for an expanding universe for case 3 with $\phi<<M_{p}$ for two sets of parameter values, $V_{0}=8{\rm x}10^{-3}M_{p}^{4},\ \alpha=0.8,\ B_{22}=M_{p}^{-2},\ \beta=0.15,\ B_{23}=1,\ \Lambda=-3.6{\rm x}10^{-2}M_{p}^{4}$ and $V_{0}=1.2{\rm x}10^{-3}M_{p}^{4},\ \alpha=10^{-8},\ B_{22}=10^{-8}M_{p}^{-2},\ \beta=-0.7,\ B_{23}=1,\ \Lambda=4.3{\rm x}10^{-3}M_{p}^{4}$ respectively. Figs. \ref{lambda31} and \ref{lambda30} have been obtained with the help of Eqn. (\ref{potential13}). Graphical analysis show that within the allowed range of the parameter values, as discussed before, for values of $\Lambda$ near the upper bound, expansion is possible only for negative values of $\beta$ and values of other parameters ($\alpha,\ V_{0},\ B_{22}$) much less than unity ($\sim 10^{-8}$). For positive value of $\beta$ expansion is possible if the value of $V_{0}>10^{-4}M_{p}^{4}$ and that of $\alpha$ is close to unity. In this case, in order to get an expanding universe, $\Lambda$ must have small negative values. Larger the magnitude of $\beta$, steeper the fall of the potential.
\end{itemize}

\section{Hysteresis from Loop Quantum gravity (LQG) model}
\label{sq5}
Loop Quantum gravity (LQG) is one of the candidate theories of quantum gravity which successfully resolves the big bang singularity. It is primarily based on the
quantum geometry effects of loop quantum gravity.

For the spatially flat case when the curvature parameter $k=0$, the modified Friedmann equations in this model are given by:
\bea
H^{2} &=&\left(\frac{\dot{a}}{a}\right)^2= \frac{\rho}{3M^{2}}\left(1 - \frac{\rho}{\rho_{c}}\right),\\
\dot{H}+H^{2}&=&\frac{\ddot{a}}{a} = -\frac{1}{6M^{2}}\left((\rho + 3p) - \frac{2\rho}{\rho_{c}}(2\rho + 3p)\right),
\eea
which is exactly same as that we get for RSII brane world cosmology (see Appendix).  Since this analysis has already been done in detail by \cite{Sahni:2012er}, we have not shown 
it here again. However the results of this analysis have been quoted in the Appendix for reference. Hence the rest of the conclusions remain same as that obtained for RSII brane world model.

Modified Friedmann equations for non flat case i.e. for $k \neq 0$ in this model \cite{Singh:2010qa,Ashtekar:2006es} are given by:
\bea
H^{2} &=&\left(\frac{\dot{a}}{a}\right)^2=  \frac{1}{3M^{2}} \, (\rho - \rho_1)\left(\frac{1}{\rho_{c}} (\rho_2 - \rho)\right)
\label{lqc1},\\
\dot{H}+H^{2}&=&\frac{\ddot{a}}{a} = \nonumber -\frac{1}{6M^{2}} (\rho + 3 p) + \frac{2}{3M^{2}} \left(\frac{\rho}{\rho_{c}} + k \chi\right)\left(\rho + \frac{3}{2} p \right) \\
&& ~~~~~~~~~~~~~~~+ \frac{k \chi}{\gamma^2 \Delta} \left(\frac{\rho}{\rho_{c}} + k \chi\right) - \frac{2 \zeta k}{\gamma^2 \Delta} \left(\frac{\rho}{\rho_{c}} + k \chi - \frac{1}{2} \right),
\label{lqc2}
\eea
where we introduce two new constants $\rho_1$ and $\rho_2$ are defined as: 
\bea
\rho_1 :&=& \frac{- 3 k \chi M^{2}}{\gamma^2 \Delta} = - k \, \chi \, \rho_{c},\\
\rho_2 :&=& \rho_{c}(1 - k \chi) ~.\eea
Here the critical density $\rho_{c}$, the model parameters $\Delta$, $\chi$ and $\zeta$ can be expressed within LQG setup as \cite{Singh:2010qa,Ashtekar:2006es}:
\bea 
\rho_{c} &=&\frac{3M^2_p}{\gamma^2 \Delta},\\
\Delta&=& 4\sqrt{3} \pi\gamma {\it l}^2_{p}=\bar{\mu}^2 p,\\
 \displaystyle \chi&=&\left\{\begin{array}{ll}
                    \displaystyle   \sin^2\bar{\mu} - \left(1+\gamma^2\right)\bar{\mu}^2,~~~~~~~ &
 \mbox{\small {\bf for {$k=+1$}}}  \\ \\
         \displaystyle  -\gamma^2 \bar{\mu}^2, & \mbox{\small {\bf for {$k=-1$}}}.
          \end{array}
\right.\\
\zeta&=& \sin^2\bar{\mu}-\frac{\bar{\mu}}{2}\sin 2\bar{\mu},
\eea
where $p$ is the triad (which without any loss of generality will be
chosen with positive orientation) and the Barbero-Immirzi
parameter $\gamma$ can be fixed by computing the black hole entropy
in LQG.
Now let us concentrate on the classical limit, $\Delta\rightarrow 0$, from LQG setup one can write:
\bea 
\chi&\rightarrow&-\gamma^2 \bar{\mu}^2,\\
\rho_{1}&\rightarrow&\frac{3k M^2_{p}}{p},\\
\frac{(\rho_{2}-\rho)}{\rho_{c}}&\rightarrow&1.
\eea
Thus we recover the Friedmann equation in classical GR
in the limit $\Delta\rightarrow 0$. In the next subsections we will discuss the detailed cosmological consequences as well the phenomena of cosmological hysteresis from this model.
\subsection{Condition for bounce} 

Bounce occurs when the following condition holds good:
\be \rho = \rho_{2},\ee where the universe
reaches its minimum radius $a_{min}$ and maximum density \cite{Ashtekar:2006es} is achieved:
\be \rho_{max} = \rho_{2}\mid_{a_{min}} \approx \rho_{c}.\ee Therefore condition for bounce is given by:
\begin{equation}
\rho_{b} = \rho_{c}.
\end{equation}
The mass at bounce (neglecting the constant factor) is given by:
\be M_{b} = \rho_{b}a_{b}^{3} = \rho_{c}a_{b}^{3}.\ee 
Therefore an infinitesimal variation in mass at bounce can be expressed as: \be \delta M =  \rho_{c}\delta(a_{b})^{3}.\ee 
By setting the following constraint condition:
\be \delta M = -\delta W = -\oint pdV,\ee we get the expression for change in amplitude of the scale factor at each successive cycle as:
\begin{equation}
\delta(a_{min})^{3} = -\frac{1}{\rho_{c}}\oint pdV
\label{lqc3}
\end{equation}
Thus, the amplitude of the scale factor increases iff: \be \oint pdV < 0\ee as
the critical density $\rho_{c}>0$ always. We also observe that now the increase in scale factor depends on the critical density 
$\rho_{c}$ but is independent of the curvature parameter $k$. Hence the result holds good for $k=\pm 1$ i.e. for both open and closed universe.

\subsection{Condition for acceleration}

From Eq.~ (\ref{lqc2}), at bounce the condition for acceleration is given by:
\begin{eqnarray}
\frac{\ddot a}{a} &=& \nonumber -\frac{1}{6M^{2}} (\rho_{c} + 3 p_{b}) + \frac{4}{6M^{2}} \left(1 + k \chi\right)\left(\rho_{c} + \frac{3}{2} p_{b} \right) \\
&& ~~~~~~~~~~~~~~~~~~~~~~~~+ \frac{k \chi}{\gamma^2 \Delta} \left(1 + k \chi\right) - \frac{2 \zeta k}{\gamma^2 \Delta} \left( k \chi + \frac{1}{2} \right)
\label{lqcaccel}
\end{eqnarray}
Thus it implies that whether the condition for acceleration violates the energy condition, 
now depends upon the values of different parameters of the LQG model present in the above expression.

For different values of the curvature parameter $k$ we get the following constraint conditions for acceleration at bounce as:
\be\begin{array}{lll}\label{accellqc}
 \displaystyle p_{b} >\left\{\begin{array}{ll}
                    \displaystyle   -\frac{2\chi M^{2}}{\gamma^{2}\Delta}\frac{(1 + 
                    \chi)}{1 + 2\chi} + \frac{2\zeta M^{2}}{\gamma^{2}\Delta} - \frac{\rho_{c}(3 + 4\chi)}{3(1 + 2\chi)}~~~~ &
 \mbox{\small {\bf for {$k=+1$}}}  \\ \\
         \displaystyle  \frac{2\chi M^{2}}{\gamma^{2}\Delta}\frac{(1 - \chi)}{1 - 2\chi} -
         \frac{2\zeta M^{2}}{\gamma^{2}\Delta} - \frac{\rho_{c}(3 - 4\chi)}{3(1 - 2\chi)}~~~~ & \mbox{\small {\bf for {$k=-1$}}}.
          \end{array}
\right.
\end{array}\ee

\begin{figure*}[htb]
\centering
\subfigure[ An illustration of the bouncing condition for a universe with$k=1, \rho_{c}=10M^{4}$ for red curve and $k=1,\ w=1/3,\ \rho_{c}=10M^{4}$ for blue curve and $k=-1,\ w=1/3,\ \rho_{c}=10M^{4}$ for green curve.]{
    \includegraphics[width=7.2cm,height=8cm] {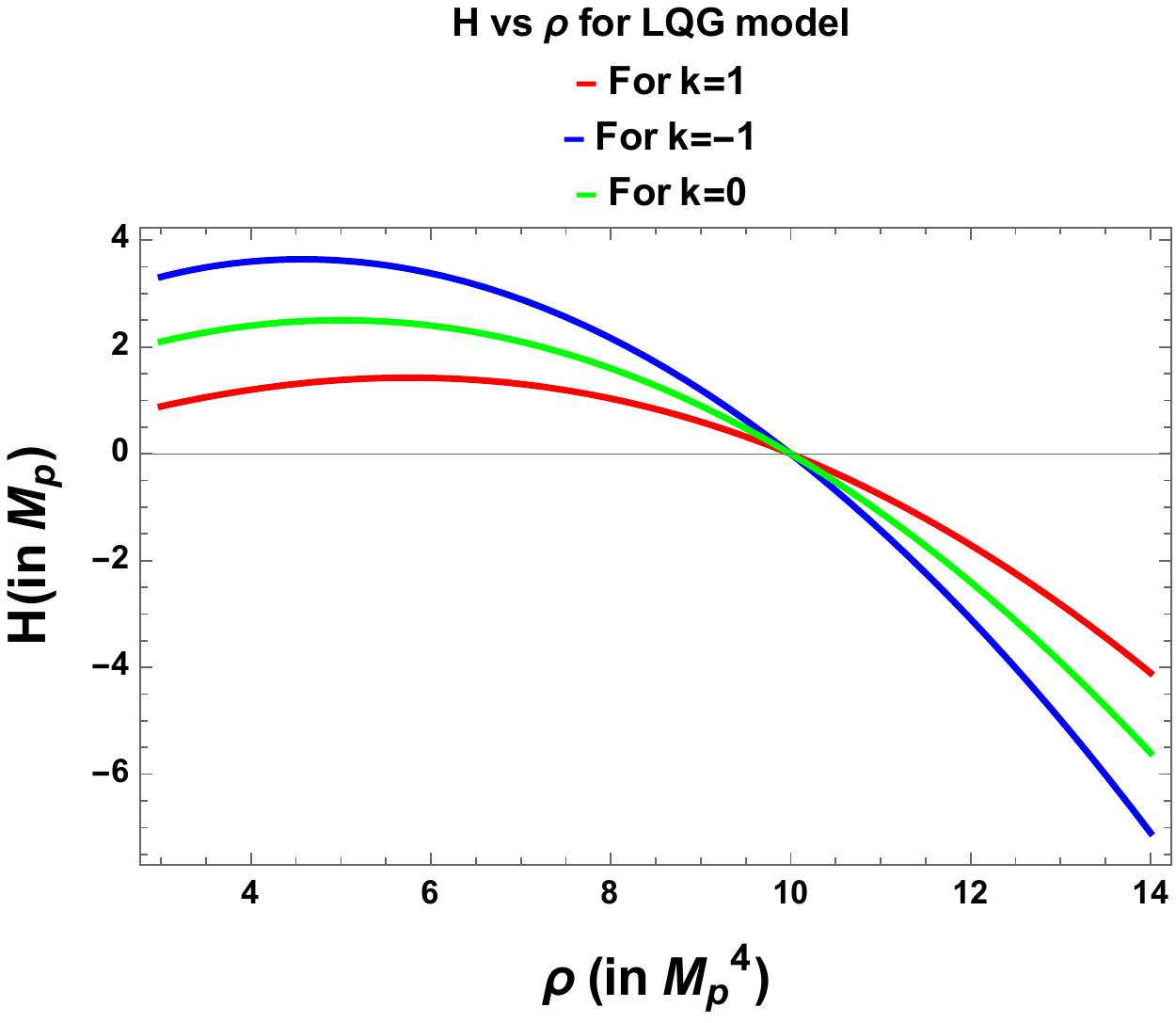}
    \label{lqg1}
}
\subfigure[An illustration of the acceleration condition at the time of bounce for a universe with an equation of state $w=0,\ k=0,\ \rho_{c}=10M^{4}$.]{
    \includegraphics[width=7.2cm,height=8cm] {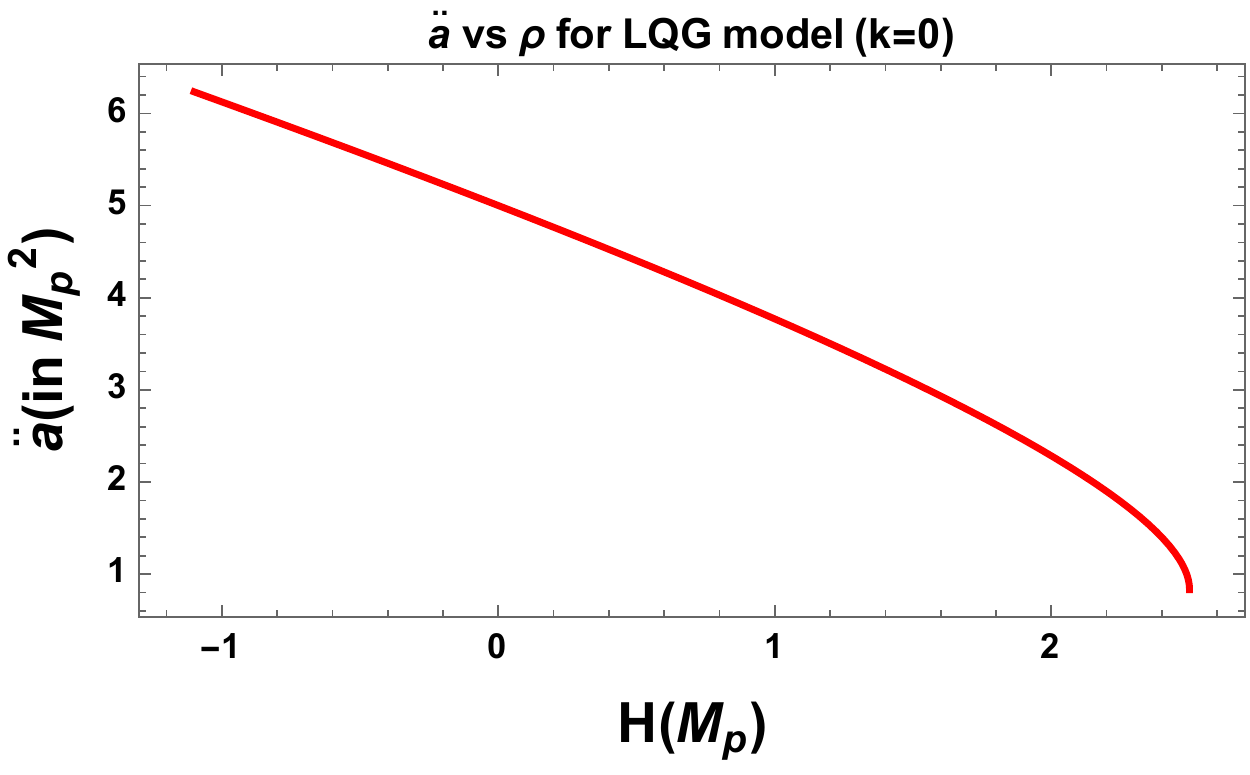}
    \label{lqg2}
}
\subfigure[An illustration of the acceleration condition at the time of bounce for a universe with an equation of state $w=0,\ k=-1,\ \rho_{c}=10M^{4}$]{
    \includegraphics[width=7.2cm,height=8cm] {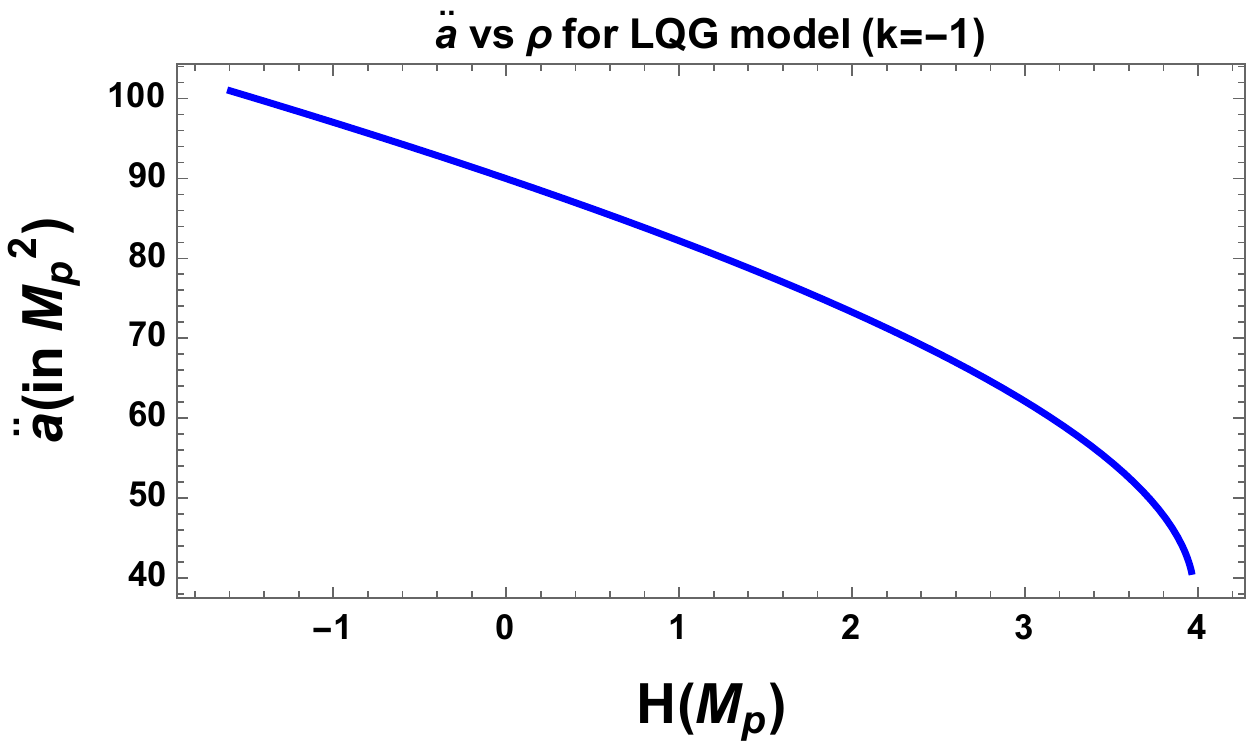}
    \label{lqg3}
}
\subfigure[An illustration of the acceleration condition at the time of bounce for a universe with an equation of state $w=0,\ k=1,\ \rho_{c}=10M^{4}$.]{
    \includegraphics[width=7.2cm,height=8cm] {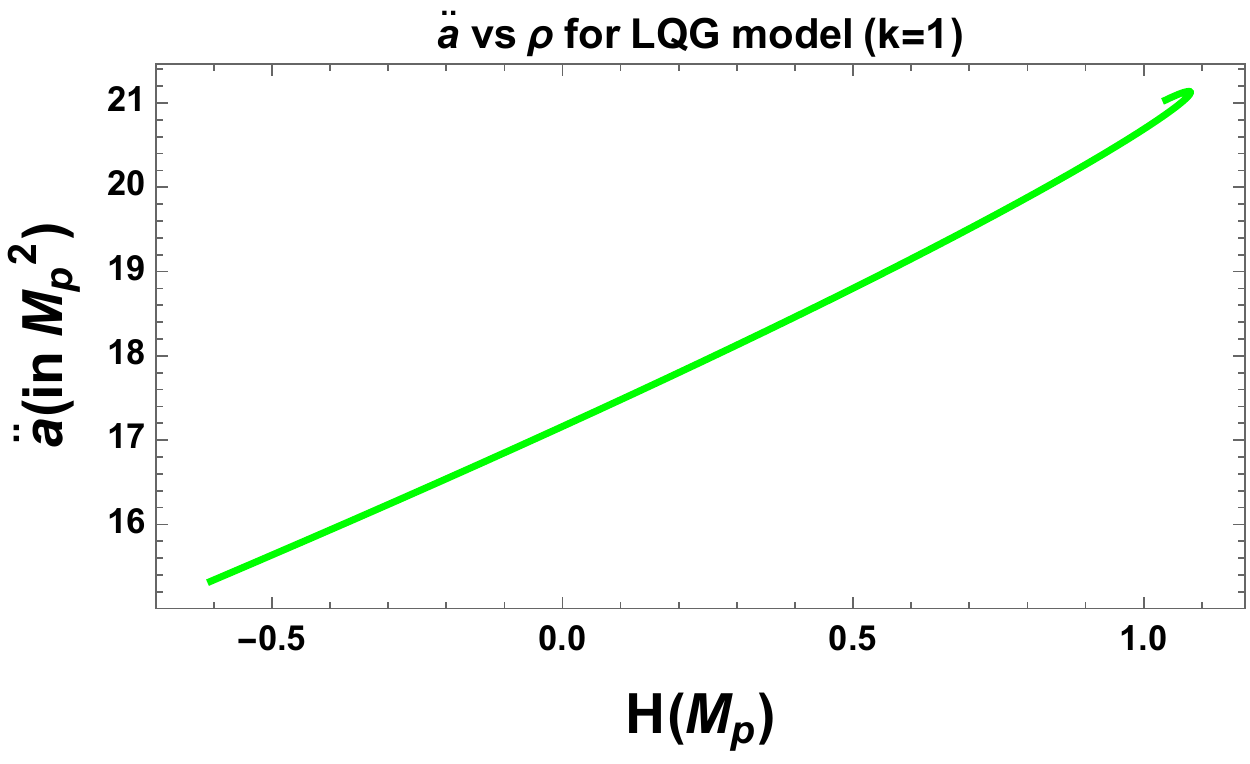}
    \label{lqg4}
}
\caption[Optional caption for list of figures]{ Graphical representation of the phenomena of bounce and acceleration for LQG model.} 
\label{fig24}
\end{figure*}

In Figs. \ref{fig24}, we have shown the phenomena of bounce and acceleration in the LQG model. We can draw the following conclusions from the above figures:
\begin{itemize}
\item In Fig. \ref{lqg1}, we have plotted the r.h.s of Eq. (\ref{lqc1}) using the relation $\rho=a^{-3(1+w)}$ with $k=1, \rho_{c}=10M^{4}$ for red curve and $k=1,\ w=1/3,\ \rho_{c}=10M^{4}$ for blue curve and $k=-1,\ w=1/3,\ \rho_{c}=10M^{4}$ for green curve. 
\item We have used $w=1/3$, since we require a soft equation of state for causing the acceleration and expansion. The case $w=0$ causes bounce equally well. The value of $\rho_{c}$ chosen is arbitrary. 
\item From Fig. \ref{lqg1}, we can also say that the behavior of the Hubble parameter w.r.t. density is nearly same for any value of $k$.
\item From Fig. \ref{lqg1}, we get the bounce at $\rho=\rho_{b}=10M^{4}$ for all the three cases. 
\item Figs. \ref{lqg2} to \ref{lqg4}, we have shown the necessary condition of acceleration ($\ddot{a}> 0$) at the time of bounce for $k=0,\ -1,\ 1$. Here we have plotted the r.h.s of Eqns. (\ref{lqc2}) and (\ref{lqc1}). These plots have also been obtained for $w=0,\ \rho_{c}=10M^{4}$.
\item Thus from Fig. \ref{fig24} we can conclude that for this model, bounce is possible for closed, open and flat universe.
\end{itemize}

\subsection{Condition for turnaround}   

Turnaround or re-collapse occurs when the following criteria is achieved:
\be \rho = \rho_{1},\ee
when the universe reaches its maximum radius $a_{max}$ and minimum density can be written as \cite{Ashtekar:2006es}:
\be\rho_{1} = \rho_{min} \approx \frac{3kM^{2}}{a_{max}^{2}}.\ee Therefore, in place of Eq.~(\ref{lqc3}), we get:
\begin{equation}
\delta a_{max} = -\frac{1}{3kM^{2}}\oint pdV
\label{lqc5}
\end{equation}
Unlike for the bounce case, the condition for an increase in expansion amplitude at turnaround now depends on the curvature parameter only (apart from the sign of the work done).
Substituting the values of curvature parameter Eq.~(\ref{lqc5}) can be recast in the following form:
\be\begin{array}{lll}\label{beg}
 \displaystyle \delta a_{max} =\left\{\begin{array}{ll}
                    \displaystyle   -\frac{1}{3M^{2}}\oint pdV~~~~ &
 \mbox{\small {\bf for {$k=+1$}}}  \\ \\
         \displaystyle  \frac{1}{3M^{2}}\oint pdV~~~~ & \mbox{\small {\bf for {$k=-1$}}}.
          \end{array}
\right.
\end{array}\ee
Let us explicitly mention the two possible physical outcomes from Eq.~(\ref{beg}) appearing in the present context:
\begin{itemize}
 \item  For $k = +1$, in order to get an increase in the amplitude of the scale factor after each successive cycle, we need $\oint pdV < 0$.
 \item  For $k = -1$, in order to get an increase in the amplitude of the scale factor after each successive cycle, we need $\oint pdV > 0$.
\end{itemize}
Rest of all the conclusions remain same as that we obtained for the case of bounce from LQG setup.

\subsection{Condition for deceleration}

Using the definition of $\rho_{1} = -k\chi\rho_{c}$ from Eq.~ (\ref{lqc2}), at turnaround time scale 
the condition for acceleration is given by:
\begin{equation}
\rho_{1} + 3p_{t} > \frac{6M^{2}\chi k}{\gamma^{2}\Delta}.
\end{equation}
Further substituting the expression for $\rho_{1}$ for different values of the curvature parameter $k$
in non-flat case we get the following constraint conditions for acceleration at bounce as:
\be\begin{array}{lll}\label{beg}
 \displaystyle p_{t} =\left\{\begin{array}{ll}
                    \displaystyle   M^{2}\left(\frac{2\chi}{\gamma^{2}\Delta} - \frac{1}{a_{t}^{2}}\right) ~~~~ &
 \mbox{\small {\bf for {$k=+1$}}}  \\ \\
         \displaystyle  M^{2}\left(\frac{2\chi}{\gamma^{2}\Delta} - \frac{1}{a_{t}^{2}}\right) ~~~~ & \mbox{\small {\bf for {$k=-1$}}}.
          \end{array}
\right.
\end{array}\ee
Just like the case of acceleration within LQG setup, the condition for deceleration
at turnaround also depends on the LQG model parameters. Additionally it is important to note that whether
this constraint violates the energy condition or not, solely governed by the numerical values of the constants.

\begin{figure*}[htb]
\centering
\subfigure[ An illustration of the turnaround condition for a universe with $k=0,\ \rho_{c}=10M^{4}$ for red curve, $k=-1,\ w=1,\ \rho_{c}=10M^{4}$ for blue curve and $k=1,\ w=1,\ \rho_{c}=10M^{4}$ for green curve .]{
    \includegraphics[width=7.2cm,height=8cm] {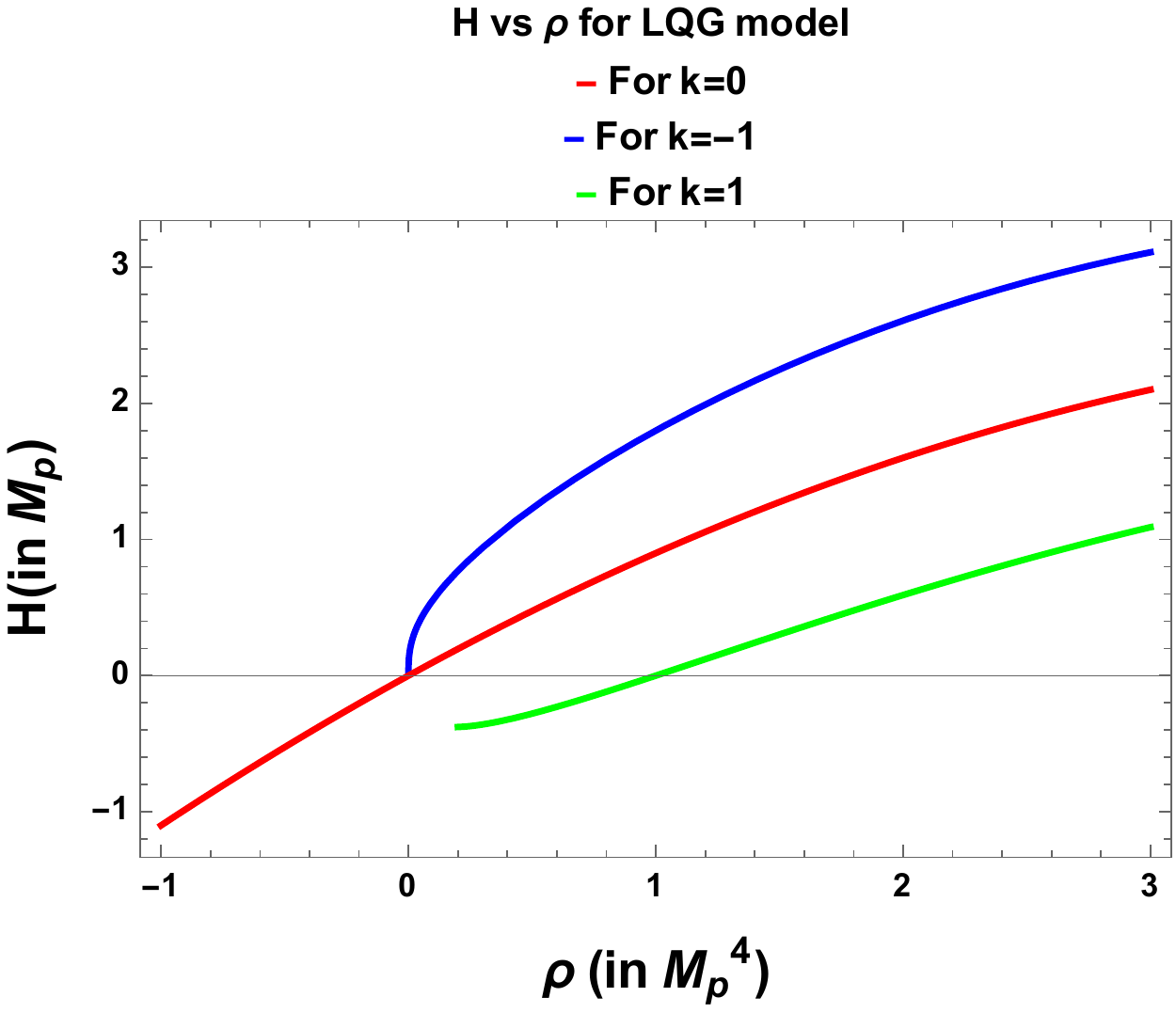}
    \label{lqg5}
}
\subfigure[An illustration of the deceleration condition at turnaround for a universe with $ k=0, \ \rho_{c}=10M^{4}$.]{
    \includegraphics[width=7.2cm,height=8cm] {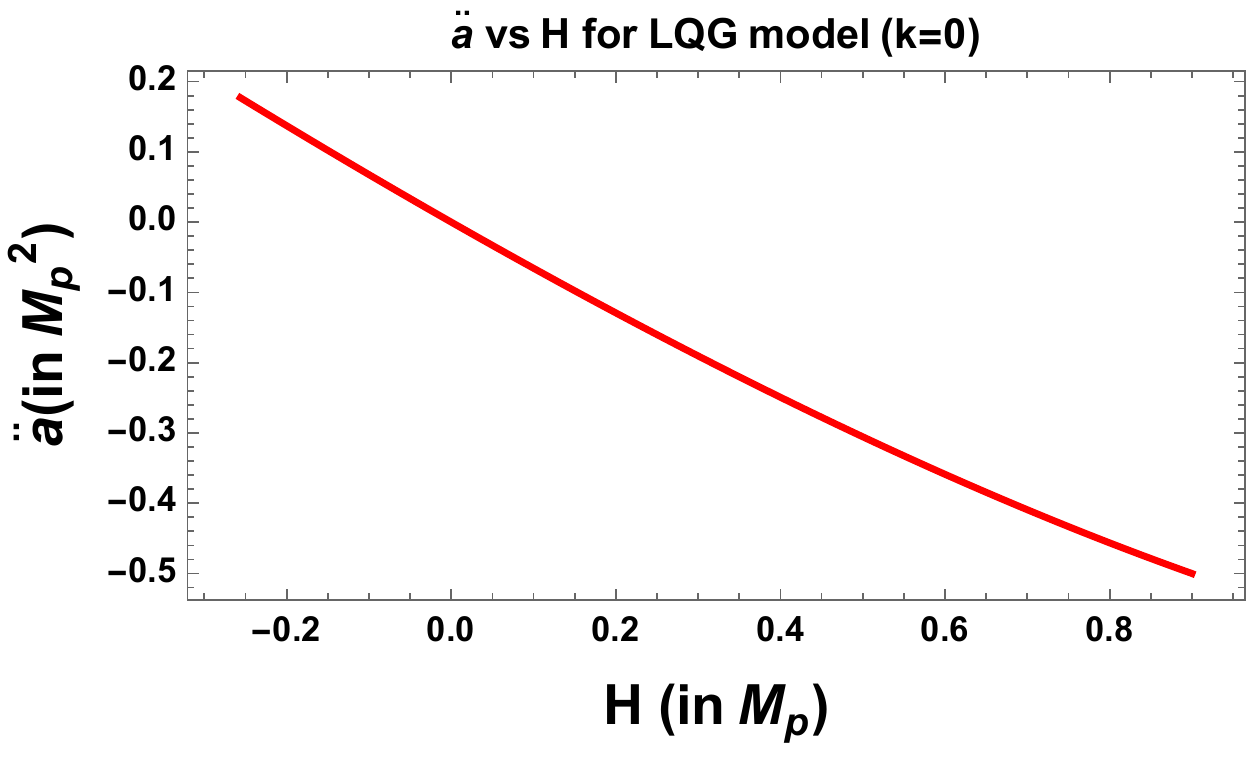}
    \label{lqg6}
}
\subfigure[An illustration of the deceleration condition at turnaround for a universe with an equation of state $w=1,\ k=-1, \rho_{c}=10M^{4}$.]{
    \includegraphics[width=7.2cm,height=8cm] {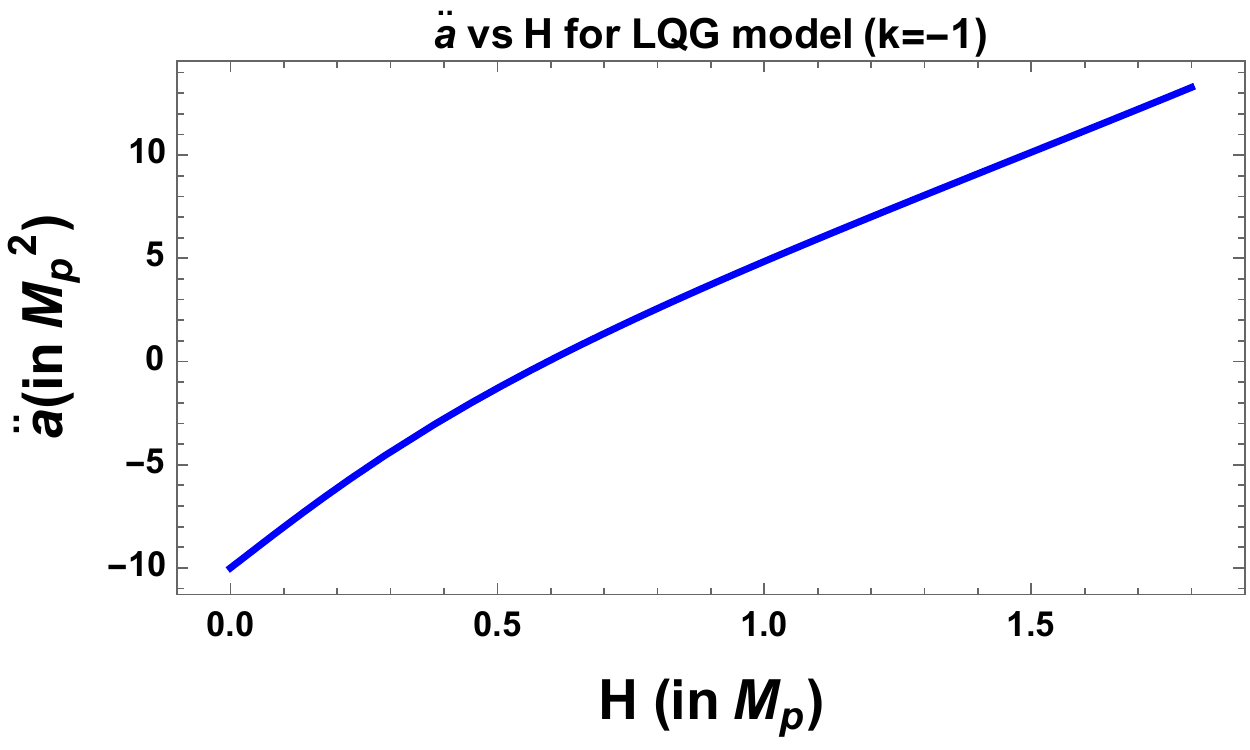}
    \label{lqg7}
}    
\subfigure[An illustration of the deceleration condition at turnaround for a universe with an equation of state $w=1,\ k=1, \rho_{c}=10M^{4}$.]{
    \includegraphics[width=7.2cm,height=8cm] {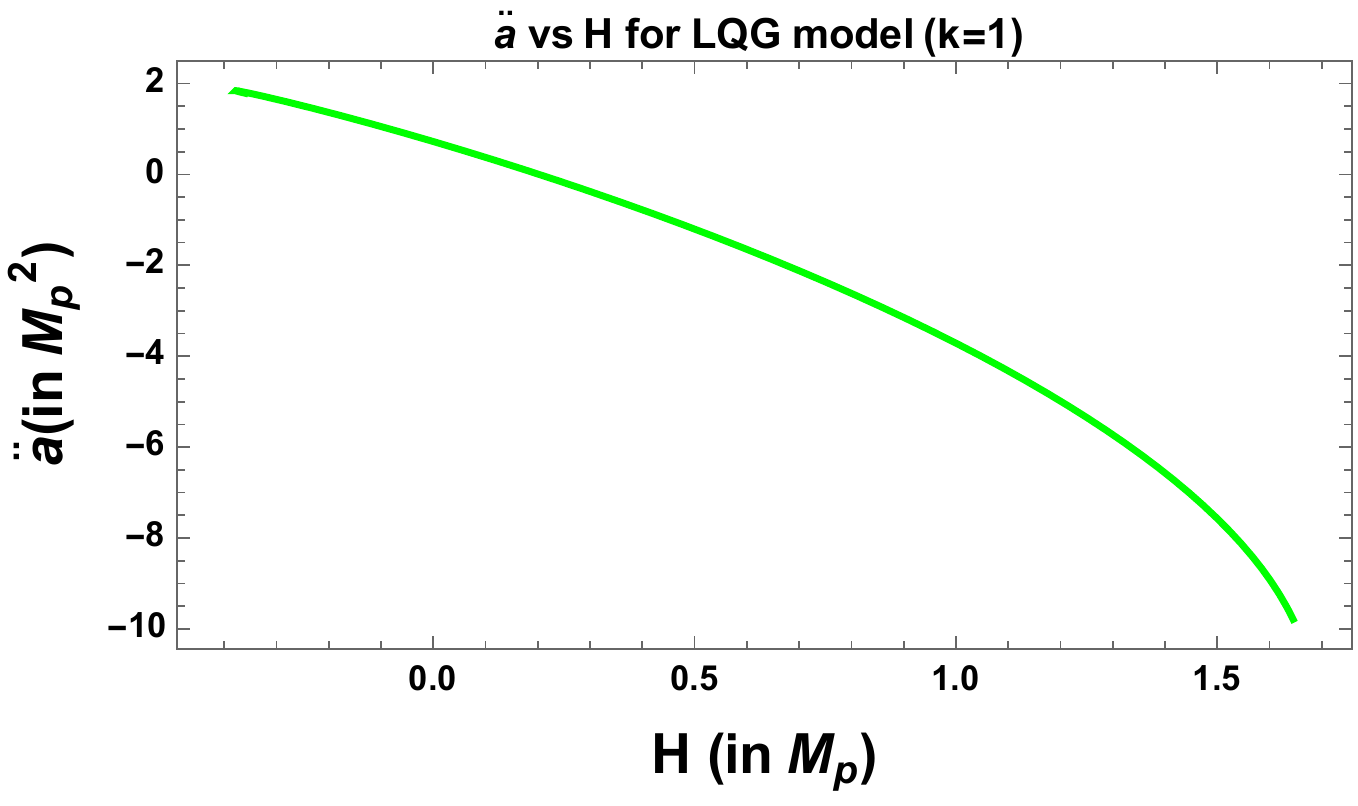}
    \label{lqg8}
}
\caption[Optional caption for list of figures]{ Graphical representation of the phenomena of turnaround and deceleration for LQG model.} 
\label{fig25}
\end{figure*}

In Fig. \ref{fig25}, we have shown the phenomena of turnaround and deceleration in LQG model. We can draw the following conclusions from the above figures:
\begin{itemize}
\item In Fig. \ref{lqg5}, we have plotted the r.h.s of Eqns. (\ref{lqc1}), using the relation $\rho=a^{-3(1+w)}$ with $k=-1,\ w=1,\ \rho_{c}=10M^{4}$ for blue curve and $k=1,\ w=1,\ \rho_{c}=10M^{4}$ for green curve. For this case we have used $w=1$, since we require a stiff equation of state for causing contraction and deceleration.  
\item From Fig. \ref{lqg5}, we get the turnaround at $\rho=\rho_{t}=0$ for $k=0,\ -1$ and at $\rho=\rho_{t}=1 M^{4}$ respectively.
\item Figs. \ref{lqg6}, \ref{lqg7} and \ref{lqg8} show the necessary condition of deceleration ($\ddot{a}< 0$) at the time of turnaround. Here we have plotted the r.h.s of Eqns. (\ref{lqc2})and (\ref{lqc1}). This plot has also been obtained for the same parameter values as the earlier graph.
\item Thus Fig. \ref{fig25} shows graphically that the phenomenon of turnaround is possible for the DGP model having $w=1$.
\end{itemize}

\subsection{Evaluation of work done in one cycle}

The expression for the total work done is same as that given by Eq.~(\ref{hyst1}) in DGP model. 
But in order to get an expression similar to Eq.~(\ref{dgphyst5}), we see that just like in DGP model,
here also the Friedmann equations given by Eq.~(\ref{lqc1}) and Eq.~(\ref{lqc2}), which are highly complicated. 
Hence to get a analytical solution, just like in the case of DGP model, here also we can use the early and late time
approximations to Eq.~(\ref{lqc1}). 

At early time, $\rho >> \rho_{1}$, hence we can neglect $\rho_{1}$ in Eq.~(\ref{lqc1}). Then we get:
\begin{equation}
\left(\frac{\dot{a}}{a}\right)^{2} = \frac{1}{3M^{2}} \, \rho\left(\frac{1}{\rho_{c}} (\rho_2 - \rho)\right)
\label{lqcnew}
\end{equation}
Solving the above equation for $\rho$ and using the following relation \cite{Singh:2010qa}:
\be \rho_{1} + \rho_{2} = \rho_{c},\ee we get the following two-fold solution for the energy density:
\begin{equation}\label{g1}
\rho = \frac{\rho_{c} \pm \sqrt{\rho_{c}^{2} - 4(3M^{2}\rho_{c}\left(\frac{\dot{a}}{a}\right)^{2} + \rho_{1}\rho_{2})}}{2}.
\end{equation}
But we know that at the time of bounce $(H = 0)$, the solution of this equation is: \be \rho = \rho_{2} = \rho_{c},\ee
which we approximately get by setting $H = 0$ in Eq.~(\ref{g1}) and taking the positive signature in the solution. 
Therefore the physically acceptable solution for the density $\rho$ is given by:
\begin{equation}
\rho_{phys}=\rho = \frac{\rho_{c} + \sqrt{\rho_{c}^{2} - 4(3M^{2}\rho_{c}\left(\frac{\dot{a}}{a}\right)^{2} + \rho_{1}\rho_{2})}}{2}.
\label{rho}
\end{equation}
Using the energy conservation equation (continuity equation):
\begin{equation}
\dot{\rho} + 3H(\rho+p) = 0
\end{equation}  
and Eq.~(\ref{lqcnew}), we get the following form of the second Friedmann equation written as:
\begin{equation}
\frac{\ddot{a}}{a} - \left(\frac{\dot{a}}{a}\right)^{2} = \frac{1}{2M^{2}\rho_{c}}(2\rho - \rho_{2})(\rho + p),
\end{equation}
from which we get the following expression for the pressure $p$ in early times as: 
\begin{equation}
p = \frac{\dot{\phi}^{2}}{2} - V(\phi) =  \frac{2M^{2}\rho_{c}}{(2\rho - \rho_{2})}\left[\left(\frac{\ddot{a}}{a}\right) - \left(\frac{\dot{a}}{a}\right)^{2}\right] - \rho.
\label{pressure}
\end{equation} 
Further using Eq.~(\ref{rho}) into Eq.~(\ref{pressure}), we get the following expression for the second and third integral of Eq.~(\ref{hyst2}) as:
\begin{eqnarray}
\int_{a'^{i-1}}^{a_{min}^{i-1}} pdV &=& \int_{a_{min}^{i-1}}^{a'^{i-1}} pdV \nonumber \\
&=& \int 3\left(\frac{2M^{2}\rho_{c}}{(2\rho - \rho_{2})}\left[\frac{\ddot{a}}{a}
- \left(\frac{\dot{a}}{a}\right)^{2}\right] - \frac{\rho_{c}}{2}\right) a^{2}\dot{a}dt \nonumber \\ &&~~~~~~~~~~~~
+ \int 3\left(\frac{\sqrt{\rho_{c}^{2} - 4\left(3M^{2}\rho_{c}\left(\frac{\dot{a}}{a}\right)^{2} + \rho_{1}\rho_{2}
\right)}}{2} \right) a^{2}\dot{a}dt.
\label{lqchystnew}
\end{eqnarray}
At late times, $\rho << \rho_{2}$, hence the following condition holds good:
\be \frac{(\rho_{2} - \rho)}{\rho_{c}} \rightarrow 1.\ee Hence using $\rho_{1} = 3kM^{2}/a^{2}_{max}$, Eq.~ (\ref{lqc1}) reduces to standard Friedmann equation given by:
\begin{equation}
\left(\frac{\dot{a}}{a}\right)^{2} = \frac{(\rho - \rho_{1})}{3M^{2}}.
\label{lqccont}
\end{equation}
Using the standard second Friedmann equation for acceleration and the above equation we get the following expression for pressure $p$ as:
\begin{equation}
p=\frac{\dot{\phi}^{2}}{2} - V(\phi) = -M^{2}\left[\frac{2\ddot{a}}{a} + \left(\frac{\dot{a}}{a}\right)^{2} + \frac{k}{a^{2}}\right].
\label{lqcp}
\end{equation}
Therefore the expressions for the first and last integral in Eq.~ (\ref{hyst2}) are given by
\begin{eqnarray}
\int_{a_{max}^{i-1}}^{a'^{i-1}} pdV &=& \int_{a'^{i-1}}^{a_{max}^{i}} pdV \nonumber \\
&=& -\int 3M^{2} \left(2\ddot{a}\dot{a}a + \dot{a}^{3} + k\dot{a}\right)dt
\label{lqchystnew2}
\end{eqnarray}
Therefore the complete expression of the work done in one single expansion-contraction cycle is obtained by substituting Eq.~(\ref{lqchystnew}) and Eq.~(\ref{lqchystnew2}) into Eq.~(\ref{hyst2}) i.e. 
\bea 
\oint pdV&=&\int 3\left(\frac{2M^{2}\rho_{c}}{(2\rho - \rho_{2})}\left[\frac{\ddot{a}}{a}
- \left(\frac{\dot{a}}{a}\right)^{2}\right] - \frac{\rho_{c}}{2}\right) a^{2}\dot{a}dt \nonumber \\ &&~~~~~~~~~~~~
+ \int 3\left(\frac{\sqrt{\rho_{c}^{2} - 4\left(3M^{2}\rho_{c}\left(\frac{\dot{a}}{a}\right)^{2} + \rho_{1}\rho_{2}
\right)}}{2} \right) a^{2}\dot{a}dt\nonumber \\ &&~~~~~~~~~~~~~~~~~~~~~~~~-\int 3M^{2} \left(2\ddot{a}\dot{a}a + \dot{a}^{3} + k\dot{a}\right)dt.
\eea
From the above expressions we can conclude that, at early times, work done depends on the LQG model parameters like the critical density $\rho_{c}$, but depends only on the curvature parameter $k$ at late times.

\subsection{Semi-analytical analysis for cosmological potentials}
As we have done for the other models, here also we  will consider the cases of three different potentials as mentioned earlier. In this subsection we will denote Planck mass by $M_{p}$.

\subsubsection{Case I: Hilltop potential}

Since we need to substitute the expression for the scale factor into Eq.~(\ref{dgpwork}), we need to find separate expressions for the scale factor for both
early and late times during expansion and contraction phases of the Universe. 
\\ \\
\textbf{A. Expansion}
\\ \\
\underline{\bf i) Early time}
\\ \\
At early time within the interval $a_{min}<a< a'$ we will do the analysis for the case when the curvature parameter $k\neq0$, because as we have already 
seen, that the curvature term is necessary for causing the turnaround at late times. To serve this purpose we can use the approximate Friedmann equation
given by Eq.~(\ref{lqcnew}), where $\rho$ is now given by the hilltop potential. Then substituting the resulting expression for the Hubble parameter $H$ from LQG setup
into Eq.~(\ref{modeqn1}), we get an integral equation of the following form:
\begin{equation}
\int d\left(\frac{\phi}{M_{4}}\right)\frac{\left(\frac{V_{0}}{3M_{p}^{2}}\right)^{1/2}\sqrt{\left(1+\beta\left(\frac{\phi}{M_{p}}\right)^{p}\right)}\sqrt{\left(1-\frac{V_{0}}{\rho_{c}}\left(1+\beta\left(\frac{\phi}{M_{p}}\right)^{p}\right)
\right)}}{\beta p\left(\frac{\phi}{M_{4}}\right)^{p-1}} = -\frac{V_{0}}{3M_{p}^{2}}\int dt.
\label{lqcearly20}
\end{equation}
Now it is important to note that the exact analytical solution of the
left hand side of the above integral equation is given in the Appendix for RSII model. But for LQG setup to get analytical result, we consider the following 
field redefinition:
\be \frac{\phi}{M_{p}}=e^{\lambda},\ee where now we will
solve for $\lambda$ instead of $\phi$. Simplified analytical expression for $\lambda$ was possible only if we consider the small field limiting case i.e. $\phi/M_{p}<<1$, which has been elaborately discussed below.

For this case we can expand the exponentials upto linear order and apply the following constraint conditions within LQG setup as: \bea \frac{p\beta\lambda}{(1+\beta)}<<1,\\
\frac{V_{0}(1+\beta(1+p\lambda))}{\rho_{c}}<<1.\eea Since we are in the small $\lambda$ limit, these conditions are possible
provided we choose the values of the parameters of the model accordingly. The solution for $\lambda$ is now given by:
\begin{equation}
\lambda=\frac{\left(-\frac{V_{0}t}{3M_{p}^{2}}+E_{0}\right)}{Q},
\label{potential26}
\end{equation}
where we introduce two constants $Q$ and $E_{0}$ given by:
\begin{eqnarray}
Q &=& \left(\frac{V_{0}}{3M_{p}^{2}}\right)^{1/2}(1+\beta)^{1/2}\left(1-\frac{V_{0}}{2\rho_{c}}(1+\beta)\right) \\
E_{0} &=& Q\lambda_{i}+\frac{V_{0}t_{i}}{3M_{p}^{2}}
\end{eqnarray}
Here $\lambda_{i}$ is the value at bounce which corresponds to the time $t=t_{i}$.

Further substituting the expression for $\lambda$ back into the approximated version of the Friedmann equation, we get the following expression for scale factor as:
\bea
a(t) &=& E_{1}\exp\left[\sqrt{\frac{V_{0}}{3M_{p}^{2}}}\left\{\frac{\left(\left(1-\frac{V_{0}}{2\rho_{c}}(1+\beta)\right)Q-\frac{E_{0}p\beta V_{0}}{2\rho_{c}}+\frac{E_{0}p\beta}
{2(1+\beta)}\left(1-\frac{V_{0}}{2\rho_{c}}(1+\beta)\right)\right)t}{Q} \nonumber 
\right.\right.\\ && \left.\left. ~~~~~~~~~~~~~~~~~~~~~~~~~~~~~~+\frac{\left(\frac{pV_{0}^{2}\beta t^{2}}{12M_{p}^{2}\rho_{c}}-\frac{V_{0}
p\beta t^{2}}{12M_{p}^{2}(1+\beta)}\left(1-\frac{V_{0}}{2\rho_{c}}(1+\beta)\right)\right)}{Q}\right\}\right],
\label{scalefactor26}
\eea
where we introduce a new constant $E_{1}$ given by:
\bea
E_{1} &=& a_{i}\exp\left[-\sqrt{\frac{V_{0}}{3M_{p}^{2}}}\left\{\frac{\left(\left(1-\frac{V_{0}}{2\rho_{c}}(1+\beta)\right)Q-\frac{E_{0}p\beta V_{0}}
{2\rho_{c}}+\frac{E_{0}p\beta}{2(1+\beta)}\left(1-\frac{V_{0}}{2\rho_{c}}(1+\beta)\right)\right)t_{i}}{Q}\nonumber 
\right.\right.\\ && \left.\left. ~~~~~~~~~~~~~~~~~~~~~~~~~~~~~~~~+\frac{\left(-\frac{pV_{0}^{2}\beta t^{2}}
{12M_{p}^{2}\rho_{c}}+\frac{V_{0}p\beta t_{i}^{2}}{12M_{p}^{2}(1+\beta)}\left(1-\frac{V_{0}}{2\rho_{c}}(1+\beta)\right)\right)}{Q}\right\}\right].
\eea
 \\ \\
\underline{ii) Late time:}
\\ \\
At late times within the interval $a'<a< a_{max}$, the Friedmann equation is given by Eq.~(\ref{lqccont}), which is the standard Friedmann equation appearing in the context of classical GR. 
Thus, in order to generate the condition for turnaround, we need to take the curvature parameter $k\neq0$ for this specific case. 
Thus, using the standard classical GR version of the Friedmann acceleration equation and considering that we have been solving
for the cases when the contribution from the canonical kinetic term is smaller compared to the potential i.e. $\rho\approx -p\approx V(\phi)$ for expansion
and the contribution from the canonical kinetic term is larger compared to the potential i.e. $\rho\approx p\approx \dot{\phi}^{2}/2$ for contraction, 
we get a differential equation for the scale factor of the form:
\bea
\frac{\ddot{a}}{a}-\left(\frac{\dot{a}}{a}\right)^{2}-\frac{k}{a^{2}}=0 
\eea
and solving this equation 
we get the two-fold expressions for the scale factor of the following form:
\begin{equation}
a(t)=\exp\left[(\sqrt{E_{2}}t+E'_{2})\pm e^{2\sqrt{E_{2}}t+E'_{2}}-2k\right]
\label{scalefactor27}
\end{equation}
and the form of $ E'_{2}$ and $E_{2}$ can be evaluated from the appropriate boundary conditions.

Substituting the above expression back into the Friedmann equation, we get the expression for the potential as 
\begin{equation}
V(\phi)=E_{2}(1+2ke^{(E_{2}'+2\sqrt{E_{2}}t)})+e^{(-2E_{2}'-2e^{(E_{2}'+2\sqrt{E_{2}}t)}+4k-2\sqrt{E_{2}}t)}.
\label{potential27}
\end{equation}
\\ \\
\textbf{B. Contraction}
\\ \\
 From Eq.~(\ref{modeqn}), we see that for the contraction phase, under the approximation which we have assumed, the analysis becomes
 independent of the choice of the potential. Hence, the final results for the LQG model holds good for any form of the potential.
 \\ \\
\underline{i) Early time:}
\\ \\
At eaarly times within the interval $a_{min}<a< a'$, following the same analysis as we have already done for the case of expansion, with the new expression for density, the integral equation is now given by:
\begin{equation}
\int d\dot{\phi}\frac{1}{\frac{3\dot{\phi}^{2}}{\sqrt{6}M_{p}}\left(1-\frac{\dot{\phi}^{2}}{2\rho_{c}}\right)^{1/2}} = -\int dt.
\end{equation}
Solving the above integrals, we get the following expression for $\dot{\phi}$ as:
\begin{eqnarray}
\dot{\phi}^{2}&=&\frac{2M_{p}^{2}}{3(t+E_{3})^{2}+\frac{M_{p}^{2}}{2\rho_{c}}} 
\end{eqnarray}
where we introduce a new constant $E_{3}$ defined as:
\begin{eqnarray}
E_{3} &=& -t_{i}\mp \frac{1}{\sqrt{3}}\left(\frac{2M_{p}^{2}}{\dot{\phi}_{i}}-\frac{M_{p}^{2}}{2\rho_{c}}\right)^{1/2}.
\end{eqnarray}
Here $\dot{\phi}_{i}$ is the value of the derivative of the scalar field at bounce.

Integrating the above equation we get the solution for $\phi$ as
\begin{equation}
\phi(t)=2\sqrt{\frac{2}{3}}\frac{\sqrt{\rho_{c}}}{M_{p}}\tan^{-1}(\sqrt{6}(E_{3}+t)\sqrt{\rho_{c}}/M_{p})+E_{3}'
\label{potential33}
\end{equation}
where 
\begin{equation}
E_{3}'=\phi_{i}-2\sqrt{\frac{2}{3}}\frac{\sqrt{\rho_{c}}}{M_{p}}\tan^{-1}(\sqrt{6}(E_{3}+t_{i})\sqrt{\rho_{c}}/M_{p}
\end{equation}

Further using the above expression in the Friedmann equation, we get the expression for scale factor as:
\begin{equation}
a(t)=E_{4}\exp\left[\frac{(-E_{3}+t)\ln[M_{p}^{2}+3(E_{3}-t)^{2}\rho_{c}]\sqrt{\frac{M_{p}^{2}\rho_{c}}{M_{p}^{2}+3(E_{3}-t)^{2}\rho_{c}}}}{6M_{p}\sqrt{\frac{(E_{3}-t)^{2}\rho_{c}}{M_{p}^{2}+3(E_{3}-t)^{2}\rho_{c}}}}\right]
\label{scalefactor33}
\end{equation}
where we introduce a new constant $E_{4}$ defined as:
\begin{equation}
E_{4}=a_{i}exp\left[-\frac{(-E_{3}+t_{i})\ln[M_{p}^{2}+3(E_{3}-t_{i})^{2}\rho_{c}]\sqrt{\frac{M_{p}^{2}\rho_{c}}{M_{p}^{2}+3(E_{3}-t_{i})^{2}\rho_{c}}}}{6M_{p}\sqrt{\frac{(E_{3}-t_{i})^{2}\rho_{c}}
{M_{p}^{2}+3(E_{3}-t_{i})^{2}\rho_{c}}}}\right],
\end{equation}
where $a_{i}$ is the value of the scale factor at the time of bounce $t_{i}$.
\\ \\
\underline{ii) Late time:}
\\ \\
At late times within the interval $a'<a<a_{max}$, the Friedmann equation given by
Eq.~(\ref{lqccont}), which is the standard Friedmann equation as appearing in the
context of classical GR. Thus, in order to generate the condition for turnaround,
we need to take $k\neq0$ for this case. Thus, using the standard Friedmann acceleration
equation and considering that we have been solving for the cases when the
$\rho\approx -p\approx V(\phi)$ for expansion and $\rho\approx p\approx \dot{\phi}^{2}/2$
for contraction, we get a differential equation for the scale factor of the form:
\begin{equation}
\ddot{a}+\frac{2\dot{a}^{2}}{a}+\frac{2k}{a}=0.
\end{equation}
Now solving the above differential equation for a(t), we get the following expression for the scale factor as:
\begin{equation}
a(t)= {\rm InverseFunction}\left[\frac{1}{-k\left(1+{\rm ProductLog}[\frac{e^{-1-\frac{E_{6}}{k}}(K_{1})^{2/k}}{k}]\right)}\right](t+E_{7})
\label{scalefactor34}
\end{equation}
where $E_{6}$, $E_{7}$ and $K_{1}$ are the arbitrary integration constants whose values are determined from the appropriate choice of boundary conditions~\footnote{
Here InverseFunction$(f)(x)$ represents the inverse of the function $f(x)$. ProductLog or Lambert $w$ function or omega function gives the principal solution
for $w$ in $z=we^{w}$,
for any complex number z. 
By implicit differentiation, one can show that all branches of $w$ satisfy the differential equation: $\frac{dw}{dz}=\frac{w}{z(1+w)}~~~{\rm for} ~~z\neq -\frac{1}{e}$.}.
\\ \\
\textbf{C. Expression for work done}
\\ \\
The expression for work done is given by the sum of Eq.~(\ref{lqchystnew2}) and Eq.~(\ref{lqchystnew}). 
Evaluation of the integrals using the corresponding scale factor for small field limiting situation as have been calculated in this section gives 
\begin{eqnarray}
\oint pdV &=& W_{1}+W_{2},\end{eqnarray} 
where $W_{1}$ and $W_{2}$ is defined as:
\bea
W_{1}&=&k\ {\rm InverseFunction}\left[\frac{1}{-k\left(1+{\rm ProductLog}[\frac{-e^{-1-\frac{E_{6}}{k}}(K_{1})^{2/k}}{k}]\right)}\right](t'-t_{max}+E_{7})\nonumber
\\ &&~~~~+ 3 \ {\rm InverseFunction}\left[\frac{1}{-k\left(1+{\rm ProductLog}[\frac{-e^{-1-\frac{E_{6}}{k}}(K_{1})^{2/k}}{k}]\right)}\right](t'-t_{max}+E_{7})^{3}
\nonumber \\ &&~~~~-d_{7}\left\{{\rm Ei}\left[\frac{d_{6}}{t_{min}}\right]-{\rm Ei}\left[\frac{d_{6}}{t'}\right]\right\}\nonumber \\ 
&&~~~~+ \frac{1}{9 d_{18}}e^{-6d_{19}k}\left[e^{3e^{d_{20}+d_{18}t_{max}}}(2-6e^{d_{20}+d_{18}t_{max}}+9e^{2(d_{20}+d_{18}t_{max})})\nonumber 
\right.\\&&\left.~~~~+ e^{3e^{d_{20}+d_{18}t'}}(-2+6e^{d_{20}+d_{18}t'}-9e^{2(d_{20}+d_{18}t')})+9e^{4d_{2}k}(e^{e^{d_{20}+d_{18}t_{max}}}-e^{e^{d_{20}+d_{18}t'}})k\right],~~~~~~~~\eea
and 
\bea
W_{2}&=&- d_{8}\left\{{\rm Ei}\left[\frac{3d_{6}}{t_{min}}\right]^{}_{}-{\rm Ei}\left[\frac{3d_{6}}{t'}\right]\right\}-d_{5}e^{d_{6}/t'}t'- d_{5}e^{3d_{6}/t'}t'-e^{d_{6}/t_{min}}(d_{9}+d_{10}e^{2d_{6}/t_{min}})t_{min}\nonumber \\ 
&&~~~~+ d_{11}\left\{{\rm Ei}\left[\frac{d_{6}}{t_{min}}\right]-{\rm Ei}\left[\frac{d_{6}}{t'}\right]\right\}+d_{12}\left\{{\rm Ei}\left[\frac{3d_{6}}{t_{min}}\right]-{\rm Ei}\left[\frac{3d_{6}}{t'}\right]\right\}\nonumber \\ 
&&~~~~+ e^{d_{6}/t'}(d_{9}+d_{10}e^{2d_{6}/t'})t'+d_{13}e^{-d_{14}}\left\{{\rm Erfi}\left[\frac{d_{15}+2d_{16}t_{min}}{2d_{13}}\right]-{\rm Erfi}\left[\frac{d_{15}+2d_{16}t'}{2d_{13}}\right]\nonumber \right.\\&&\left. 
~~~~- \left(d_{17}{\rm Erfi}\left[\frac{d_{15}+2d_{16}t_{min}}{2d_{13}}\right]-d'_{17}{\rm Erfi}\left[\frac{d_{15}+2d_{16}t'}{2d_{13}}\right]\right)\right\}\nonumber \\&& 
~~~~+\left(d_{17}{\rm Erfi}\left[\frac{d_{15}+2d_{16}t_{min}}{2d_{13}}\right]-d'_{17}{\rm Erfi}\left[\frac{d_{15}+2d_{16}t'}{2d_{13}}\right]\right)\nonumber\\ &&
~~~~~~~+ d_{5}e^{d_{6}/t_{min}}t_{min}+d_{5}e^{3d_{6}/t_{min}}t_{min},
\eea
where $d_{1}...d_{20}$ are constants that depends on the parameters of the model whose explicit forms have been given in the appendix.~\footnote{Here ${\rm Ei}(z)$ represents the exponential integral function, where ${\rm Ei}(z)=\int \frac{e^{-z}}{z}
dz$ and ${\rm Erfi}$ represents the imaginary error function given by, ${\rm Erfi}(z)=\frac{{\rm Erf}(iz)}{i}$.}.
\\ \\

\subsubsection{Case II: Natural potential}

\textbf{A. Expansion}
\\ \\
\underline{\bf i) Early time:}\\ \\
At early time, within the interval $a_{min}<a< a'$, we will do the analysis for
the case when k$\neq$0, because as we have seen, the curvature term is necessary for causing the turnaround at late times. We can use the approximate Friedmann equation given by Eq.~ (\ref{lqcnew}), where $\rho$ is now given by the natural potential. Then substituting the resulting expression of H into Eq.~ (\ref{modeqn1}), we get an integral equation given by
\begin{equation}
\int\left(\frac{V_{0}}{3M_{p}^{2}}\right)^{1/2}\frac{\left(1+\cos\left(\frac{\phi}{f}\right)\right)^{1/2}}{\sin\left(\frac{\phi}{f}\right)}\left(1-\frac{V_{0}}{\rho_{c}}-\frac{V_{0}}{\rho_{c}}\cos\left(\frac{\phi}{f}\right)\right)^{1/2}= \frac{V_{0}}{3f^{2}}\int dt
\label{lqcearly6}
\end{equation}
The exact solutions of the above integrals are given in the Appendix. In order to simplify the expressions, we solve the above integrals for two limiting cases:
\\ \\
\underline{\bf a) $\phi/f<<1$:}
\\ \\
For this case we take small argument approximations of the trigonometric functions after which we get the solution for $\phi$ as
\begin{equation}
\frac{\phi(t)}{f} = E_{8}exp\left[\frac{V_{0}t}{3\sqrt{2}f^{2}\left(1-\frac{2V_{0}}{\rho_{c}}\right)^{1/2}\left(\frac{V_{0}}{3M_{p}^{2}}\right)^{1/2}}\right]
\label{potential30}
\end{equation}
where
\begin{equation}
E_{8}=\frac{\phi_{i}}{f}exp\left[-\frac{V_{0}t_{i}}{3\sqrt{2}f^{2}\left(1-\frac{2V_{0}}{\rho_{c}}\right)^{1/2}\left(\frac{V_{0}}{3M_{p}^{2}}\right)^{1/2}}\right]
\end{equation}
Here $\phi_{i}$ is the value of $\phi$ at the initial time of bounce $t_{i}$. 

Substituting the expression for $\phi$ back into the Friedmann equation, we get the expression for scale factor as
\begin{equation}
a(t)=E_{9}exp\left[\sqrt{2}\left(\frac{V_{0}}{3M_{p}^{2}}\left(1-\frac{2V_{0}}{\rho_{c}}\right)^{1/2}\right)^{1/2}t\right]
\label{scalefactor30}
\end{equation}
where
\begin{equation}
E_{9}=a_{i}exp\left[-\sqrt{2}\left(\frac{V_{0}}{3M_{p}^{2}}\left(1-\frac{2V_{0}}{\rho_{c}}\right)^{1/2}\right)^{1/2}t_{i}\right]
\end{equation}
Here $a_{i}$ is the value of the scale factor at bounce. 
\\ \\
\underline{\bf b) $\phi/f>>1$:}
\\ \\
For this case, considering the fact that for large argument the cosine function is small and the sine function can be approximated to unity, the solution for $\phi$ in this case  simplifies to
\begin{equation}
\frac{\phi(t)}{f}=\frac{\left(\frac{V_{0}t}{2f^{2}}+E_{10}\right)}{\left(1-\frac{V_{0}}{2\rho_{c}}\right)\left(\frac{V_{0}}{3M_{p}^{2}}\right)^{1/2}}
\label{potential31}
\end{equation}
where 
\begin{equation}
E_{10}=\left(1-\frac{V_{0}}{2\rho_{c}}\right)\left(\frac{V_{0}}{3M_{p}^{2}}\right)^{1/2}\left(\phi_{i}/f\right)-\frac{V_{0}t_{i}}{3f^{2}}
\end{equation}
Here $\phi_{i}$ is the value of the scalar field at the time of bounce.

Substituting back the expression for $\phi$ into the Friedmann equation and integrating, we get the expression for the scale factor as
\begin{eqnarray}
a(t) &=& E_{11}exp\left[-\frac{\sqrt{\frac{V_{0}}{3M_{p}^{2}}}(V_{0}-2\rho_{c})}{8V_{0}\rho_{c}^{2}}\left(-6\sqrt{\frac{V_{0}}{3M_{p}^{2}}}\sin(\frac{2(3E_{10}f^{2}+tV_{0})\rho_{c}}{3\sqrt{\frac{V_{0}}{3M_{p}^{2}}}f^{2}(V_{0}-2\rho_{c})})\rho_{c}\right)\right]\nonumber \\ && exp\left[-\frac{\sqrt{\frac{V_{0}}{3M_{p}^{2}}}(V_{0}-2\rho_{c})}{8V_{0}\rho_{c}^{2}}\left(V_{0}(9\sqrt{\frac{V_{0}}{3M_{p}^{2}}}f^{2}\sin(\frac{2(3E_{10}f^{2}+tV_{0})\rho_{c}}{3\sqrt{\frac{V_{0}}{3M_{p}^{2}}}f^{2}(V_{0}-2\rho_{c})})+4t\rho_{c})\right)\right] \nonumber \\
\label{scalefactor31}
\end{eqnarray}
where
\begin{eqnarray}
E_{11} &=& a_{i}exp\left[\frac{\sqrt{\frac{V_{0}}{3M_{p}^{2}}}(V_{0}-2\rho_{c})}{8V_{0}\rho_{c}^{2}}\left(-6\sqrt{\frac{V_{0}}{3M_{p}^{2}}}\sin(\frac{2(3E_{10}f^{2}+t_{i}V_{0})\rho_{c}}{3\sqrt{\frac{V_{0}}{3M_{p}^{2}}}f^{2}(V_{0}-2\rho_{c})})\rho_{c}\right)\right]\nonumber \\ && exp\left[\frac{\sqrt{\frac{V_{0}}{3M_{p}^{2}}}(V_{0}-2\rho_{c})}{8V_{0}\rho_{c}^{2}}\left(V_{0}(9\sqrt{\frac{V_{0}}{3M_{p}^{2}}}f^{2}\sin(\frac{2(3E_{10}f^{2}+t_{i}V_{0})\rho_{c}}{3\sqrt{\frac{V_{0}}{3M_{p}^{2}}}f^{2}(V_{0}-2\rho_{c})})+4t_{i}\rho_{c})\right)\right] \nonumber \\
\end{eqnarray}
\\ \\
\underline{\bf ii) Late time:}
\\ \\
At late times, within the interval $a' <a< a_{max}$ the analysis is independent of the choice of the
potential. Consequently the results
shown in this subsection hold good for natural potential.
\\ \\
\textbf{B. Contraction}
 \\ \\
As has already been mentioned that while performing the analysis for expansion, the conclusions for contraction phase is independent of any potential, hence remains same.
\\ \\
\textbf{C. Expression for work done}
\\ \\
The expression for work done is given by the sum of Eq.~(\ref{lqchystnew2}) and Eq.~(\ref{lqchystnew}). 
Evaluation of the integrals using the corresponding scale factor analytically has been possible for small field limiting situation, which as have been calculated in this section as shown below: 
\bea
\oint pdV &=& D_{1}+D_{2},\eea 
where $D_{1}$ and $D_{2}$ is defined as:
\bea
D_{1}&=& k\ {\rm InverseFunction}\left[\frac{1}{-k\left(1+{\rm ProductLog}[\frac{-e^{-1-\frac{E_{6}}{k}}(K_{1})^{2/k}}{k}]\right)}\right](t'-t_{max}+E_{7})\nonumber
\\ &&~~~~+ 3 \ {\rm InverseFunction}\left[\frac{1}{-k\left(1+{\rm ProductLog}[\frac{-e^{-1-\frac{E_{6}}{k}}(K_{1})^{2/k}}{k}]\right)}\right](t'-t_{max}+E_{7})^{3}\nonumber \\
&&~~~~~~~+ \frac{1}{9 d_{18}}e^{-6d_{19}k}
\left\{e^{3e^{d_{20}+d_{18}t_{max}}}(2-6e^{d_{20}+d_{18}t_{max}}+9e^{2(d_{20}+d_{18}t_{max})})\nonumber 
\right.\\&&\left.~~~~~~~+ e^{3e^{d_{20}+d_{18}t'}}(-2+6e^{d_{20}+d_{18}t'}-9e^{2(d_{20}+d_{18}t')})+9e^{4d_{2}k}(e^{e^{d_{20}+d_{18}t_{max}}}-e^{e^{d_{20}+d_{18}t'}})k\right\},\nonumber\\
\eea
and 
\bea 
D_{2}&=& d_{5}e^{d_{6}/t_{min}}t_{min}+d_{5}e^{3d_{6}/t_{min}}t_{min}\nonumber 
\\&&~~~~~-d_{7}\left\{{\rm Ei}\left[\frac{d_{6}}{t_{min}}
\right]-{\rm Ei}\left[\frac{d_{6}}{t'}
\right]\right\}-d_{8}\left\{{\rm Ei}\left[\frac{3d_{6}}{t_{min}}\right]-{\rm Ei}\left[\frac{3d_{6}}{t'}\right]\right\}\nonumber 
\\ &&~~~~~- d_{5}e^{d_{6}/t'}t'- d_{5}e^{3d_{6}/t'}t'-e^{d_{6}/t_{min}}(d_{9}+d_{10}e^{2d_{6}/t_{min}})t_{min}
\nonumber 
\\&&~~~~~~+d_{11}\left\{{\rm Ei}\left[\frac{d_{6}}{t_{min}}\right]-{\rm Ei}\left[\frac{d_{6}}{t'}\right]\right\}+d_{12}\left\{{\rm Ei}
\left[\frac{3d_{6}}{t_{min}}\right]-{\rm Ei}
\left[\frac{3d_{6}}{t'}\right]\right\}\nonumber \\ &&~~~~~+ 3d_{21}(e^{d_{22}t_{min}}-e^{d_{22}t'})-d_{23}(e^{3d_{22}t_{min}}
-e^{3d_{22}t'})\nonumber \\ &&~~~~~+\frac{d_{24}}{3}(e^{3d_{22}t_{min}}-e^{3d_{22}t'}),
\eea
where $d_{1}...d_{24}$ are constants that depends on the parameters of the model whose explicit forms have been given in the appendix. Hence from our analysis we can see that the phenomenon of hysteresis is achieved due to non-zero work done for natural potential in LGG setup.
\\ \\
\\ \\

\subsubsection{Case III:  Coleman-Weinberg potential}

\textbf{A. Expansion}
\\ \\
\underline{\bf i) Early time:}
\\ \\
At early time within the interval, $a_{min}<a< a'$ to do the analysis we use non-vanishing curvature parameter $k\neq0$,
because as we have seen, the curvature term is necessary for causing the turnaround at late times. 
We can use the approximated version of the Friedmann equation as given by Eq.~(\ref{lqcnew}), where the energy density $\rho$ is now characterized
by the Coleman-Weinberg potential. Then substituting the resulting expression of H into Eq.~(\ref{modeqn1}), we get an integral equation given by
\be\begin{array}{lll}
\int  d\left(\frac{\phi}{M_{p}}\right)\frac{\sqrt{\left(\frac{V_{0}}{3M_{p}^{2}}\right)\left[1+\left\{\alpha+\beta \ln\left(\frac{\phi}{M_{p}}\right)\right\}\left(\frac{\phi}{M_{p}}\right)^{4}\right]\left(1-\frac{V_{0}}{3M_{p}^{2}\rho_{c}}
\left[1+\left\{\alpha+\beta \ln\left(\frac{\phi}{M_{p}}\right)\right\}\left(\frac{\phi}{M_{p}}\right)^{4}\right]\right)} }{\left(\frac{\phi}{M_{p}}\right)^{3}\left[4\alpha+
\beta+4\beta \ln\left(\frac{\phi}{M_{p}}\right)\right]}= -\frac{V_{0}}{3M_{p}^{2}} \int dt.
\label{lqcearly10}
\end{array}\ee
To compute the left hand side of the above integral equation, we consider the following field redefinition:
\be \frac{\phi}{M_{p}}=e^{\lambda},\ee as we
had done for the case of hilltop potential. Here now we will solve for the transformed or redefined field $\lambda$ instead of $\phi$. Simplified analytical expressions for the above integral
was possible only in the small field limiting case, $\phi/M_{p}<<1$, which has been elaborately discussed below.

For this case we can expand the exponentials upto linear order after which we get the following solution for $\lambda$ as:
\begin{equation}
\lambda=\frac{\left(\frac{V_{0}}{3M_{p}^{2}}t-E_{12}\right)(4\alpha+\beta)}{(2+\alpha)\left(\frac{V_{0}}{2\rho_{c}}\alpha+\frac{V_{0}}{2\rho_{c}}-1\right)\left(\frac{V_{0}}{3M_{p}^{2}}\right)^{1/2}},
\label{potential32}
\end{equation}
where $E_{12}$ is the arbitrary integration constant given in terms of the model parameters as:
\begin{equation}
E_{12}=\frac{V_{0}}{3M_{p}^{2}}t_{i}-\lambda_{i}(2+\alpha)\left(\frac{V_{0}}{2\rho_{c}}\alpha+\frac{V_{0}}{2\rho_{c}}-1\right)\left(\frac{V_{0}}{3M_{p}^{2}}\right)^{1/2}.
\end{equation}
Here $\lambda_{i}$ is the initial value at the time of bounce.

While obtaining the above expression we have assumed the following three constraint conditions: 
\bea \frac{V_{0}(1+(\alpha+\beta\lambda)(1+4\lambda)}{\rho_{c}}<<1,\\
(\alpha+\beta\lambda)(1+4\lambda)<<1,\\
\frac{4\beta\lambda}{(4\alpha+\beta)}<<1.\eea 
Since we are in the small $\lambda$ limit, these conditions are satisfied if we also choose the parameters of the model appropriately.

Further substituting the expression for $\lambda$ back into the Friedmann equation, we get the following expression for scale factor as:
\begin{equation}
a(t)= E_{13}\exp\left[\left(\frac{V_{0}}{3M_{p}^{2}}\right)^{1/2}\left(E't+t^{2}\frac{\left(\frac{V_{0}}{3M_{p}^{2}}\right)}{2(\frac{V_{0}}{2\rho_{c}}+4\alpha+\beta)
\left(\frac{V_{0}}{3M_{p}^{2}}\right)^{1/2}}\right)\right],\nonumber \\
\label{scalefactor32}
\end{equation}
where $E_{13}$ is the arbitrary integration constant given in terms of the model parameters as:
\begin{equation}
E_{13}= a_{i}exp\left[-\left(\frac{V_{0}}{3M_{p}^{2}}\right)^{1/2}\left(E't_{i}-t_{i}^{2}
\frac{\left(\frac{V_{0}}{3M_{p}^{2}}\right)}{2(\frac{V_{0}}{2\rho_{c}}+4\alpha+\beta)\left(\frac{V_{0}}{3M_{p}^{2}}\right)^{1/2}}\right)\right]\nonumber \\
\end{equation}
where we introduce two constants $E'$ and $E''$ defined as:
\bea
E'&=&\left(1+\frac{\alpha}{2}\right)\left(1-\frac{V_{0}}{2\rho_{c}}(1+\alpha)\right)-E_{12}\frac{E''}{\frac{V_{0}}{2\rho_{c}}+4\alpha+\beta},\\
E''&=&\left(1+\frac{\alpha}{2}\right)\left(1-\frac{V_{0}}{2\rho_{c}}(1+\alpha)\right)+\left(\frac{V_{0}}{2\rho_{c}}+4\alpha+\beta\right)\left(1+\frac{\alpha}{2}\right)\nonumber\\&&~~~~~~~~~~~~~~~~~~~~~~~~~~~~~~-\left(2\alpha
+\frac{\alpha}{2}\right)\left(1-\frac{V_{0}}{2\rho_{c}}(1+\alpha)\right).\nonumber\\
\eea
Here $a_{i}$ is the value of the scale factor at the time of bounce.
\\ \\
 \underline{\bf ii) Late time:}\\ \\
 At late time, within the interval $a'<a<a_{max}$, the result is independent of any specific choice of potential.
 We have already seen that for late times for the hilltop potential did not depend on the form of the potential, hence the results shown in that section holds good for Coleman-Weinberg potential.
\\ \\
\textbf{B. Contraction}
\\ \\
As has already been mentioned that while performing the analysis for expansion, the conclusions for contraction phase is independent of any choice of potential, hence the conclusion remains the same as mentioned earlier for other 
potentials within LQG setup.
\\ \\
\textbf{C. Expression for work done}
\\ \\
The expression for work done is given by the sum of Eq.~(\ref{lqchystnew2}) and Eq.~(\ref{lqchystnew}).
Since the evolution of the scale factor with time is same as that for the hilltop potential,
the final expression for the work done also remains the same as that obtained for the hilltop potential. But now the constants
depend on the parameters of this model under consideration. Hence we can see that the phenomenon of hysteresis is possible for Coleman-Weinberg potential in LQG setup.
\\ \\
\subsection{Graphical Analysis}

\subsubsection{Case I: Hilltop potential}
All the graphs in this section and in the following sections have been plotted in units of $M_{p}=1,\ H_{0}=1,\ c=1$, where $M_{p}$ is the Planck mass, $H_{0}$ is the present value of the Hubble parameter and $c$ is the speed of light. 
\\ \\
\begin{figure*}[htb]
\centering
\subfigure[ An illustration of the behavior of the scale factor with time during the early  expansion phase for $\phi<<M_{p}$ with $V_{0}=1.8{\rm x}10^{-3}M_{p}^{4},\ p=3,\ \beta=1,\ E_{0}=10M_{p},\ \rho_{c}=0.86M_{p}^{4},\ E_{1}=1$.]{
    \includegraphics[width=7.2cm,height=8cm] {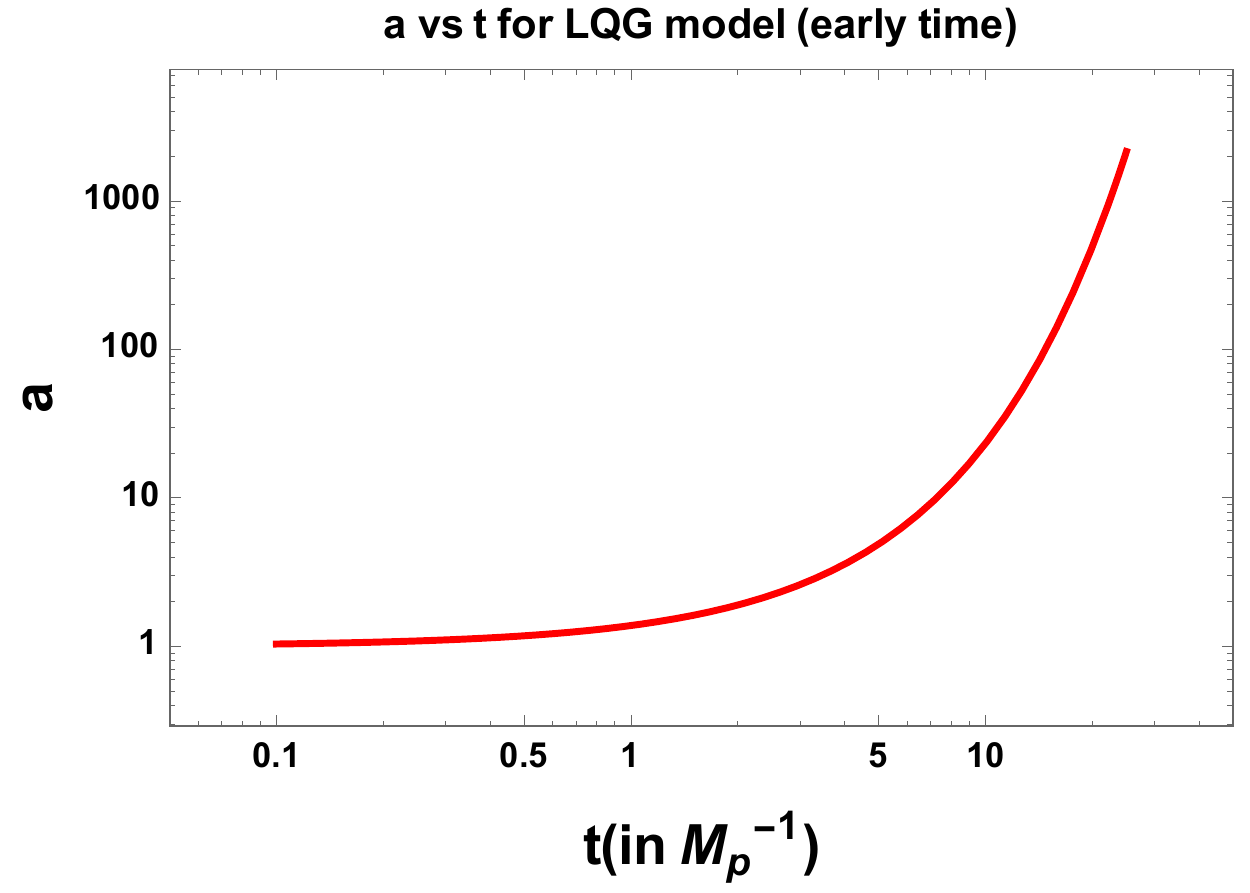}
    \label{lqg9}
}
\subfigure[An illustration of the behavior of the potential during early expansion phase for $\phi<<M_{p}$ with $V_{0}=1.8{\rm x}10^{-3}M_{p}^{4},\ p=3,\ \beta=1,\ E_{0}=0.38M_{p},\ \rho_{c}=0.86M_{p}^{4}$ .]{
    \includegraphics[width=7.2cm,height=8cm] {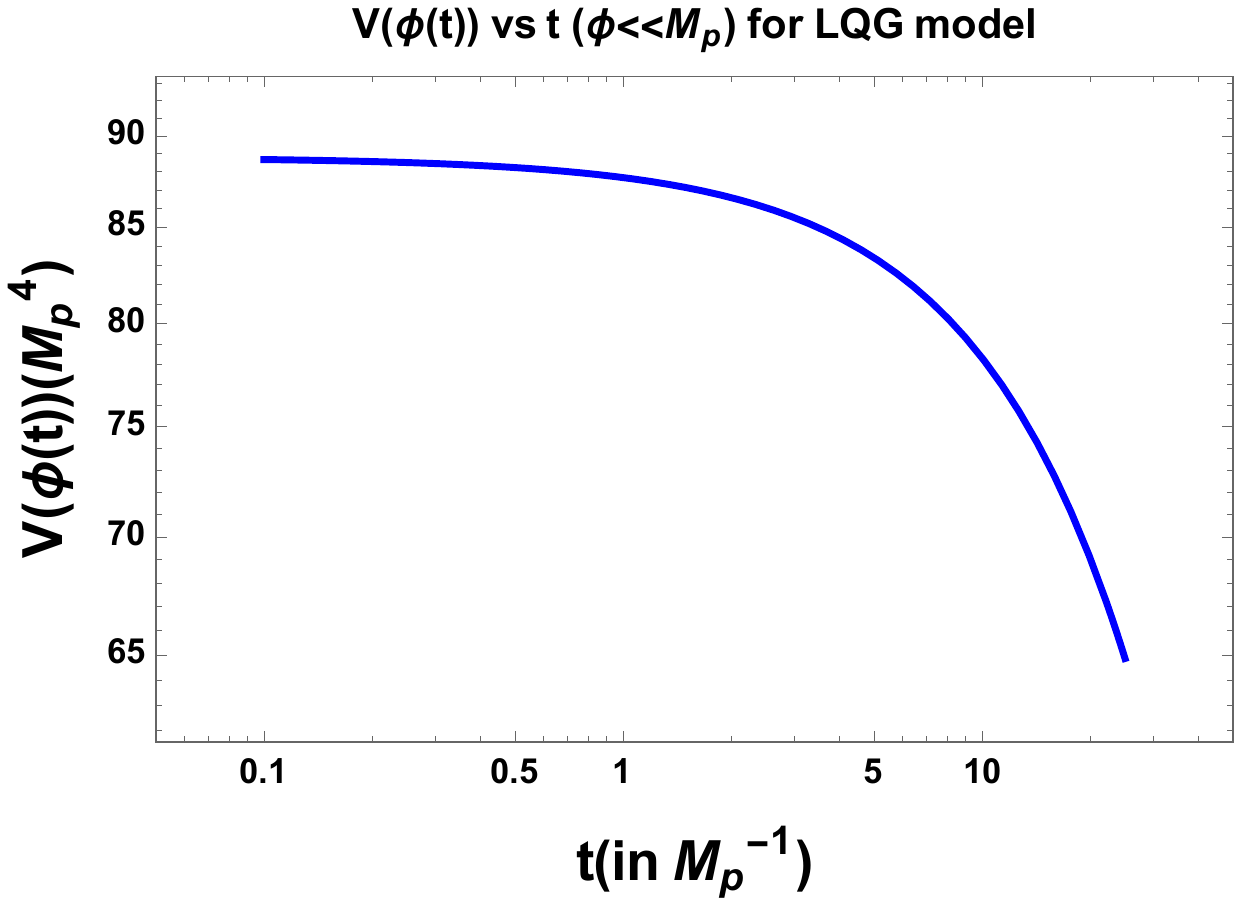}
    \label{lqg10}
}
\subfigure[An illustration of the behavior of the scale factor with time during late time expansion phase with $E_{2}=M_{p}^{2},\ E_{2}'=0.6,\ k=-1.0$.]{
    \includegraphics[width=7.2cm,height=8cm] {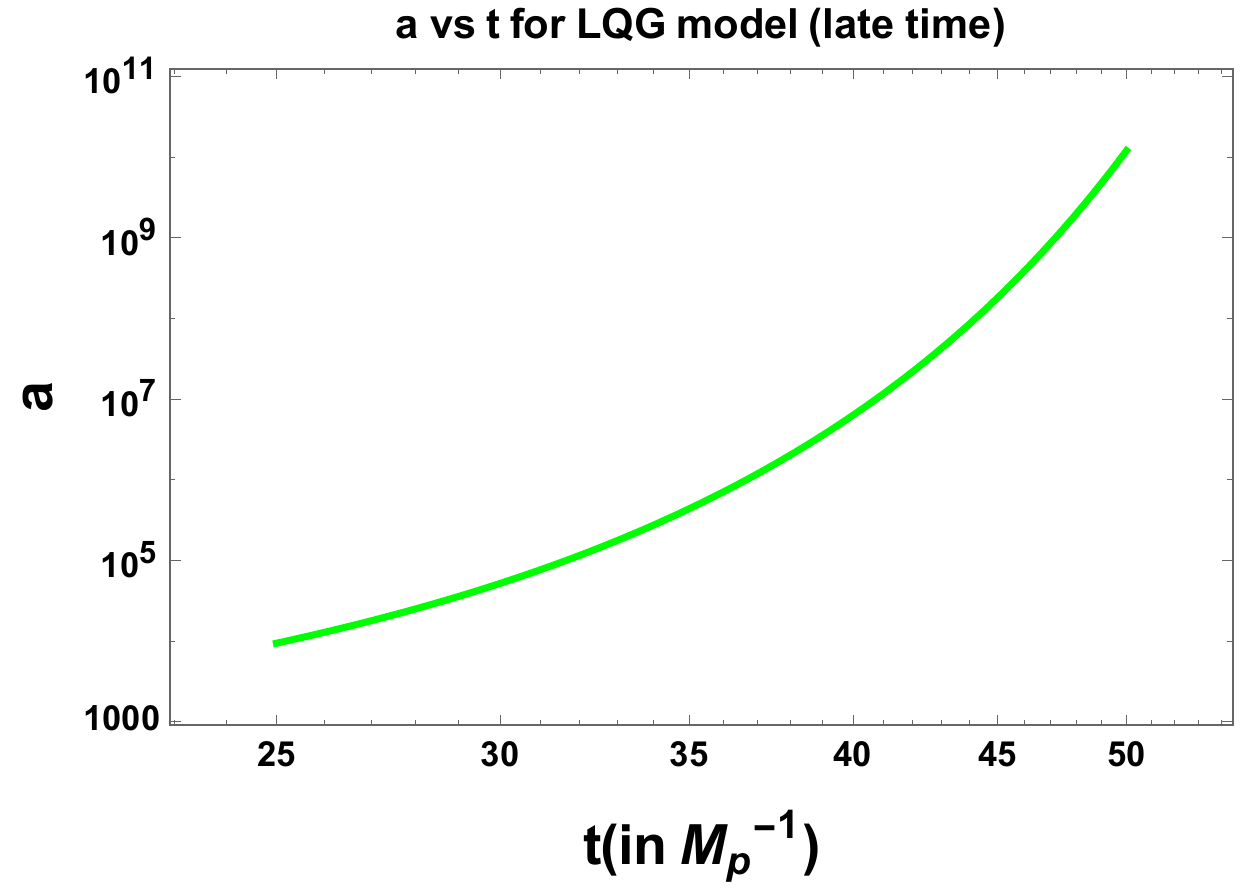}
    \label{lqg11}
}
\subfigure[An illustration of the behavior of the potential during late time expansion phase with $E_{2}=0.1M_{p}^{2},\ E_{2}'=-30,\ k=1.0$ .]{
    \includegraphics[width=7.2cm,height=8cm] {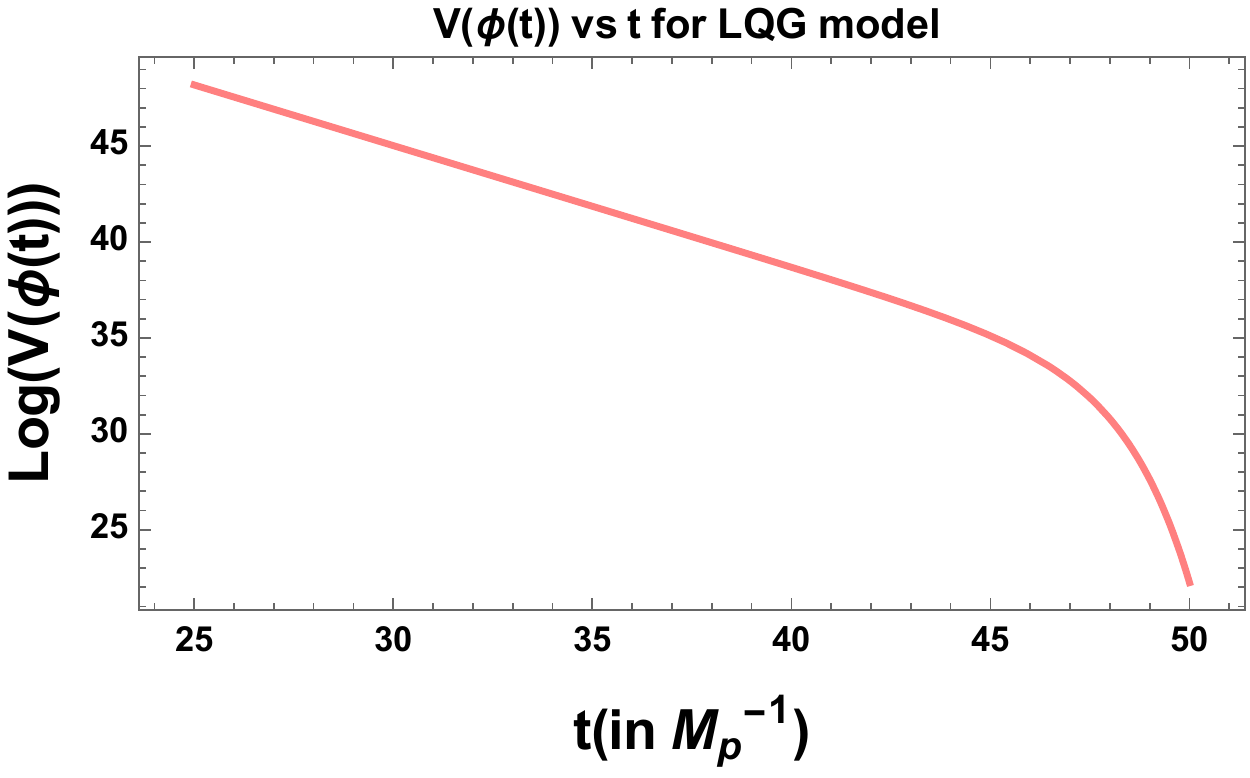}
    \label{lqg12}
}
\caption[Optional caption for list of figures]{ Graphical representation of the evolution of the scale factor and the potential during the expansion phase for LQG model.} 
\label{fig26}
\end{figure*}
In Fig. \ref{fig26} we have shown the evolution of the scale factor and the potential for early and late time expansion phase. From the above plots, we can draw the following conclusions:
\begin{itemize}
\item Fig. \ref{lqg9}, shows the plot of the scale factor in the small field limit for hilltop potential given by Eq. (\ref{scalefactor26}) with the parameter values $V_{0}=1.8{\rm x}10^{-3}M_{p}^{4},\ p=3,\ \beta=1,\ E_{0}=10M_{p},\ \rho_{c}=0.86M_{p}^{4},\ E_{1}=1$. 
\item Detail graphical analysis show that expansion is possible for both positive and negative values of $\beta$. However, for large positive values of $\beta$ (such as $>2$) expansion is possible only if the values of $V_{0}$ and $E_{0}$ are large. For $\beta<-1$, expansion is not possible. And for $\beta$ close to $-1$, expansion is obtained only if $V_{0}$ and $E_{0}$ is O($10^{-8}$). Increase in $p$ only results in an increase in the amplitude, keeping the pattern same. The plot is almost independent of the value of $\rho_{c}$ 
\item Fig. \ref{lqg10} shows the plot of the behavior of the potential with time for small field hilltop potential. This graph has been obtained with the help of Eqn. (\ref{potential26}) with parameter values $V_{0}=1.8{\rm x}10^{-3}M_{p}^{4},\ p=3,\ \beta=1,\ E_{0}=0.38 M_{p},\ \rho_{c}=0.86M_{p}^{4}$. The evolution of the potential is not much affected by the value of $p$ and $\rho_{c}$ (only the amplitude changes). $\beta<0$ do not give the required evolution of the potential for causing expansion. Expansion is possible for larger positive values of $\beta$ only if $V_{0}$ is close to $10^{-8}M_{p}^{4}$. Larger the value of $V_{0}$, more linear the fall of the potential becomes. Expansion is possible only if the value of $E_{0}<1 M_{p}$.
\item Fig. \ref{lqg11} shows the plot of the scale factor in the late time expansion phase for hilltop potential given by Eqn. (\ref{scalefactor27}). This plot has been obtained for $E_{2}=M_{p}^{2},\ E_{2}'=0.6,\ k=-1.0$. Since the solution is true for any potential, Fig. \ref{lqg11} is also true for all the three potentials. Detail graphical analysis have shown that the pattern of the graph remains same for $k=1$, only its amplitude decreases. Expansion is possible only for $E_{2}'<1$. Larger values of $E_{2}$ increases the amplitude of expansion. Expansion is possible for both negative and positive values of the integration constants.
\item Fig. \ref{lqg12} shows the evolution of the potential in the late time expansion phase for hilltop potential. Fig. \ref{lqg12} has been obtained by with the help of Eqn. (\ref{potential27}) with the parameter values $E_{2}=0.1M_{p}^{2},\ E_{2}'=-30,\ k=1.0$. Expansion is possible for negative values and slightly positive values of $E_{2}'$ only and for getting the correct form of the potential with positive $E_{2}'$, we need to make $E_{2}$ much less than unity. Very large values of $E_{2}$ do not give the required nature of potential. Larger the magnitude of $E_{2}'$, larger is the height of the potential. In this case, we get a potential decreasing with time, which is required for expansion, only for $k=1$ and not for $k=-1$.
\end{itemize}
\begin{figure*}[htb]
\centering
\subfigure[ An illustration of the behavior of the scale factor with time during the late contraction phase with $A_{4}=100M_{p}^{-1},\ r=6$.]{
    \includegraphics[width=7.2cm,height=8cm] {dgp13.pdf}
    \label{dgp13}
}
\subfigure[An illustration of the behavior of the potential during late contraction phase with $V_{0}=6.1{\rm x}10^{-3}M_{p}^{4},\ A_{3}=152M_{p}^{-1},\ A_{4}=730M_{p},\ \beta=0.52,\ r=1.62,\ p=1$ .]{
    \includegraphics[width=7.2cm,height=8cm] {dgp14.pdf}
    \label{dgp14}
}
\subfigure[An illustration of the behavior of the scale factor with time during early time contraction phase with $E_{3}=10^{-8}M_{p}^{-1},\ E_{4}=10^{7},\ \rho_{c}=0.86M_{p}^{4}$.]{
    \includegraphics[width=7.2cm,height=8cm] {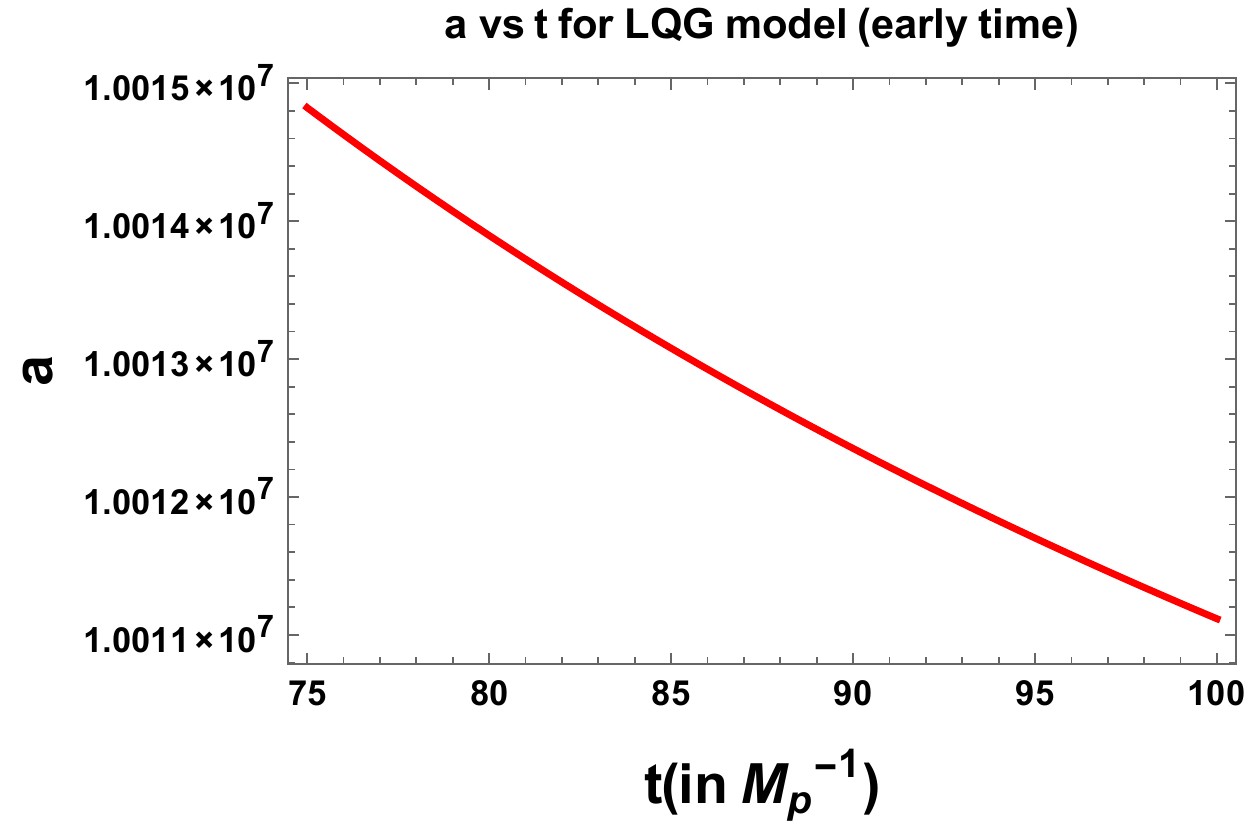}
    \label{lqg15}
}
\subfigure[An illustration of the behavior of the potential during early time contraction phase with $V_{0}=6.3{\rm x}10^{-3}M_{p}^{4},\ p=5,\ \beta=0.35,\ E_{3}'=10.2M_{p},\ E_{3}=5M_{p}^{-1},\ \rho_{c}=0.86M_{p}^{4}$ .]{
    \includegraphics[width=7.2cm,height=8cm] {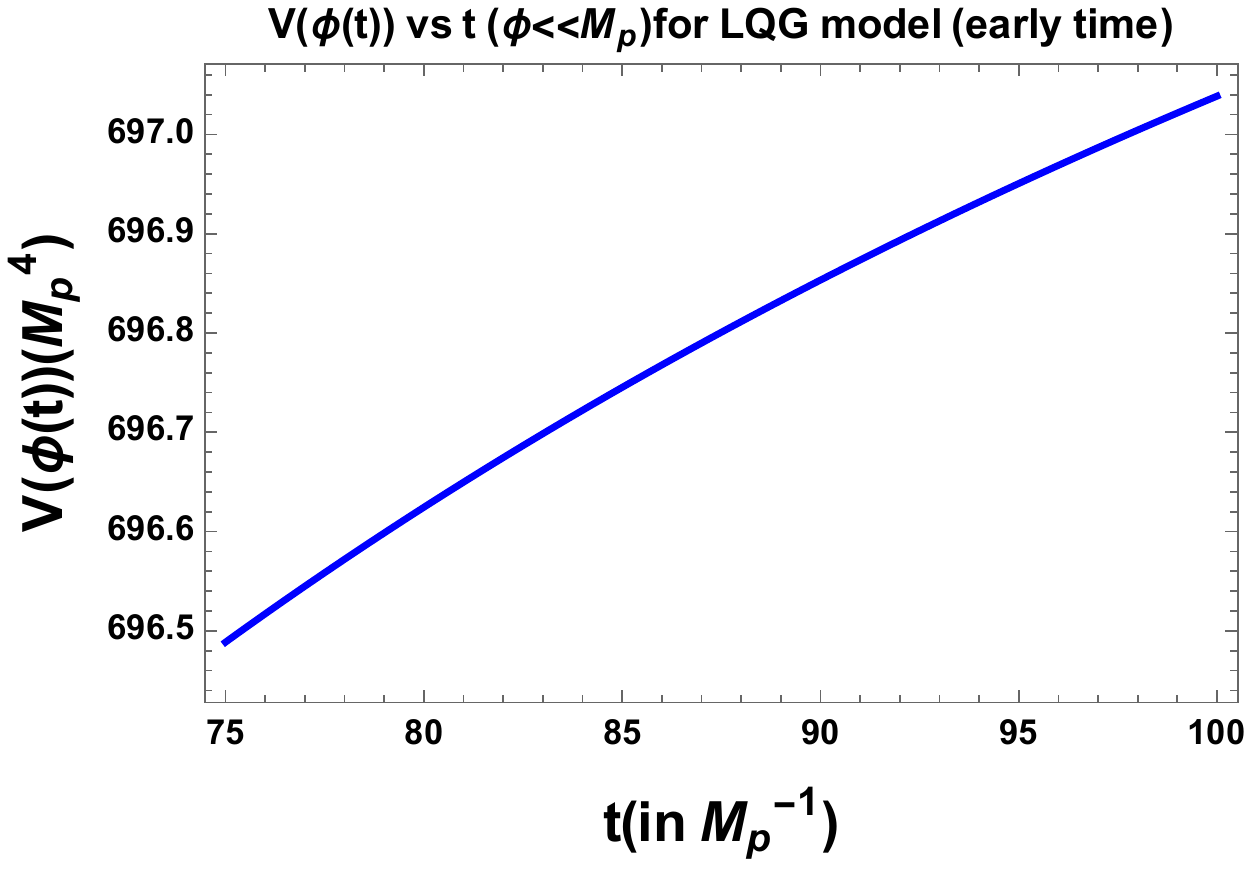}
    \label{lqg17}
}
\caption[Optional fig caption for list of figures]{ Graphical representation of the evolution of the scale factor and the potential during the contraction phase for LQG model.} 
\label{fig31}
\end{figure*}
In Fig. \ref{fig31} we have shown the evolution of the scale factor and the potential for early and late time contraction phase. From the above plots, we can draw the following conclusions:
\begin{itemize} 
\item Fig. \ref{lqg15} shows the plot of the scale factor in the early time contraction phase for hilltop potential given by Eqn. (\ref{scalefactor33}). This plot has been obtained for $E_{3}=10^{-8}M_{p}^{-1},\ E_{4}=10^{7},\ \rho_{c}=0.86M_{p}^{4}$. Detail graphical analysis have shown that as we increase the value of $E_{4}$ and $E_{3}$, the amplitude of expansion increases but the nature of the graph remains the same. Both the amplitude and nature is almost independent of any change in the value of $\rho_{c}$.
\item Fig. \ref{lqg17} shows the evolution of the potential in the early time contraction phase for hilltop potential. Fig. \ref{lqg17} has been obtained by with the help of Eqn. (\ref{potential33} with the parameter values $V_{0}=6.3{\rm x}10^{-3}M_{p}^{4},\ p=5,\ \beta=0.35,\ E_{3}'=10.2M_{p},\ E_{3}=5M_{p}^{-1},\ \rho_{c}=0.86M_{p}^{4}=1.4$. Detail graphical analysis show that in order to get the correct nature of the potential which will result in contraction, we need $\beta>0$. The other parameters do not affect the nature of the graph, but only increases its amplitude. 
\item The solution for the late time contraction phase given by Eqn. (\ref{scalefactor34}), contains inverse functions. Hence graphically solutions can be obtained only after solving the equation numerically.
\end{itemize}

\subsubsection{Case II: Natural potential}

\begin{figure*}[htb]
\centering
\subfigure[ An illustration of the behavior of the scale factor with time during the early  expansion phase for $\phi<<f$ with $V_{0}=4.1{\rm x}10^{-3}M_{p}^{4},\ E_{9}=10^{2}$.]{
    \includegraphics[width=7.2cm,height=7.7cm] {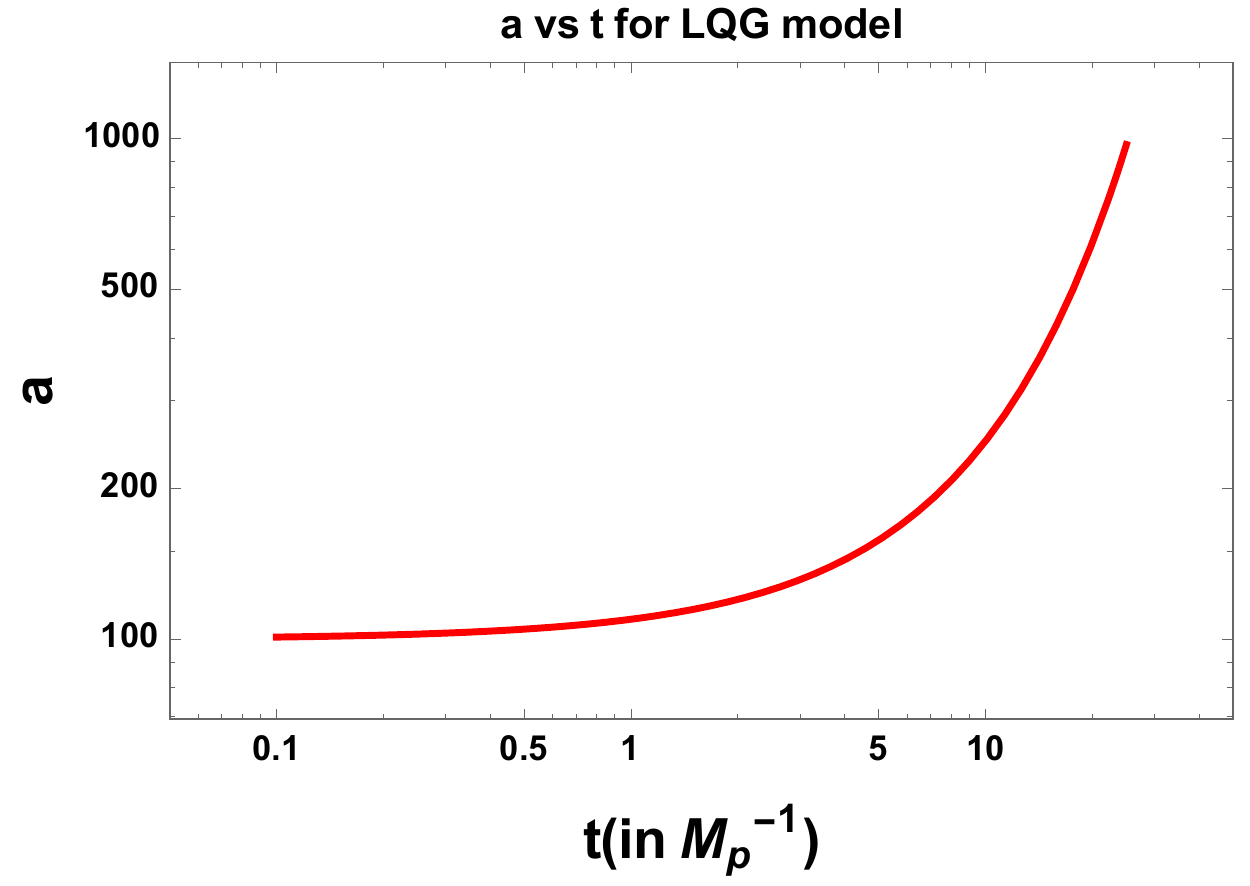}
    \label{lqg18}
}
\subfigure[An illustration of the behavior of the potential during early expansion phase for $\phi<<f$ with $V_{0}=1.3{\rm x}10^{-1}M_{p}^{4},\ f=20M_{p},\ E_{8}=33,\ \rho_{c}=0.86M_{p}^{4}$ .]{
    \includegraphics[width=7.2cm,height=7.7cm] {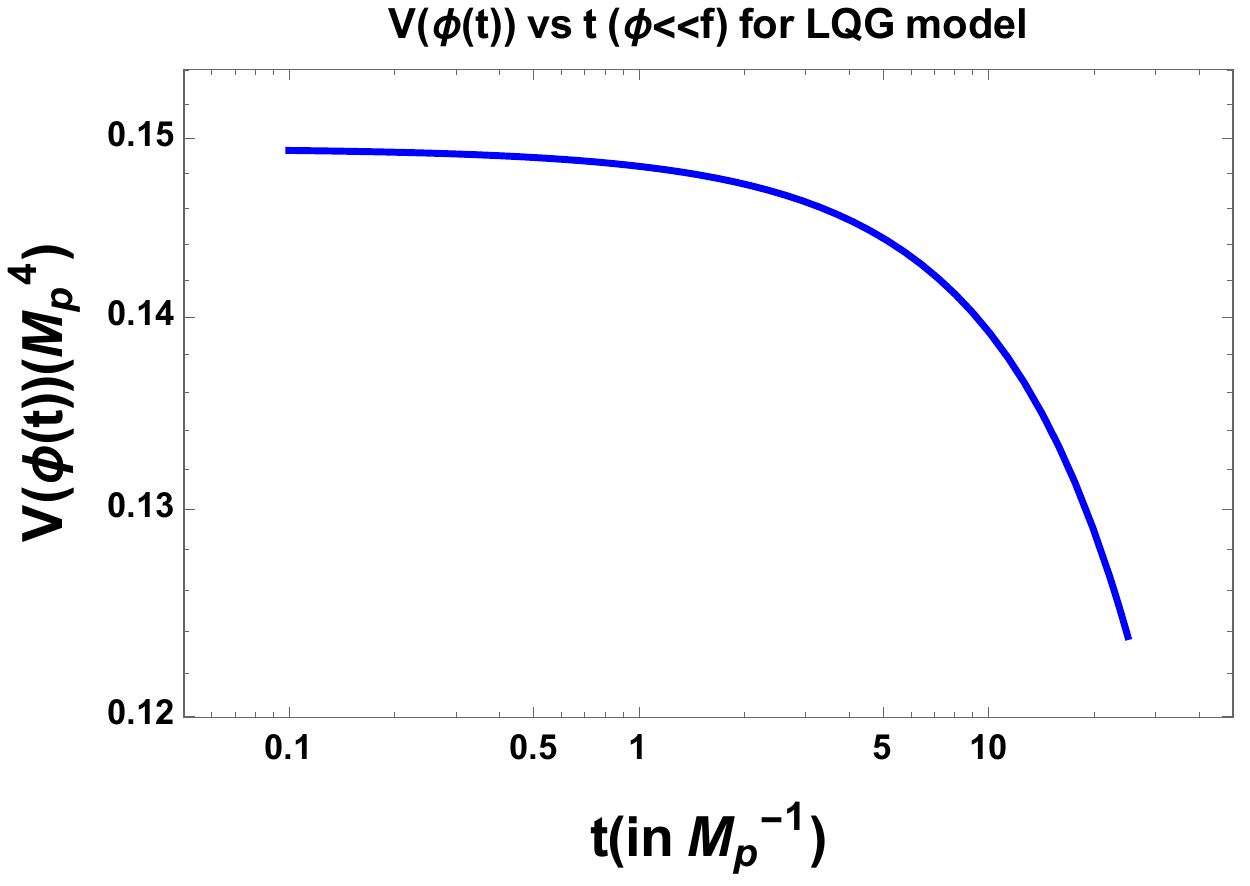}
    \label{lqg19}
}
\subfigure[An illustration of the behavior of the scale factor with time during early time expansion phase for $\phi>>f$ with $V_{0}=1.5{\rm x}10^{-1}M_{p}^{4},\ f=1M_{p},\ E_{10}=50M_{p},\ E_{11}=1,\ \rho_{c}=0.86M_{p}^{4}$.]{
    \includegraphics[width=7.2cm,height=7.7cm] {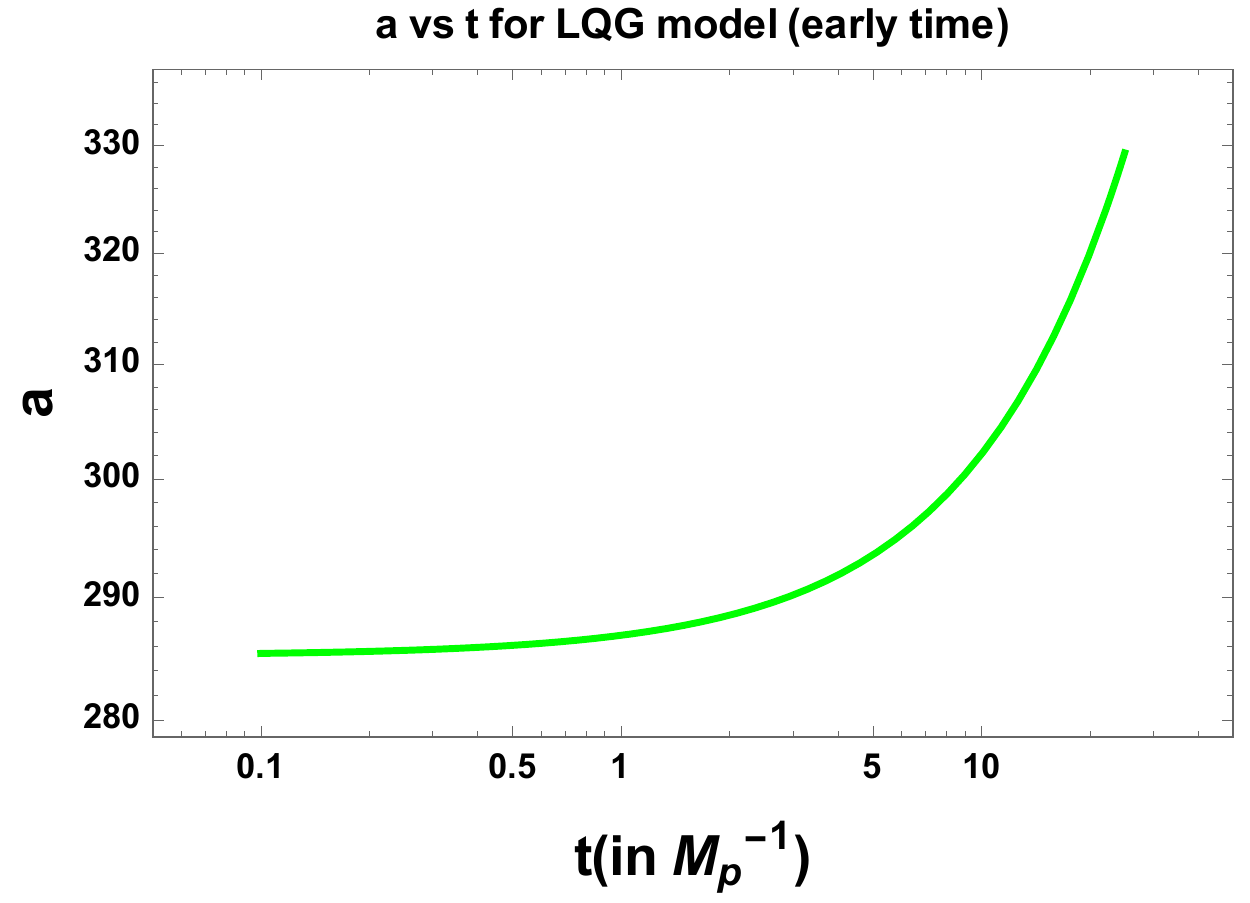}
    \label{lqg20}
}
\subfigure[An illustration of the behavior of the potential during early time expansion phase for $\phi>>f$ with $V_{0}=4.4{\rm x}10^{-2}M_{p}^{4},\ f=1M_{p},\ E_{10}=6.4M_{p},\ \rho_{c}=0.86M_{p}^{4}$ .]{
    \includegraphics[width=7.2cm,height=7.7cm] {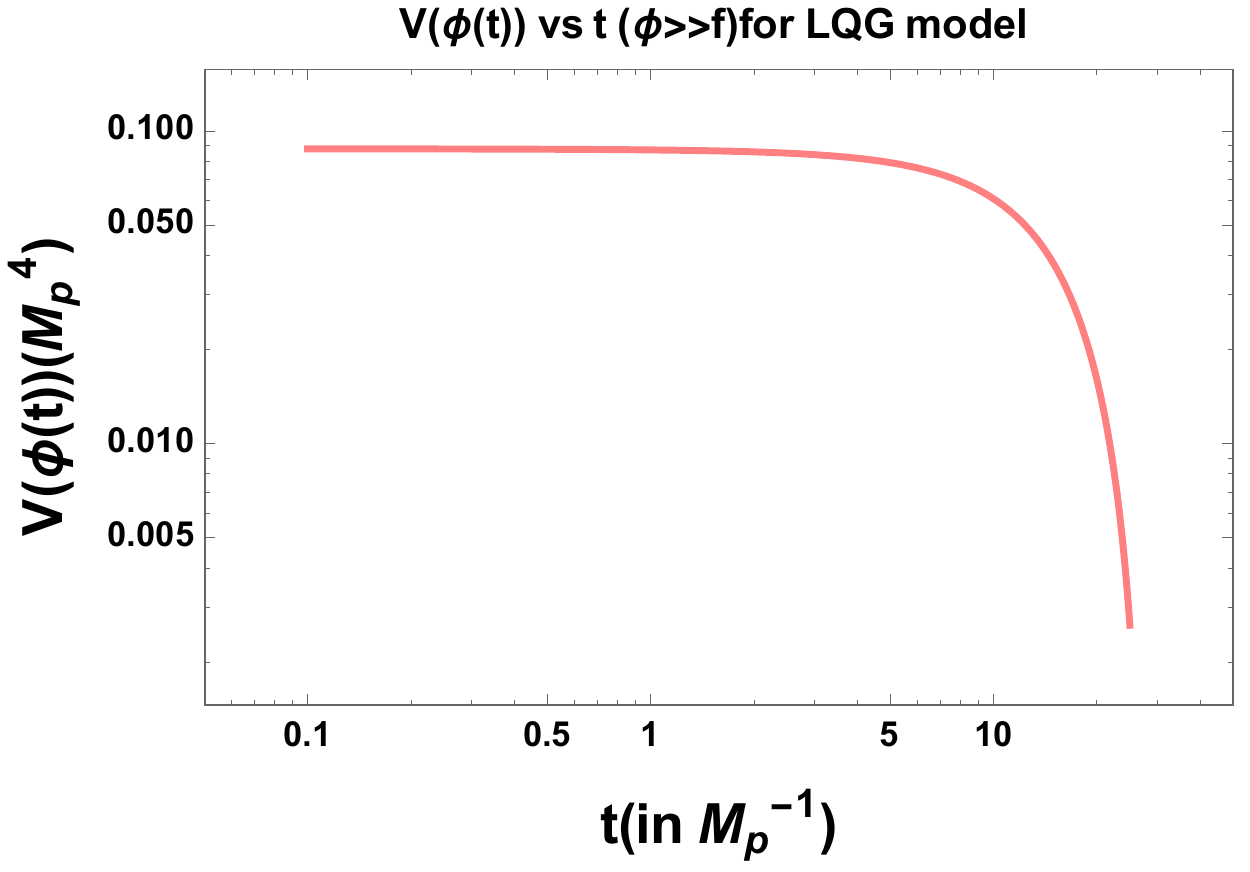}
    \label{lqg21}
}
\caption[Optional caption for list of figures]{ Graphical representation of the evolution of the scale factor and the potential during the early time expansion phase for the LQG model.} 
\label{fig28}
\end{figure*}
In Fig. \ref{fig28} we have shown the evolution of the scale factor and the potential for early expansion phase. From the above plots, we can draw the following conclusions:
\begin{itemize}
\item Fig. \ref{lqg18}, shows the plot of the scale factor in the small field limit for natural  potential given by Eq. (\ref{scalefactor30}) with the parameter values $V_{0}=4.1{\rm x}10^{-3}M_{p}^{4},\ E_{9}=10^{2}$ during the early phase of expansion. 
\item Higher values of $V_{0}$ and $E_{9}$ only increase the amplitude of expansion, keeping the nature of the plot unchanged. The change in the amplitude and nature of the plot is almost independent of the value of $\rho_{c}$.
\item Fig. \ref{lqg19} shows the plot of the behavior of the potential with time for small field natural potential during the early phase of expansion. This graph has been obtained with the help of Eq. (\ref{potential30}) with parameter values $V_{0}=1.3{\rm x}10^{-1}M_{p}^{4},\ f=20M_{p},\ E_{8}=33,\ \rho_{c}=0.86M_{p}^{4}$. Large values of $V_{0}$ and $f$ increases the height of the potential. Only for certain range of values of $E_{8}$, we get the required evolution of the potential just like in case of the behavior of natural potential in DGP model. The evolution is almost independent of the value of $\rho_{c}$.
\item The late time expansion plots remain same as given in Figs. \ref{lqg11} and \ref{lqg12}. The amplitudes can be adjusted accordingly by changing the magnitudes of different parameters.
\item Fig. \ref{lqg20}, shows the plot of the scale factor in the large field limit for natural  potential given by Eq. (\ref{scalefactor31}) with the parameter values $V_{0}=1.5{\rm x}10^{-1}M_{p}^{4},\ f=1M_{p},\ E_{10}=50M_{p},\ E_{11}=1,\ \rho_{c}=0.86M_{p}^{4}$ during the early phase of expansion. 
\item Higher values of the $\rho_{c},\ V_{0}$ and $f$ only increase the amplitude of expansion, keeping the nature of the plot unchanged.
\item Very large values of $V_{0}$ (close to $1M_{p}^{4}$ makes the expansion uneven. Expansion is possible for both negative and positive values of $E_{10}$. Increasing the value of $E_{10}$ can both increase and decrease the amplitude and makes the expansion more non linear and linear alternately.
\item Fig. \ref{lqg21} shows the plot of the behavior of the potential with time for large field natural potential. This graph has been obtained with the help of Eqn. (\ref{potential31}) with parameter values $V_{0}=4.4{\rm x}10^{-2}M_{p}^{4},\ f=1M_{p},\ E_{10}=6.4M_{p},\ \rho_{c}=0.86M_{p}^{4}$. Detail analysis show that larger values of $f$ gives smoother evolution of the potential. There are certain ranges of values for $V_{0}$ and $E_{10}$, appearing after certain intervals, which gives rise to the correct nature of potential. Larger values of $\rho_{c}$ may decrease or increase the height of the potential. 
\item If we compare the amplitudes of the scale factor in Fig. \ref{lqg15} with Fig. \ref{lqg18} or \ref{lqg20}, we find that after one cycle of expansion and contraction, we can get a net increase in amplitude of the scale factor provided the parameters are chosen properly. 
\end{itemize}

\subsubsection{Case III:  Coleman-Weinberg potential}

\begin{figure*}[htb]
\centering
\subfigure[ An illustration of the behavior of the scale factor with time during the early  expansion phase for $\phi<<M_{p}$ with $V_{0}=3{\rm x}10^{-4}M_{p}^{4},\ \alpha=0.08,\ \beta=0.7,\ E'=18.4,\ \rho_{c}=0.86,\ E_{13}=1$.]{
    \includegraphics[width=7.2cm,height=8cm] {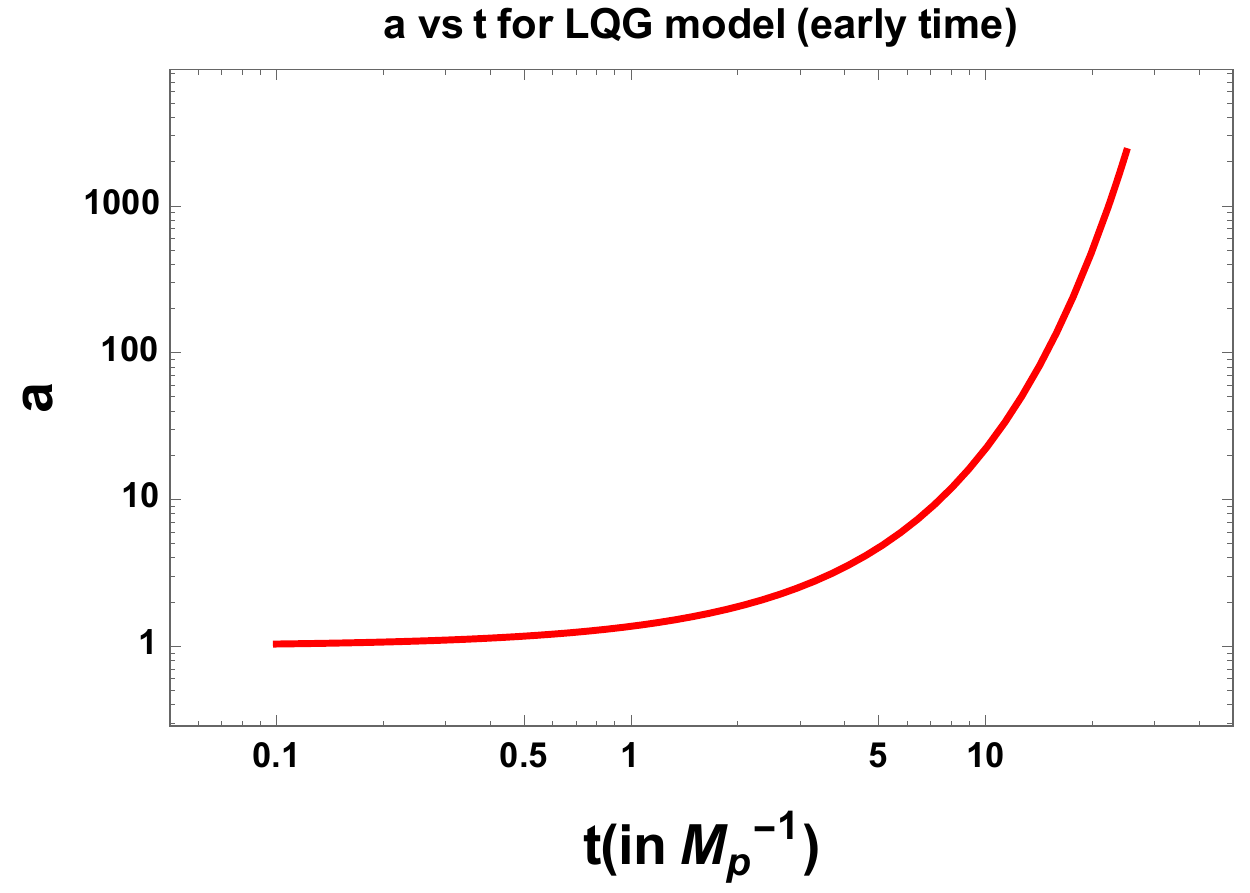}
    \label{lqg22}
}
\subfigure[An illustration of the behavior of the potential during early expansion phase for $\phi<<M_{p}$ with $V_{0}=2.7{\rm x}10^{-3}M_{p}^{4},\ \beta=0.4,\ E_{12}=0.61M_{p},\ \alpha=0.02,\ \rho_{c}=0.86$ .]{
    \includegraphics[width=7.2cm,height=8cm] {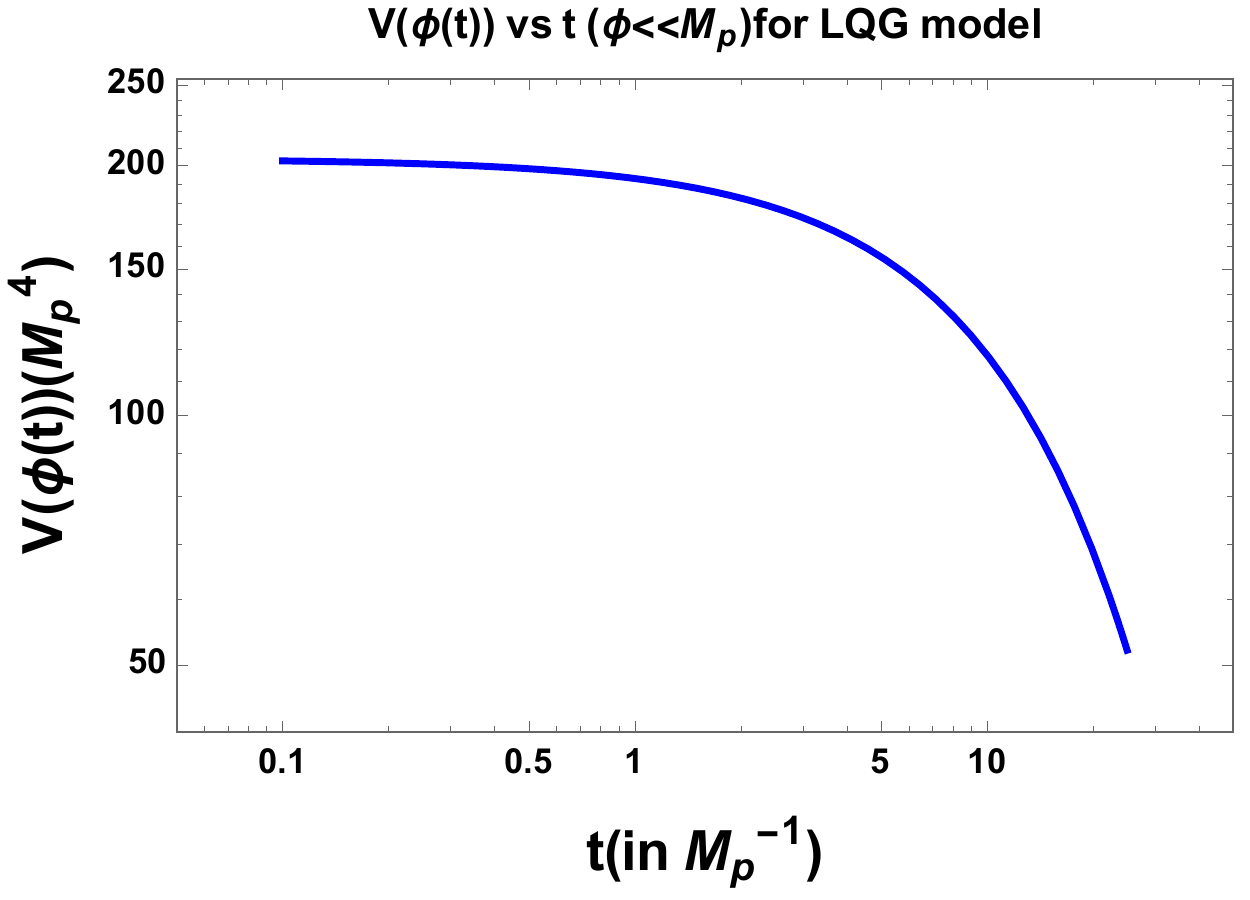}
    \label{lqg23}
}
\caption[Optional caption for list of figures]{ Graphical representation of the evolution of the scale factor and the potential during the early expansion phase for the LQG model.} 
\label{fig30}
\end{figure*}
In Fig. \ref{fig30} we have shown the evolution of the scale factor and the potential for early small field expansion phase for supergravity potential. From the above plots, we can draw the following conclusions:
\begin{itemize}
\item Fig. \ref{lqg22}, shows the plot of the scale factor in the small field limit for supergravity  potential given by Eqn. (\ref{scalefactor32}) with the parameter values $V_{0}=3{\rm x}10^{-4}M_{p}^{4},\ \alpha=0.08,\ \beta=0.7,\ E'=18.4,\ \rho_{c}=0.86,\ E_{13}=1$ during the early phase of expansion. 
\item Expansion is possible for both negative and positive values of $\beta$. However, for negative values of $\beta$, expansion is possible only if $\alpha>0.5$ and $E'>1$. Larger values of $V_{0}$ and $E'$ increases the amplitude of expansion. The expansion is almost independent of the value of $\rho_{c}$.
\item Fig. \ref{lqg23} shows the plot of the behavior of the potential with time for small field supergravity potential during the early phase of expansion. This graph has been obtained with the help of Eqn. (\ref{potential32}) with parameter values $V_{0}=2.7{\rm x}10^{-3}M_{p}^{4},\ \beta=0.4,\ E_{12}=0.61M_{p},\ \alpha=0.02,\ \rho_{c}=0.86$. Correct form of the potential is obtained for both negative and positive values of $\beta$. However, for negative values of $\beta$, $E_{12}$ must be much less than unity. Larger the value of $V_{0}$, smaller the value of $E_{12}$ allowed. For large positive $\beta$, proper evolution of the potential is possible only for very small value of $V_{0}$. $\rho_{c}$ has very negligible effect on the evolution of the potential. 
\item If we compare the amplitudes of the scale factor in Fig. \ref{lqg22} with Fig. \ref{lqg15}, we find that after one cycle of expansion and contraction, we can get a net increase in amplitude of the scale factor provided the parameters are chosen properly. 
\end{itemize}
\section{Hysteresis from Einstein-Gauss-Bonnet brane world gravity model}
\label{sq6}
Einstein Gauss-Bonnet gravity is based on the theory of higher (five) dimensional brane world scenario where the simplest
RS brane world action for single brane model (RSII) is modified by the addition of a Gauss Bonnet term  (${\cal L}_{GB}$) which is a combination of the Ricci scalar $\tilde{R}$, the Ricci tensor $\tilde{R}_{AB}$
and the Riemann tensor $\tilde{R}_{ABCD}$ and is given by:
\begin{equation}
{\cal L}_{GB} = \tilde{R}^{2} -4\tilde{R}_{AB}\tilde{R}^{AB}+\tilde{R}_{ABCD}\tilde{R}^{ABCD}.
\end{equation}
Consequently the action in five dimensional space-time takes the following form~\footnote{One can generalize this action in arbitrary dimension $D\geq 5$. It is important to note that 
at $D=4$ the contribution from Gauss-Bonnet term is
a topological invariant and does not contribute to the field
equations. For rest of the analysis we fix the space-time dimension to be $D=5$ for the sake of simplicity.}:
\bea
S&=&\frac{1}{2\kappa_{5}^{2}}\int d^{5}X\sqrt{-\tilde{g}}\,
\left[\tilde{R}-2\Lambda+\alpha\left(\tilde{R}^{2} -4\tilde{R}_{AB}\tilde{R}^{AB}+\tilde{R}_{ABCD}\tilde{R}^{ABCD}\right)\right]\nonumber\\
&&~~~~~~~~~~~~~~~~~~~~~~~~~~~~~~~~~~~~~~~+\int d^{4}x\sqrt{-g}\,\left({\cal L}_{M}^{\mathrm{brane}}-\sigma\right)\,,
\label{DGPaction}
\eea
where $\tilde{g}_{AB}$ is the metric in the 5D bulk and 
\be g_{\mu\nu}=\partial_{\mu}X^{A}\partial_{\nu}X^{B}\tilde{g}_{AB}\ee
is the induced metric on the brane with $X^{A}(x^{c})$ being the
coordinates of an event on the brane labeled by $x^{c}$.

In the present context the modified Friedmann equation is given by  \cite{Maeda:2007cb}:
\begin{equation}
\frac{\kappa^4_5}{36}(\rho+\sigma)^2 = \left(\frac{h(a)}{a^2}+\varepsilon
H^2\right)\left[1+\frac{4\alpha}{3}
\left(\frac{3k-\varepsilon h(a)}{a^2}+2H^2\right)\right]^2\, , \label{GB-F}
\end{equation}
where $\sigma$ is the single brane tension, $\alpha$ is the Gauss-Bonnet coupling,
$\Lambda$ is the 5-D cosmological constant, $\varepsilon = +1,-1$ for space-like or time-like extra dimension respectively.
Here the the function $h(a)$ is given by the following expression:
\begin{equation}
h(a) = \varepsilon k+\frac{a^2}{4\alpha}\Biggl(\varepsilon\mp
\sqrt{1+\frac{\alpha \mu}{a^{4}}+\frac43\alpha\Lambda}\Biggr) \,
,\label{horizon-2}
\end{equation}
where $\mu$ is a constant.

Since the above equations are highly complicated, following the analysis of \cite{Maeda:2007cb}, we can write the above equation as:
\begin{equation} \label{plus1}
C (\rho + \sigma)^{2} = \left(A \pm H^{2}\right) \left( B + H^{2} \right)^{2} \, ,
\end{equation}
where $\pm$ corresponds to $\varepsilon = +1$ and $\varepsilon = -1$ respectively. Here $A,B$ and $C$ are defined as:
\bea
A :&=& \frac{k}{a^{2}} + \frac{h(a)}{a^{2}} \, ,\\
 B :&=& \frac{3k}{2a^{2}} + \frac{3}{8\alpha} - \frac{\varepsilon h(a)}{2a^{2}} = {3 \over 8 \alpha} + \frac{3k}{2a^{2}}
- \frac{\varepsilon A}{2} \, ,\\
C :&=& {\kappa_5^4 \over 36} \left( {3 \over 8 \alpha} \right)^2 > 0 \,.
\label{ABC}
\eea

\subsection{Condition for bounce}

{\bf A. Space-like extra dimension with $\varepsilon=1$}\\
\\
Let $\rho = \rho_{b}$ is the corresponding energy density at which bounce occurs i.e. the Hubble parameter $H = 0$ condition achieved. Hence substituting the Hubble parameter $H = 0$ in Eq.~(\ref{plus1}), we get:
\begin{equation}
\rho_{b} = \frac{\sqrt{A}B}{C'} - \sigma
\label{rhob}
\end{equation} 
where $C' = \sqrt{C}$ is also a constant, $A, B$ are now fixed at the bounce.

Total mass at the bounce is given by \be M_{b} = \rho_{b} a_{b}^{3} =  \left(\frac{\sqrt{A}B}{C'} - \sigma\right)a_{b}^{3}\ee
and infinitisimal change in mass at bounce can be expressed as:
\begin{equation}
\delta M_{b} = \frac{\sqrt{A}}{C'}\left(\frac{ B\delta Aa_{b}^{3}}{2A} + \delta Ba_{b}^{3} + 3Ba_{b}^{2}\delta a_{b} \right), 
\label{gb1}
\end{equation}
where infinitisimal change in $A,B$ i.e. $\delta A, \delta B$ can be expressed using Eq.~(\ref{ABC}) with $\varepsilon = +1$ as: 
\begin{eqnarray}
\delta A &=& \frac{-2k}{a^{3}}\delta a + \frac{\mu a^{-5}}{(1-\left(A - \frac{k}{a^{2}}\right)4\alpha)}\delta a \label{gb2.1}, \\
\delta B &=& -\frac{2k}{a^{3}}\delta a - \frac{\mu a^{-5}}{(1-\left(A - \frac{k}{a^{2}}\right)4\alpha)}\delta a. 
\label{gb2}
\end{eqnarray}
Further substituting Eq.(\ref{gb2.1}) and Eq.~(\ref{gb2}) into Eq.~(\ref{gb1}) and using the energy conservation equation (or continuity equation) 
we get the following expression for the change of the amplitude of the scale factor after one cycle as:
\begin{equation}
\delta a_{min} = \frac{\oint pdV}{\left(3\sigma a_{min}^{2} - X'\right)}
\label{gbamin}
\end{equation}
where we introduce a new model dependent parameter $X'$ defined as:
\bea X'& =& \frac{\sqrt{A}}{C'}\left[\left(-2k + \frac{\mu a_{min}^{-2}}{\left(1-\left(A - \frac{k}{a_{min}^{2}}\right)4\alpha\right)}\right)\frac{B}{2A}
\right.\nonumber\\&&\left.~~~~~~~~~~~~~~~~~~~~~~~- \left(2k + \frac{\mu a_{min}^{-2}}{\left(1-\left(A - \frac{k}{a_{min}^{2}}\right)4\alpha\right)}\right) + 3 Ba_{min}^{2}\right].\eea
Thus we see that the condition for an increase in the amplitude of the scale factor depends on $A$ through $X'$, which in turn depends on the curvature parameter $k$ and the 
model parameters like $\mu,\ \alpha$ etc.

For $k = +1,0,-1$, Eq.~(\ref{gbamin}) can be rewritten as:

\be\begin{array}{lll}\label{gbamin1}
 \displaystyle \delta a_{min} =\left\{\begin{array}{lll}
                    \displaystyle   \frac{\oint pdV}{\left(3\sigma a_{min}^{2} - Y'\right)}~~~~ &
 \mbox{\small {\bf for {$k=+1$}}}  \\ \\
 \displaystyle   \frac{\oint pdV}{\left(3\sigma a_{min}^{2} - J'\right)}~~~~ &
 \mbox{\small {\bf for {$k=0$}}}  \\ \\
         \displaystyle  \frac{\oint pdV}{\left(3\sigma a_{min}^{2} - Z'\right)}~~~~ & \mbox{\small {\bf for {$k=-1$}}}.
          \end{array}
\right.
\end{array}\ee

where $Y'$, $J'$ and $Z'$ defined as:
\be\begin{array}{lll}\label{zxzcvb}
 \displaystyle X' =\left\{\begin{array}{lll}
                    Y' = \frac{\sqrt{A}}{C'}\left[\left(-2 + \frac{\mu a_{min}^{-2}}{\left(1-\left(A - \frac{1}{a_{min}^{2}}\right)4\alpha\right)}\right)\frac{B}{2A} 
                    - \left(2 + \frac{\mu a_{min}^{-2}}{\left(1-\left(A - \frac{1}{a_{min}^{2}}\right)4\alpha\right)}\right) + 3 Ba_{min}^{2}\right], &
 \mbox{\small {\bf for {$k=+1$}}}  \\ \\
 J'=\frac{\sqrt{A}}{C'}\left[ \left( \frac{\mu a_{min}^{-2}}{(1 - 4A\alpha)}\right)\frac{B}{2A} - \left( \frac{\mu a_{min}^{-2}}{(1- 4A\alpha)}\right) + 3 Ba_{min}^{2}\right], &
 \mbox{\small {\bf for {$k=0$}}}  \\ \\
         Z' = \frac{\sqrt{A}}{C'}\left[\left(2 + \frac{\mu a_{min}^{-2}}{\left(1-\left(A + \frac{1}{a_{min}^{2}}\right)4\alpha\right)}\right)\frac{B}{2A}
         - \left(-2 + \frac{\mu a_{min}^{-2}}{\left(1-\left(A + \frac{1}{a_{min}^{2}}\right)4\alpha\right)}\right) + 3 Ba_{min}^{2}\right], & \mbox{\small {\bf for {$k=-1$}}}.
          \end{array}
\right.
\end{array}\ee
and now the expressions for $A$ and $B$ for different values of curvature parameter $k=+1,0,-1$ are given by:
\be\begin{array}{lll}\label{gb3}
 \displaystyle A =\left\{\begin{array}{lll}
                    \displaystyle  
                    \frac{1}{a_{min}^{2}} + \frac{h(a_{min})}{a_{min}^{2}} \,, &
 \mbox{\small {\bf for {$k=+1$}}}  \\ \\
 \displaystyle  
 \frac{h(a_{min})}{a_{min}^{2}} \,, &
 \mbox{\small {\bf for {$k=0$}}}  \\ \\
         \displaystyle  
         -\frac{1}{a_{min}^{2}} + \frac{h(a_{min})}{a_{min}^{2}} \,, & \mbox{\small {\bf for {$k=-1$}}}.
          \end{array}
\right.
\end{array}\ee
and 
\be\begin{array}{lll}\label{gb3v}
 \displaystyle B =\left\{\begin{array}{lll}
                    \displaystyle  
                    \frac{3}{2a_{min}^{2}} + \frac{3}{8\alpha} - \frac{h(a_{min})}{2a_{min}^{2}} = {3 \over 8 \alpha} + \frac{3}{2a_{min}^{2}}- \frac{A}{2}, &
 \mbox{\small {\bf for {$k=+1$}}}  \\ \\
 \displaystyle  
 \frac{3}{8\alpha} - \frac{h(a_{min})}{2a_{min}^{2}} = {3 \over 8 \alpha} - \frac{A}{2} \,, &
 \mbox{\small {\bf for {$k=0$}}}  \\ \\
         \displaystyle  
         -\frac{3}{2a_{min}^{2}} + \frac{3}{8\alpha} - \frac{h(a_{min})}{2a_{min}^{2}} = {3 \over 8 \alpha} - \frac{3}{2a_{min}^{2}}- \frac{A}{2}, & \mbox{\small {\bf for {$k=-1$}}}.
          \end{array}
\right.
\end{array}\ee
For the various values of the curvature parameter $k=+1,0,-1$ one can point out the following characteristics from our analysis:
\begin{itemize}
 \item For $k=+1$ case if we consider the situation where $3\sigma a_{min}^{2} > Y'$, i.e. if the brane tension to be large,
 then an increase in the amplitude of the scale factor is possible for a positive sign of the work done i.e. $\oint pdV >0$.
 And if the brane tension is small enough then a negative sign of the integral gives expansion with increasing amplitude.
 \item If we follow the analysis of \cite{Maeda:2007cb}, it shows pictorially that bounce does not occur when the extra dimension is space-like along with $k=0$. We can also do
similar pictorial study for the case when $k\neq 0$ and see whether this statement remains true for $k\neq 0$.
 \item We also see that for $k=-1$ case depending on the relative signs of $A$, $B$ and $\sigma$, $\delta a_{min}$ can be positive for both $\oint pdV > or < 0$ cases.
 Thus just like for $k=+1$, if $3\sigma a_{min}^{2} > Z'$, i.e. if we consider the brane tension to be large,
 then an increase in the amplitude of the scale factor is possible for a positive sign of the work done i.e. $\oint pdV >0$. Additionally if the brane tension
 is small enough then a negative sign of the integral  i.e. $\oint pdV <0$ gives expansion with increasing amplitude.
\end{itemize}
{\bf B. Time-like extra dimension with $\varepsilon=-1$}\\
 \\ \\
In this context the condition for bounce is given by Eq.~(\ref{rhob})
with the expression of $A$ and $B$ now given by Eq.~(\ref{ABC}) with $\varepsilon=-1$. 
Hence the expression for infinitesimal change in mass at bounce $\delta M_{b}$ is again given by Eq.~(\ref{gb1}), where the expressions for $\delta A$ and $\delta B$ are now given by:
\begin{eqnarray}
\delta A &=& +\frac{2k}{a^{3}}\delta a - \frac{\mu a^{-5}}{(1 + \left(A + \frac{k}{a^{2}}\right)4\alpha)}\delta a, \label{gb2.1new} \\
\delta B &=& -\frac{2k}{a^{3}}\delta a - \frac{\mu a^{-5}}{(1 + \left(A + \frac{k}{a^{2}}\right)4\alpha)}\delta a. 
\label{gb2new}
\end{eqnarray}
Further substituting Eq.~(\ref{gb2.1new}) and Eq.~(\ref{gb2new}) into Eq.~(\ref{gb1}) and using the energy conservation equation (or continuity equation) 
we get the following expression for the change of the amplitude of the scale factor after one cycle as:
\begin{equation}
\delta a_{min} = \frac{\oint pdV}{\left(3\sigma a_{min}^{2} - X'\right)}
\label{gbaminnew}
\end{equation}
where the parameter $X'$ is defined as:
\bea 
X'&=& \frac{\sqrt{A}}{C'}\left[\left(2k + \frac{\mu a_{min}^{-2}}{\left(1+\left(A + \frac{k}{a_{min}^{2}}\right)4\alpha\right)}\right)\frac{B}{2A}
\right.\nonumber\\&&\left.~~~~~~~~~~~~~~~~~~~~~~~- \left(2k + \frac{\mu a_{min}^{-2}}{\left(1+\left(A + \frac{k}{a_{min}^{2}}\right)4\alpha\right)}\right) + 3 Ba_{min}^{2}\right].\eea

For $k = +1,0,-1$, Eq.~(\ref{gbaminnew}) can be rewritten as:

\be\begin{array}{lll}\label{gbamin1new}
 \displaystyle \delta a_{min} =\left\{\begin{array}{lll}
                    \displaystyle   \frac{\oint pdV}{\left(3\sigma a_{min}^{2} - Y'\right)}~~~~ &
 \mbox{\small {\bf for {$k=+1$}}}  \\ \\
 \displaystyle   \frac{\oint pdV}{\left(3\sigma a_{min}^{2} - J'\right)}~~~~ &
 \mbox{\small {\bf for {$k=0$}}}  \\ \\
         \displaystyle  \frac{\oint pdV}{\left(3\sigma a_{min}^{2} - Z'\right)}~~~~ & \mbox{\small {\bf for {$k=-1$}}}.
          \end{array}
\right.
\end{array}\ee

where $Y'$, $J'$ and $Z'$ defined as:
\be\begin{array}{lll}\label{zxzcvb}
 \displaystyle X' =\left\{\begin{array}{lll}
                    Y' = \frac{\sqrt{A}}{C'}\left[\left(2 + \frac{\mu a_{min}^{-2}}{\left(1+\left(A + \frac{1}{a_{min}^{2}}\right)4\alpha\right)}\right)\frac{B}{2A}
                    - \left(2 + \frac{\mu a_{min}^{-2}}{\left(1+\left(A + \frac{1}{a_{min}^{2}}\right)4\alpha\right)}\right) + 3 Ba_{min}^{2}\right], &
 \mbox{\small {\bf for {$k=+1$}}}  \\ \\
 J'=\frac{\sqrt{A}}{C'}\left[\left(\frac{\mu a_{min}^{-2}}{(1+ 4A\alpha)}\right)\frac{B}{2A} - \left(\frac{\mu a_{min}^{-2}}{(1+ 4A\alpha)}\right) + 3 Ba_{min}^{2}\right], &
 \mbox{\small {\bf for {$k=0$}}}  \\ \\
         Z' = \frac{\sqrt{A}}{C'}\left[\left(-2 + \frac{\mu a_{min}^{-2}}{\left(1+\left(A - \frac{1}{a_{min}^{2}}\right)4\alpha\right)}\right)\frac{B}{2A}
         - \left(-2 + \frac{\mu a_{min}^{-2}}{\left(1+\left(A - \frac{1}{a_{min}^{2}}\right)4\alpha\right)}\right) + 3 Ba_{min}^{2}\right], & \mbox{\small {\bf for {$k=-1$}}}.
          \end{array}
\right.
\end{array}\ee
and now the expressions for $A$ and $B$ for differfent values of curvature parameter $k=+1,0,-1$ are given by:
\be\begin{array}{lll}\label{gb3}
 \displaystyle A =\left\{\begin{array}{lll}
                    \displaystyle  
                    -\frac{1}{a_{min}^{2}} + {1 \over 4 \alpha} \left(\sqrt{-1 + {\alpha \mu \over a_{min}^4} + \frac43 \alpha
\Lambda} - 1 \right) \,, &
 \mbox{\small {\bf for {$k=+1$}}}  \\ \\
 \displaystyle  
 \frac{h(a)}{a_{min}^{2}} \,, &
 \mbox{\small {\bf for {$k=0$}}}  \\ \\
         \displaystyle  
          \frac{1}{a_{min}^{2}} + {1 \over 4 \alpha} \left(\sqrt{-1 + {\alpha \mu \over a_{min}^4} + \frac43 \alpha
\Lambda} - 1 \right) \,, & \mbox{\small {\bf for {$k=-1$}}}.
          \end{array}
\right.
\end{array}\ee
and 
\be\begin{array}{lll}\label{gb3v}
 \displaystyle B =\left\{\begin{array}{lll}
                    \displaystyle  
                    {3 \over
8 \alpha} +\frac{3}{2a_{min}^{2}} + \frac{A}{2} \,, &
 \mbox{\small {\bf for {$k=+1$}}}  \\ \\
 \displaystyle  
 \frac{3}{8\alpha} + \frac{h(a_{min})}{2a_{min}^{2}} = {3 \over 8 \alpha} + \frac{A}{2} \,, &
 \mbox{\small {\bf for {$k=0$}}}  \\ \\
         \displaystyle  
         {3 \over
8 \alpha} -\frac{3}{2a_{min}^{2}} + \frac{A}{2} \,, & \mbox{\small {\bf for {$k=-1$}}}.
          \end{array}
\right.
\end{array}\ee
For the various values of the curvature parameter $k=+1,0,-1$ one can point out the following characteristics from our analysis:
\begin{itemize}
 \item Here for $k=+1$ case we see that depending on the relative signs of $A$, $B$ and $\sigma$, $\delta a_{min}$ can be positive for both $\oint pdV > or < 0$. 
 Thus, if $3\sigma a_{min}^{2} > Y'$, i.e. if we consider the brane tension to be
 large, then an increase in the amplitude of the scale factor is possible for a positive sign of the work done i.e. $\oint pdV >0$. Additionally
 if the brane tension is small enough then a negative sign of the integral i.e. $\oint pdV <0$ gives expansion with increasing amplitude.
 \item In \cite{Maeda:2007cb}, the authors show pictorially that the bouncing condition is reached if either $0<B<2A$ or $B>2A$ and $\sigma <0$ for $k = 0$ case.
  We can also do
similar pictorial study for the case when $k\neq 0$ and see whether this statement remains true for $k\neq 0$.
 \item Also for $k=-1$ case we see that depending on the relative signs of A, B and $\sigma$, $\delta a_{min}$ can be positive for both $\oint pdV > or < 0$. Thus
 just like for k=+1, if $3\sigma a_{min}^{2} > Z'$, i.e if we consider the brane tension to be
 large, then an increase in the amplitude of the scale factor is possible for a positive sign of the work done i.e. $\oint pdV >0$.
 And if the brane tension is small enough then a negative sign of the integral  i.e. $\oint pdV <0$ gives expansion with increasing amplitude.
\end{itemize}

 \subsection{Condition for acceleration}
 
 { \bf A. Space-like extra dimension with $\varepsilon=1$}\\
 \\  
 In order to get the condition for acceleration, we need to use the second Friedmann equation
 which we get after differentiating Eq.~(\ref{plus1}) w.r.t. time $t$ for $\varepsilon = +1$ and using the energy conservation equation to get rid of $\dot{\rho}$. The resulting equation is given by:
 \begin{equation}
 \dot{H}+H^2=\frac{\ddot{a}}{a}  = -\frac{3C'^{2}(\rho + p)(\rho + \sigma)}{Y} + \left(\frac{\dot{a}}{a}\right)^{2} - \frac{Z}{Y}
 \label{gbacceleration}
 \end{equation}
 where $Z$ and $Y$ are functions of $A, B$ and their time derivatives, which are in turn functions of scale factor $a(t)$ and its time derivative $\dot{a}$ respectively.
 Their explicit expressions in terms of $A, B$ and their derivatives are given by:
 \begin{eqnarray}
Y &=& 4AH^{2} + 4AB + 2B^{2} + 6H^{4} + 8BH^{2}, \\
Z &=& \frac{\dot{A}B^{2}}{H} + 2\frac{AB\dot{B}}{H} + \dot{A}H^{3} + 2H(\dot{A}B + A\dot{B} + B\dot{B} + \dot{B}H^{2}). \nonumber \\
 \end{eqnarray}
 where $H$ is the Hubble parameter.
 
The general condition for acceleration is given by the following expression:
\begin{equation}
(\rho + p)(\rho + \sigma) < \frac{1}{3C'^{2}}(H^{2}Y - Z)
\label{gbaccel}
\end{equation}
Using Eq.~(\ref{rhob}) and setting $H^{2} = 0$ in the above equation, we get the following condition for acceleration at bounce as:
\begin{equation}
p_{b} < \frac{\sqrt{A}B}{C'}\left(-\frac{Z}{3AB^{2}} - 1 + \frac{\sigma C'}{\sqrt{A}B}\right)
\label{accelbounce}
\end{equation} 
where $A, B, Z$ are fixed at bounce. Thus, whether this condition violates the energy condition or not, depends upon the values of the different parameters of the model.

Finally, in terms of the scalar field and its potential, the condition for acceleration can be written as:
\begin{equation}
\frac{\dot{\phi}^{4}}{2} < \frac{1}{3C'^{2}}(H^{2}Y - Z) - \frac{1}{2}\dot{\phi}^{2}(V(\phi) + \sigma)
\label{accelphi}
\end{equation}
Thus we see that the condition for acceleration in terms of the scalar field now not only depends on the kinetic term, but also on the fourth power of the time derivative of $\phi$.
\\ \\
{\bf B. Time-like extra dimension with $\varepsilon=-1$}\\
\\
The second Friedmann equation is again given by Eq.~(\ref{gbacceleration}) i.e.
\begin{equation}
 \dot{H}+H^2=\frac{\ddot{a}}{a}  = -\frac{3C'^{2}(\rho + p)(\rho + \sigma)}{D} + \left(\frac{\dot{a}}{a}\right)^{2} - \frac{E}{D},
 \label{gbacceleration1}
 \end{equation}
where now the expression for $D$ and $E$ are given by:
 \begin{eqnarray}
D &=& 4AH^{2} + 4AB - 2B^{2} - 6H^{4} - 8BH^{2}, \label{ED}\\
E &=& \frac{\dot{A}B^{2}}{H} + 2\frac{AB\dot{B}}{H} + \dot{A}H^{3} + 2H(\dot{A}B + A\dot{B} - B\dot{B} - \dot{B}H^{2}). \nonumber \\
\label{DE}
 \end{eqnarray}
Rest all the conditions for acceleration in terms of the pressure and the scalar field remains the same with expressions for $D$ and $E$ now given by Eq.~(\ref{ED}) and Eq.~(\ref{DE}) respectively.
\\ \\
\begin{figure*}[htb]
\centering
\subfigure[ An illustration of the bouncing condition for a universe with$k=0, A=10M_{p}^{2},\ B=40M_{p}^{2},\ \sigma=-10^{-9}M_{4}^{4},\ C=100$.]{
    \includegraphics[width=7.2cm,height=8cm] {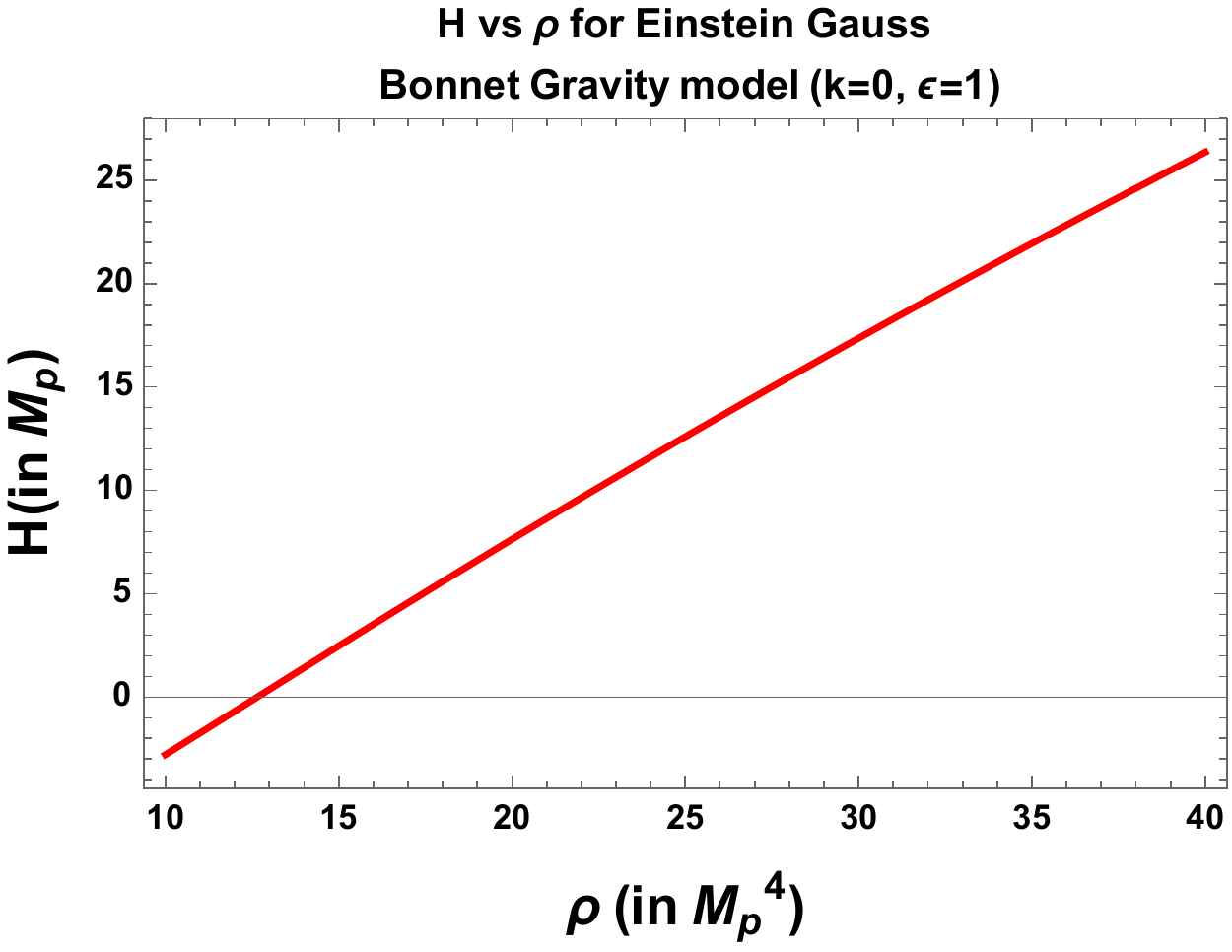}
    \label{egb1}
}
\subfigure[An illustration of the bouncing condition for a universe with$k=1,\ \varepsilon=1,\ w=1/3,\ C=100,\ A=10M_{p}^{2},\ B=40M_{p}^{2},\ \sigma=-10^{-9}M_{4}^{4},\ C=100,\ \Lambda=6{\rm x}10^{-3}M_{p}^{2}$.]{
    \includegraphics[width=7.2cm,height=7.5cm] {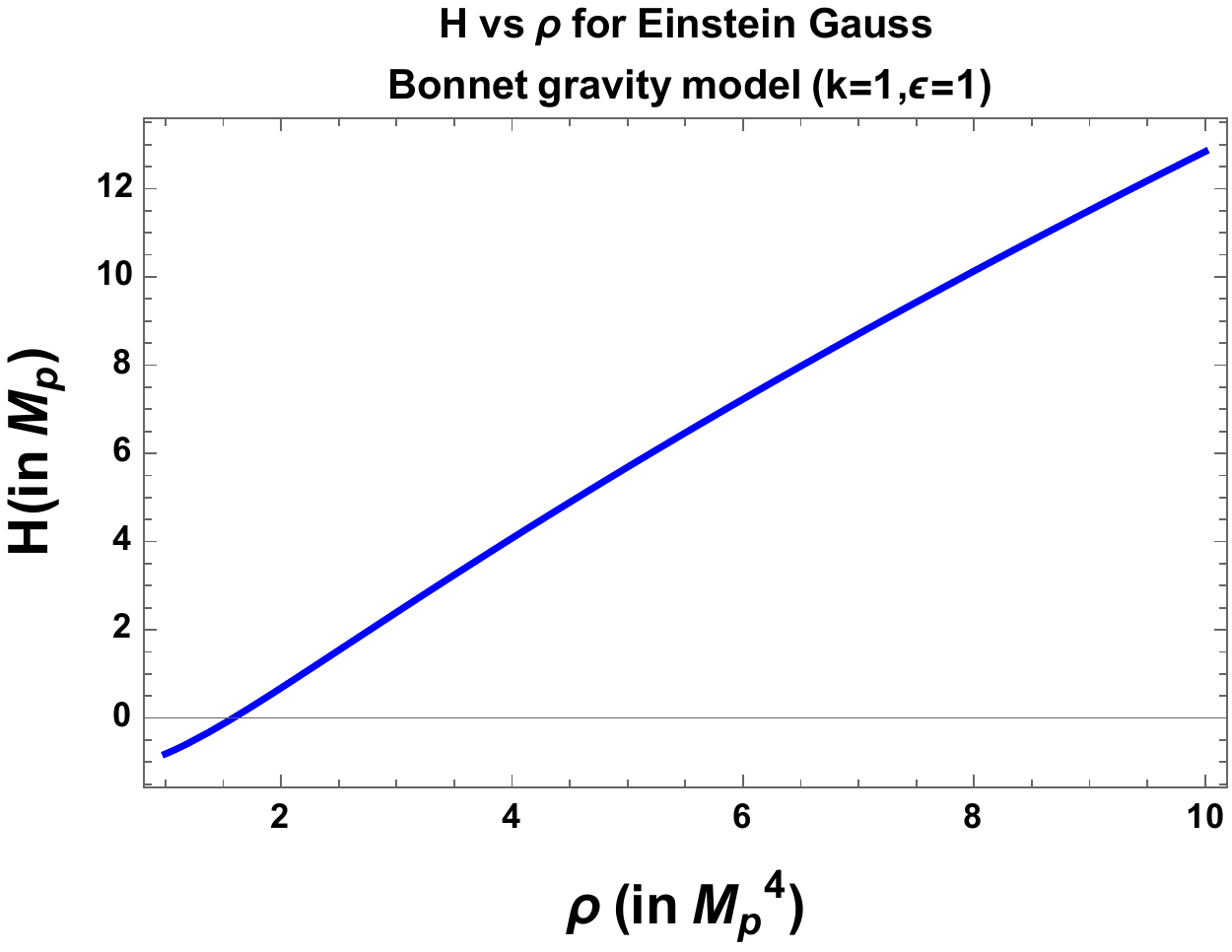}
    \label{egb2}
}
\subfigure[An illustration of the acceleration condition at the time of bounce for a universe with an equation of state $w=1/3,\ k=-0,\ A=-10M_{p}^{2},\ B=10M_{p}^{2},\ C'=10,\ \sigma=1M_{p}^{4}$]{
    \includegraphics[width=11.2cm,height=7.8cm] {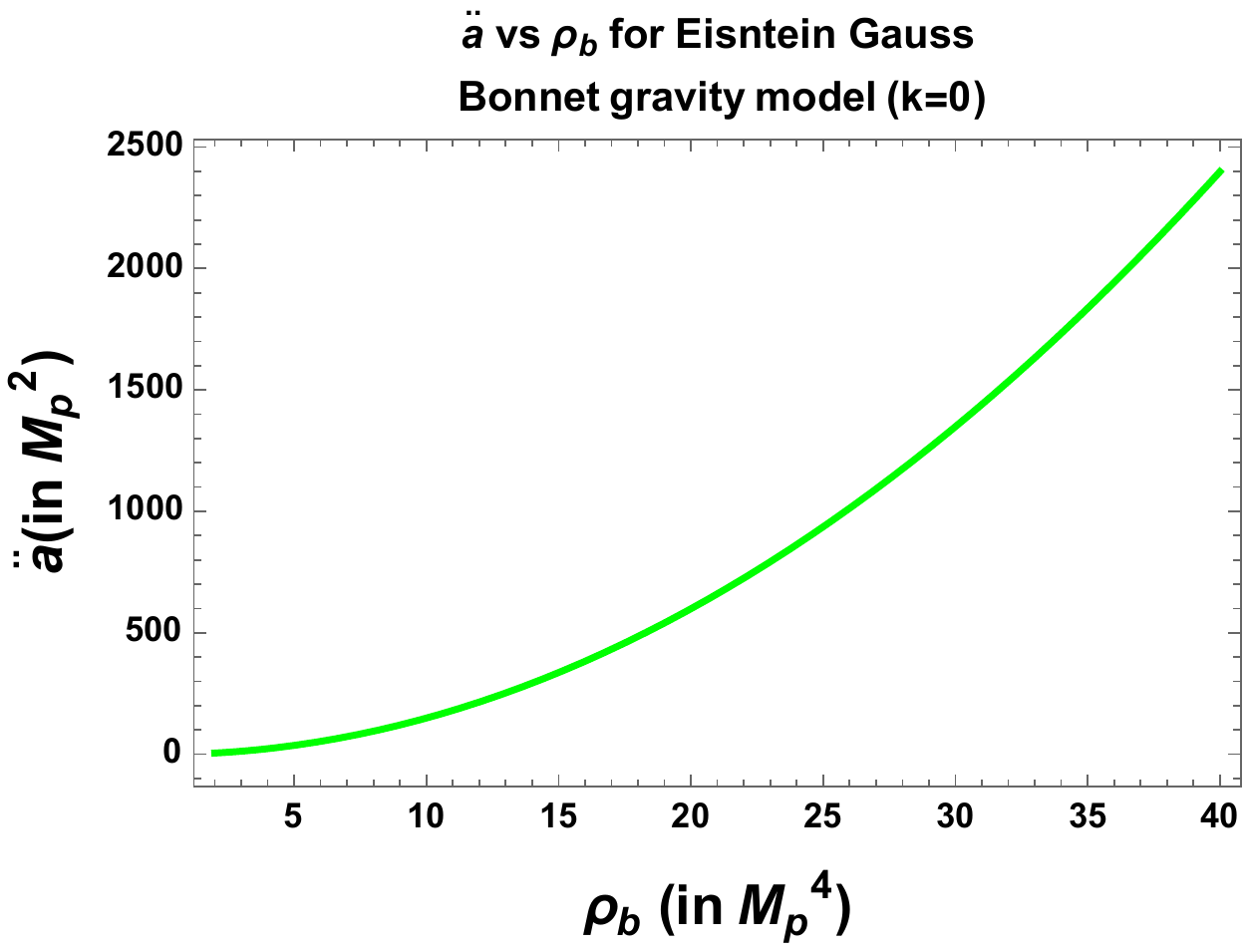}
    \label{egb3}
}
\caption[Optional caption for list of figures]{ Graphical representation of the phenomena of bounce and acceleration for Einstein Gauss Bonnet gravity model.} 
\label{fig35}
\end{figure*}

In Figs. \ref{fig35}, we have shown the phenomena of bounce and acceleration in the Einstein Gauss Bonnet gravity model. While performing this analysis, we have considered that $\mu$ is very small and hence can be neglected. We can draw the following conclusions from the above figures:
\begin{itemize}
\item Figs. \ref{egb1} and \ref{egb2}, have been plotted with the help of Eqns. (\ref{GB-F}), (\ref{horizon-2}), (\ref{plus1}) and (\ref{ABC}) using the relation $\rho=a^{-3(1+w)}$ with $k=0, A=10M_{p}^{2},\ B=40M_{p}^{2},\ \sigma=-10^{-9}M_{4}^{4},\ C=100$. Bounce occurs for both negative and positive brane tension. Detail analysis of the scenarios under which bouncing is possible has been discussed extensively in \cite{Maeda:2007cb}.  
\item From Fig. \ref{egb1}, we get the bounce at $\rho=\rho_{b}=12.78M_{p}^{4}$.
\item In Fig. \ref{egb2}, we have shown the bouncing condition for $k=1,\ \varepsilon=1,\ w=1/3,\ C=100,\ A=10M_{p}^{2},\ B=40M_{p}^{2},\ \sigma=-10^{-9}M_{4}^{4},\ C=100,\ \Lambda=6{\rm x}10^{-3}M_{p}^{2}$. The condition remains valid for the other combinations like ($k=-1,\ \varepsilon=-1$), ($k=1,\ \varepsilon=-1$) and ($k=-1,\ \varepsilon=1$). But for the last case, bouncing occurs at a very small value of $\rho$, while for the other cases, it occurs at some larger value of$\rho$, which depends upon the choice of our parameters.
\item For the cases ($k=1,\ \varepsilon=\pm 1$), bounce is possible only for negative brane tension. But for ($k=-1,\ \varepsilon=\pm 1$), it is possible for both positive and negative brane tension. Bounce is also possible for $w=0$. From Fig. \ref{egb2}, we get the bounce at $\rho=\rho_{b}=1.9M_{p}^{4}$.
\item Fig. \ref{egb3}, we have shown the necessary condition of acceleration ($\ddot{a}> 0$) at the time of bounce for $w=1/3,\ k=0,\ A=-10M_{p}^{2},\ B=10M_{p}^{2},\ C'=10,\ \sigma=1M_{p}^{4}$. This plot has been obtained with the help of Eqn. (\ref{gbacceleration}). But since Eqn. (\ref{gbacceleration}) is highly complicated, in order to know the parameter space for which we get acceleration at bounce, we have instead set $H^{2}=0$ and $\rho=\rho_{b}$ in Eqn. (\ref{gbacceleration}) and analysed for what values of the parameters do we get acceleration at bounce. From detail graphical analysis, we have found that this possible if the condition $4AB+2B^{2}<0$ is satisfied for ($\rho+\sigma$)$>0$, and vice versa. This is possible for both positive and negative values of $A,\ B,\ \sigma$ provided the above conditions are satisfied. Similar analysis can be done for $k=1,-1$ also. The results remains nearly same.
\item Thus from Fig. \ref{fig35} we can conclude that for this model, bounce is possible for closed, open and flat universe.
\end{itemize}
\subsection{Condition for turnaround}

{ \bf A. Space-like extra dimension with $\varepsilon=1$}\\
\\
Following the same line of treatment as for bounce we get the condition for turnaround for $\varepsilon = +1$ as: 
\begin{equation}
\rho_{t} = \frac{\sqrt{A}B}{C'} - \sigma
\label{turn}
\end{equation}
where $A$ and $B$ are fixed at turnaround.

Following the same analysis as for bounce we get the following expression for change in the amplitude of the scale factor after each successive cycle is given by:
\be\begin{array}{lll}\label{gbmax1}
 \displaystyle \delta a_{max} = \frac{\oint pdV}{\left(3\sigma a_{max}^{2} - X\right)}=\left\{\begin{array}{lll}
                    \displaystyle  \frac{\oint pdV}{\left(3\sigma a_{max}^{2} - Y\right)}~~~~ &
 \mbox{\small {\bf for {$k=+1$}}}  \\ \\
 \displaystyle   \frac{\oint pdV}{\left(3\sigma a_{max}^{2} - J\right)}~~~~ &
 \mbox{\small {\bf for {$k=0$}}}  \\ \\
         \displaystyle  \frac{\oint pdV}{\left(3\sigma a_{max}^{2} - Z\right)} ~~~~ & \mbox{\small {\bf for {$k=-1$}}}.
          \end{array}
\right.
\end{array}\ee

where where the expressions for $Y, J, Z$ remains same as in the bounce section for $Y', J', Z'$, only with $a_{min}$ replaced by $a_{max}$ i.e.
\be\begin{array}{lll}\label{zxzcvbgg}
 \displaystyle X =\left\{\begin{array}{lll}
                    Y = \frac{\sqrt{A}}{C'}\left[\left(-2 + \frac{\mu a_{max}^{-2}}{\left(1-\left(A - \frac{1}{a_{max}^{2}}\right)4\alpha\right)}\right)\frac{B}{2A} 
                    - \left(2 + \frac{\mu a_{max}^{-2}}{\left(1-\left(A - \frac{1}{a_{max}^{2}}\right)4\alpha\right)}\right) + 3 Ba_{max}^{2}\right], &
 \mbox{\small {\bf for {$k=+1$}}}  \\ \\
 J=\frac{\sqrt{A}}{C'}\left[ \left( \frac{\mu a_{max}^{-2}}{(1 - 4A\alpha)}\right)\frac{B}{2A} 
- \left( \frac{\mu a_{max}^{-2}}{(1- 4A\alpha)}\right) + 3 Ba_{max}^{2}\right], &
 \mbox{\small {\bf for {$k=0$}}}  \\ \\
         Z = \frac{\sqrt{A}}{C'}\left[\left(2 + \frac{\mu a_{max}^{-2}}{\left(1-\left(A + \frac{1}{a_{max}^{2}}\right)4\alpha\right)}\right)\frac{B}{2A}
         - \left(-2 + \frac{\mu a_{max}^{-2}}{\left(1-\left(A + \frac{1}{a_{max}^{2}}\right)4\alpha\right)}\right) + 3 Ba_{max}^{2}\right], & \mbox{\small {\bf for {$k=-1$}}}.
          \end{array}
\right.
\end{array}\ee
and now the expressions for $A$ and $B$ for differfent values of curvature parameter $k=+1,0,-1$ are given by:
\be\begin{array}{lll}\label{gb3}
 \displaystyle A =\left\{\begin{array}{lll}
                    \displaystyle  
                    \frac{1}{a_{max}^{2}} + \frac{h(a_{max})}{a_{max}^{2}} \,, &
 \mbox{\small {\bf for {$k=+1$}}}  \\ \\
 \displaystyle  
 \frac{h(a_{max})}{a_{max}^{2}} \,, &
 \mbox{\small {\bf for {$k=0$}}}  \\ \\
         \displaystyle  
         -\frac{1}{a_{max}^{2}} + \frac{h(a_{max})}{a_{max}^{2}} \,, & \mbox{\small {\bf for {$k=-1$}}}.
          \end{array}
\right.
\end{array}\ee
and 
\be\begin{array}{lll}\label{gb3v}
 \displaystyle B =\left\{\begin{array}{lll}
                    \displaystyle  
                    \frac{3}{2a_{max}^{2}} + \frac{3}{8\alpha} - \frac{h(a_{max})}{2a_{max}^{2}} = {3 \over 8 \alpha} + \frac{3}{2a_{max}^{2}}- \frac{A}{2}, &
 \mbox{\small {\bf for {$k=+1$}}}  \\ \\
 \displaystyle  
 \frac{3}{8\alpha} - \frac{h(a_{max})}{2a_{max}^{2}} = {3 \over 8 \alpha} - \frac{A}{2} \,, &
 \mbox{\small {\bf for {$k=0$}}}  \\ \\
         \displaystyle  
         -\frac{3}{2a_{max}^{2}} + \frac{3}{8\alpha} - \frac{h(a_{max})}{2a_{max}^{2}} = {3 \over 8 \alpha} - \frac{3}{2a_{max}^{2}}- \frac{A}{2}, & \mbox{\small {\bf for {$k=-1$}}}.
          \end{array}
\right.
\end{array}\ee

Therefore the conclusions also remain same as for bounce case.

In \cite{Maeda:2007cb}, the authors show pictorially that for the case when $k = 0$, re-collapse occurs
when $A>0 , B>0$ and $\sigma <0$ or is small but positive. We can also do similar pictorial study for the
case when $k \neq 0$ and find similar conditions on $A, B$ and $\sigma$ which gives re-collapse.
\\ \\
{\bf B. Time-like extra dimension with $\varepsilon=-1$}\\
\\ 

The condition for turnaround remains same as Eq.~ (\ref{turn}).

Following the same analysis as for bounce we get the expression for change in the amplitude of the scale factor after each successive cycle as:

\be\begin{array}{lll}\label{gbmax1new11}
 \displaystyle \delta a_{max} = \frac{\oint pdV}{\left(3\sigma a_{max}^{2} - X\right)}=\left\{\begin{array}{lll}
                    \displaystyle   \frac{\oint pdV}{\left(3\sigma a_{max}^{2} - Y\right)}~~~~ &
 \mbox{\small {\bf for {$k=+1$}}}  \\ \\
 \displaystyle   \frac{\oint pdV}{\left(3\sigma a_{max}^{2} - J\right)}~~~~ &
 \mbox{\small {\bf for {$k=0$}}}  \\ \\
         \displaystyle  \frac{\oint pdV}{\left(3\sigma a_{max}^{2} - Z\right)}~~~~ & \mbox{\small {\bf for {$k=-1$}}}.
          \end{array}
\right.
\end{array}\ee
where the expressions for $Y, J, Z$ remains same as in the bounce section for $Y', J', Z'$, only with $a_{min}$ replaced by $a_{max}$ i.e.
\be\begin{array}{lll}\label{zxzcvb}
 \displaystyle X =\left\{\begin{array}{lll}
                    Y = \frac{\sqrt{A}}{C'}\left[\left(2 + \frac{\mu a_{max}^{-2}}{\left(1+\left(A + \frac{1}{a_{max}^{2}}\right)4\alpha\right)}\right)\frac{B}{2A}
                    - \left(2 + \frac{\mu a_{max}^{-2}}{\left(1+\left(A + \frac{1}{a_{max}^{2}}\right)4\alpha\right)}\right) + 3 Ba_{max}^{2}\right], &
 \mbox{\small {\bf for {$k=+1$}}}  \\ \\
 J=\frac{\sqrt{A}}{C'}\left[\left(\frac{\mu a_{max}^{-2}}{(1+ 4A\alpha)}\right)\frac{B}{2A} - \left(\frac{\mu a_{max}^{-2}}{(1+ 4A\alpha)}\right) + 3 Ba_{max}^{2}\right], &
 \mbox{\small {\bf for {$k=0$}}}  \\ \\
         Z = \frac{\sqrt{A}}{C'}\left[\left(-2 + \frac{\mu a_{max}^{-2}}{\left(1+\left(A - \frac{1}{a_{max}^{2}}\right)4\alpha\right)}\right)\frac{B}{2A}
         - \left(-2 + \frac{\mu a_{max}^{-2}}{\left(1+\left(A - \frac{1}{a_{max}^{2}}\right)4\alpha\right)}\right) + 3 Ba_{max}^{2}\right], & \mbox{\small {\bf for {$k=-1$}}}.
          \end{array}
\right.
\end{array}\ee
and now the expressions for $A$ and $B$ for different values of curvature parameter $k=+1,0,-1$ are given by:
\be\begin{array}{lll}\label{gb3}
 \displaystyle A =\left\{\begin{array}{lll}
                    \displaystyle  
                    -\frac{1}{a_{max}^{2}} + {1 \over 4 \alpha} \left(\sqrt{-1 + {\alpha \mu \over a_{max}^4} + \frac43 \alpha
\Lambda} - 1 \right) \,, &
 \mbox{\small {\bf for {$k=+1$}}}  \\ \\
 \displaystyle  
 \frac{h(a)}{a_{max}^{2}} \,, &
 \mbox{\small {\bf for {$k=0$}}}  \\ \\
         \displaystyle  
          \frac{1}{a_{max}^{2}} + {1 \over 4 \alpha} \left(\sqrt{-1 + {\alpha \mu \over a_{max}^4} + \frac43 \alpha
\Lambda} - 1 \right) \,, & \mbox{\small {\bf for {$k=-1$}}}.
          \end{array}
\right.
\end{array}\ee
and 
\be\begin{array}{lll}\label{gb3v}
 \displaystyle B =\left\{\begin{array}{lll}
                    \displaystyle  
                    {3 \over
8 \alpha} +\frac{3}{2a_{max}^{2}} + \frac{A}{2} \,, &
 \mbox{\small {\bf for {$k=+1$}}}  \\ \\
 \displaystyle  
 \frac{3}{8\alpha} + \frac{h(a_{max})}{2a_{max}^{2}} = {3 \over 8 \alpha} + \frac{A}{2} \,, &
 \mbox{\small {\bf for {$k=0$}}}  \\ \\
         \displaystyle  
         {3 \over
8 \alpha} -\frac{3}{2a_{max}^{2}} + \frac{A}{2} \,, & \mbox{\small {\bf for {$k=-1$}}}.
          \end{array}
\right.
\end{array}\ee
Therefore the conclusions also remain same as for bounce case.

In \cite{Maeda:2007cb}, the authors have shown pictorially that re-collapse occurs if $0<B<2A$ for $k = 0$.
We can also do similar pictorial study for the
case when $k \neq 0$ and find similar conditions on $A, B$ and $\sigma$ which gives re-collapse.

\subsection{Condition for deceleration}

{ \bf A. Space-like extra dimension with $\varepsilon=1$}\\
\\ 
In this subsection the analysis again remains the same giving in place of Eq.~(\ref{gbaccel}), the governing constraint condition is given by:
\begin{equation}
(\rho + p)(\rho + \sigma) > \frac{1}{3C'^{2}}(H^{2}Y - Z),
\label{gbdecel}
\end{equation}
and in place of Eq.~(\ref{accelbounce}) and Eq.~(\ref{accelphi}), we get:
\bea
\label{decelbounce1}p_{t} &>& \frac{\sqrt{A}B}{C'}\left(-\frac{Z}{3AB^{2}} - 1 +  \frac{\sigma C'}{\sqrt{A}B}\right),\\
\frac{\dot{\phi}^{4}}{2} &>& \frac{1}{3C'^{2}}(H^{2}Y - Z) - \frac{1}{2}\dot{\phi}^{2}(V(\phi) + \sigma).
\label{decelphi}
\eea
Thus we again see that this condition now depends on the fourth power of the time derivative of the scalar field as mentioned earlier.
\\ \\
{\bf B. Time-like extra dimension with $\varepsilon=-1$}\\
\\ \\
In the presesnt context the analysis again remains the same as the final governing equations are exactly same as Eq.~(\ref{gbdecel}), Eq.~(\ref{decelbounce1}) and Eq.~(\ref{decelphi})
in which only the signatures of $Y, Z, A, B$ changes accordingly due to $\varepsilon=-1$.

\begin{figure*}[htb]
\centering
\subfigure[ An illustration of the turnaround condition for a universe with $k=0,\ A=1M_{p}^{2},\ B=10M_{p}^{2},\ \sigma=-10^{-9}M_{p}^{4},\ C=200$.]{
    \includegraphics[width=7.2cm,height=8cm] {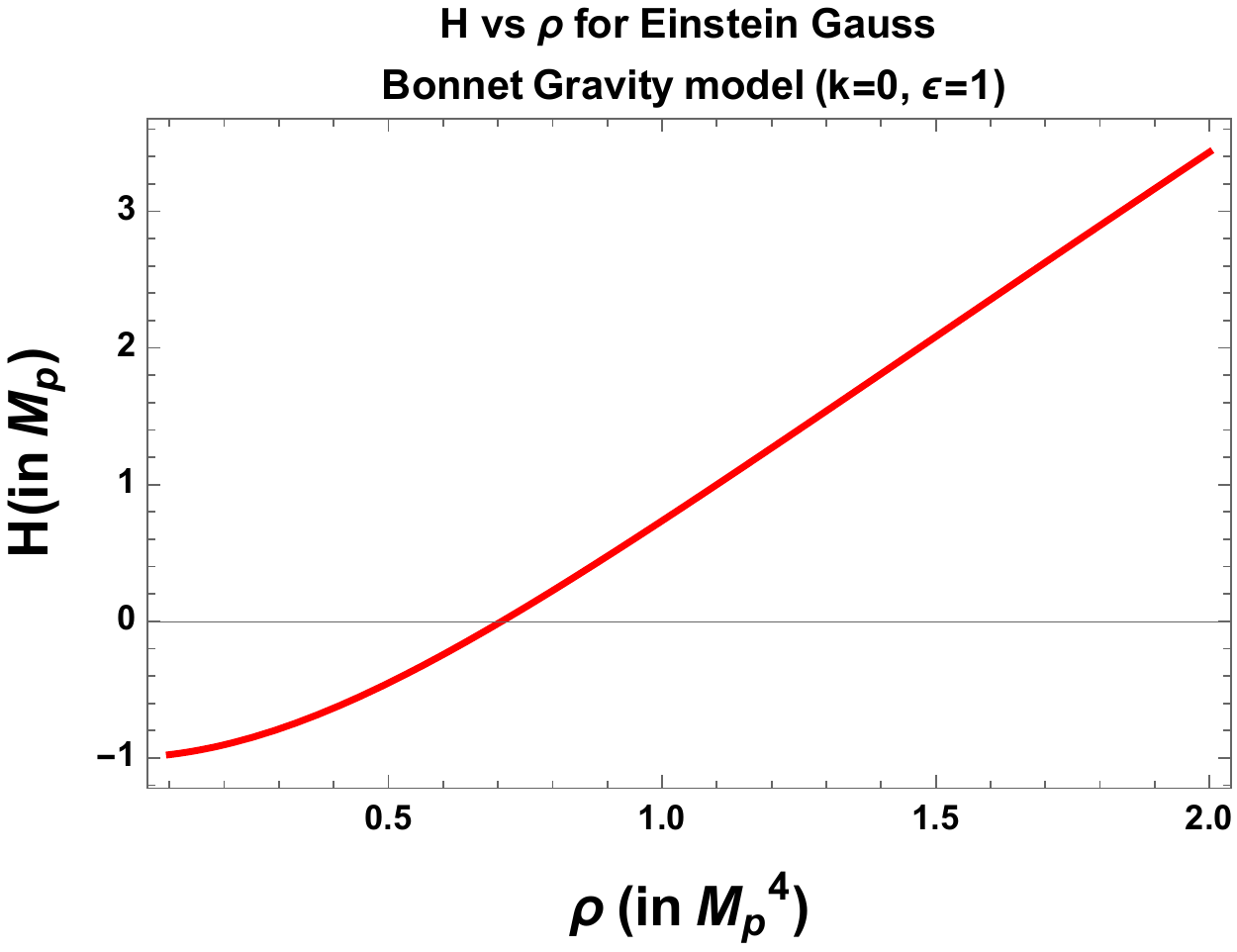}
    \label{egb5}
}
\subfigure[ An illustration of the turnaround condition for a universe with an equation of state $w=1,\  k=1,\ \varepsilon=1,\ \sigma=-10^{-9}M_{p}^{4},\ C=200$.]{
    \includegraphics[width=7.2cm,height=8cm] {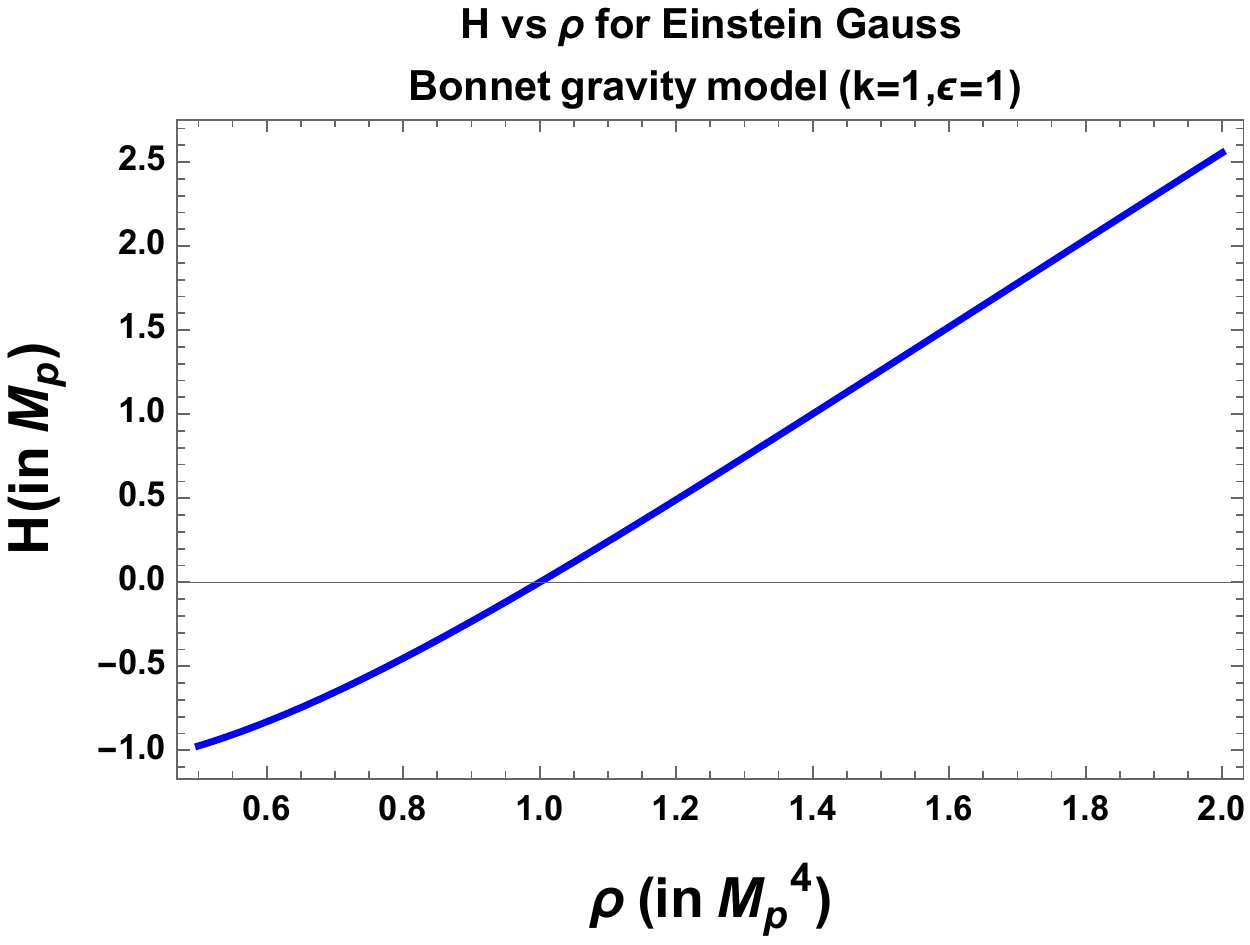}
    \label{egb6}
}
\subfigure[An illustration of the deceleration condition at turnaround for a universe with an equation of state $w=1,\ k=0,\ A=10,\ B=15,\ \sigma=10^{-9}M_{p}^{4},\ C'=10$.]{
    \includegraphics[width=11.2cm,height=8.2cm] {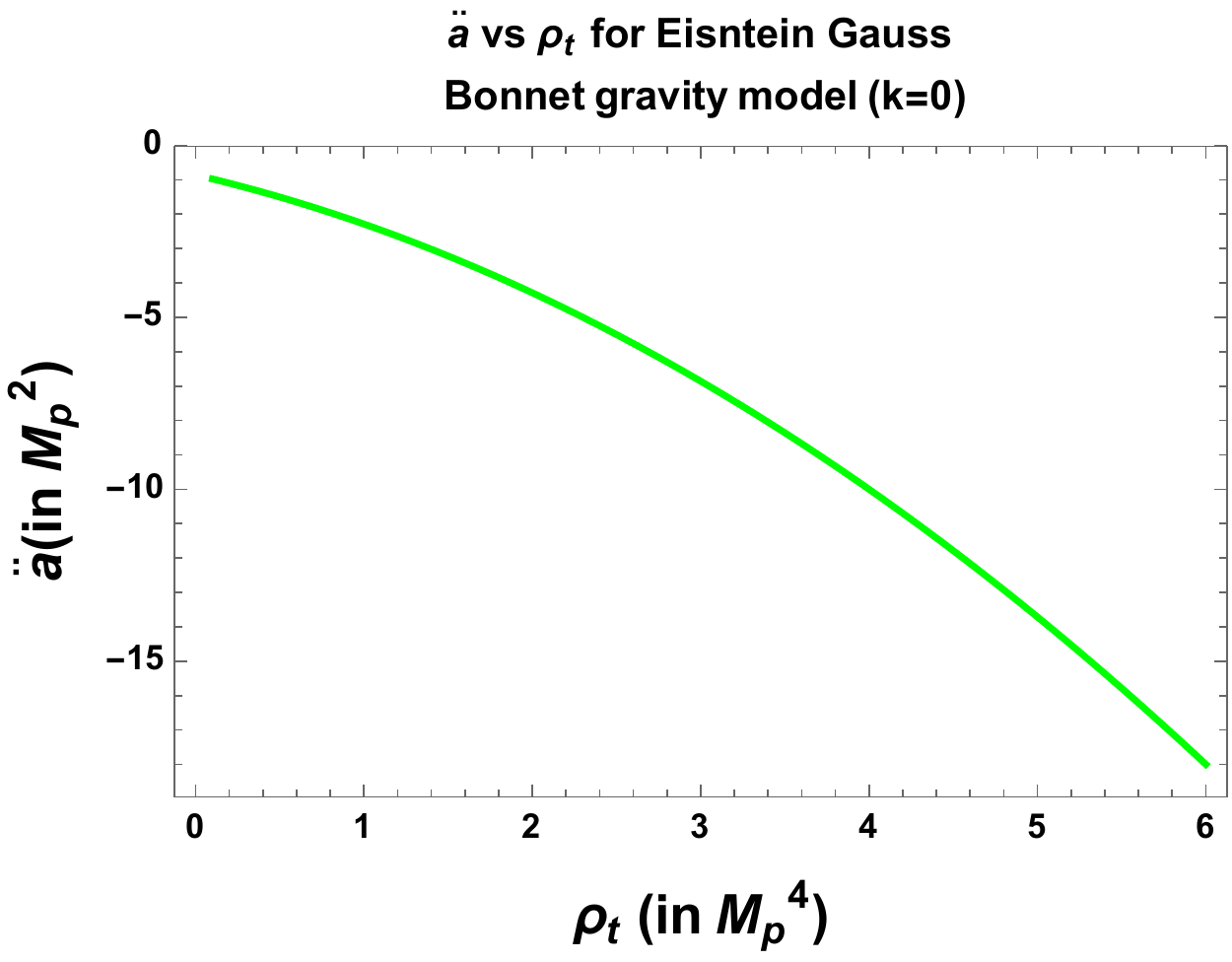}
    \label{egb66}
}
\caption[Optional caption for list of figures]{ Graphical representation of the phenomena of turnaround and deceleration for Einstein Gauss Bonnet gravity model.} 
\label{fig37}
\end{figure*}

In Fig. \ref{fig37}, we have shown the phenomena of turnaround and deceleration in Einstein Gauss Bonnet gravity model. We can draw the following conclusions from the above figures:
\begin{itemize}
\item Fig. \ref{egb5}, has been plotted with the help of Eqns. (\ref{GB-F}), (\ref{horizon-2}), (\ref{plus1}) and (\ref{ABC}) using the relation $\rho=a^{-3(1+w)}$ with $k=0,\ A=1,\ B=10,\ \sigma=-10^{-9}M_{p}^{4},\ C=200$. Detail graphical analysis show that the condition of turnaround is achieved only when $A>0,\ B>0,\ \sigma<0$ or is small positive. For further detail analysis of this case, one may refer to \cite{Maeda:2007cb}.  
\item From Fig. \ref{egb5}, we get the turnaround at $\rho=\rho_{t}=0.7M_{p}^{4}$.
\item Fig. \ref{egb6} shows turnaround using Eqns. (Eqns. (\ref{GB-F}), (\ref{horizon-2}), (\ref{plus1}) and (\ref{ABC})) with $w=1,\  k=1,\ \varepsilon=1,\ \sigma=-10^{-9}M_{p}^{4},\ C=200$. For this case we have used $w=1$, since we require a stiff equation of state for causing contraction and deceleration. The parameter space causing turnaround remains same as that for bounce case. From Fig. \ref{egb6}, we get the turnaround at $\rho=\rho_{t}=1.0M_{p}^{4}$.
\item Fig. \ref{egb66}, we have shown the necessary condition of deceleration ($\ddot{a}< 0$) at the time of turnaround for $w=1,\ k=0,\ A=10,\ B=15,\ \sigma=1M_{p}^{4},\ C'=10$. This plot has been obtained with the help of Eqn. (\ref{gbacceleration}). But since Eqn. (\ref{gbacceleration}) is highly complicated, in order to know the parameter space for which we get deceleration at turnaround, we have instead set $H^{2}=0$ and $\rho=\rho_{b}$ in Eqn. (\ref{gbacceleration}) and analysed for what values of the parameters do we get deceleration at turnaround (similar to the analysis done for bounce). From detail graphical analysis, we have found that this possible if the condition $4AB+2B^{2}>0$ is satisfied for ($\rho+\sigma$)$>0$, and vice versa. This is possible for both positive and negative values of $A,\ B,\ \sigma$ provided the above conditions are satisfied. Similar analysis can be done for $k=1,-1$ also. The results remains nearly same. 
\item Thus from Fig. \ref{fig37} we can conclude that for this model, turnaround is possible for closed, open and flat universe.
\end{itemize}
\subsection{C. Evaluation of work done for one cycle}
The expression for the total work done is same as that given by Eq.~ (\ref{hyst1}) in DGP model.
But we can also express it in terms of the scale factor by using Eq.~ (\ref{gbacceleration}) and
Eq.~ (\ref{plus1}) in order to get an expression of the pressure in terms of the scale factor,
which we can then substitute into the integral for the work done to get:
\\ \\
{\bf A. Space-like extra dimension with $\varepsilon=1$}
\\ \\
\begin{eqnarray}
\oint pdV &=& \oint 3 \dot{a}\frac{a^{2}}{C'}(A+H^{2})^{1/2}(B+H^{2})\left(\pm \frac{\left(H^{2}Y -
Z - \frac{\ddot{a}}{a}Y\right)}{3(A + H^{2})(B + H^{2})^{2}} \mp 1\right) dt \nonumber \\ && ~~~~~~~+
\oint 3 \dot{a}\frac{a^{2}}{C'}(A+H^{2})^{1/2}(B+H^{2})\left(\frac{\sigma C'}{(A +  H^{2})^{1/2}(B + H^{2})} \right) dt. \nonumber \\
\label{egbworkdone}
\end{eqnarray}
\\ \\
{\bf B. Time-like extra dimension with $\varepsilon=-1$}
\\ \\
\begin{eqnarray}
\oint pdV &=& \oint 3 \dot{a}\frac{a^{2}}{C'}(A-H^{2})^{1/2}(B+H^{2})\left(\pm \frac{\left(H^{2}Y 
- Z - \frac{\ddot{a}}{a}Y\right)}{3(A - H^{2})(B + H^{2})^{2}} \mp 1\right) dt \nonumber \\&& ~~~~~~~+
\oint 3 \dot{a}\frac{a^{2}}{C'}(A-H^{2})^{1/2}(B+H^{2})\left(\frac{\sigma C'}{(A - H^{2})^{1/2}(B + H^{2})} \right) dt. \nonumber \\
\end{eqnarray}
Thus the work done now depends not only on the scale factor, but also on the different parameters of the model like the coupling constant, brane tension etc within the present setup.
\\
\subsection{RSII limit}
It has been shown in \cite{Shtanov:2002mb}, that in the limit, $\alpha\rightarrow 0$, Eq.~(\ref{GB-F}) reduces to RSII scenario. 
Then all the results shown in \cite{Sahni:2012er}, for the case when all the other parameters are made zero and the extra dimension
is time-like, will also hold true for the above model. These results have been shown once again in the appendix for the sake of clarity.

\subsection{Semi-analytical analysis for cosmological potentials}
In the analysis below, we will denote Planck mass by $M_{p}$.
Here we will perform the analysis for $k=0$ case. One can repeat the analysis for $k=\pm 1$ as well.

\subsubsection{Case I: Hilltop potential} 

\textbf{A. Expansion}
\\ \\
{\bf A1. Space-like extradimension ($\varepsilon=+1$)}
\\ \\
Using Eq.~(\ref{modeqn1}) and Eq.~(\ref{plus1}) for $\varepsilon=+1$ and using the explicit form of $\rho$, which is specified by the hilltop potential as mentioned earlier, we get an integral equation of the following form:
\be\begin{array}{lll}
\displaystyle\int d\left(\frac{\phi}{M_{p}}\right)\frac{\sqrt{\left[-\left(\sqrt{A}+\frac{1}{2}\frac{B}{\sqrt{A}}\right)2\sqrt{A}
+\sqrt{A}\sqrt{\left\{R\pm 2\sqrt{\frac{C}{A}}V_{0}\left(1+\beta\left(\frac{\phi}{M_{p}}\right)^{p}\right)\right\}}\right]}}{\sqrt{2}\beta p \left(\frac{\phi}{M_{p}}\right)^{p-1}}\\ \displaystyle~~~~~~~~~~~~~~~~~~~~~
= -\frac{V_{0}}{3M_{p}^{2}}\int dt,
\end{array}\ee
where
\begin{equation}
R= A+\frac{B^{2}}{4A}-3B\pm 2\sqrt{\frac{C}{A}}\sigma
\end{equation}
While arriving at the above integral, we have considered that $\mu$ is small
enough such that the term $\alpha\mu/a^{4}$ can be neglected and
$\alpha$ should be small such that $A$ is a large number hence the following constraint holds good:
\be \frac{H^{2}}{A}<<1.\ee
By neglecting $\mu$, the quantities A and B becomes independent of scale factor,
hence allowing us to attain some analytical approximate solutions. To compute the left hand side of the above integral equation, we will again use the following 
redefinition of the field:
\be\frac{\phi}{M_{p}}=e^{\lambda}.\ee
Analytical solution was possible only for the case in the small filed limiting case i.e. when $\phi/M_{p}<<1$ which has been discussed below.

In this limit Expanding the exponentials upto linear order and applying two conditions:
\bea \frac{\sqrt{C}\beta V_{0}\left(\frac{\phi}{M_{p}}\right)^{p}}{\left(A+\frac{B}{2A}-3B\pm 4\sqrt{\frac{C}{A}}\sigma+4\sqrt{\frac{C}{A}}V_{0}\right)}&<<&1,\\
\frac{\sqrt{C}\beta V_{0}}{\left(A+\frac{B}{2A}-3B\pm 4\sqrt{\frac{C}{A}}\sigma+4\sqrt{\frac{C}{A}}V_{0}-\left(A+\frac{B}{2}\right)\right)}&<<&1,\eea we get the solution for $\lambda$ as:
\begin{equation}
\lambda=\frac{-\frac{V_{0}p\beta\sqrt{2}}{3M_{p}^{2}}t+F_{0}}{(\sqrt{R+2\sqrt{C}V_{0}}-2(A+\frac{B}{2}))^{1/2}S}
\label{potential39}
\end{equation}
where we introduce a new constant $F_{0}$,  which is defined in terms of the model parameters as:
\begin{equation}
F_{0}=\lambda_{i}(\sqrt{R+4\sqrt{\frac{C}{A}}V_{0}}-(A+\frac{B}{2}))^{1/2}S+\frac{V_{0}p}{3M_{p}^{2}}t_{i},
\end{equation}
where $S$ is given by:
\begin{equation}
S=1\pm \frac{1}{2}\left(\frac{C}{\left(R+2\sqrt{C}V_{0}\right)\left(\sqrt{R+2\sqrt{C}V_{0}}-2\left(A+\frac{B}{2}\right)\right)}\right)^{1/2}\beta.
\end{equation}
Here $\lambda_{i}$ is the value at bounce.

Substituting the above expression in the Friedmann equation, we get the expression for scale factor as
\begin{eqnarray}
a(t)&=& F_{1}\exp\left[\left\{\sqrt{R+4\sqrt{\frac{C}{A}}V_{0}}\right.\right.\nonumber\\&&\left.\left.~~~~~~~~~~~-(A+\frac{B}{2})\right\}^{1/2}\left(St+\frac{tp(S-1)F_{0}}{
\sqrt{\left(\sqrt{R+4\sqrt{\frac{C}{A}}V_{0}}-\left(A+\frac{B}{2}\right)\right)}S}\right)\right.\nonumber \\&&\left.
 ~~~~~~~~~~~~~~\mp\left\{\sqrt{R+4\sqrt{\frac{C}{A}}V_{0}}\right.\right.\nonumber\\&&\left.\left.~~~~~~~~~~~-(A+\frac{B}{2})\right\}^{1/2}\frac{\frac{t^{2}}{2}p(S-1)\frac{V_{0}p}{3M_{p}^{2}}
 }{\sqrt{\left(\sqrt{R+4\sqrt{\frac{C}{A}}V_{0}}-\left(A+\frac{B}{2}\right)\right)}S}\right],
 \label{scalefactor39}
\end{eqnarray}
where the overall factor $F_{1}$ can be expressed in terms of the model parameters as: 
\begin{eqnarray}
F_{1}&=& a_{i}\exp\left[\left\{\sqrt{R+2\sqrt{C}V_{0}}\right.\right.\nonumber\\&&\left.\left.~~~~~~~~~~~-2(A+\frac{B}{2})\right\}^{1/2}\left(S+\frac{p(S-1)F_{0}}{
\sqrt{\left(\sqrt{R+4\sqrt{\frac{C}{A}}V_{0}}-\left(A+\frac{B}{2}\right)\right)}S}\right)t_{i}\right.\nonumber \\&&\left.
 ~~~~~~~~~~~~~~\pm\left\{\sqrt{R+2\sqrt{C}V_{0}}\right.\right.\nonumber\\&&\left.\left.~~~~~~~~~~~-2(A+\frac{B}{2})\right\}^{1/2}
 \frac{\frac{t_{i}^{2}}{2}p(S-1)\frac{V_{0}p\beta\sqrt{2}}{3M_{p}^{2}}}{\sqrt{\left(\sqrt{R+4\sqrt{\frac{C}{A}}V_{0}}-\left(A+\frac{B}{2}\right)\right)}S}\right].
\end{eqnarray}
Here $a_{i}$ is the value of the scale factor at the time of bounce. Thus we see that the scale factor varies exponentially with time in the small field limit during expansion. 
\\ \\
{\bf A2. Time-like extra dimension}
\\ \\
The results remain same as for space-like extra dimension with the expression for R now given by:
\begin{equation}
R= A+\frac{B^{2}}{4A}+3B\pm 2\sqrt{\frac{C}{A}}\sigma.
\end{equation}
\\ \\
\textbf{B. Contraction}
\\ \\
Following the same procedure as for the expansion phase, but with the expression for density is now given by $\dot{\phi}^{2}/2$, we get the following solutions:
\bea
\dot{\phi}&=&\frac{1+\left[1-\left\{\left(-2\left(A+\frac{B}{2}\right)+\sqrt{AR}\right)(-t+F_{2})+1\right\}\frac{1}{\sqrt{R}\left(-2\left(A+\frac{B}{2}\right)+\sqrt{AR}\right)}\right]}{\frac{1}{\sqrt{R}
\left(-2\left(A+\frac{B}{2}\right)+\sqrt{AR}\right)}},~~~~~~\\
\phi(t)&=&2t\sqrt{R}\left(-2\left(A+\frac{B}{2}\right)+\sqrt{AR}\right)-F_{2}\left(-2\left(A+\frac{B}{2}\right)+\sqrt{AR}\right)t\nonumber \\ &&~~~~~~~~~~~~~~~~~~~~~~~~~~~~~~~~~~+\frac{\left(-2\left(A+\frac{B}{2}\right)+\sqrt{AR}\right)t^{2}}{2}+F_{3},
\label{potential40}
\eea
where we introduce two arbitrary integration constants $F_{2}$ and $F_{3}$, which is given in terms of model parameters as: 
\bea
F_{2}&=&\frac{\dot{\phi}_{f}-2\sqrt{R}\left(-2\left(A+\frac{B}{2}\right)+\sqrt{AR}\right)}{2A+B-\sqrt{AR}}-\frac{t_{f}(\sqrt{AR}-B-2A)}{2A+B-\sqrt{AR}},\\
F_{3}&=&\phi_{f}-2t_{f}\sqrt{R}\left(-2\left(A+\frac{B}{2}\right)+\sqrt{AR}\right)+F_{2}\left(-2\left(A+\frac{B}{2}\right)+\sqrt{AR}\right)t_{f}\nonumber\\&&~~~~~~~~~~~~~~~~~~~~~~~~~~~~~~~~~~~~~~~~~~~~~~~~~
+\frac{\left(-2\left(A+\frac{B}{2}\right)+\sqrt{AR}\right)t_{f}^{2}}{2}.
\eea
Further substituting back into the Friedmann equation, we get the following solution for the scale factor as:
\begin{equation}
a(t)=F_{4}\exp\left[\sqrt{\frac{D''}{2}}\left(t-\frac{(-2+D'+D'F_{3}D''-D'D''t)^{3}}{3D'^{2}D''}\right)\right],
\label{scalefactor40}
\end{equation}
where we introduce a arbitrary integration constant $F_{4}$, which is given in terms of model parameters as: 
\begin{equation}
F_{4}=a_{f}\exp\left[-\sqrt{\frac{D''}{2}}\left(t_{f}-\frac{(-2+D'+D'F_{3}D''-D'D''t_{f})^{3}}{3D'^{2}D''}\right)\right].
\end{equation}
Here $D'$ and $D''$ is given by:
\bea
D'&=&\frac{1}{\sqrt{R}(-2(A+\frac{B}{2})+\sqrt{AR})},\\
D''&=&-2(A+\frac{B}{2})+\sqrt{AR},
\eea
and $a_{f}$ is the value of the scale factor at turnaround. Here also the dependence is exponential but includes higher powers of time. 
\\ \\
\textbf{C. Expression for work done}
\\ \\
The expression for work done is obtained by substituting the expressions for scale factor into the Eq.~(\ref{egbworkdone}),
taking care of the assumptions that we have made earlier in the context of Einstein-Gauss-Bonnet brane world model. 
\\ \\
{\bf A1. Space-like extra dimension:}
\\ \\
\begin{eqnarray}
\oint pdV &=& f_{8}(-f_{1}t_{max}^{4}+f_{2}t_{max}^{5}-f_{3}t_{max}^{6}+f_{4}t_{max}^{7}-f_{5}t_{max}^{8} \nonumber \\
&&~~~~~~~~~~~~~~~~~~~+ f_{6}t_{max}^{9}-f_{7}t_{max}^{10}+f_{1}t_{min}^{4}-f_{2}t_{min}^{5}+f_{3}t_{min}^{6}\nonumber \\
&&~~~~~~~~~~~~~~~~~~~ -f_{4}t_{min}^{7}+f_{5}t_{min}^{8}-f_{6}t_{min}^{9}+f_{7}t_{min}^{10}) \nonumber \\
&&~~~~~~~~~~~~~~~~~~~+ f_{9}(Erf[(-f_{10}+f_{11}t_{max})]-Erf[(-f_{10}+f_{11}t_{min})])\nonumber \\
\end{eqnarray}
where $f_{1}....f_{11}$ are the constants whose functional form depends on the parameters as appearing in the expression for the scale factor. Their explicit forms have been given in the appendix.
\\ \\
{\bf B. Time like extra dimensions:}
\\ \\
Since the solutions for the scale factor in the expansion and contraction
phase remain same baring the constants, the final results for work done in this case also remain same.

Thus we see that since the work done after a complete cycle is non-zero,
at least for the approximated expressions for the scale factor which
we have evaluated in this paper. Thus hysteresis phenomenon can explained for
hilltop potential in the small field limit for Einstein Gauss Bonnet brane world gravity.
\\ \\

\subsubsection{Case II: Natural potential}

\textbf{A. Expansion}
\\ \\
Repeating the analysis as we have done for the hilltop potential, we get the expressions for the scalar field and scale factor for the two limiting cases-i) small field limit $\phi/f<<1$
and ii) large field limit $\phi/f>>1$, which we discuss elaborately in the following subsections.
\\ \\ \\
\underline{\bf i) $\phi/f<<1$}
\\ \\
{\bf A1. Space-like extra dimension ($\varepsilon=+1$)}
\\ \\
Taking the small argument approximations of the trigonometric functions, we get the solutions for the scale factor and scalar field as
\bea
a(t)&=&F_{5}\exp\left[\frac{1}{\sqrt{2}}\left(-2\left(A+\frac{B}{2}\right)+\sqrt{U}\left(1\pm \frac{2\sqrt{C}V_{0}}{U}\right)^{1/2}\right)^{1/2}t\right],\\
\frac{\phi(t)}{f}&=&F_{6}\exp\left[\frac{V_{0}t}{\frac{3f^{2}}{\sqrt{2}}\left(-2\left(A+\frac{B}{2}\right)+\sqrt{U}\left(1\pm \frac{2\sqrt{C}V_{0}}{U}\right)^{1/2}\right)^{1/2}}\right],
\label{scalefactor41}
\eea
where we introduce three arbitrary integration constants $F_{5}$, $F_{6}$ and $U$, which are given in terms of model parameters as:
\bea
F_{5}&=&a_{i}\exp\left[-\frac{1}{\sqrt{2}}\left(-2(A+\frac{B}{2})+\sqrt{U}\left(1\pm \frac{2\sqrt{C}V_{0}}{U}\right)^{1/2}\right)^{1/2}t_{i}\right],\\
F_{6}&=&\frac{\phi_{i}}{f}\exp\left[-\frac{V_{0}t_{i} }{\frac{3f^{2}}{\sqrt{2}}\left(-2(A+\frac{B}{2})+\sqrt{U}\left(1\pm \frac{2\sqrt{C}V_{0}}{U}\right)^{1/2}\right)^{1/2}}\right],\\
U&=&\sqrt{A}R\pm 2\sqrt{C}V_{0},
\eea
with the definition of $R$ is same as appearing in the previous case i.e. for hilltop potential.

Thus the scale factor depends on time exponentially with only first power or linear power of time. 
\\ \\ \\
{\bf A2. Time-like extra dimension ($\varepsilon=-1$):} 
\\ \\
The conclusion remain same as was for the hilltop potential.
\\ \\
\underline{\bf ii) $\phi/f>>1$}
\\ \\
{\bf A1. Space-like extra dimension ($\varepsilon=+1$):}
\\ \\
Using the large argument approximations of the trigonometric functions and applying the conditions:
\bea\frac{\sqrt{\frac{U}{W}}}{W}&<<&1,\\ 
W&>>&\sqrt{\frac{C}{WU}},\eea i.e. $W$ is a large quantity where it is defined as:
\begin{equation}
W=\frac{1}{2}\left(\sqrt{U}-\left(\sqrt{A}+\frac{1}{2}\frac{B}{\sqrt{A}}\right)\right),
\end{equation}
and $U$ has already been defined before, we hence get the solution of the scalar field as:
\begin{equation}
\frac{\phi(t)}{f}=2\tan^{-1}\left[\exp\left(\frac{\frac{V_{0}t}{3f^{2}}+F_{7}}{f}\right)\frac{f}{W}\right],
\label{potential42}
\end{equation}
where we introduce three arbitrary integration constant $F_{7}$, which is given in terms of model parameters as:
\begin{equation}
F_{7}=f\ln\left(\frac{W}{f}\tan\left(\frac{\phi_{i}}{2f}\right)\right)-\frac{V_{0}t_{i}}{3f^{2}}
\end{equation}
Substituting back into the approximate Friedmann equation we get the following expression for the scale factor as:
\begin{equation}
a(t)=F_{8}\exp\left[\sqrt{W}\left(\left(1-\frac{1}{W}\sqrt{\frac{C}{U}}\right)t-\frac{\frac{2}{W}\sqrt{\frac{C}{U}}\ln(1+e^{\tilde{G}})}{\left(\frac{2V_{0}}{3f^{2}W}\right)}+\frac{4}{W}\sqrt{\frac{C}{U}}t\right)\right]
\label{scalefactor42}
\end{equation}
where the functions $G$ and $F_{8}$ are given by: 
\bea
\tilde{G}&=&\frac{2F_{7}}{W}+\frac{2V_{0}t}{3f^{2}W},\\
F_{8}&=&a_{i}\exp\left[-\sqrt{W}\left(\left(1-\frac{1}{W}\sqrt{\frac{C}{U}}\right)t_{i}-\frac{\frac{2}{W}\sqrt{\frac{C}{U}}\ln(1+e^{G})}{\frac{2V_{0}}{3f^{2}W}}+\frac{4}{W}\sqrt{\frac{C}{U}}t_{i}\right)\right].~~~~~
\eea
\\ \\
{\bf A2. Time-like extra dimension ($\varepsilon=-1$):}
\\ \\
The conclusion remain same as was for the previous case.
\\ \\
\textbf{B. Contraction}
\\ \\
\\ \\
{\bf A1. Space-like extra dimension ($\varepsilon=+1$):}
\\ \\
Since the contraction phase is independent of any potential, the conclusions remain same as for hilltop potential
\\ \\
{\bf A2. Time-like extra dimensions ($\varepsilon=-1$):}
\\ \\
In this case also the conclusions remain same as for hilltop potential
\\ \\
\textbf{C. Expression for work done}
\\ \\
The expression for work done is obtained by substituting the expressions for scale factor into the Eq.~(\ref{egbworkdone}), taking care of the assumptions that we have made earlier in the present context. 
\\ \\
{\bf A1. Space-like extra dimension ($\varepsilon=+1$):}
\\ \\
\begin{eqnarray}
\oint pdV &=& f_{8}(-f_{1}t_{max}^{4}+f_{2}t_{max}^{5}-f_{3}t_{max}^{6}+f_{4}t_{max}^{7}-f_{5}t_{max}^{8} \nonumber \\
&&~~~~~~~~~~~~~~~~~~~+ f_{6}t_{max}^{9}-f_{7}t_{max}^{10}+f_{1}t_{min}^{4}-f_{2}t_{min}^{5}+f_{3}t_{min}^{6}\nonumber \\
&&~~~~~~~~~~~~~~~~~~~ -f_{4}t_{min}^{7}+f_{5}t_{min}^{8}-f_{6}t_{min}^{9}+f_{7}t_{min}^{10}) \nonumber \\ &&~~~~~~~~~~~~~~~~~~~~~~+f_{12}(e^{f_{13}t_{min}}-e^{f_{13}t_{max}}),
\end{eqnarray}
where $f_{1}....f_{13}$ are the constants whose form depends on the parameters in the expression for the scale factor. Explicit forms of the constants are given in the appendix.
\\ \\
{\bf A2. Time-like extra dimensions ($\varepsilon=-1$):}
\\ \\
Since the solutions for the scale factor in the expansion and contraction phase remains same baring the constants, the results for work done also remains same.

Thus we see that since the work done after a complete cycle is non-zero, at least for
the approximate expressions for the scale factor which we have evaluated in this subsection earlier.
Thus hysteresis phenomenon is possible for the natural potential
in the small field limit for Einstein Gauss Bonnet gravity.
\\ \\

\subsubsection{Case III:  Coleman-Weinberg potential}

\textbf{A. Expansion}
\\ \\
{\bf A1. Space-like extra dimension ($\varepsilon=+1$):}\\ \\
The original integral equation being complicated, we follow the same procedure
as has been done before i.e. use the field redefinition:
\be \phi/M_{p}=e^{\lambda}.\ee
Analytical solutions were possible only for small field limit $\phi/M_{p}<<1$ which we have discussed below.

In this limit expanding the exponentials upto linear order and applying the following constraint 
conditions: \bea 
\frac{2\sqrt{C}V_{0}(4\alpha+\beta)\lambda}{(\sqrt{U}+2\sqrt{C}V_{0}\alpha)}&<<&1,\\
\frac{2\sqrt{C}V_{0}(4\alpha+\beta)}{\sqrt{H'}\left(\sqrt{AH'}-2\left(A+\frac{B}{2}\right)\right)}&<<&1,\\
\frac{4\beta\lambda}{(4\alpha+\beta)}&<<&1,\eea we get the solution for redefined or transformed field $\lambda$ as:
\begin{equation}
\lambda=(4\alpha+\beta)\left(-\frac{V_{0}t}{3M_{p}^{2}}+F_{9}\right)\left(\frac{2}{\sqrt{AH'}-2\left(A+\frac{B}{2}\right)}\right)^{1/2},
\label{potential43}
\end{equation}
where we introduce two new constants $H'$ and $F_{9}$ defined in terms of the model parameters as:
\bea
H'&=&\sqrt{U}+2\sqrt{C}V_{0}\alpha,\\
F_{9}&=&\frac{\lambda_{i}}{(4\alpha+\beta)\left(\frac{2}{\sqrt{AH'}-2\left(A+\frac{B}{2}\right)}\right)^{1/2}}-\frac{V_{0}t_{i}}{3M_{p}^{2}}.
\eea
Here $\lambda_{i}$ is the value at bounce.

The solution for the scale factor turns out to be:
\begin{equation}
a(t)=F_{10}\exp\left[\left\{\left(\sqrt{AH'}-2\left(A+\frac{B}{2}\right)\right)^{1/2}+F_{9}I(4\alpha+\beta)\right\}t-\frac{(4\alpha+\beta)t^{2}\frac{V_{0}}{3M_{p}^{2}}}{2}\right]
\label{scalefactor43}
\end{equation}
where we introduce two new constants $I$ and $F_{10}$ defined in terms of the model parameters as:
\bea
I&=&\frac{\sqrt{2}V_{0}(4\alpha+\beta)}{\sqrt{H'}\left(\sqrt{AH'}-2\left(A+\frac{B}{2}\right)\right)},\\
F_{10}&=&a_{i}\exp\left[-\left\{\left(\sqrt{AH'}-2\left(A+\frac{B}{2}\right)\right)^{1/2}+F_{9}I(4\alpha+\beta)\right\}t_{i}+\frac{(4\alpha+\beta)t_{i}^{2}\frac{V_{0}}{3M_{p}^{2}}}{2}\right].\nonumber\\
\eea
\\ \\
{\bf A2. Time-like extra dimension ($\varepsilon=-1$):}
\\ \\
The conclusions remain same as for $\varepsilon=1$ case, only the expression for different parameters change with $\varepsilon=1$ replaced by $\varepsilon=-1$.\\ \\
\textbf{B. Contraction}\\ \\
{\bf A1. Space-like extra dimension ($\varepsilon=+1$):}\\ \\
Conclusions for hilltop potential, holds true for this case also.\\ \\
{\bf A2. Time-like extra dimension ($\varepsilon=-1$):}\\ \\
Conclusions remain same as for the case of hilltop potential.
\\ \\
\textbf{C. Expression for work done}
\\ \\
{\bf A1. Space-like extra dimension ($\varepsilon=+1$):}\\ \\
The general expression for the scale factor i.e its time dependence is same for
supergravity potential as for hilltop potential in the small field limit, hence
the conclusions for work done remains valid for this case also, thus giving rise
to the phenomenon of hysteresis.\\ \\
{\bf A2. Time-like extra dimension ($\varepsilon=-1$):}\\ \\
The conclusion is same as above.\\ \\

\subsection{Graphical Analysis}

\subsubsection{Case I: Hilltop potential}
All the graphs in this section and in the following sections have been plotted in units of $M_{p}=1,\ H_{0}=1,\ c=1$, where $M_{p}$ is the Planck mass, $H_{0}$ is the present value of the Hubble parameter and $c$ is the speed of light.
\begin{figure*}[htb]
\centering
\subfigure[ An illustration of the behavior of the scale factor with time  during expansion phase for $\phi<<M_{p}$ with $V_{0}=10^{-8}M_{p}^{4},\ p=3,\ F_{1}=1,\ F_{0}=0.1M_{p},\ A=0.1M_{p}^{2},\ B=0.1M_{p}^{2},\ C=10,\ S=0.1,\ R=0.1M_{p}^{2}$.]{
    \includegraphics[width=7.2cm,height=7.5cm] {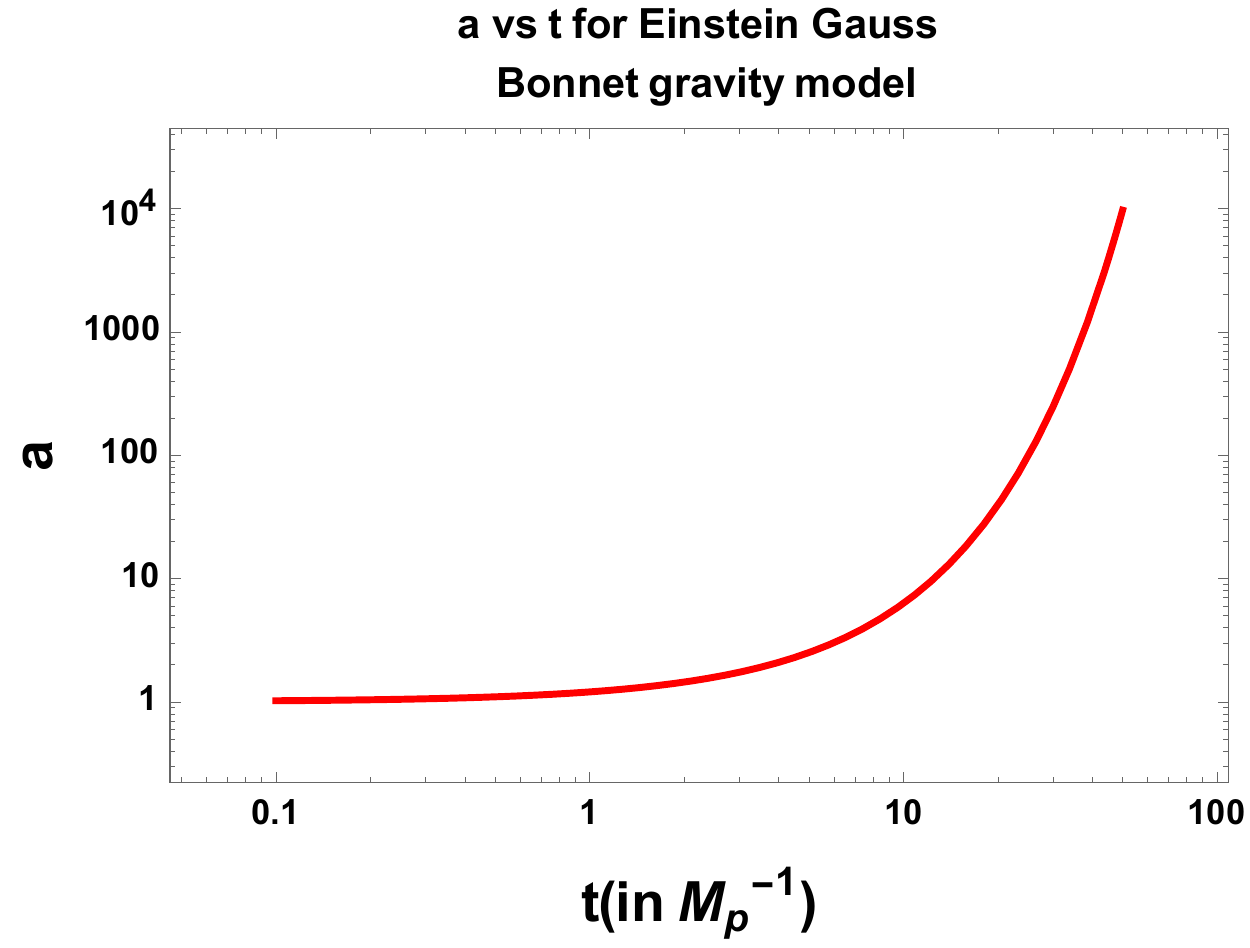}
    \label{egb7}
}
\subfigure[An illustration of the behavior of the potential during expansion phase for $\phi<<M_{p}$ with $V_{0}=2.7{\rm x}10^{-3}M_{p}^{4},\ p=4,\ F_{0}=54M_{p},\ A=0.1M_{p}^{2},\ B=0.1M_{p}^{2}, C=500,\ R=30M_{p}^{2},\ S=0.3,\ \beta=0.05$ .]{
    \includegraphics[width=7.2cm,height=7.5cm] {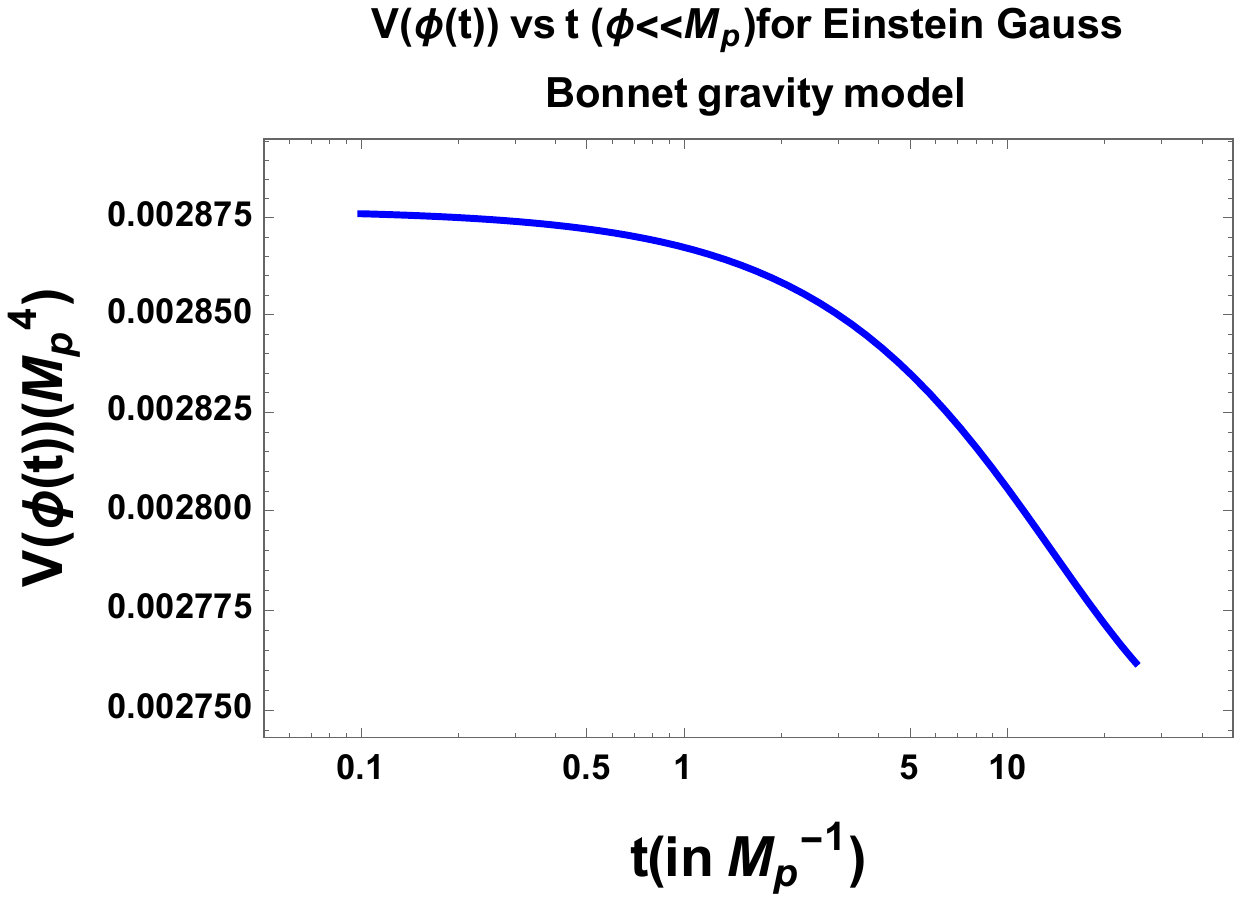}
    \label{egb8}
}
\subfigure[An illustration of the behavior of the scale factor with time  during contraction phase with $A=10M_{p}^{2},\ B=20M_{p}^{2},\ F_{3}=150M_{p},\ F_{4}=1 $.]{
    \includegraphics[width=7.2cm,height=7.5cm] {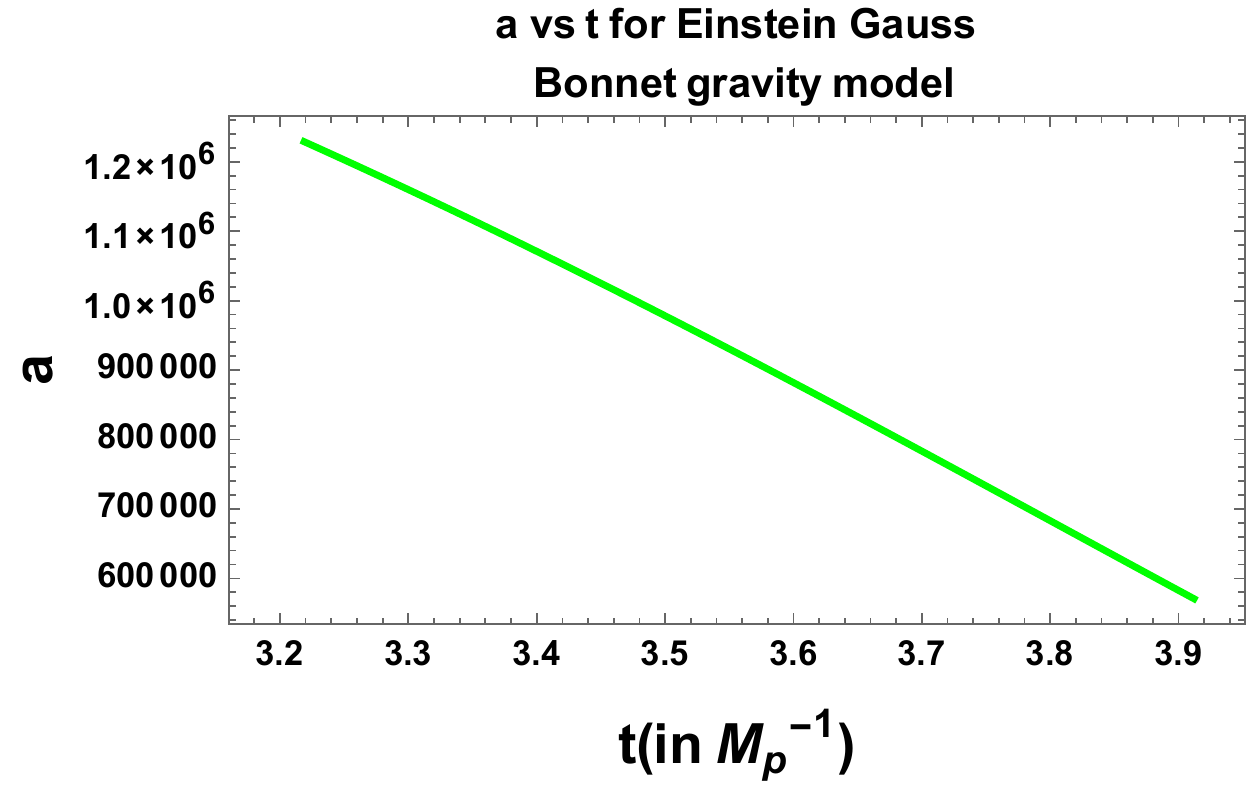}
    \label{egb10}
}
\subfigure[An illustration of the behavior of the potential during contraction phase with $V_{0}=10^{-6}M_{p}^{4},\ p=1,\ F_{2}=24M_{p}^{-1},\ F_{3}=1M_{p},\ A=2M_{p}^{2},\ B=-2M_{p}^{2},\ R=10^{-3}M_{p}^{2},\ \beta=-0.9$ .]{
    \includegraphics[width=7.2cm,height=7.5cm] {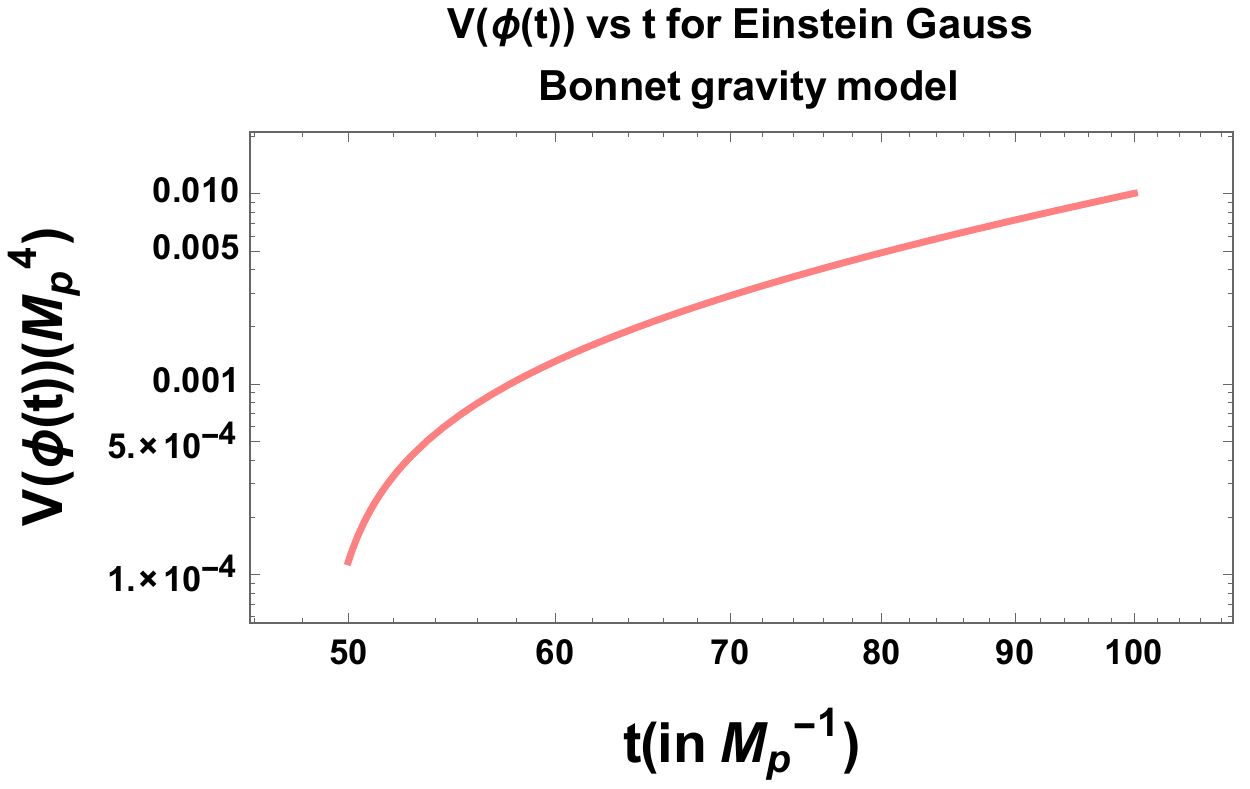}
    \label{egb11}
}    
\caption[Optional caption for list of figures]{ Graphical representation of the evolution of the scale factor and the potential during the expansion and contraction phase for Einstein Gauss Bonnet gravity model.} 
\label{fig40}
\end{figure*}

Fig. \ref{fig40} shows the evolution of scale factor and potential during expansion and contraction phase for Einstein Gauss Bonnet gravity model. We can draw the following conclusions:
\begin{itemize}
\item Fig. \ref{egb7}, shows the plot of the scale factor in the small field limit for hilltop potential given by Eqn. (\ref{scalefactor39}) with $V_{0}=10^{-8}M_{p}^{4},\ p=3,\ F_{1}=1,\ F_{0}=0.1M_{p},\ A=0.1M_{p}^{2},\ B=0.1M_{p}^{2}, C=10,\ S=0.1,\ R=0.1M_{p}^{2}$. 
\item Detail graphical analysis show that the amplitude of expansion increases with increase in the value of $C$, decrease in $A$ and $B$. Expansion is not possible for $A<0$. For $R,S<0$, expansion is possible for large values of other parameters. For $B<0$, expansion is possible if $V_{0}$ and $A$ are small, $R$ and $F$ are negative. In this case, $S$ can be both negative and positive. Again for expansion to occur, the parameter space ($R,F,S<0$), ($A,B>0$) and ($A,B,R>0$), ($S,F<0$) not allowed. But if we look at the expression of $F_{0}$, then $R,A$ cannot take negative values. This will make the expression imaginary, which is unphysical. For $F<0, S>0$ but small, expansion is possible if $p$ lies between $1$ and $3$.
\item Fig. \ref{egb8} shows the plot of the behavior of the potential with time for small field hilltop potential. This graph has been obtained with the help of Eqn. (\ref{potential39}) with parameter values $V_{0}=2.7{\rm x}10^{-3}M_{p}^{4},\ p=4,\ F_{0}=54M_{p},\ A=0.1M_{p}^{2},\ B=0.1M_{p}^{2}, C=500,\ R=30M_{p}^{2},\ S=0.3\ \beta=0.05$. Correct form of potential is possible for both negative and positive values of $\beta$. In this case, the condition \be \sqrt{R+2\sqrt{C}V_{0}}>2(A+B/2)\ee should always be satisfied for getting real outputs which we can conclude from Eqn. (\ref{potential39}). Larger values of $S$ makes the potential flatter for longer time. Larger values of other parameters make the potential fall more linearly and decreases its slope. Both negative and positive values of $B$ are allowed provided the above condition is satisfied. $S<0$ not allowed because we do not get the correct nature of  the potential.
\item In Fig. \ref{egb10}, we have plotted Eqn. (\ref{scalefactor40}), which results in the contraction phase of the universe provided we choose the value of the constants accordingly. This plot has been obtained for $A=10M_{p}^{2},\ B=20M_{p}^{2},\ F_{3}=150M_{p},\ F_{4}=1 $. Output is possible provided \be \sqrt{AR}/2>2(A+B/2),\ee
which we can also conclude from Eqn. (\ref{scalefactor40}). Contraction is possible provided \be B,R>A \ee and \be A,R,B>0.\ee Higher values of the parameters increases the amplitude of expansion. If we compare Fig. \ref{egb7} with Fig. \ref{egb10}, we find that there occurs a net increase in the amplitude of the scale factor after one expansion-contraction cycle.
\item Fig. \ref{egb11} shows the plot of the potential during the contraction phase given by Eqn. (\ref{potential40}). This plot has been obtained for $V_{0}=10^{-6}M_{p}^{4},\ p=1,\ F_{2}=24M_{p}^{-1},\ F_{3}=1M_{p},\ A=2M_{p}^{2},\ B=-2M_{p}^{2},\ R=10^{-3}M_{p}^{2},\ \beta=-0.9$. Larger values of the integration constants $F_{2}$ and $F_{3}$ makes the rise more nonlinear. For $\beta>0$ rise of the potential is possible only for very large values of $F_{2}$.  Larger values of other parameters increases the amplitude of the potential.

\end{itemize}

\subsubsection{Case II: Natural potential}
All the graphs in this section and in the following sections have been plotted in units of $M_{p}=1,\ H_{0}=1,\ c=1$, where $M_{p}$ is the Planck mass, $H_{0}$ is the present value of the Hubble parameter and $c$ is the speed of light.
\begin{figure*}[htb]
\centering
\subfigure[ An illustration of the behavior of the scale factor with time  during expansion phase for $\phi<<f$ with $V_{0}=8{\rm x}10^{-4}M_{p}^{4},\ F_{5}=1,\ A=0.8M_{p}^{2},\ B=4.2M_{p}^{2},\ C=1,\ R=40.6M_{p}^{2}$.]{
    \includegraphics[width=7.2cm,height=8.2cm] {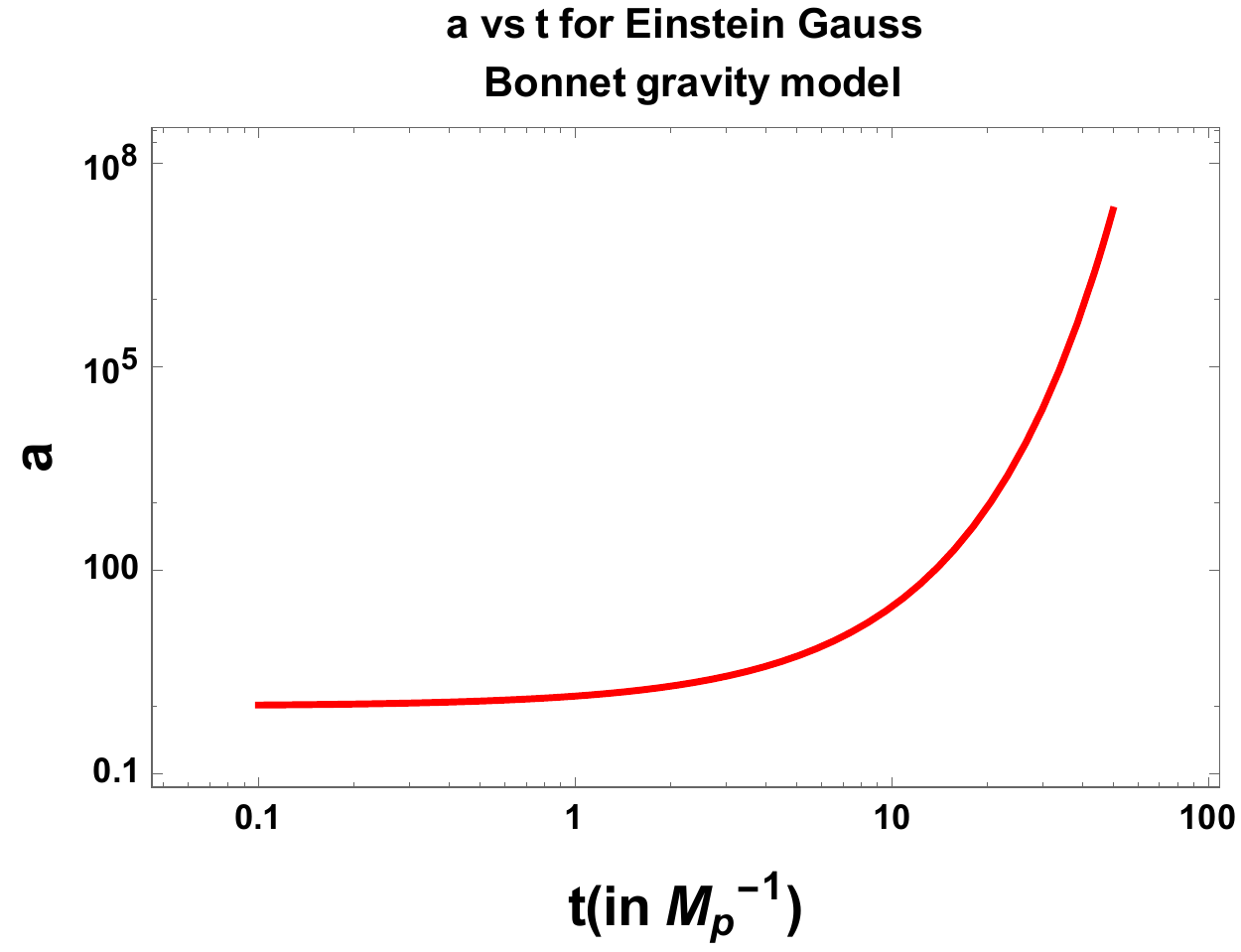}
    \label{egb13}
}
\subfigure[An illustration of the behavior of the potential during expansion phase for $\phi<<f$ with $V_{0}=8{\rm x}10^{-4}M_{p}^{4},\ F_{6}=1,\ A=0.1M_{p}^{2},\ B=0.1M_{p}^{2}, C=10,\ R=40.6M_{p}^{2},\ f=0.1M_{p}$ .]{
    \includegraphics[width=7.2cm,height=8.2cm] {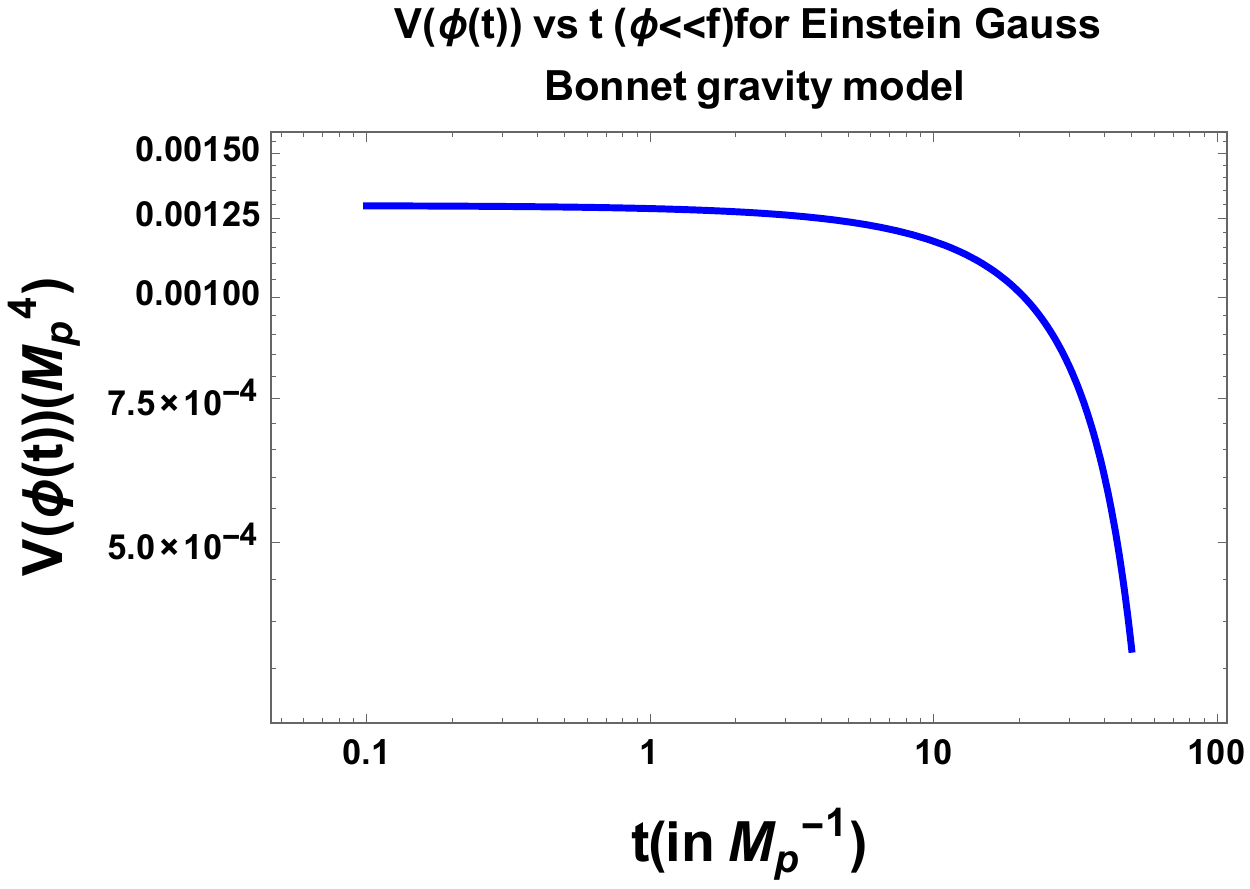}
    \label{egb14}
}
\caption[Optional caption for list of figures]{ Graphical representation of the evolution of the scale factor and the potential during the expansion for Einstein Gauss Bonnet gravity model.} 
\label{fig41}
\end{figure*}

Fig. \ref{fig41} shows the evolution of scale factor and potential during expansion phase for Einstein Gauss Bonnet gravity model. We can draw the following conclusions:
\begin{itemize}
\item Fig. \ref{egb13}, shows the plot of the scale factor in the small field limit for natural potential given by Eqn. (\ref{scalefactor41}) with $V_{0}=8{\rm x}10^{-4}M_{p}^{4},\ F_{5}=1,\ A=0.8M_{p}^{2},\ B=4.2M_{p}^{2},\ C=1,\ R=40.6M_{p}^{2}$. 
\item Detail graphical analysis show that the amplitude of expansion increases with increase in the value of $R$. From Eqn. (\ref{scalefactor41}), we can conclude that $R,A<0$ not possible. Expansion is possible for $B<0$. But large negative values of $B$ expansion is not possible. The nature of the graph is almost independent of the value of $C$ and $V_{0}$.
\item Fig. \ref{egb14} shows the plot of the behavior of the potential with time for small field natural potential. This graph has been obtained with the help of Eqn. (\ref{scalefactor41}) with parameter values $V_{0}=8{\rm x}10^{-4}M_{p}^{4},\ F_{6}=1,\ A=0.1M_{p}^{2},\ B=0.1M_{p}^{2}, C=10,\ R=40.6M_{p}^{2},\ f=0.1M_{p}$. The conclusions regarding the allowed parameter space for the expansion of the scale factor holds true for this case also. Very large values of $V_{0}$ and $A$ are not allowed since they cause uneven oscillations in the potential.
\item For large field case given by Eqns. (\ref{scalefactor42}) and (\ref{potential42}), the behavior is similar, hence have not been shown here explicitly.
\item If we compare Fig. \ref{egb13} with Fig. \ref{egb10}, we find that there occurs a net increase in the amplitude of the scale factor after one expansion-contraction cycle.
\end{itemize}

\subsubsection{Case III:  Coleman-Weinberg potential}

All the graphs in this section and in the following sections have been plotted in units of $M_{p}=1,\ H_{0}=1,\ c=1$, where $M_{p}$ is the Planck mass, $H_{0}$ is the present value of the Hubble parameter and $c$ is the speed of light.
\begin{figure*}[htb]
\centering
\subfigure[ An illustration of the behavior of the scale factor with time  during expansion phase for $\phi<<M_{p}$ with $V_{0}=10^{-8}M_{p}^{4},\ F_{9}=1M_{p},\ F_{10}=1,\ A=0.1M_{p}^{2},\ B=0.1M_{p}^{2},\ C=1,\ R=4.4M_{p}^{2},\ \beta=1.5,\ \alpha=0.33$.]{
    \includegraphics[width=7.2cm,height=8.2cm] {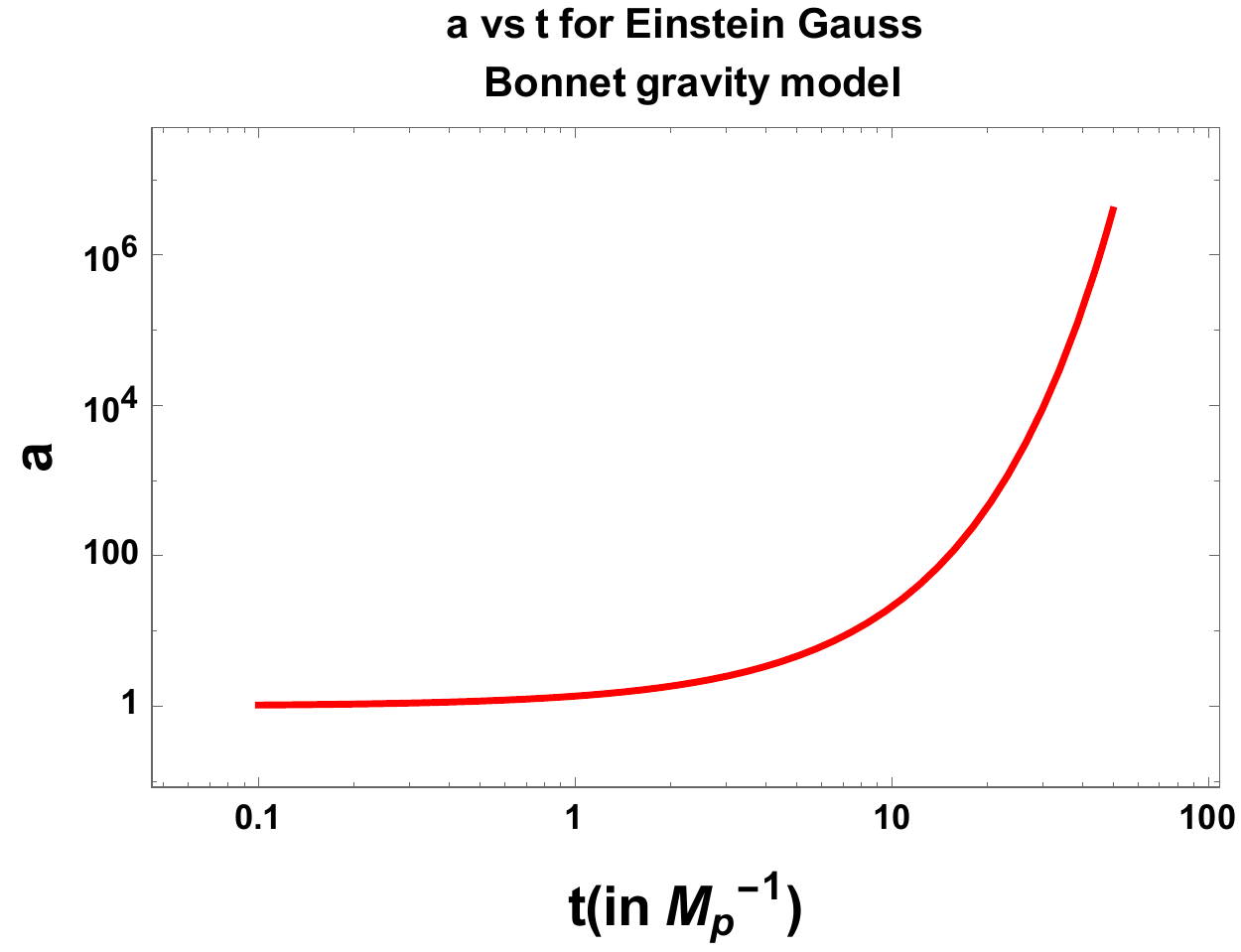}
    \label{egb15}
}
\subfigure[An illustration of the behavior of the potential during expansion phase for $\phi<<M_{p}$ with $V_{0}=1.2{\rm x}10^{-4}M_{p}^{4},\ F_{9}=6.6M_{p},\ A=0.1M_{p}^{2},\ B=0.1M_{p}^{2}, C=10,\ R=20M_{p}^{2},\ \alpha=0.38,\ \beta=0.85$ .]{
    \includegraphics[width=7.2cm,height=8.2cm] {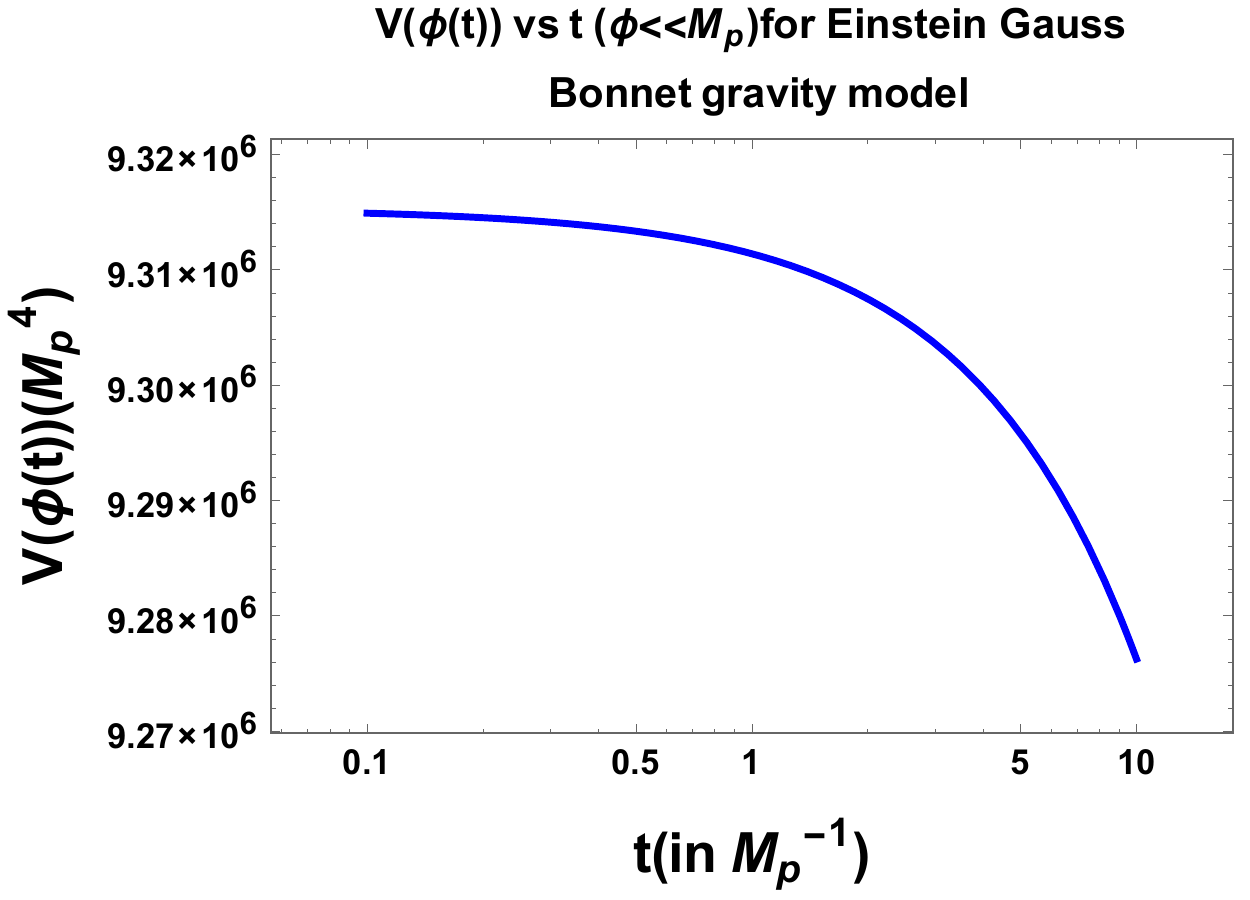}
    \label{egb16}
}
\caption[Optional caption for list of figures]{ Graphical representation of the evolution of the scale factor and the potential during the expansion for Einstein Gauss Bonnet gravity model.} 
\label{fig42}
\end{figure*}

Fig. \ref{fig42} shows the evolution of scale factor and potential during expansion phase for Einstein Gauss Bonnet gravity model. We can draw the following conclusions:
\begin{itemize}
\item Fig. \ref{egb15}, shows the plot of the scale factor in the small field limit for supergravity potential given by Eqn. (\ref{scalefactor43}) with $V_{0}=10^{-8}M_{p}^{4},\ F_{9}=1M_{p},\ F_{10}=1,\ A=0.1M_{p}^{2},\ B=0.1M_{p}^{2},\ C=1,\ R=4.4M_{p}^{2},\ \beta=1.5,\ \alpha=0.33$. 
\item Detail graphical analysis show that the amplitude of expansion increases with increase in the value of $R$. But the expansion is almost independent of $\alpha, \beta, F_{9}$. Negative values of $B$ gives large expansion. For larger values of $R$, $F_{9}$ cannot take large values. Expansion occurs for both negative and positive values of $\beta$. Larger values of $V_{0}$ makes the graph more linear. For $A<0$, expansion is not possible. However, $B$ can take negative values, but both $B$ and $R$ taking negative values not allowed.
\item Fig. \ref{egb16} shows the plot of the behavior of the potential with time for small field supergravity potential. This graph has been obtained with the help of Eqn. (\ref{potential43}) with parameter values $V_{0}=1.2{\rm x}10^{-4}M_{p}^{4},\ F_{9}=6.6M_{p},\ A=0.1M_{p}^{2},\ B=0.1M_{p}^{2}, C=10,\ R=20M_{p}^{2},\ \alpha=0.38,\ \beta=0.85$. The conclusions regarding the allowed parameter space for the expansion of the scale factor holds true for this case also. Very large negative values of $\beta$ are not allowed. Expansion is possible only for small negative and positive values of $\beta$. Other parameter values only increases the amplitude, keeping the nature of the graph unchanged. The plot is almost independent of the value of $C$ and $R$.
\item If we compare Fig. \ref{egb15} with Fig. \ref{egb10}, we find that there occurs a net increase in the amplitude of the scale factor after one expansion-contraction cycle.
\end{itemize}
\section{Conclusion}
\label{sq7}
In this paper, we have further explored the role of hysteresis in making cyclic models
of universe as an alternative proposal to inflationary paradigm. The idea, originally proposed by the authors in refs.~\cite{Kanekar:2001qd,Sahni:2012er},
have been studied by us for a wide variety of cosmological models, especially, higher dimensional gravity setup. The basic analysis
for getting the conditions which lead to an increase in expansion maximum, remains the same as discussed in refs.~\cite{Kanekar:2001qd,Sahni:2012er}.
The most interesting outcome of this analysis is the dependence of the bouncing and turnaround conditions on the various model parameters.
Through this analysis we find that the phenomenon of hysteresis is very robust and is eternal for all of these models. 
We study essentially those effective field theoretic models which can give rise to both the conditions for bounce and turnaround and at the same
time also satisfies the observed features of the present universe. Like in the previous analysis as performed in the ref.~\cite{Sahni:2012er},
here also we find that inflationary conditions are not necessary for causing cosmological hysteresis. Hence the models analyzed in
this paper can also be treated as an alternative prescription to inflation. Most of the results indicate that the value of the scale
factor maximum and minimum after each cycle, depends not only on the signature of the hysteresis loop integral but also on the relative
amplitudes of the model parameters, or in other words, we can fine tune the model parameters and get an amplitude
increase after each cycle even if the signature of the hysteresis loop integral is positive. While doing the analysis we have
not constrained the potential by any particular form. The potential can have any general form like a power series,
oscillatory, etc, but with well defined minimum/minima which is essential for generating the required randomness or
mixing of the field in the phase space ($\dot\phi, \phi$) so that its value during contraction and expansion
are uncorrelated. 

We have also tried to find the nature of dependence of the scale factor on time by directly
solving the equation of motion for the scalar field under different scenarios and varying
forms of the cosmological potential. Exact solutions of the equation of motion for the scalar field being
highly complicated, we applied certain valid and possible approximations while doing the
analysis by redefining the field in terms of other dynamical parameter, in order to get an approximate analytical expression for the dependence of the
scale factor on time. This in turn helps us to directly guess the nature of the behavior of the
scale factor as well as the scalar field during each cycle for different models. We have also derived
an explicit expression for work done per cycle, hence have shown that its value comes out
to be non zero for the cases which we have studied explicitly, provided we choose the parameters of
the models accordingly. It shows that the phenomenon of hysteresis and a cyclic universe
with an ever increasing scale factor, can be produced by a wide variety of models and
potentials. This will also help us to constrain the parameters of the models and the
potentials, such that we get the required results compatible with observations.\\

The future prospects of our work are mentioned below:
\begin{itemize}
 \item  Through this analysis we have seen the beautiful correlation between purely
thermodynamical principle and relativistic models and how the former can be
used for extracting interesting results from the later. But the models that
we have considered are the variants of minimally coupled gravity frameworks. It would therefore
be interesting to investigate what new features arises once we relax this constraint.
Also we can check whether modified gravity models like for variants of $f(R)$ gravity, two brane-world model in presence of the Einsitein-Hillbert term and the 
Einstein-Hillbert-Gauss-Bonnet gravity setup succeeds
in generating a cyclic universe with increasing amplitude of expansion.
 
 \item One
would also like to ask what other observational signatures we can get from
such models which can be tested using CMB.  We have also not yet verified
whether these models produces a universe with the right amount of anisotropy
and inhomogeneity present in the universe. In future we plan to connect these analyses with CMB observations,
by rigorous study of the cosmological perturbation theory in various orders of metric fluctuations
and computation of two point correlations to get the expressions for scalar and tensor power
spectrum in this context. Hence we extend the study of this paper to compute the primordial
non-Gaussianity in CMB from
three and four point correlations. We also plan to derive the explicit
expression for various modified consistency relations between the non-Gaussian
as well as other cosmological parameters in the present context.

\item Also, our present analysis have
been performed for three different potentials. It would be interesting to
check whether we can formulate any general form of a potential, which can
be used to repeat the analysis of this paper for all the models. 

\item We also carry forward our analysis in the development of density inhomogeneities, which is the prime component to form large scale structures at late times. Also 
the specific role of cosmological hysteresis in the study of cosmological perturbations i.e. for interacting/decoupled dark matter and dark energy have not been explored at all earlier. We have some future plan to do
some computations from this setup.
\item Further using the reconstruction techniques we want to study the generic features of scalar field potentials in the framework of cosmological hysteresis.
\end{itemize}


\section*{Acknowledgments}
SC would like to thank Department of Theoretical Physics, Tata Institute of Fundamental
Research, Mumbai for providing me Visiting (Post-Doctoral) Research Fellowship.  SC take this opportunity to thank sincerely to Prof. Soumitra SenGupta,
Prof. Sayan Kar, Prof. Sandip P. Trivedi, Prof. Shiraz Minwalla, Dr. Subhabrata Majumdar and Dr. Supratik Pal
for their constant support and inspiration. SC take this opportunity to thank all the active members and the
regular participants of weekly student discussion meet ``COSMOMEET'' from Department of Theoretical Physics and Department of Astronomy and Astrophysics, Tata Institute of Fundamental
Research for suggesting various crucial issues at the initial stage of the work, which finally helped us to improve 
the quality of the work. SC also thanks 
Indian Association for the Cultivation of Science (IACS), Kolkata and Physics and Applied Mathematics Unit (PAMU), Indian Statistical Institute (ISI), Kolkata for
extending hospitality during the work. Additionally SC take this opportunity to thank the organizers of 28th IAGRG Meeting, 2015, Raman Research Institute (RRI) and Indian Institute of Science (IISC)
for providing the local hospitality during the work. SB sincerely thanks Prof. T. P. Singh for constant support and inspiration. Last but not the least, we would all like to acknowledge our debt to the people of
India for their generous and steady support for research in natural sciences, especially for theoretical physics.

\section{Appendix}
\subsection{Hysteresis from RSII brane world model}
\label{sq8}
Below we have quoted the results for RSII brane world cosmology with time like extra dimension necessary for causing the bounce, which was studied in detail by the authors of \cite{Sahni:2012er}. 

The Friedmann equations in this model are given by:
\begin{eqnarray}
H^2 &=&\left(\frac{\dot{a}}{a}\right)^{2}= \frac{8\pi G}{3}\rho\left\lbrace 1 - \frac{\rho}{\rho_c}\right\rbrace
-\frac{k}{a^2}~,\\
\dot{H}+H^{2}&=&\frac{\ddot a}{a} = -\frac{4\pi G}{3}\left\lbrace (\rho+3p) -
\frac{2\rho}{\rho_c}(2\rho+3p)\right\rbrace~.
\label{bouncenew}
\end{eqnarray}
\subsubsection{Condition for bounce}

At bounce, by setting the Hubble parameter $H = 0$, we get:
\begin{equation}
\rho_{b} = \rho_{c}
\end{equation}
The mass at bounce (neglecting the constant factor) is given by:
\be M_{b} = \rho_{b} a_{b}^{3}.\ee
Therefore, the infinitesimal change in the mass content at bounce is given by:
\be \delta M = \delta(\rho_{c} a_{b})^{3}.\ee
From energy conservation, we get: \be \delta M + \delta W = 0,\ee
where $\delta W$ is the work done during each expansion-contraction cycle which is given by $\oint pdV $ which includes contribution from the area of the hysteresis loop.

Further setting \be \delta M = -\delta W = -\oint pdV,\ee we get the expression for change in amplitude of the scale factor at each successive cycle as: 
\begin{equation}
\delta (a_{\rm min})^3 \equiv \left\lbrace a_{min}^{(i)}\right\rbrace^3 - 
\left\lbrace a_{min}^{(i-1)}\right\rbrace^3
= -\frac{1}{\rho_c}\oint pdV~,
\label{eq:work4}
\end{equation}
Thus we see that the change in minimum of the expansion factor depends upon the work done and the density at bounce.

\subsubsection{Condition for acceleration}

From Eq.~(\ref{bouncenew}), condition for acceleration is given by:
\begin{equation}
4\pi G(\rho_{c} + p_{c}) > 0
\label{dgpaccel}
\end{equation}
i.e. bounce can be obtained without violating the strong energy condition.

\subsubsection{Condition for turnaround}   

Though the authors of  \cite{Sahni:2012er} have suggested different ways of getting the condition for turnaround,
but in there analysis they have used the presence of density which take negative values at late times to be the
cause for turnaround. The authors of  \cite{Sahni:2012er} have assumed that the late time behavior of the universe is governed by:
\begin{equation}
H^2 \simeq \kappa \rho - \frac{A}{a^n}, 
\label{eq:late}
\end{equation}
where $\kappa = \frac{8\pi G}{3}$, and $A>0$, $n \leq 2$. Here $A \equiv \Lambda < 0, ~ n=0$ corresponds to 
a {\em negative} cosmological constant,
whereas $A=1, ~n=2$ describes a universe which is spatially closed.
Setting the Hubble parameter $H = 0$, we get the condition for turnaround as:
\begin{equation}
\kappa \rho_{t} = \frac{A}{a_{t}^n}
\label{eq:ta}
\end{equation}
where $\rho_{t}$ is the density and $a_{t}$ the expansion factor at turnaround. They are connected via the following equation:
\begin{equation}
\rho_{t} = \frac{3M_{4}^{2}}{2}\left(\frac{k}{a_{t}^{2}}-\frac{1}{r_{c}^{2}}\right)
\end{equation}
where $\rho_{t}$ and $a_{t}$ are the density and scale factor at turnaround respectively.

The change in amplitude of the scale factor after each successive cycle is given by:
\begin{equation}
\delta\left ( a_{\rm max}\right )^{3-n} 
\equiv \Big\lbrace a_{max}^{(i)}\Big\rbrace^{3-n} - 
\Big\lbrace a_{max}^{(i-1)}\Big\rbrace^{3-n}
= - \frac{\kappa}{A} \oint pdV~,
\label{eq:work1}
\end{equation}
which is responsible for turnaround. Here two extreme physical situation correspond to:
\begin{itemize}
 \item  (i) the negative
cosmological constant ($n=0$) for which
\begin{equation}
\delta a^3_{\rm max} = - \frac{\kappa}{\Lambda} 
\oint pdV ~,
\label{eq:amax_lambda}
\end{equation}
 
 \item (ii) the spatially closed universe ($n=2$) for which
\begin{equation}
\delta a_{\rm max} \equiv a_{max}^{(i)} - a_{max}^{(i-1)}
 = - \kappa \oint pdV ~.
\label{eq:amax}
\end{equation}
\end{itemize}

\subsubsection{Condition for deceleration}
Turnaround occurs only if:
\begin{equation}
\rho + 3p \geq 0
\end{equation}
Therefore, turnaround can be obtained without violating the energy condition in RSII setup. 

\subsection{Exact expressions for $\phi$ integrals}

\underline{\textbf{DGP model-Hilltop potential}}\\ \\
i) The exact solution to the left hand side of the integral in Eq.~(\ref{dgpearly1}) for early times is given by
\begin{eqnarray}
{\cal I}_{1}&=&\frac{(\phi/M_{4})^{2-p}\left(-4(\frac{1}{2r_{c}}+\sqrt{\frac{V_{0}}{3M_{4}^{2}}(1+(\phi/M_{4})^{p}
\beta}\right)}{4(-2+p)\beta p}\nonumber \\ &&+\frac{(\phi/M_{4})^{2-p}
\left(\sqrt{\frac{V_{0}}{3M_{p}^{2}}}p(\phi/M_{4}^{2})^{p}\beta \ {\rm Hypergeometric2F}1\left[
\frac{1}{2},\frac{2}{p},\frac{2+p}{p},-(\phi/M_{4})^{p}\beta\right]\right)}{4(-2+p)\beta p}~~~~~~~~~~~
\end{eqnarray}
ii) The exact solution to the left hand side of Eq.~(\ref{dgplate}) for late times is given by
\begin{eqnarray}
{\cal I}_{2}&=&\frac{(\phi/M_{4})^{2}\left(-\frac{8V_{0}\beta}{3M_{4}^{2}}-4D_{G}
(\phi/M_{4})^{-p}\right)}{4(-2+p)\left(D_{G}+\frac{2V_{0}\beta}{3M_{4}^{2}t^{p}}\right)^{1/2}}\nonumber 
\\ &&+\frac{(\phi/M_{4})^{2}\left(\frac{2V_{0}\beta}{3M_{4}^{2}}p
\sqrt{1+\frac{\frac{2V_{0}\beta}{3M_{4}^{2}}(\phi/M_{4})^{p}}{D_{G}}}
{\rm Hypergeometric2F}1\left[\frac{1}{2},\frac{2}{p},\frac{2+p}{p},
-\frac{\frac{2V_{0}\beta}{3M_{4}^{2}}}{D_{G}}(\phi/M_{4})^{p}\beta\right]
\right)}{4(-2+p)\left(D_{G}+\frac{2V_{0}\beta}{3M_{4}^{2}t^{p}}\right)^{1/2}} \nonumber \\
\end{eqnarray}
where \be D_{G}=\frac{2}{3}\frac{V_{0}}{M_{4}^{2}}+\frac{1}{r_{c}^{2}}.\ee
\\ \\
\underline{\textbf{DGP model- Natural potential}}\\ \\
i) Exact solution to the integral on left hand side of Eq.~(\ref{dgpearly6}) for early times is given by
\begin{equation}
{\cal I}_{3}=\left(\frac{V_{0}}{3M_{4}^{2}}\right)^{1/2}\left[1+\cos\left(\frac{\phi}{f}\right)
\right]^{1/2}\left[-\ln\left(\cos\left(\frac{\phi}{4f}\right)\right)+\ln\left(\sin\left(\frac{\phi}{4f}\right)\right)\right]\sec\left(\frac{\phi}{2f}\right)
\end{equation}
ii) Exact solution to the integral on left hand side of Eq.~(\ref{dgplate1}) for late times is given by
\begin{equation}
{\cal I}_{4}=\sqrt{D_{P}-\frac{2V_{0}}{3M_{4}^{2}}}\tan^{-1}\left[\frac{\sqrt{\frac{2V_{0}}{3M_{4}^{2}}+D_{P}
\cos\left(\frac{\phi}{f}\right)}}{\sqrt{D_{P}-\frac{2V_{0}}{3M_{4}^{2}}}}\right]-\sqrt{D_{P}+\frac{2V_{0}}{3M_{4}^{2}}}\tan^{-1}
\left[\frac{\sqrt{\frac{2V_{0}}{3M_{4}^{2}}+D_{P}\cos\left(\frac{\phi}{f}\right)}}{\sqrt{D_{P}+\frac{2V_{0}}{3M_{4}^{2}}}}\right]
\end{equation}
where \be D_{P}=\frac{2V_{0}}{3M_{4}^{2}}+\frac{1}{r_{c}^{2}}.\ee
\\ \\
\underline{\textbf{Cosmological constant model- Hilltop potential}}
\\ \\
 Exact solution to the integral on left hand side of Eq.~(\ref{dgpearly6}) for expansion is given by 
\begin{eqnarray}
{\cal I}_{5}&=&\frac{(\phi/M_{4})^{2}\left(-\frac{4V_{0}\beta}{3M_{4}^{2}}
-4D_{\Lambda}(\phi/M_{4})^{-p}\right)}{4(-2+p)
\left(D_{\Lambda}+\frac{V_{0}\beta}{3M_{4}^{2}t^{p}}\right)^{1/2}}\nonumber \\ 
&&+\frac{(\phi/M_{4})^{2}\left(\frac{V_{0}\beta}{3M_{4}^{2}}
p\sqrt{1+\frac{\frac{V_{0}\beta}{3M_{4}^{2}}(\phi/M_{4})^{p}}
{D_{\Lambda}}}{\rm Hypergeometric2F}1\left[\frac{1}{2},\frac{2}{p},
\frac{2+p}{p},-\frac{\frac{V_{0}\beta}{3M_{4}^{2}}}{D_{\Lambda}}
(\phi/M_{4})^{p}\beta\right]\right)}{4(-2+p)\left(D_{\Lambda}+\frac{V_{0}\beta}{3M_{4}^{2}t^{p}}\right)^{1/2}} \nonumber \\
\end{eqnarray}
where \be D_{G}=\frac{V_{0}}{3M_{4}^{2}}+\frac{\Lambda}{3}.\ee
\\ \\
\underline{\textbf{Cosmological constant model- Natural potential}}
\\ \\
 Exact solution to the integral on left hand side of Eq.~(\ref{lambdanatural}) for expansion is given by 
 \begin{equation}
{\cal I}_{6}=\sqrt{D_{f}-\frac{V_{0}}{3M_{4}^{2}}}\tan^{-1}
\left[\frac{\sqrt{\frac{V_{0}}{3M_{4}^{2}}+D_{f}\cos\left(\frac{\phi}{f}\right)}}{\sqrt{D_{f}-\frac{V_{0}}{3M_{4}^{2}}}}\right]
-\sqrt{D_{f}+\frac{V_{0}}{3M_{4}^{2}}}\tan^{-1}\left[\frac{\sqrt{\frac{V_{0}}{3M_{4}^{2}}+D_{f}\cos\left(\frac{\phi}{f}\right)}}{\sqrt{D_{f}+\frac{V_{0}}{3M_{4}^{2}}}}\right]
\end{equation}
where \be D_{f}=\frac{V_{0}}{3M_{4}^{2}}+\frac{\Lambda}{3}.\ee
\\ \\
\underline{\textbf{LQG model- Hilltop potential}}
 \\ \\
  Exact solution to the integral on left hand side of Eq.~(\ref{lqcearly20}) for early time expansion is given by: 
 \begin{eqnarray}
 {\cal I}_{7}&=&-\left(\frac{V_{0}}{3M_{p}^{2}}\right)^{1/2}\\&&\times
 \frac{(\phi/M_{p})^{2-p}\sqrt{1-\frac{V_{0}}{\rho_{c}}(1+(\phi/M_{p})^{p}\beta)}
 {\rm AppellF}1\left[-1+\frac{2}{p},-\frac{1}{2},-\frac{1}{2},\frac{2}{p},
 -(\phi/M_{p})^{p}\beta,\frac{V_{0}(\phi/M_{p})^{p}\beta}{\rho_{c}
 (1-(\frac{V_{0}}{\rho_{c}}))}\right]}{(-2+p)\left(\frac{-1+\frac{V_{0}}{\rho_{c}}
 +\frac{V_{0}(\phi/M_{p})^{p}\beta}{\rho_{c}}}{-1+(\frac{V_{0}}{\rho_{c}})}\right)^{1/2}\beta p}\nonumber
 \end{eqnarray}
 \underline{\textbf{LQC model- Natural potential}}
 \\ \\
  Exact solution to the integral on left hand side of Eq.~(\ref{lqcearly6}) for early time expansion is given by 
  \begin{eqnarray}
 {\cal I}_{8}&=&\left(\frac{V_{0}}{3M_{p}^{2}}\right)^{1/2} \sec\left(\frac{\phi}{2f}\right)\sqrt{1+\cos\left(\frac{\phi}{f}\right)}\left\{ 
 -\sqrt{1-2\frac{V_{0}}{\rho_{c}}}\tanh^{-1}\left[\frac{\sqrt{1-2\frac{V_{0}}{\rho_{c}}}\cos\left(\phi/2f\right)}{\sqrt{1-\frac{V_{0}}{\rho_{c}}
 -\frac{V_{0}}{\rho_{c}}\cos(\frac{\phi}{f})}}\right]\nonumber \right.\\ &&\left.+ \sqrt{-\frac{2V_{0}}{\rho_{c}}}\ln\left[\sqrt{-\frac{2V_{0}}
 {\rho_{c}}}\cos(\phi/2f)+\sqrt{1-\frac{V_{0}}{\rho_{c}}-\frac{V_{0}}{\rho_{c}}\cos\left(\frac{\phi}{f}\right)}\right]\right\} \nonumber \\
  \end{eqnarray}
  
  \subsection{Expressions of the constants in work done analysis}
  \label{sq9}
  \subsubsection{Dvali-Gabadadze-Porrati (DGP) brane world model}
  \underline{\textbf{Hilltop potential}}
  \begin{eqnarray}
  a_{6}&=&\left(\left(\frac{V_{0}}{3M_{4}^{2}}\right)^{1/2}(1+\beta)^{1/2}+\frac{1}{2r_{c}}+\beta p\frac{\left(\frac{V_{0}}{3M_{4}^{2}}\right)^{1/2}}{2(1+\beta)^{1/2}}
\left\{\lambda_{i}+\frac{\frac{V_{0}}{3M_{4}^{2}}}{\left(\frac{V_{0}}{3M_{4}^{2}}\right)^{1/2}\frac{(1+\beta)^{1/2}}{\beta p}+\frac{1}{2r_{c}}}t_{i}\right\}\right)\nonumber \\
  a_{7}&=&\left(\beta p\frac{\left(\frac{V_{0}}{3M_{4}^{2}}\right)^{1/2}}{2(1+\beta)^{1/2}}\right)\frac{\frac{V_{0}}{3M_{4}^{2}}}{2 \left(\frac{V_{0}}{3M_{4}^{2}}\right)^{1/2}\frac{(1+\beta)^{1/2}}{\beta p}+\frac{1}{2r_{c}}}\nonumber \\ 
   a_{4}&=&e^{\frac{3a_{6}^{2}}{4a_{7}}}a_{i}^{3}\sqrt{\frac{\pi}{3}}\frac{(2+(1-\frac{1}{2r_{c}})^{2})}{2\sqrt{a_{7}}}\nonumber \\
  a_{5}&=&\frac{1}{2}\sqrt{\frac{3}{a_{7}}}\nonumber \\
  a_{1}&=&e^{A_{2}/r_{c}}\nonumber \\
  a_{2}&=&a_{1}^{1/2}\nonumber \\
  a_{3}&=&a_{1}^{3/2}/r_{c}\nonumber \\
  a'_{1}&=&\frac{1}{2r_{c}}\nonumber \\
  a'_{2}&=& A_{2}''^{3}\left(2+\left(1-\frac{1}{2r_{c}}\right)^{2}\right)\frac{r_{c}}{3}\nonumber \\
  \end{eqnarray}
  \underline{\textbf{Natural potential}}
  \begin{eqnarray}
  b_{2}&=&\left(\sqrt{\frac{V_{0}}{3M_{4}^{2}}}\sqrt{2}+\frac{1}{2r_{c}}\right)\nonumber \\
  b_{1}&=&=\frac{A_{7}^{3}}{3b_{2}}(2+(1-\frac{1}{2r_{c}})^{2})
  \end{eqnarray}
  \subsubsection{Cosmological constant dominated Einstein gravity model}
  \underline{\textbf{Hilltop potential}}
  \begin{eqnarray}
  l_{1}&=&\sqrt{\left(\frac{V_{0}}{3M_{p}^{2}}(1+\beta)+\frac{\Lambda}{3}\right)}\nonumber \\
  l_{2}&=&\frac{\frac{V_{0}}{3M_{p}^{2}}\beta^{2}p^{2}}{2\left(\frac{V_{0}}{3M_{p}^{2}}(1+\beta)+\frac{\Lambda}{3}\right)}B_{0}\nonumber \\
  l_{3}&=&\frac{\left(\frac{V_{0}}{3M_{p}^{2}}\right)^{2}\beta^{2}p^{2}}{4\left(\frac{V_{0}}{3M_{p}^{2}}(1+\beta)+\frac{\Lambda}{3}\right)}\nonumber \\
  b_{5}&=& 2/p^{2}\nonumber \\
  b_{6}&=&\frac{\beta^{1/2}p^{2}B_{2}}{2(\frac{V_{0}}{3M_{p}^{2}})^{1/2}}\nonumber \\
  b_{7}&=&\beta^{1/2}\frac{p^{2}}{2}(\frac{V_{0}}{3M_{p}^{2}})^{1/2}\nonumber \\
  b_{1}&=&B_{1}\frac{e^{3(l_{1}+l_{2})^{2}/(2l_{3})}}{\sqrt{l_{3}}}\nonumber \\
   b_{2}&=&l_{1}\sqrt{3/(2l_{3})}\nonumber \\
   b_{3}&=&\sqrt{3l_{3}/2}
  \end{eqnarray}
  \underline{\textbf{Natural Potential}}\\ \\
  \underline{\textbf{a) $\phi<<f$}}
 \begin{eqnarray}
  b_{8}&=&B_{7}^{3}\nonumber \\
  b_{9}&=&\left(1+\frac{\frac{V_{0}}{3M_{p}^{2}}}{\frac{V_{0}}{3M_{p}^{2}}+\frac{\Lambda}{3}}\right)^{1/2}\left(\frac{V_{0}}{3M_{p}^{2}}+\frac{\Lambda}{3}\right)^{1/2}
  \end{eqnarray}
 \underline{\textbf{b) $\phi>>f$}} 
 \begin{eqnarray}
 b_{10}&=&\left(\frac{V_{0}}{3M_{p}^{2}}+\frac{\Lambda}{3}\right)^{1/2}\nonumber \\
 b_{13}&=&\frac{\frac{V_{0}}{3M_{p}^{2}}}{2(\frac{V_{0}}{3M_{p}^{2}}+\frac{\Lambda}{3})}\nonumber \\ l_{6}&=&\frac{V_{0}}{3M_{p}^{2}}\nonumber \\
 l_{7}&=&\frac{\frac{V_{0}}{3M_{p}^{2}}}{\left(\frac{V_{0}}{3M_{p}^{2}}+\frac{\Lambda}{3}\right)^{1/2}}\nonumber \\
 b_{12}&=&1,\ b_{11}=l_{7}/b_{10},\ b_{14}=B_{8}(-3+\Lambda),\nonumber \\
  b_{15}&=&\frac{l_{6}l_{7}-l_{5}l_{6}b_{10}+b_{10}l_{6}}{b_{10}l_{6}}-1
\end{eqnarray}
\subsubsection{Loop Quantum Gravity model}
\underline{\textbf{Hilltop potential}}
\begin{eqnarray}
q_{1}&=&\sqrt{\frac{V_{0}}{3M_{p}^{2}}}\left(\frac{\left(\left(1-\frac{V_{0}}{2\rho_{c}}(1+\beta)\right)Q-\frac{E_{0}p\beta V_{0}}{2\rho_{c}}+\frac{E_{0}p\beta}
{2(1+\beta)}\left(1-\frac{V_{0}}{2\rho_{c}}(1+\beta)\right)\right)}{Q}\right)\nonumber \\
q_{2}&=&\sqrt{\frac{V_{0}}{3M_{p}^{2}}}\left(\frac{\left(\frac{pV_{0}^{2}\beta}{12M_{p}^{2}\rho_{c}}-\frac{V_{0}
p\beta}{12M_{p}^{2}(1+\beta)}\left(1-\frac{V_{0}}{2\rho_{c}}(1+\beta)\right)\right)}{Q}\right)\nonumber \\
d_{2}&=&1/M_{p}^{2}\nonumber \\
d_{5}&=& (1+M_{p}^{2})/\sqrt{6}\nonumber \\
d_{6}&=&\sqrt{\frac{1+M^{2}}{2}}\nonumber \\
d_{7}&=&1\nonumber \\
d_{8}&=&\frac{E_{4}^{3}e^{3\rho_{c}/2}}{4\sqrt{1+M_{p}^{2}}}.
\end{eqnarray}
\begin{eqnarray}
d_{11}&=&3E_{4}^{2}e^{\rho_{c}}(1+M_{p}^{2})\sqrt{2\pi}(-12M_{p}^{2}+\rho_{c})\nonumber \\
d_{12}&=&4k(1+M_{p}^{2})\sqrt{6\pi}\nonumber \\
d_{19}&=&1 \nonumber \\
d_{18}&=&2\sqrt{E_{2}}\nonumber \\
d_{20}&=&E'_{2}\nonumber \\
d_{13}&=&\sqrt{\frac{\pi}{q_{2}}}\frac{E_{1}^{3}\rho_{c}}{4}\nonumber \\
d_{14}&=&\frac{3q_{1}^{2}}{4q_{2}}\nonumber \\
d_{15}&=&\frac{q_{1}}{2}\sqrt{\frac{3}{q_{2}}}\nonumber \\
d_{16}&=&\frac{\sqrt{3q_{2}}}{2}\nonumber \\
d_{17}&=&12e^{q_{1}^{2}/(2q_{2})}k\nonumber \\
d'_{17}&=&\sqrt{3}E_{1}^{2}(12M_{p}^{2}-\rho_{c})
\end{eqnarray}
\underline{\textbf{Natural potential}}
\begin{eqnarray}
d_{22}&=&\sqrt{2}\left(\frac{V_{0}}{3M_{p}^{2}}\left(1-\frac{2V_{0}}{\rho_{c}}\right)^{1/2}\right)^{1/2}\nonumber \\
d_{21}&=&\frac{6 E_{9}k}{d_{22}}\nonumber \\
d_{23}&=&\frac{E_{9}^{3}}{2 d_{22}}\rho_{c}(12M_{p}^{2}-\rho_{c})\nonumber \\
d_{24}&=&\frac{E_{9}^{3}\rho_{c}}{2d_{22}}
\end{eqnarray}
\subsubsection{Einstein Gauss Bonnet Gravity brane world model}
\underline{\textbf{Hilltop potential}}
\begin{eqnarray}
g_{1}&=&\left(\sqrt{R+4\sqrt{\frac{C}{A}}V_{0}}-(A+B/2)^{1/2}\right)^{1/2}S\nonumber \\
g_{2}&=&\frac{p(S-1)F_{0}}{\sqrt{(\sqrt{R+4\sqrt{\frac{C}{A}}V_{0}}-(A+B/2))}S}g_{1}\nonumber \\
g_{3}&=&\frac{p^{2}(S-1)\frac{V_{0}}{3M_{p}^{2}}}{\sqrt{(\sqrt{R+4\sqrt{\frac{C}{A}}V_{0}}-(A+B/2))}S}g_{1}
\end{eqnarray}
\begin{eqnarray}
g_{4}&=&\sqrt{\frac{D''}{2}}\nonumber \\
g_{5}&=&-2+D'+D'F_{3}D''\nonumber \\
g_{6}&=&D'D''\nonumber\\ g_{7}&=&D^{2}D''\nonumber \\
f_{1}&=&210g_{1}(28g_{4}^{2}g_{5}^{6}g_{6}^{3}+15g_{4}g_{5}^{3}(3g_{4}g_{5}-4g_{6})g_{6}^{2}g_{7}+9g_{6}(3g_{4}^{2}g_{5}^{2}-6g_{4}g_{5}g_{6}+g_{6}^{2})g_{7}^{2}+9g_{4}^{2}g_{7}^{3}\nonumber \\
f_{2}&=&504g_{4}^{2}g_{6}^{2}(14g_{4}g_{5}^{5}g_{6}^{2}+5g_{4}^{2}(4g_{4}g_{5}-3g_{6})g_{6}g_{7}+ 3(3g_{4}g_{5}-2g_{6})g_{7}^{2}\nonumber \\
f_{3}&=& 420g_{4}^{2}g_{6}^{3}(14g_{4}g_{5}^{4}g_{6}^{2}+3g_{4}(5g_{4}g_{5}-2g_{6})g_{6}g_{7}+3g_{4}g_{7}^{2})\nonumber \\
f_{4}&=& -120g_{4}^{2}g_{6}^{5}(28g_{4}g_{5}^{3}g_{6}+18g_{4}g_{5}g_{7}-3g_{4}g_{7})\nonumber \\
f_{5}&=& 3 15g_{4}^{3}g_{6}^{6}(4g_{4}^{2}g_{6}+g_{7})\nonumber \\
f_{6}&=& 280g_{4}^{3}g_{5}g_{6}^{8}\nonumber \\
f_{7}&=& 28g_{4}^{3}g_{6}^{9}\nonumber \\
f_{8}&=& \frac{1}{2520g_{5}g_{7}}\sqrt{1+A}(1+B)F_{4}^{3}\left(-1+\frac{\sigma C'}{\sqrt{1+A}(1+B)}\right)\nonumber \\
f_{9} &=& F_{1}^{3}e\frac{^{3(g_{1}+g_{2})^{2}/(2g_{3})}}{\sqrt{g_{3}}}\nonumber \\
f_{10}&=& g_{1}\sqrt{3/(2g_{3})}\nonumber \\
f_{11} &=& \sqrt{3g_{3}/2}
\end{eqnarray}
\underline{\textbf{Natural potential}}
\begin{eqnarray}
f_{13}&=& \frac{1}{\sqrt{2}}\left(-2\left(A+\frac{B}{2}\right)+\sqrt{U}\left(1\pm \frac{2\sqrt{C}V_{0}}{U}\right)^{1/2}\right)^{1/2}\nonumber \\
f_{12}&=&F_{5}^{3}\frac{(\sqrt{1+A}+\sqrt{1+A}B-\sigma C')}{f_{13}C'}
\end{eqnarray}



\begin{thebibliography}{}
\bibitem{eliade}
M. Eliade, \textcolor{blue}{\it The Myth of the Eternal Return}, \textcolor{purple}{Pantheon, New York (1934)}.

\bibitem{jaki}
S. L. Jaki, \textcolor{blue}{\it Science and Creation: From eternal Cycles to an Oscillating Universe}, \textcolor{purple}{Science History Publ., New York (1974)}.

\bibitem{starobinsky}
A. A. Starobinsky, \textcolor{purple}{JETP Lett. {\bf 30}, 719 (1979)}.

\bibitem{tolman}
R. C. Tolman, \textcolor{blue}{\it Relativity, Thermodynamics and Cosmology}, \textcolor{purple}{ Clarendon Press, Oxford (1934)}.

\bibitem{Steinhardt:2001st}
  P.~J.~Steinhardt and N.~Turok,
  \textcolor{blue}{\it``Cosmic evolution in a cyclic universe,''}
  \textcolor{purple}{Phys.\ Rev.\ D {\bf 65} (2002) 126003
  [hep-th/0111098]}.
  
  \bibitem{Lehners:2008vx}
  J.~L.~Lehners,
  \textcolor{blue}{\it``Ekpyrotic and Cyclic Cosmology,''}
  \textcolor{purple}{Phys.\ Rept.\  {\bf 465} (2008) 223
  [arXiv:0806.1245 [astro-ph]]}.


  \bibitem{Baumann:2014nda}
  D.~Baumann and L.~McAllister,
  \textcolor{blue}{\it``Inflation and String Theory,''}
  \textcolor{purple}{arXiv:1404.2601 [hep-th]}.
  
  \bibitem{Baumann:2009ds} 
  D.~Baumann,
  \textcolor{blue}{\it``TASI Lectures on Inflation,''}
  \textcolor{purple}{arXiv:0907.5424 [hep-th]}.
  
  \bibitem{Lyth:1998xn}
  D.~H.~Lyth and A.~Riotto,
  \textcolor{blue}{\it``Particle physics models of inflation and the cosmological density perturbation,''}
  \textcolor{purple}{Phys.\ Rept.\  {\bf 314} (1999) 1
  [hep-ph/9807278]}.
  
\bibitem{Kanekar:2001qd}
  N.~Kanekar, V.~Sahni and Y.~Shtanov,
  \textcolor{blue}{\it``Recycling the universe using scalar fields,''}
  \textcolor{purple}{Phys.\ Rev.\ D {\bf 63} (2001) 083520
  [astro-ph/0101448]}.
  
\bibitem{Sahni:2012er}
  V.~Sahni and A.~Toporensky,
  \textcolor{blue}{\it``Cosmological Hysteresis and the Cyclic Universe,''}
  \textcolor{purple}{Phys.\ Rev.\ D {\bf 85} (2012) 123542
  [arXiv:1203.0395 [gr-qc]]}.

  
  \bibitem{Cai:2013kja}
  Y.~F.~Cai, E.~McDonough, F.~Duplessis and R.~H.~Brandenberger,
   \textcolor{blue}{\it``Two Field Matter Bounce Cosmology,''}
  \textcolor{purple}{JCAP {\bf 1310} (2013) 024
  [arXiv:1305.5259 [hep-th]]}.
  
  \bibitem{Cai:2013vm}
  Y.~F.~Cai, R.~Brandenberger and P.~Peter,
  \textcolor{blue}{\it``Anisotropy in a Nonsingular Bounce,''}
  \textcolor{purple}{Class.\ Quant.\ Grav.\  {\bf 30} (2013) 075019
  [arXiv:1301.4703 [gr-qc]]}.
  
  \bibitem{Cai:2012ag}
  Y.~F.~Cai, C.~Gao and E.~N.~Saridakis,
  \textcolor{blue}{\it``Bounce and cyclic cosmology in extended nonlinear massive gravity,''}
  \textcolor{purple}{JCAP {\bf 1210} (2012) 048
  [arXiv:1207.3786 [astro-ph.CO]]}.
  
  \bibitem{Cai:2012va}
  Y.~F.~Cai, D.~A.~Easson and R.~Brandenberger,
  \textcolor{blue}{\it``Towards a Nonsingular Bouncing Cosmology,''}
  \textcolor{purple}{JCAP {\bf 1208} (2012) 020
  [arXiv:1206.2382 [hep-th]]}.
  
  \bibitem{Li:2014era}
  C.~Li, R.~H.~Brandenberger and Y.~K.~E.~Cheung,
  \textcolor{blue}{\it``Big Bounce Genesis,''}
  \textcolor{purple}{Phys.\ Rev.\ D {\bf 90} (2014) 12,  123535
  [arXiv:1403.5625 [gr-qc]]}.
  
  \bibitem{Brandenberger:2012zb}
  R.~H.~Brandenberger,
  \textcolor{blue}{\it``The Matter Bounce Alternative to Inflationary Cosmology,''}
  \textcolor{purple}{arXiv:1206.4196 [astro-ph.CO]}.
  
  \bibitem{Cai:2011zx}
  Y.~F.~Cai, R.~Brandenberger and X.~Zhang,
  \textcolor{blue}{\it``The Matter Bounce Curvaton Scenario,''}
  \textcolor{purple}{JCAP {\bf 1103} (2011) 003
  [arXiv:1101.0822 [hep-th]]}.
  
  \bibitem{Lilley:2015ksa}
  M.~Lilley and P.~Peter,
  \textcolor{blue}{\it``Bouncing alternatives to inflation,''}
  \textcolor{purple}{arXiv:1503.06578 [astro-ph.CO]}.
  
  \bibitem{Falciano:2008gt}
  F.~T.~Falciano, M.~Lilley and P.~Peter,
  \textcolor{blue}{\it``A Classical bounce: Constraints and consequences,''}
  \textcolor{purple}{Phys.\ Rev.\ D {\bf 77} (2008) 083513
  [arXiv:0802.1196 [gr-qc]]}.
  
  \bibitem{Lilley:2010av}
  M.~Lilley, F.~Di Marco, J.~Martin and P.~Peter,
  \textcolor{blue}{\it``Nonabelian Bosonic Currents in Cosmic Strings,''}
  \textcolor{purple}{Phys.\ Rev.\ D {\bf 82} (2010) 023510
  [arXiv:1003.4601 [hep-th]]}.
  
  \bibitem{Lilley:2011ag}
  M.~Lilley, L.~Lorenz and S.~Clesse,
  \textcolor{blue}{\it``Observational signatures of a non-singular bouncing cosmology,''}
  \textcolor{purple}{JCAP {\bf 1106} (2011) 004
  [arXiv:1104.3494 [gr-qc]]}.
  
  \bibitem{Battefeld:2014uga}
  D.~Battefeld and P.~Peter,
  \textcolor{blue}{\it``A Critical Review of Classical Bouncing Cosmologies,''}
  \textcolor{purple}{Phys.\ Rept.\  {\bf 571} (2015) 1
  [arXiv:1406.2790 [astro-ph.CO]]}.
  
  
 \bibitem{Graham:2011nb}
  P.~W.~Graham, B.~Horn, S.~Kachru, S.~Rajendran and G.~Torroba,
  \textcolor{blue}{\it``A Simple Harmonic Universe,''}
  \textcolor{purple}{JHEP {\bf 1402} (2014) 029
  [arXiv:1109.0282 [hep-th]]}.
  
  \bibitem{Koehn:2013upa}
  M.~Koehn, J.~L.~Lehners and B.~A.~Ovrut,
  \textcolor{blue}{\it``Cosmological super-bounce,''}
  \textcolor{purple}{Phys.\ Rev.\ D {\bf 90} (2014) 2,  025005
  [arXiv:1310.7577 [hep-th]]}.
  
  \bibitem{Vilenkin:2013rza}
  A.~Vilenkin,
  \textcolor{blue}{\it``Arrows of time and the beginning of the universe,''}
  \textcolor{purple}{Phys.\ Rev.\ D {\bf 88} (2013) 043516
  [arXiv:1305.3836 [hep-th]]}.
  
  \bibitem{Choudhury:2014hja}
  S.~Choudhury, B.~K.~Pal, B.~Basu and P.~Bandyopadhyay,
  \textcolor{blue}{\it``Measuring CP violation within Effective Field Theory of inflation from CMB,''}
  \textcolor{purple}{arXiv:1409.6036 [hep-th]}.
  
  \bibitem{Choudhury:2014sxa}
  S.~Choudhury, A.~Mazumdar and E.~Pukartas,
  \textcolor{blue}{\it``Constraining ${\cal N}=1$ supergravity inflationary framework with non-minimal Kähler operators,''}
  \textcolor{purple}{JHEP {\bf 1404} (2014) 077
  [arXiv:1402.1227 [hep-th]]}.
  
  \bibitem{Choudhury:2013woa}
  S.~Choudhury and A.~Mazumdar,
  \textcolor{blue}{\it``Primordial blackholes and gravitational waves for an inflection-point model of inflation,''}
   \textcolor{purple}{Phys.\ Lett.\ B {\bf 733} (2014) 270
  [arXiv:1307.5119 [astro-ph.CO]]}.
  
  \bibitem{Choudhury:2013zna}
  S.~Choudhury, T.~Chakraborty and S.~Pal,
  \textcolor{blue}{\it``Higgs inflation from new Kähler potential,''}
   \textcolor{purple}{Nucl.\ Phys.\ B {\bf 880} (2014) 155
  [arXiv:1305.0981 [hep-th]]}.
  
  \bibitem{Choudhury:2011jt}
  S.~Choudhury and S.~Pal,
  \textcolor{blue}{\it``Fourth level MSSM inflation from new flat directions,''}
  \textcolor{purple}{JCAP {\bf 1204} (2012) 018
  [arXiv:1111.3441 [hep-ph]]}.
  
  
  \bibitem{Biswas:2015kha}
  T.~Biswas, R.~Mayes and C.~Lattyak,
  \textcolor{blue}{\it``Perturbations in Bouncing and Cyclic Models, a General Study,''}
  \textcolor{purple}{arXiv:1502.05875 [gr-qc]}.
  
\bibitem{Biswas:2012bp}
  T.~Biswas, A.~S.~Koshelev, A.~Mazumdar and S.~Y.~Vernov,
  \textcolor{blue}{\it``Stable bounce and inflation in non-local higher derivative cosmology,''}
  \textcolor{purple}{JCAP {\bf 1208} (2012) 024
  [arXiv:1206.6374 [astro-ph.CO]]}.
  
  \bibitem{Battarra:2014tga}
  L.~Battarra, M.~Koehn, J.~L.~Lehners and B.~A.~Ovrut,
  \textcolor{blue}{\it``Cosmological Perturbations Through a Non-Singular Ghost-Condensate/Galileon Bounce,''}
  \textcolor{purple}{JCAP {\bf 1407} (2014) 007
  [arXiv:1404.5067 [hep-th]]}.
  
  
  \bibitem{Sahni:2015kga}
  V.~Sahni, Y.~Shtanov and A.~Toporensky,
  \textcolor{blue}{\it``Arrow of time in dissipationless cosmology,''}
  \textcolor{purple}{arXiv:1506.01247 [gr-qc]}.
  
  \bibitem{Choudhury:2015yna}
  S.~Choudhury and S.~Pal,
 \textcolor{blue}{\it``Primordial non-Gaussian features from DBI Galileon inflation,''}
  \textcolor{purple}{Eur.\ Phys.\ J.\ C {\bf 75} (2015) 6,  241 [arXiv:1210.4478 [hep-th]]}.
  
  
  \bibitem{Choudhury:2014uxa}
  S.~Choudhury,
  \textcolor{blue}{\it``Constraining ${\cal N} = 1$ supergravity inflation with non-minimal Kaehler operators using $\delta$N formalism,''}
  \textcolor{purple}{JHEP {\bf 1404} (2014) 105
  [arXiv:1402.1251 [hep-th]]}.
  
  \bibitem{Gao:2014hea}
  X.~Gao, M.~Lilley and P.~Peter,
  \textcolor{blue}{\it``Production of non-gaussianities through a positive spatial curvature bouncing phase,''}
  \textcolor{purple}{JCAP {\bf 1407} (2014) 010
  [arXiv:1403.7958 [gr-qc]]}.
  
  \bibitem{Gao:2014eaa}
  X.~Gao, M.~Lilley and P.~Peter,
  \textcolor{blue}{\it``Non-Gaussianity excess problem in classical bouncing cosmologies,''}
  \textcolor{purple}{Phys.\ Rev.\ D {\bf 91} (2015) 2,  023516
  [arXiv:1406.4119 [gr-qc]]}.
  
  \bibitem{Maldacena:2002vr}
  J.~M.~Maldacena,
  \textcolor{blue}{\it``Non-Gaussian features of primordial fluctuations in single field inflationary models,''}
  \textcolor{purple}{JHEP {\bf 0305} (2003) 013
  [astro-ph/0210603]}.
  
  \bibitem{Maldacena:2011nz}
  J.~M.~Maldacena and G.~L.~Pimentel,
  \textcolor{blue}{\it``On graviton non-Gaussianities during inflation,''}
  \textcolor{purple}{JHEP {\bf 1109} (2011) 045
  [arXiv:1104.2846 [hep-th]]}.
  
  \bibitem{Arkani-Hamed:2015bza}
  N.~Arkani-Hamed and J.~Maldacena,
  \textcolor{blue}{\it``Cosmological Collider Physics,''}
  \textcolor{purple}{arXiv:1503.08043 [hep-th]}.
  
  \bibitem{Mata:2012bx}
  I.~Mata, S.~Raju and S.~Trivedi,
  \textcolor{blue}{\it``CMB from CFT,''}
  \textcolor{purple}{JHEP {\bf 1307} (2013) 015
  [arXiv:1211.5482 [hep-th]]}.
  
  \bibitem{Ghosh:2014kba}
  A.~Ghosh, N.~Kundu, S.~Raju and S.~P.~Trivedi,
  \textcolor{blue}{\it``Conformal Invariance and the Four Point Scalar Correlator in Slow-Roll Inflation,''}
  \textcolor{purple}{JHEP {\bf 1407} (2014) 011
  [arXiv:1401.1426 [hep-th]]}.
  
  \bibitem{Kundu:2014gxa}
  N.~Kundu, A.~Shukla and S.~P.~Trivedi,
  \textcolor{blue}{\it``Constraints from Conformal Symmetry on the Three Point Scalar Correlator in Inflation,''}
  \textcolor{purple}{JHEP {\bf 1504} (2015) 061
  [arXiv:1410.2606 [hep-th]]}.
  
  \bibitem{Choudhury:2014hua}
  S.~Choudhury,
  \textcolor{blue}{\it``Inflamagnetogenesis redux: Unzipping sub-Planckian inflation via various cosmoparticle probes,''}
  \textcolor{purple}{Phys.\ Lett.\ B {\bf 735} (2014) 138
  [arXiv:1403.0676 [hep-th]]}.
  
   
\bibitem{Choudhury:2015jaa}
  S.~Choudhury,
  \textcolor{blue}{\it``Braneflamagnetogenesis from Cosmoparticle Physics after Planck,''}
  \textcolor{purple}{arXiv:1504.08206 [astro-ph.CO]}.
  
  \bibitem{Subramanian:2015lua}
  K.~Subramanian,
  \textcolor{blue}{\it``The origin, evolution and signatures of primordial magnetic fields,''}
  \textcolor{purple}{arXiv:1504.02311 [astro-ph.CO]}.
  
  
  \bibitem{Choudhury:2014kma}
  S.~Choudhury and A.~Mazumdar,
  \textcolor{blue}{\it``Reconstructing inflationary potential from BICEP2 and running of tensor modes,''}
  \textcolor{purple}{arXiv:1403.5549 [hep-th]}.
  
  \bibitem{Choudhury:2013iaa}
  S.~Choudhury and A.~Mazumdar,
  \textcolor{blue}{\it``An accurate bound on tensor-to-scalar ratio and the scale of inflation,''}
  \textcolor{purple}{Nucl.\ Phys.\ B {\bf 882} (2014) 386
  [arXiv:1306.4496 [hep-ph]]}.
  
  \bibitem{Choudhury:2014wsa}
  S.~Choudhury and A.~Mazumdar,
  \textcolor{blue}{\it``Sub-Planckian inflation \& large tensor to scalar ratio with $r\geq 0.1$,''}
  \textcolor{purple}{arXiv:1404.3398 [hep-th]}.
  
  \bibitem{Choudhury:2014sua}
  S.~Choudhury,
  \textcolor{blue}{\it``Can Effective Field Theory of inflation generate large tensor-to-scalar ratio within Randall–Sundrum single braneworld?,''}
  \textcolor{purple}{Nucl.\ Phys.\ B {\bf 894} (2015) 29
  [arXiv:1406.7618 [hep-th]]}.
  
  \bibitem{Lidsey:1995np}
  J.~E.~Lidsey, A.~R.~Liddle, E.~W.~Kolb, E.~J.~Copeland, T.~Barreiro and M.~Abney,
  \textcolor{blue}{\it``Reconstructing the inflation potential : An overview,''}
  \textcolor{purple}{Rev.\ Mod.\ Phys.\  {\bf 69} (1997) 373
  [astro-ph/9508078]}.
  
  \bibitem{Sotiriou:2008rp}
  T.~P.~Sotiriou and V.~Faraoni,
  \textcolor{blue}{\it``f(R) Theories Of Gravity,''}
  \textcolor{purple}{Rev.\ Mod.\ Phys.\  {\bf 82} (2010) 451
  [arXiv:0805.1726 [gr-qc]]}.
  
  \bibitem{DeFelice:2010aj}
  A.~De Felice and S.~Tsujikawa,
  \textcolor{blue}{\it``f(R) theories,''}
  \textcolor{purple}{Living Rev.\ Rel.\  {\bf 13} (2010) 3
  [arXiv:1002.4928 [gr-qc]]}.
  
  \bibitem{Faraoni:2008mf}
  V.~Faraoni,
  \textcolor{blue}{\it``f(R) gravity: Successes and challenges,''}
  \textcolor{purple}{arXiv:0810.2602 [gr-qc]}.
  
  \bibitem{Capozziello:2011et}
  S.~Capozziello and M.~De Laurentis,
  \textcolor{blue}{\it``Extended Theories of Gravity,''}
  \textcolor{purple}{Phys.\ Rept.\  {\bf 509} (2011) 167
  [arXiv:1108.6266 [gr-qc]]}.
  
  \bibitem{Choudhury:2012yh}
  S.~Choudhury and S.~Pal,
  \textcolor{blue}{\it``DBI Galileon inflation in background SUGRA,''}
  \textcolor{purple}{Nucl.\ Phys.\ B {\bf 874} (2013) 85
  [arXiv:1208.4433 [hep-th]]}.
  
  \bibitem{Choudhury:2013yg}
  S.~Choudhury and S.~Sengupta,
  \textcolor{blue}{\it``Features of warped geometry in presence of Gauss-Bonnet coupling,''}
  \textcolor{purple}{JHEP {\bf 1302} (2013) 136
  [arXiv:1301.0918 [hep-th]]}.
  
  \bibitem{Choudhury:2013eoa}
  S.~Choudhury, S.~Sadhukhan and S.~SenGupta,
  \textcolor{blue}{\it``Collider constraints on Gauss-Bonnet coupling in warped geometry model,''}
  \textcolor{purple}{arXiv:1308.1477 [hep-ph]}.
  
  \bibitem{Choudhury:2013qza}
  S.~Choudhury and A.~Dasgupta,
  \textcolor{blue}{\it``Galileogenesis: A new cosmophenomenological zip code for reheating through R-parity violating coupling,''}
  \textcolor{purple}{Nucl.\ Phys.\ B {\bf 882} (2014) 195
  [arXiv:1309.1934 [hep-ph]]}.
  
  \bibitem{Choudhury:2013aqa}
  S.~Choudhury and S.~SenGupta,
  \textcolor{blue}{\it``A step toward exploring the features of Gravidilaton sector in Randall–Sundrum scenario via lightest Kaluza–Klein graviton mass,''}
  \textcolor{purple}{Eur.\ Phys.\ J.\ C {\bf 74} (2014) 11,  3159
  [arXiv:1311.0730 [hep-ph]]}.
  
\bibitem{Choudhury:2014hna}
  S.~Choudhury, J.~Mitra and S.~SenGupta,
  \textcolor{blue}{\it``Modulus stabilization in higher curvature dilaton gravity,''}
  \textcolor{purple}{JHEP {\bf 1408} (2014) 004
  [arXiv:1405.6826 [hep-th]]}.
  
  \bibitem{Choudhury:2015wfa}
  S.~Choudhury, J.~Mitra and S.~SenGupta,
  \textcolor{blue}{\it``Fermion localization and flavour hierarchy in higher curvature spacetime,''}
   \textcolor{purple}{arXiv:1503.07287 [hep-th]}.
  
  \bibitem{Randall:1999vf}
  L.~Randall and R.~Sundrum,
  \textcolor{blue}{\it``An Alternative to compactification,''}
  \textcolor{purple}{Phys.\ Rev.\ Lett.\  {\bf 83} (1999) 4690
  [hep-th/9906064]}.
  
  \bibitem{Maartens:2010ar}
  R.~Maartens and K.~Koyama,
  \textcolor{blue}{\it``Brane-World Gravity,''}
  \textcolor{purple}{Living Rev.\ Rel.\  {\bf 13} (2010) 5
  [arXiv:1004.3962 [hep-th]]}.
  
  \bibitem{Randall:1999ee}
  L.~Randall and R.~Sundrum,
  \textcolor{blue}{\it``A Large mass hierarchy from a small extra dimension,''}
  \textcolor{purple}{Phys.\ Rev.\ Lett.\  {\bf 83} (1999) 3370
  [hep-ph/9905221]}.
  
  
  \bibitem{Dvali:2000hr}
  G.~R.~Dvali, G.~Gabadadze and M.~Porrati,
  \textcolor{blue}{\it``4-D gravity on a brane in 5-D Minkowski space,''}
  \textcolor{purple}{Phys.\ Lett.\ B {\bf 485} (2000) 208
  [hep-th/0005016]}.
  

\bibitem{Copeland:2006wr}
  E.~J.~Copeland, M.~Sami and S.~Tsujikawa,
  \textcolor{blue}{\it``Dynamics of dark energy,''}
  \textcolor{purple}{Int.\ J.\ Mod.\ Phys.\ D {\bf 15} (2006) 1753
  [hep-th/0603057]}.
 
\bibitem{Freese:2004un}
  K.~Freese and W.~H.~Kinney,
  \textcolor{blue}{\it``On: Natural inflation,''}
  \textcolor{purple}{Phys.\ Rev.\ D {\bf 70} (2004) 083512
  [hep-ph/0404012]}.
  
 \bibitem{Choudhury:2011sq}
  S.~Choudhury and S.~Pal,
  \textcolor{blue}{\it``Brane inflation in background supergravity,''}
  \textcolor{purple}{Phys.\ Rev.\ D {\bf 85} (2012) 043529
  [arXiv:1102.4206 [hep-th]]}.
 
 
 \bibitem{Choudhury:2011rz}
  S.~Choudhury and S.~Pal,
  \textcolor{blue}{\it``Reheating and leptogenesis in a SUGRA inspired brane inflation,''}
  \textcolor{purple}{Nucl.\ Phys.\ B {\bf 857} (2012) 85
  [arXiv:1108.5676 [hep-ph]]}.
  
  \bibitem{Choudhury:2012ib}
  S.~Choudhury and S.~Pal,
  \textcolor{blue}{\it``Brane inflation: A field theory approach in background supergravity,''}
  \textcolor{purple}{J.\ Phys.\ Conf.\ Ser.\  {\bf 405} (2012) 012009
  [arXiv:1209.5883 [hep-th]]}.
  
  
  
\bibitem{Gumjudpai:2003vv}
  B.~Gumjudpai,
  \textcolor{blue}{\it``Brane cosmology dynamics with induced gravity,''}
  \textcolor{purple}{Gen.\ Rel.\ Grav.\  {\bf 36} (2004) 747
  [gr-qc/0308046]}.
  
  
  \bibitem{Ade:2015xua}
  P.~A.~R.~Ade {\it et al.}  [Planck Collaboration],
  \textcolor{blue}{\it``Planck 2015 results. XIII. Cosmological parameters,''}
  \textcolor{purple}{arXiv:1502.01589 [astro-ph.CO]}.
  
\bibitem{Singh:2010qa}
  P.~Singh and F.~Vidotto,
  \textcolor{blue}{\it``Exotic singularities and spatially curved Loop Quantum Cosmology,''}
   \textcolor{purple}{Phys.\ Rev.\ D {\bf 83} (2011) 064027
  [arXiv:1012.1307 [gr-qc]]}.

\bibitem{Ashtekar:2006es}
  A.~Ashtekar, T.~Pawlowski, P.~Singh and K.~Vandersloot,
  \textcolor{blue}{\it``Loop quantum cosmology of $k=1$ FRW models,''}
  \textcolor{purple}{Phys.\ Rev.\ D {\bf 75} (2007) 024035
  [gr-qc/0612104]}.
  
  \bibitem{Maeda:2007cb}
  H.~Maeda, V.~Sahni and Y.~Shtanov,
  \textcolor{blue}{\it``Braneworld dynamics in Einstein-Gauss-Bonnet gravity,''}
  \textcolor{purple}{Phys.\ Rev.\ D {\bf 76} (2007) 104028
   [Phys.\ Rev.\ D {\bf 80} (2009) 089902]
  [arXiv:0708.3237 [gr-qc]]}.
  
\bibitem{Shtanov:2002mb}
  Y.~Shtanov and V.~Sahni,
  \textcolor{blue}{\it``Bouncing brane worlds,''}
  \textcolor{purple}{Phys.\ Lett.\ B {\bf 557} (2003) 1
  [gr-qc/0208047]}.

\end{thebibliography}
\end{document}